\PassOptionsToPackage{dvipsnames}{xcolor}
\documentclass[a4paper,11pt]{article}
\usepackage{jheppub} 

\usepackage{tabularx}
\makeatletter
\def\hlinewd#1{%
\noalign{\ifnum0=`}\fi\hrule \@height #1 %
\futurelet\reserved@a\@xhline}
\makeatother
\usepackage{scrextend}
\usepackage{epsfig}
\usepackage[normalem]{ulem}

\usepackage[T1]{fontenc}
\usepackage[utf8]{inputenc}
\usepackage{csquotes}
\usepackage{amsmath,amssymb,amsthm}
\usepackage{graphicx}
\usepackage{orcidlink} 
\usepackage{subfigure}
\usepackage{longtable}
\usepackage{bm}
\usepackage{multirow}
\usepackage[export]{adjustbox}
\usepackage[section]{placeins}
\usepackage{siunitx}

\numberwithin{equation}{section}
\renewcommand{\theequation}{\arabic{section}.\arabic{equation}}

\usepackage{xcolor}

\hypersetup{
    colorlinks=true,
    linkcolor=MidnightBlue,
    citecolor=ForestGreen,
    urlcolor=RoyalBlue
}
 
\numberwithin{equation}{section}
\renewcommand{\theequation}{\arabic{section}.\arabic{equation}}

\newcommand{\dd}[0]{\mathrm{d}}

\newcommand{\github}[2]{\href{https://github.com/#1/#2}{\includegraphics[width=9pt, valign=c]{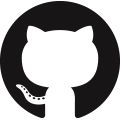}\ #2}}

\renewcommand\[{\left[}

\newcommand{\be}{\begin{equation}}
\newcommand{\ee}{\end{equation}}
\newcommand{\bea}{\begin{eqnarray}}
\newcommand{\eea}{\end{eqnarray}}

\newcommand{\bq}{\begin{eqnarray}}
\newcommand{\eq}{\end{eqnarray}}

\begin{document}

\title{Circular Axion String Loop Simulations:\\ Field Theory and Effective Description}

\author[a,b,c]{Mathieu Kaltschmidt~\orcidlink{0000-0002-6470-5371},}
\author[c]{Javier Redondo~\orcidlink{0000-0002-1044-8197},}
\author[d,c]{Ivan Rybak~\orcidlink{0000-0001-8501-8629},}
\author[e]{Ken'ichi Saikawa~\orcidlink{0000-0002-9205-9813},}
\author[c]{Iker Sanz-González~\orcidlink{0009-0006-2152-3320}}

\affiliation[a]{Dipartimento di Fisica e Astronomia “Galileo Galilei”, Università di Padova, \\ Via F. Marzolo 8, 35131 Padova, Italy}

\affiliation[b]{Istituto Nazionale di Fisica Nucleare (INFN), Sezione di Padova,\\ Via F. Marzolo 8, 35131 Padova, Italy}

\affiliation[c]{CAPA \& Departamento de F\'isica Te\'orica, Universidad de Zaragoza,\\
C. Pedro Cerbuna 12, 50009 Zaragoza, Spain}

\affiliation[d]{Cosmology, Gravity, and Astroparticle Physics Group,\\
Center for Theoretical Physics of the Universe,
Institute for Basic Science (IBS),\\
Daejeon 34126, Korea}

\affiliation[e]{Institute for Theoretical Physics, Kanazawa University,\\ Kakuma-machi, Kanazawa, Ishikawa 920-1192, Japan}

\emailAdd{mathieu.kaltschmidt@unipd.it, jredondo@unizar.es, irybak@ibs.re.kr, saikawa@hep.s.kanazawa-u.ac.jp, iker.sanz@unizar.es}

\preprint{KANAZAWA-26-06}

\abstract{
An accurate description of axion strings and their radiation is crucial for understanding their cosmological implications. Motivated by the differing conclusions of previous works, we study a simple, controlled setup: the collapse of an initially static circular loop in a Minkowski background. We compare field-theory simulations of the conventional complex scalar and high-tension Klaer--Moore models with solutions of the renormalised Nambu--Goto--Kalb--Ramond (NGKR) equation of motion, covering the range $3 \lesssim \mathcal{T} \lesssim 100$ in dimensionless string tension. Both the field theory and NGKR approach the Nambu--Goto (NG) limit with corrections scaling as $1/\mathcal{T}$. However, the leading correction to the halving-time is $20-35\%$ larger in NGKR than in the field-theory simulations over the same range of tensions, while for large tensions the value agrees with an independent
perturbative calculation to 0.2\%. The late-time axion spectrum agrees with NGKR in the infrared, but the collapse generates additional structure in the ultraviolet part of the spectrum, which we attribute to different events such as an outgoing wave front that carries a sizable fraction of the energy or the rebound of the initial loop. These results provide a useful benchmark for a more accurate understanding of axion radiation from cosmological string networks.
}

\maketitle

\section{Introduction}
\label{sec:intro}

The hypothetical quantum chromodynamics (QCD) axion~\cite{Weinberg:1977ma, Wilczek:1977pj}, pseudo-Nambu-Goldstone boson of the spontaneous breaking of the Peccei-Quinn (PQ) symmetry~\cite{Peccei:1977hh, Peccei:1977ur}, remains one of the best-motivated solutions to the strong CP problem and a viable candidate for cold dark matter (CDM). In the post-inflationary scenario, the PQ phase transition produces a network of global axion strings as a consequence of the Kibble mechanism~\cite{Kibble:1976sj, Vilenkin:2000jqa}. The decay of this string network could be the dominant source of the present-day axion dark matter abundance.
Turning this into a prediction for the QCD axion dark matter mass requires precise knowledge of the spectrum of axions radiated by the string network, a quantity that is only accessible via modern high-performance-computing (HPC) simulations.

Despite three decades of theoretical and numerical progress~\cite{Hagmann:1990mj, Hagmann:2000ja, Battye:1993jv, Battye:1994au}, the axion radiation spectrum from strings remains heavily debated in the literature, e.\,g. Refs.~\cite{Arza:2026rsl, Cicoli:2026fqp, OHare:2024nmr}. Recent
large-scale field-theory simulations of axion string networks disagree on whether the instantaneous
emission spectrum is sharply peaked in the infrared (IR) with a spectral index
$q\gg1$, that grows logarithmically with the string tension, see Refs.~\cite{
Gorghetto:2018myk, Gorghetto:2020qws, Kim:2024wku}, or approaches $q\sim1$, see Refs.~\cite{Buschmann:2019icd, Buschmann:2021sdq, Saikawa:2024bta,
Benabou:2024msj,  Correia:2024cpk, Correia:2025nns,
Battye:2026whd}. The two scenarios translate into an order-of-magnitude
disagreement in the predicted QCD axion mass, and pushing network
simulations to yet larger dynamical ranges has so far not settled the
question.

In this work we take a complementary approach. Rather than simulating
larger networks, we isolate single
string loops and study them in as much detail as possible.
We focus on circular loops, arguably the simplest loop configuration,
because they allow us to probe the largest dynamical range
and provide a controlled setting in which field-theory simulations can be
directly compared with the effective Nambu--Goto--Kalb--Ramond (NGKR)
description of the string~\cite{Vilenkin:1986ku, Davis:1988rw}. Although the NGKR action appears far simpler than the underlying field theory, the renormalisation connecting the two is subtle, particularly in the large-tension, weakly coupled regime relevant for QCD axion strings. The literature has not reached full agreement on this issue since the pioneering work of Ref.~\cite{Dabholkar:1989ju}, with later studies refining the analysis using different renormalisation schemes~\cite{Copeland:1990qu, Battye:1995hw, Buonanno:1998kx, Buonanno:1998is}. Understanding this correspondence in detail is a
necessary step towards a first-principles prediction not only of the axion
dark matter mass, but also of the gravitational wave and other radiation
signatures of cosmic string loops more
generally, see Refs.~\cite{Blanco-Pillado:2023sap, Drew:2019mzc, Drew:2022iqz,
Drew:2023ptp, Drew:2025txu,
Aurrekoetxea:2026xrl}.

To this end, we compare four independent numerical approaches: three-dimensional field-theory simulations with \github{veintemillas}{jaxions}~\cite{
Jaxions:2026}, three-dimensional simulations with adaptive mesh refinement (AMR)
using \github{}{axioNyx}~\cite{Schwabe:2020eac}, axisymmetric two-dimensional simulations that exploit the loop's symmetry to reach a
larger dynamical range (also realised with \texttt{jaxions}), and three-dimensional simulations of the
high-tension Klaer--Moore model~\cite{Klaer:2017qhr,
Klaer:2019fxc}, which allows us to access string tensions far beyond the \enquote{standard} complex-scalar model simulations~\cite{Klaer:2017ond}. 

This paper is organised as follows. In Sec.~\ref{sec:theory} we introduce the field-theory description of global strings via the conventional complex scalar and the high-tension Klaer--Moore models, together with their corresponding straight-string solutions. In Sec.~\ref{sec:eft} we introduce the \enquote{Kalb--Ramond-Axion Duality} and the NGKR equations of motion for the circular loop, discuss the renormalisation of the string tension, and compute the axion spectrum in the effective theory. In Sec.~\ref{sec:simu} we present the four simulation setups, the loop initial conditions and the anatomy of a typical circular loop collapse. We then describe how we measure the loop radius, construct its continuum limit, and extract the axion spectrum from the simulations. In Sec.~\ref{sec:compa} we compare the field-theory and effective-theory results for the collapse dynamics and the axion spectrum, before concluding in Sec.~\ref{sec:conclusions}.

Further material is collected in the appendices: details on the numerical setups (App.~\ref{app:numerics}), the relevant timescales during the collapse (App.~\ref{app:collapse_scalings}), the extraction of the halving and collapse times (App.~\ref{app:collapse_times}), the limitations on the validity of the effective description (App.~\ref{app:limitations}), the first-order correction to the halving time (App.~\ref{app:halving_coefficient}), the construction of the loop initial conditions (App.~\ref{app:string_ics}), details on the retarded Kalb--Ramond (KR) field of a circular loop and its local singularity (App.~\ref{app:KR-force-derivation}) and a comparison with previous studies (App.~\ref{app:context}).

Throughout this work we use natural units such that $c=\hbar=1$ and assume a flat Minkowski background with metric signature $(+, -,-,-)$. Greek indices run
over spacetime, lower-case Latin indices label the string worldsheet, and capital Latin indices label the two directions normal to it.

\section{Field Theories of Global Strings}
\label{sec:theory}
In this section, we develop the field theory description for global axion strings. We focus on two distinct UV completions that cover the phenomenologically relevant range in string tension.
\subsection{Complex Scalar Model}
\label{subsec:complex-scalar-model}

The conventional field theory of a global string consists of a single complex scalar field $\phi$, and the action reads
\begin{equation}
 \label{eq:lagrangian}
 S=\int \dd^4x\left[
\frac12\partial_\mu\phi^\dagger\partial^\mu\phi
 -\frac{\lambda}{4}\left(|\phi|^2-v^2\right)^2\right],
\end{equation}
where we denote the vacuum expectation value $\langle\phi\rangle\equiv v$. Writing
\begin{equation}
 \phi=v f e^{i\theta},\qquad m_s=\sqrt{2\lambda}\,v,
\end{equation}
yields
\begin{equation}
 \label{eq:lagrangianJ}
 \mathcal L=\frac{v^2}{2}\left[(\partial_\mu f)^2
 +f^2(\partial_\mu\theta)^2
 -\frac{m_s^2}{4}(f^2-1)^2\right].
\end{equation}
The radial excitation (sometimes called \emph{saxion}) has mass $m_s$, whereas the angular field is the
massless \emph{axion} outside the string core. Its equation of motion is
\begin{equation}
 \label{eq:eom_theta}
\partial_\mu\left(f^2\partial^\mu\theta\right)=0.
\end{equation}

\paragraph{Straight-string Solution.}
\label{subsec:straight-global-string}

For a static unit-winding string along the $z$-axis we use cylindrical
coordinates $(\rho,\varphi,z)$, where $\rho$, $\varphi$, and $z$ are
the radial, angular, and height coordinates, respectively. We set
$\theta=\varphi$ and $f=f(\rho)$. In terms of $x=m_s\rho$, the radial profile
obeys
\begin{equation}
 \label{eq:eom_f}
 f''+\frac{f'}{x}-\frac{f}{x^2}
 -\frac{f}{2}(f^2-1)=0,
 \qquad f(0)=0,\quad f(\infty)=1.
\end{equation}
A fit to the numerical solution that incorporates both  the
linear behaviour at the origin and the correct large-radius expansion is 
\begin{equation}
 \label{eq:solution_field}
 f(x)=\frac{x P(x)}{\sqrt{1+x^2P^2(x)}},
 \qquad
  P(x)=\frac{p_1+p_2x^2}{1+p_3x+\sqrt2\,p_2x^2},
\end{equation}
where
\begin{equation}
 p_1=0.4107\ldots,\qquad p_2=0.0302\ldots,\qquad p_3=0.0047\ldots.
\end{equation}

\begin{figure}[t]
    \centering
    \includegraphics[width=0.49\linewidth]{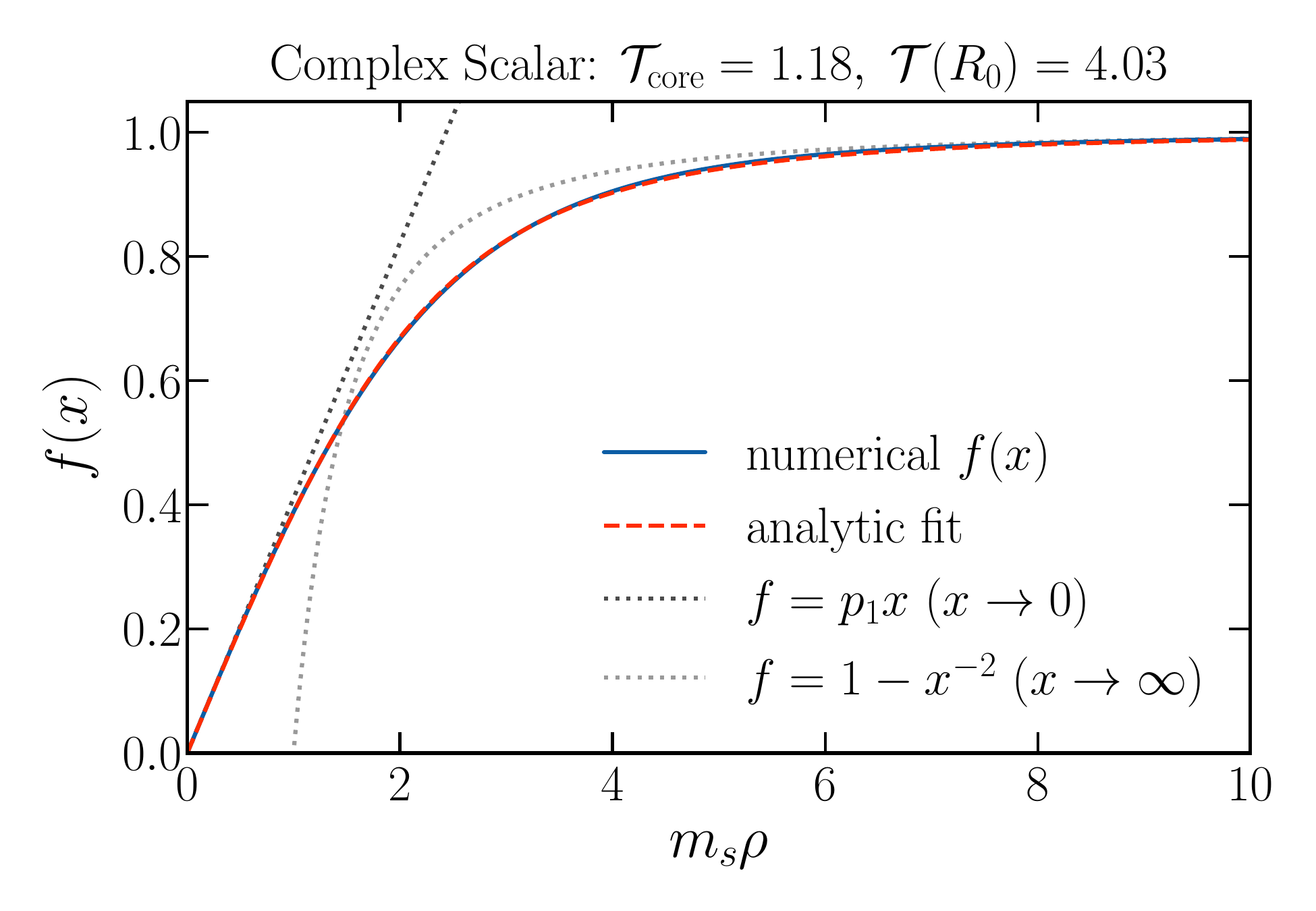}
    \includegraphics[width=0.49\linewidth]{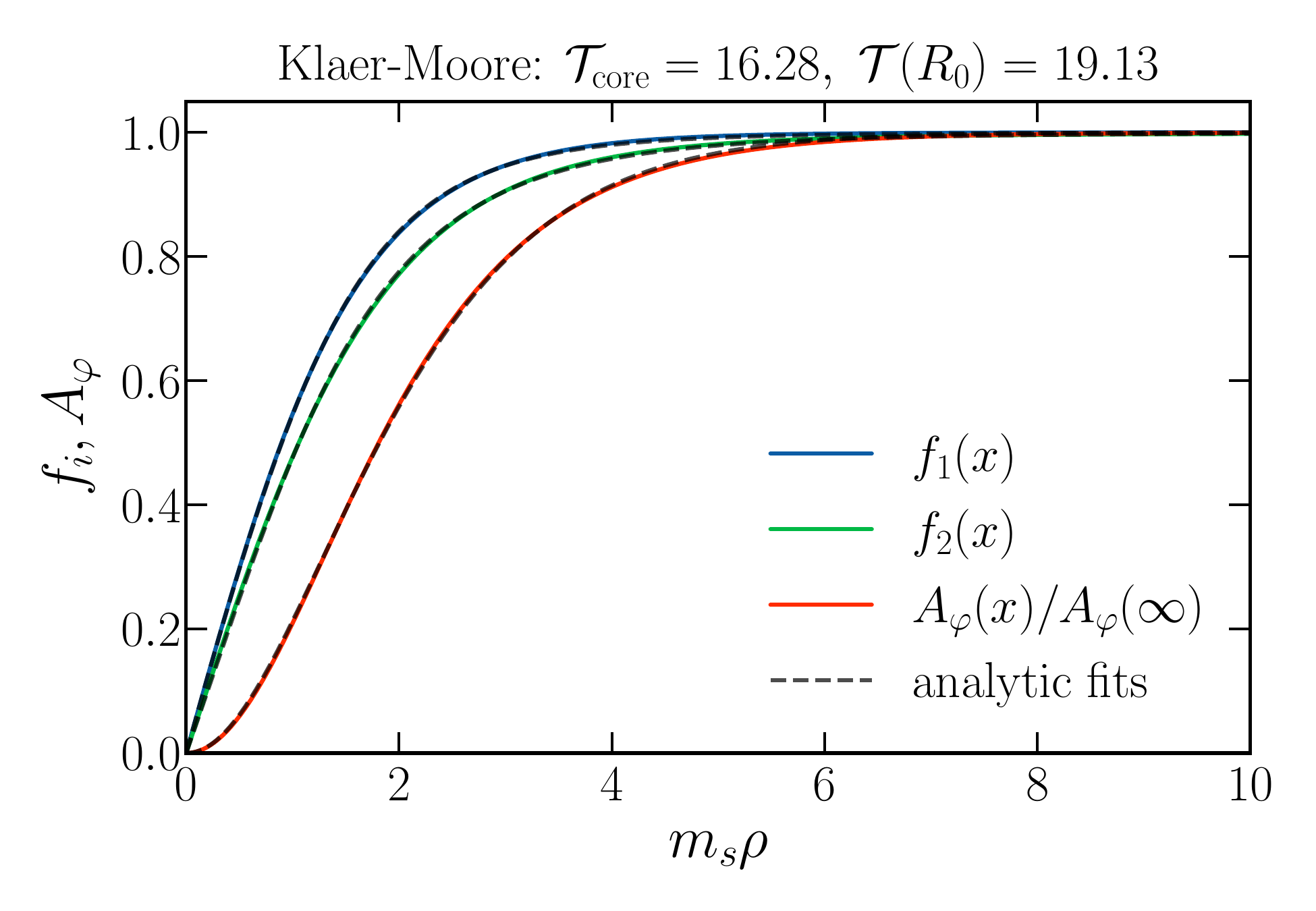}
   \caption{Radial profiles of a straight string. \emph{Left:} complex scalar model, with the numerical solution of Eq.~\eqref{eq:eom_f} (blue solid), the analytic fit of Eq.~\eqref{eq:solution_field} (red dashed), and the $x\to0$ and $x\to\infty$ asymptotics (gray dotted). \emph{Right:} Klaer--Moore $(1,1)$ string at the BPS point ($m_1=m_2=m_V\equiv m_s$) for $v_1^2=2.66$ and $(q_1,q_2)=(2,1)$, showing the numerical scalar profiles $f_1$, $f_2$ and the normalised gauge field $A_\varphi/A_\varphi(\infty)$ (solid), together with analytic fits (black dashed).}
    \label{fig:profile}
\end{figure}

Thus, $f(x)=p_1 x+\mathcal O(x^2)$ close to the core and
$f(x)=1-x^{-2}+\mathcal O(x^{-3})$ far from it, see the left panel of Fig.~\ref{fig:profile}. The fit reproduces the numerical profile to better than $4\times10^{-3}$ everywhere. The numerical values shown below assume the fitted form Eq.~\eqref{eq:solution_field}, but the results from the fit and the numerical solution differ by less than one percent.

The energy per unit length up to a radial cutoff is 
\begin{equation}
 \mu(\rho_{\rm cut})=\pi v^2\int_0^{m_s\rho_{\rm cut}}\dd x\,x
 \left[(f')^2+\frac{(f^2-1)^2}{4}+\frac{f^2}{x^2}\right].
 \label{eq:global-string-tension}
\end{equation}

As it turns out the physics of the collapse is not sensitive to $\mu$, but rather to its value normalised by $\pi v^2$. It is therefore very convenient to define the dimensionless tension parameter,
\begin{equation}
    {\mathcal{T}} = \frac{\mu}{\pi v^2}.
    \label{eq:calT_def}
\end{equation}

We can now separate the tension directly into the fraction of energy carried by the
massive radial field in the core and the fraction carried by the regulated axion phase, 
\begin{equation}
 {\mathcal{T}}(\rho_{\rm cut})
 ={\mathcal{T}}_{\rm core}+{\mathcal{T}}_\theta(\rho_{\rm cut})
 +\mathcal O\!\left((m_s\rho_{\rm cut})^{-2}\right).
 \label{eq:tension-core-axion-split}
\end{equation}
The core contribution contains only the radial-gradient and potential
energies and is IR finite, 
\begin{equation}
 {\mathcal{T}}_{\rm core}
 =\int_0^\infty \dd x\,x
 \left[(f')^2+\frac{(f^2-1)^2}{4}\right]
 =0.7843\ldots\,.
 \label{eq:core-tension}
\end{equation}
The axion contribution is the phase-gradient energy.  The radial profile
$f(x)$ provides its physical core regulator,
\begin{equation}
 {\mathcal{T}}_\theta(\rho_{\rm cut})
 =\int_0^{m_s\rho_{\rm cut}}\frac{\dd x}{x}\,f^2(x)
 \approx \log\!\left(
0.4725 
\,m_s\rho_{\rm cut}\right) 
 +\mathcal O\!\left((m_s\rho_{\rm cut})^{-2}\right)
 .
 \label{eq:axion-tension}
\end{equation}

\subsection{Klaer--Moore Two-Scalar Model}
\label{subsec:Klaer--Moore-model}
To simulate strings whose core tension is parametrically
large compared with their long-range axion coupling, Klaer and Moore
introduced a weakly coupled $U(1)$-theory containing \emph{two} complex scalars
$(\phi_1,\phi_2)$ with integer charges $(q_1,q_2)$~\cite{Klaer:2017qhr,Klaer:2019fxc}. We use the normalisation 
\begin{align}
 \label{eq:KM-lagrangian}
 &\mathcal L_{\rm KM}={}-\frac{1}{4e^2}F_{\mu\nu}F^{\mu\nu}
 +\sum_{i=1}^2\frac{1}{2}|D_\mu\phi_i|^2
 -\sum_{i=1}^2\frac{m_i^2}{8v_i^2}
 \left(|\phi_i|^2-v_i^2\right)^2, \\
& D_\mu\phi_i=(\partial_\mu-iq_iA_\mu)\phi_i,
\end{align}
and neglect direct interactions between the two scalars. Writing
$\phi_i=(v_i+s_i)e^{i\theta_i}$, the spectrum consists of two radial
modes of masses $m_i$, a massive vector, and one massless axion. The vector
eats the phase combination parallel to $q_1 v_1^2 \theta_1+q_2 v_2^2 \theta_2$, getting a mass
\begin{equation}
\label{eq:mv_sq}
 m_V^2=e^2(q_1^2v_1^2+q_2^2v_2^2).
\end{equation}
The orthogonal, gauge-invariant angular variable and its
canonical normalisation are
\begin{equation}
 \label{eq:KM-axion}
 \theta_A=q_2\theta_1-q_1\theta_2=\frac{a}{v},
 \qquad
 v^2=\frac{v_1^2v_2^2}{q_1^2v_1^2+q_2^2v_2^2},
\end{equation}
where $a$ is the canonically normalised axion field: integrating out the
massive vector leaves $\tfrac12v^2(\partial\theta_A)^2=\tfrac12(\partial a)^2$,
so that $v$ is the axion decay constant. 

We take the charges to be coprime. The points of the $(\theta_1,\theta_2)$ torus that
are gauge-equivalent to the origin are
$(2\pi n_1+q_1\lambda,\,2\pi n_2+q_2\lambda)$ with $n_i\in\mathbb Z$ and
$\lambda$ arbitrary, where
$\theta_A=2\pi(q_2n_1-q_1n_2)$. Since
$\{q_2n_1-q_1n_2\}=\gcd(q_1,q_2)\,\mathbb Z$, the axion has period $2\pi$ for
coprime charges, and the canonical field $a$ therefore has period $2\pi v$.
The same combination appears as the winding $Q_A$ of
Eq.~\eqref{eq:KM-axion-winding}, so coprimality also ensures that every
integer winding is realisable.

\paragraph{Straight-string Solution.}
\label{subsec:straight-KM-string}
A cylindrically symmetric vortex can be written as
\begin{equation}
\phi_i=v_i f_i(\rho)e^{in_i\varphi},\qquad
 A_\varphi=A_\varphi(\rho),\qquad n_i\in\mathbb Z,
 \label{eq:KM-string-ansatz}
\end{equation}
where $A_\varphi$ denotes the coordinate component of the gauge field. Its axion winding is the gauge-invariant integer quantity
\begin{equation}
 Q_A=q_2n_1-q_1n_2.
 \label{eq:KM-axion-winding}
\end{equation}
Minimising the angular-gradient energy at large radius fixes
\begin{equation}
 A_\varphi(\infty)=\frac{q_1v_1^2n_1+q_2v_2^2n_2}{q_1^2v_1^2+q_2^2v_2^2}.
 \label{eq:g_infty}
\end{equation}
For $Q_A\neq0$ a single gauge field cannot cancel the phase gradients of both scalars, and each field retains a residual winding
\begin{equation}
 \Delta_i\equiv n_i-q_iA_\varphi(\infty),\qquad
 \Delta_1=\frac{q_2\,Q_A}{q_1^2r+q_2^2},\qquad
 \Delta_2=-\frac{q_1r\,Q_A}{q_1^2r+q_2^2},\qquad r\equiv\frac{v_1^2}{v_2^2}.
 \label{eq:KM-residual-winding}
\end{equation}
These satisfy $q_1v_1^2\Delta_1+q_2v_2^2\Delta_2=0$ and $q_2\Delta_1-q_1\Delta_2=Q_A$. The gauge field screens all winding except the unit winding of the massless axion, which is shared between the two phases.

In analogy with the complex scalar model, the profile functions and tension of the Klaer--Moore string can be obtained by solving the corresponding field equations. In terms of $x=m_2\rho$, the Euler--Lagrange equations derived from Eq.~\eqref{eq:KM-lagrangian} are
\begin{align}
 f_1''+\frac{f_1'}{x}-\frac{f_1}{x^2}\left(n_1-q_1A_\varphi\right)^2-\frac{\beta_m}{2}f_1\left(f_1^2-1\right)&=0,
 \label{eq:KM_f1_profile}\\
 f_2''+\frac{f_2'}{x}-\frac{f_2}{x^2}\left(n_2-q_2A_\varphi\right)^2-\frac{1}{2}f_2\left(f_2^2-1\right)&=0,
 \label{eq:KM_f2_profile}\\
 A_\varphi''-\frac{A_\varphi'}{x}+C\left[r\,q_1f_1^2\left(n_1-q_1A_\varphi\right)+q_2f_2^2\left(n_2-q_2A_\varphi\right)\right]&=0,
 \label{eq:KM_A_profile}
\end{align}
with $\beta_m\equiv m_1^2/m_2^2$ and $C\equiv e^2v_2^2/m_2^2$. Subject to the boundary conditions 
\begin{equation}
    f_i(0) = 0,\quad f_i(\infty) = 1,\quad 
    A_\varphi(0) = 0,
\end{equation}
and $A_\varphi(\infty)$ of Eq.~\eqref{eq:g_infty}, 
the system of coupled differential equations \eqref{eq:KM_f1_profile}--\eqref{eq:KM_A_profile} is integrated numerically, see the right panel of Fig.~\ref{fig:profile} for an example. Far from the core the residual winding gives the scalars a power-law tail, $f_i=1-\Delta_i^2x^{-2}+\mathcal O(x^{-4})$, whereas the massive gauge field approaches $A_\varphi(\infty)$ exponentially.

An ansatz analogous to Eq.~\eqref{eq:solution_field} for the scalar profiles is
\begin{equation}
 f_i(x)=\frac{xP_i(x)}{\sqrt{1+x^2P_i^2(x)}},\qquad
 P_i(x)=\frac{p_1+p_2x^2}{1+p_3x+s_i\,p_2x^2},\qquad s_i=\sqrt2\,|\Delta_i|,
 \label{eq:KM_fit_f}
\end{equation}
which generalises $s=\sqrt2$ of the single-scalar model, where $|\Delta|=1$. It reproduces the behaviour near the core, $f_i(x)=p_1x+\mathcal O(x^2)$, and the leading large-distance tail, $f_i(x)=1-\Delta_i^2x^{-2}+\mathcal O(x^{-3})$, exactly. For the gauge field we use
\begin{equation}
 \frac{A_\varphi(x)}{A_\varphi(\infty)}=1-\left(1+bx+cx^2\right)e^{-bx},
 \label{eq:KM_fit_A}
\end{equation}
which is regular at the origin, $A_\varphi(x)/A_\varphi(\infty)=(b^2/2-c)\,x^2+\mathcal O(x^3)$. For the example shown in Fig.~\ref{fig:profile}, with $v_1^2=2.66$, $v_2^2=1.00$ and $(q_1,q_2)=(2,1)$, the residual windings are
\begin{equation}
 \Delta_1=-0.0859,\qquad \Delta_2=0.4570,
\end{equation}
and the fits give
\begin{alignat}{4}
 f_1:&\quad & p_1&=0.6166\ldots, &\qquad p_2&=0.0545\ldots, &\qquad p_3&=\phantom{-}0.0250\ldots,\\
 f_2:&\quad & p_1&=0.4938\ldots, &\qquad p_2&=0.0127\ldots, &\qquad p_3&=-0.0729\ldots,\\
 A_\varphi:&\quad & b&=1.3037\ldots, &\qquad c&=0.5955\ldots, & &
\end{alignat}
reproducing the numerical profiles to better than $6\times10^{-3}$. The coefficients for all runs used in this work are listed in Table~\ref{tab:KM_fit_coefficients} in App.~\ref{app:numerics}.

The normalised energy per unit length up to a radial cutoff is
\begin{equation}
\begin{aligned}
 {\mathcal{T}}(\rho_{\rm cut})=\frac{q_1^2r+q_2^2}{r}\int_0^{m_2\rho_{\rm cut}}\dd x\,x\biggl[
 &\frac{(A_\varphi')^2}{Cx^2}
 +\frac{r\,f_1^2\left(n_1-q_1A_\varphi\right)^2+f_2^2\left(n_2-q_2A_\varphi\right)^2}{x^2}\\
 &+r\,f_1'^2+f_2'^2
 +\frac{r\,\beta_m\left(f_1^2-1\right)^2+\left(f_2^2-1\right)^2}{4}\biggr].
\end{aligned}
\label{eq:KM_tension}
\end{equation}
The prefactor converts from units of $\pi v_2^2$ to $\pi v^2$, allowing for a direct comparison between the models. The second term, corresponding to the angular gradient energy, is the only one requiring regularisation since at large $x$ it approaches $(r\Delta_1^2+\Delta_2^2)/x$, which diverges logarithmically, in close analogy with the single-scalar model. Furthermore, the quadratic behaviour of the gauge field near the core keeps the first term finite at the origin.

Separating the contribution of the long-range axion field outside the core from the local contributions gives 
\begin{equation}
 {\mathcal{T}}_\theta(\rho_{\rm cut})= Q_A^2
 \log\!\left(m_2\rho_{\rm cut}\right)+\mathcal{O}(1),
 \label{eq:KM-log-tension}
\end{equation}
since inserting Eq.~\eqref{eq:KM-residual-winding} gives $(q_1^2r+q_2^2)(r\Delta_1^2+\Delta_2^2)/r=Q_A^2$.
We choose adjacent charges $q_2=q_1-1$ and the
$(n_1,n_2)=(1,1)$ vortex, for which $Q_A=-1$. It therefore has exactly the
same long-distance axion field as the unit-winding global string. 

However, the local gauge-string tension can be much larger in units of $\pi v^2$. The enhancement is confined to the UV string core and does not modify the coefficient of the IR axion logarithm. 
We adopt the BPS case, $m_1=m_2=m_V\equiv m_s$, for which $\beta_m=1$, $C=1/(q_1^2r+q_2^2)$ and $x=m_s\rho$. The core energy is then approximately that of an Abelian-Higgs string with vev $v_{\mathrm{AH}}^2=v_1^2+v_2^2$~\cite{Klaer:2017qhr}, 
\begin{equation}
{\mathcal{T}}_{\rm core}=\frac{\mu_{\rm core}}{\pi v^2}\simeq
\frac{(v_1^2+v_2^2)(q_1^2v_1^2+q_2^2v_2^2)}{v_1^2v_2^2}\,,
 \label{eq:KM-core-tension}
\end{equation}
where $\mu_{\rm core}\equiv\mu(\rho_{\rm cut})-\pi v^2Q_A^2\log(m_s\rho_{\rm cut}/\pi)$ is independent of $\rho_{\rm cut}$ up to $\mathcal O((m_s\rho_{\rm cut})^{-2})$. With this separation scale, ${\mathcal{T}}_{\rm core}$ coincides with $\bar\kappa$
of Ref.~\cite{Klaer:2017qhr}. Equation~\eqref{eq:KM-core-tension} is not exact, because the residual $\Delta_i^2/x^2$ tails of the scalar profiles are not captured by the Abelian-Higgs solution. Integrating Eq.~\eqref{eq:KM_tension} numerically, we find that the true core tension exceeds Eq.~\eqref{eq:KM-core-tension} by an almost constant $0.26$--$0.29$ for all parameters used in this work, see Table~\ref{tab:simparams_moore}.

The total tension is 
\begin{equation}
 \label{eq:KM-dimensionless-tension}
 \mathcal T(\rho_{\rm cut})\equiv\frac{\mu(\rho_{\rm cut})}{\pi v^2}
 =Q_A^2\log\!\left(\frac{m_s\rho_{\rm cut}}{\pi}\right)+{\mathcal{T}}_{\rm core}\,,
\end{equation}
where the first term is the ordinary long-distance axion
contribution. The enhancement $\mathcal T_{\rm core}$ is instead purely UV: it
is the local scalar--vector core energy, made parametrically large relative
to $\pi v^2$ by the massive-vector construction. Increasing the charges
raises $\mathcal T_{\rm core}$ without
requiring an exponentially large ratio between the lattice box and the
string core. This allows the Klaer--Moore simulations to reproduce the
dynamics of global strings at otherwise inaccessible effective scale
separations.

\section{Effective Theory for Global Strings}
\label{sec:eft}
In this section we develop the Nambu--Goto--Kalb--Ramond (NGKR) effective field theory (EFT) description for global axion strings.
\subsection{Kalb--Ramond-Axion Duality}
\label{subsec:kr-axion-duality}
Outside the string core, the only light field is the axion phase.  In four
dimensions a massless scalar is dual to a two-form gauge field
$B_{\mu\nu}$ with field strength
$H_{\mu\nu\rho}=3\partial_{[\mu}B_{\nu\rho]}$.  We choose
$\varepsilon^{0123}=+1$ and write the duality as \cite{Vilenkin:1986ku, Davis:1988rw}
\begin{equation}
H^{\mu\nu\rho}=v\varepsilon^{\mu\nu\rho\lambda}
 \partial_\lambda\theta,
 \qquad
 v\partial_\mu\theta=\frac{1}{3!}
 \varepsilon_{\mu\nu\rho\sigma}H^{\nu\rho\sigma}.
 \label{eq:duality}
\end{equation}
In this normalisation $B_{\mu\nu}$ and $H_{\mu\nu\rho}$ have canonical
mass dimensions one and two, respectively.  The duality maps the vacuum
axion action into $\int H^2/12$.

A unit string is introduced through an antisymmetric worldsheet current
\cite{Vilenkin:2000jqa},
\begin{equation}
 \partial_\lambda H^{\lambda\mu\nu}= 2\pi v J^{\mu\nu}, 
 \label{eq:Rsource}
\end{equation}
such that the circulation integral of the $\theta$-field around the source gives precisely $2\pi$,
\begin{equation}
 \oint_C \dd\theta = \frac{1}{6v} \oint_C \dd x^{\mu}
\varepsilon_{\mu\nu\rho\sigma} H^{\nu\rho\sigma}  = 2\pi ,
 \label{eq:winding}
\end{equation}
where $\dd x^{\mu}$ is the coordinate line element along the contour $C$; here, we assume that the winding number is $1$.

\subsection{Thin-string Effective Action}

For $m_s\mathcal R\gg1$, where $\mathcal R$ denotes a curvature or
worldsheet-variation scale, or distance to the string core, the finite core profile can be treated in a multipolar expansion, see Appendix \ref{app:limitations} for details. To leading order in this derivative expansion the effective action is \cite{Kalb:1974yc}\footnote{We note that our definition of the effective action slightly differs from those used in \cite{Vilenkin:1986ku,Davis:1988rw,Dabholkar:1989ju,Battye:1995hw,Copeland:1990qu}. This difference arises from our convention for $\mu_0$, which is defined as one half of the value used in the literature. }
\begin{equation}
 S_{\rm EFT}=-\mu_0
 \int_\Sigma \dd^2\sigma\sqrt{-\gamma}
 +\frac{1}{12}\int \dd^4x\,H_{\mu\nu\rho}H^{\mu\nu\rho}
 +\pi v \int_\Sigma \dd\sigma^{\mu\nu} B_{\mu\nu}\,.
 \label{eq:EFT_NG+KR}
\end{equation}
Here, the first term is the Nambu-Goto (NG) action for the infinitely thin string with tension $\mu_0$, the induced metric is $\gamma_{ab}=\partial_aX^\mu\partial_bX_\mu$, with
$\partial_a$ the derivative with respect to the worldsheet coordinates
$\sigma^a$ and $X^\mu(\sigma)$ the embedding of $\Sigma$ in spacetime. Note  that $\dd\sigma^{\mu\nu}\equiv\epsilon^{ab}\partial_aX^\mu\partial_bX^\nu\,\dd^2\sigma$
is the worldsheet area element. Equivalently, the spacetime current is
\begin{equation}
 J^{\mu\nu}(x)=\int_\Sigma \dd\sigma^{\mu\nu}\,
 \delta^{(4)}\!\left(x-X(\sigma)\right),
 \label{eq:worldsheet-current}
\end{equation}
whose coupling to $B_{\mu\nu}$ reproduces
Eq.~\eqref{eq:Rsource}.
The coefficients of the
KR kinetic term and of the string source are fixed by
Eq.~\eqref{eq:duality} and the unit-winding condition; this is the usual NGKR action, see Refs.~\cite{Vilenkin:1986ku,Davis:1988rw, Dabholkar:1989ju, Battye:1995hw, Copeland:1990qu, Buonanno:1998kx}.  The only leading-order Wilson
coefficient still to be matched is the local worldsheet coefficient
$\mu_0$.  Operators containing additional powers of the extrinsic curvature
describe finite-core effects and are suppressed by powers of
$(m_s\mathcal R)^{-1}$.

\subsection{Straight-string Matching}
\label{subsec:straight-string-matching}

The value of $\mu_0$ in the effective theory can be matched to the field theory by considering the string tension as observable. 

Consider an infinite static string along the $z$-axis, with worldsheet coordinates $X^\mu=(t,0,0,\sigma)$ in cylindrical coordinates $(\rho,\varphi,z)$. We can identify the coordinate along the string as $\sigma=z$ and exclude a small region $\delta_\sigma$ around the point of interest, which we take to be at $z=0$. We use $\delta_\sigma$ as a regulator that will be taken to 0 at the end of the calculation. 
The $\partial_i\theta$ field is nonzero in the azimuthal direction and can be obtained directly from the Biot--Savart integral form of Eq.~\eqref{eq:Rsource},
\begin{align}
 H^{103}(\rho)
 &=\frac{v}{2}\int_{|\sigma|>\delta_\sigma}\dd\sigma\,
 \frac{\rho}{(\rho^2+\sigma^2)^{3/2}} =\frac{v}{\rho}
 \left(1-\frac{\delta_\sigma}
 {\sqrt{\rho^2+\delta_\sigma^2}}\right).
 \label{eq:straight-H-regulated}
\end{align}

In the limit $\delta_\sigma/\rho\to 0$ the result approaches the well known result $v/\rho$, singular at $\rho=0$.  
The energy per unit length of the KR field inside a cylinder of radius
$\rho_{\rm cut}$ follows from integrating the energy density, 
$\tfrac12(H^{103})^2$, 
\begin{equation}
 \mu_{\rm KR}(\rho_{\rm cut};\delta_\sigma)=
 \pi v^2\,
 \log\left[
 \frac{(\delta_\sigma+\sqrt{\delta_\sigma^2+\rho_{\rm cut}^2})^2}{4\delta_\sigma\sqrt{\delta_\sigma^2+\rho_{\rm cut}^2}}
 \right]
 = \pi v^2 \log\left(\frac{\rho_{\rm cut}}{4\delta_\sigma}\right)+
 \mathcal O(v^2\delta_\sigma/\rho_{\rm cut}),
 \label{eq:muKR-exact}
\end{equation}
while the energy per unit length corresponding to the NG string is $\mu_0$.

Let us match this result to the field theories discussed in Sec.~\ref{sec:theory}. In both cases we write the tension parameter as, 
\begin{equation}
 {\mathcal{T}}(\rho_{\rm cut})
 =c_{\rm FT}+\log(m_s\rho_{\rm cut})
 +\mathcal O((m_s\rho_{\rm cut})^{-2}),
 \label{eq:FT-tension-match}
\end{equation}
where,  
\begin{equation}
    c_{\rm FT} = 
    \begin{cases}
    \frac{\mu_{\rm core}}{\pi v^2}+\log(0.4725)=0.0347 & \quad \text{Complex scalar,} \\[.3em]
    {\mathcal{T}}_{\rm core}-\log(\pi) & \quad
    \text{Klaer--Moore,}\\
    \end{cases}
\end{equation}
see Eqs.~\eqref{eq:core-tension}
and~\eqref{eq:axion-tension}. 

Evaluating both equations at any $\rho_{\rm cut}$ outside the core fixes the NG tension,
\begin{equation}
 \frac{\mu_0(\delta_\sigma)}{\pi v^2}={\mathcal{T}}(\rho_{\rm cut})-\log\left(\frac{\rho_{\rm cut}}{4\delta_\sigma}\right) = 
 c_{\rm FT} + \log(4 m_s\delta_\sigma)\,.
 \label{eq:mu0-exact}
\end{equation}
up to $\delta_\sigma/\rho_{\rm cut}$ corrections, which will be irrelevant in the $\delta_\sigma\to 0 $ limit. 

Note that the right-hand side is independent of $\rho_{\rm cut}$ but depends on the choice of the regulator $\delta_\sigma$. Thus, $\mu_0$ is a regulator-dependent Wilson coefficient, not a separately
measurable ``core energy'';  its running precisely compensates the short distance part of the KR field. 

We note also that the result depends on the sum of the field theory core energy (both in saxion and axion fields) and a $\log(m_s)$ piece. Thus, the effective theory as described here cannot differentiate core tension from the core size (here represented by $1/m_s$) regulating the tension from the axion field, it is sensitive only to the sum of both. 
This is particularly relevant for Klaer--Moore strings, which have more fields and energy in the core, and therefore behave like strings of the single complex scalar field with much thinner cores.  

The effective theory of Eq.~\eqref{eq:EFT_NG+KR} seems to have three parameters, ($\mu_0,v,\delta_\sigma$). The latter will be sent to zero as soon as we regulate the equations of motion. We can normalise length and time scales with $1/v$ and fields with $v$ and obtain an action written in terms of dimensionless variables, where the only free parameter is $\mu_0/v^2$. 
It is thus quite natural to justify the choice of the tension parameter normalised by $\pi v^2$ as for ${\mathcal{T}}$, see Eq.~\eqref{eq:calT_def}. In the NGKR EFT, the evolution of strings with different tensions can be rescaled onto each other when they have the same $\mathcal{T}(R_0)$. Thus, it is the only relevant parameter for the dynamics studied in this work. 

\subsection{Renormalised Worldsheet Equation}
\label{subsec:renormalisation}
Varying Eq.~\eqref{eq:EFT_NG+KR} yields the regulated worldsheet equation \cite{Battye:1995hw}
\begin{equation}\mu_0(\delta_\sigma)K^\mu=F^\mu(X;\delta_\sigma),
 \label{eq:bare-worldsheet-eom}
\end{equation}
where the complete curvature vector $K^\mu$ (including the string acceleration) and the force $F^\mu$ due to the coupling to the KR field are given by
\begin{eqnarray}
 K^\mu &=&\frac{1}{\sqrt{-\gamma}}
 \partial_a\left(\sqrt{-\gamma}\gamma^{ab}\partial_bX^\mu\right), \\   
F^{\mu}(X;\delta_\sigma)
&=&
\frac{\pi v}{\sqrt{-\gamma}}\,
H^{\mu}{}_{\nu\rho}(X;\delta_\sigma)\,
\epsilon^{ab}\partial_aX^\nu\partial_bX^\rho .
\end{eqnarray}

The KR field strength driving the force is evaluated on the worldsheet, where it is finite and well defined, since the regularisation removes the self-field produced in the $\sigma$-interval $(-\delta_\sigma,\delta_\sigma)$ responsible for the divergences.   
Locality and worldsheet covariance require this UV-sensitive part of the
self-force to be proportional to the complete curvature vector $K^\mu$,
in direct analogy with classical self-force renormalisation
\cite{Dirac:1938nz}.  
Choosing an arbitrary intermediate proper length $\Delta$ such that
$\delta_\sigma\ll\Delta\ll\mathcal R$, we define the renormalised tension and force as \cite{Dabholkar:1989ju, Copeland:1990qu, Battye:1995hw}
\begin{align}
 \bar\mu(\Delta)
&\equiv\lim_{\delta_\sigma\to0}\left[\mu_0(\delta_\sigma)
 +\pi v^2\log\left(\frac{\Delta}{4\delta_\sigma} \right)\right],
 \label{eq:tens_renorm}\\[.5em]
 \bar F^\mu(X;\Delta)
&\equiv\lim_{\delta_\sigma\to0}\left[
 F^\mu(X;\delta_\sigma)
 +\pi v^2\log\left(\frac{\Delta}{4\delta_\sigma}\right)K^\mu
 \right].
 \label{eq:force_renorm}
\end{align}
Both limits exist since $\log(\delta_\sigma)$ in $\mu_0$ cancels the
one in the second term, and the remaining $\mathcal O(\delta_\sigma/\Delta)$
terms of Eq.~\eqref{eq:muKR-exact} vanish. The renormalised equation is therefore
\begin{equation}
 \bar\mu(\Delta)K^\mu=\bar F^\mu(X;\Delta),
 \qquad
 \bar\mu(\Delta)=\pi v^2\left( c_{\rm FT}+\log(m_s\Delta)\right),
 \label{eq:eom_renorm}
\end{equation}
 
All logarithmic dependence on $\delta_\sigma$ has cancelled.  Moreover,
\begin{equation}
 \frac{\dd\bar\mu}{\dd\log(\Delta)}=\pi v^2,
 \qquad
 \frac{\dd\bar F^\mu}{\dd\log(\Delta)}=\pi v^2K^\mu,
 \label{eq:Delta-running}
\end{equation}
so that $\bar\mu K^\mu$ and $\bar F^\mu$ shift by the same amount and
the equation is independent of the arbitrary splitting scale $\Delta$.  Changing $\Delta$  does not alter the physical
trajectory.  Once Eq.~\eqref{eq:mu0-exact} has been imposed, the microscopic
theory enters the leading EFT only through $m_s$ and $c_{\mathrm{FT}}$ in the combination $c_{\mathrm{FT}}+\log m_s$. The distinction between core energy and near-core gradient energy of the axion (KR) field is lost.  Genuine
core dynamics are represented by the power-suppressed operators omitted from
Eq.~\eqref{eq:EFT_NG+KR}.

\subsection{Circular Loop}
\label{subsec:eoms_circular_loop}

For a circular loop in the $z=0$ plane we use the parametrisation $\sigma^a=\left( t,\sigma \right)$, where $\sigma$ is the arc-length along the loop at the initial time, equivalent to the polar angle $\varphi=\sigma/R_0\in (-\pi,\pi]$\footnote{Not to be confused with the azimuthal angle around the string of
Sec.~\ref{subsec:straight-string-matching}; this angle is the coordinate of the string itself.} , 
\begin{equation}
 X^\mu(t,\sigma)=
(t, \, R(t)\cos(\varphi), \, R(t)\sin(\varphi), \, 0),
 \qquad t \in (-\infty,\infty),
 \label{eq:loop}
\end{equation}
where $R_0 \equiv R(0)$.

The component of $K^\mu$ along the radial unit normal\footnote{
$\hat n^\mu = \gamma_R(\dot R,\cos\varphi,\sin\varphi,0)$
from  $\hat n^\mu \partial_t X_\mu = 0, \hat n^\mu \partial_\varphi X_\mu = 0,\hat n^\mu \hat n_\mu=-1$.} $\hat n^\mu$, the only
one that survives for a planar circular loop, allows us to define the combination, 
\begin{equation}
 \mathcal{K} \equiv -\frac{1}{\gamma_R^3}\hat n^\mu K_\mu = \ddot R+\frac{1-\dot R^2}{R},
 \label{eq:circle-curvature}
\end{equation}
with $\gamma_R = 1/\sqrt{1-\dot R^2}$ the relativistic boost factor of the string motion. 
The unrenormalised radial equation then reads
\begin{equation}
 {\mu_0(\delta_\sigma)}\mathcal{K} =-\frac{2\pi v}{\gamma_R^3}H_{012}(t,R;\delta_{\sigma}). 
 \label{eq:eom_circular}
\end{equation}

The regulated KR field strength  component entering the radial equation is 
$H_{012}$, which would correspond to the gradient of the axion field in the $z$ direction, $\partial_z\theta$. Let us remember that it is here evaluated at the string position, $R$, but the regulation makes it finite at non-zero $\delta_\sigma$. 
Since $H_{012}$ is sourced by the whole string, it can be calculated\footnote{See Appendix~\ref{app:KR-force-derivation} for details of the derivation and of the local singularity.} as an integral over the retarded time $t_r$ given by the intersection of the worldsheet and the past light-cone of our reference point\footnote{Because of symmetry, all points of the string are equivalent, so we choose $\varphi=0$ for our calculations.} at $(t,R(t),0,0)$
\begin{equation}
 H_{012}(t,R(t);\delta_\sigma)=-2v\int_{t_\pi}^{t-\epsilon(\delta_\sigma)}
 \frac{\dd t_r}{|\partial_\varphi\mathfrak f|}
 \frac{\dd}{\dd t_r}\left[
 \frac{
 2R_r\left[\dot R_r(t-t_r)+R_r-R(t)\cos(\varphi)\right]
 }{\partial_{t_r}\mathfrak f}
 \right]_{\mathfrak f=0},
 \label{eq:forceKR}
\end{equation}
where the light cone function is 
\begin{eqnarray}
 \mathfrak f &=&(t-t_r)^2-R(t)^2-R_r^2
 +2R(t)R_r\cos(\varphi),
 \qquad R_r\equiv R(t_r).
 \label{eq:lightcone-function}
\end{eqnarray}
In \eqref{eq:forceKR}, the derivatives are taken before imposing ${\mathfrak f} = 0$ and the integration limits correspond to the extremes of the loop: $t_\pi$ corresponds to the time taken from the antipode $\varphi=\pi$ for which $t-t_\pi=R(t)+R(t_\pi)$, and  $t_\delta=t-\epsilon(\delta_\sigma)$ corresponds to the closest point to $\varphi=0$ compatible with our  worldsheet regularisation \emph{proper-distance} = $R(t)|\varphi|>\delta_\sigma$. 
Using the near light-cone approximation $\varphi\simeq(t-t_r)/(R\gamma_R)$ we get $\epsilon(\delta_\sigma) = t-t_\delta = \gamma_R \delta_\sigma$. 

Near the coincidence, $t_r\to t$, the integrand behaves as
$-\gamma_R^3 {\mathcal K}(t)/4(t-t_r)$, so that $H_{012}$ diverges
logarithmically contributing a term  $\pi v^2 {\mathcal K}(t)\log \delta_\sigma$ to the equation of motion \eqref{eq:eom_circular}.  This is precisely the circular-loop projection of the covariant counterterm in Eq.~\eqref{eq:force_renorm}, which cancels the regulator dependence on $\delta_\sigma$. Like before, we can write, 
\begin{eqnarray}
 \bar\mu(\Delta)\,
 {\mathcal K}
 =
  \lim_{\delta_\sigma\to0}\left[
 -\frac{2\pi v}{\gamma_R^3}H_{012}(t,R;\delta_\sigma)
 +\pi v^2\log\left(\frac{\Delta}{4\delta_\sigma}\right)\mathcal K
 \right]
 \equiv \bar F[R;\Delta]
 \label{eq:circle-renormalised}
\end{eqnarray}
The $\delta_\sigma$-dependence cancels between the two terms in the RHS of  Eq.~\eqref{eq:circle-renormalised}, so that their sum depends only
on $R$ and $\Delta$ and the $\Delta$-dependence then cancels with that of $\bar\mu$ via Eq.~\eqref{eq:Delta-running}. Note that
$\delta_\sigma$ defines the EFT and is \emph{not} to be identified with the
physical core size. 
Note also that a factor $\pi v^2$ factors out of all terms of the equation, which depends on the tension only through $c_{\rm FT}+\log m_s$. 
At fixed macroscopic $R$, increasing the UV parameter 
$c_{\rm FT}+\log m_s$ makes $\bar\mu/(\pi v^2)$ large and the trajectory approaches the
NG evolution, ${\mathcal K}=0$.  

\begin{figure*}[t]
 \centering
 \includegraphics[width=\linewidth]{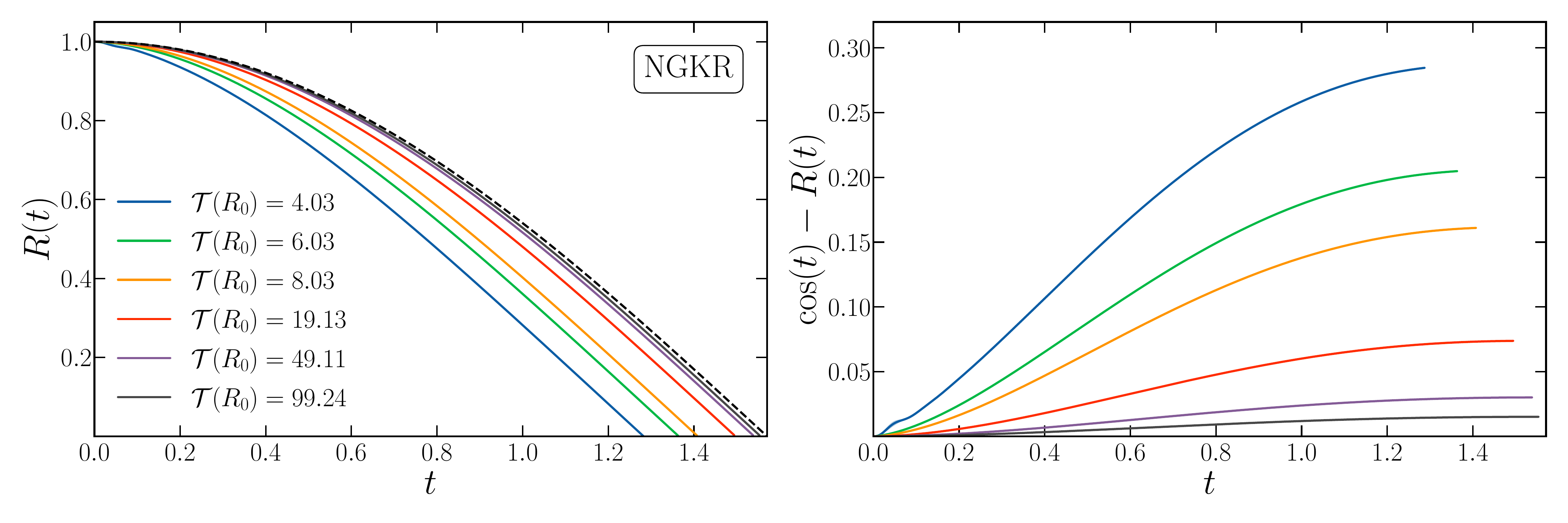}
 \caption{Selected NGKR trajectories computed with the \github{mathieukaltschmidt}{NGKR-circular-loop} module. \emph{Left:} loop radius $R(t)$, with the NG trajectory $R=\cos(t)$ shown as a black dashed line. \emph{Right:} deviation from the NG trajectory, $\cos(t)-R(t)$. The curves are labelled by their dimensionless tension $\mathcal T(R_0)=c_{\rm FT}+\log(m_sR_0)$, given by Eq.~\eqref{eq:FT-tension-match}. The curves are extrapolated to $\Delta t\to 0$ and  $\delta_\sigma\to0$, see App.~\ref{sec:ngkr-evolution} for details. Note that the computation stops once the loop has collapsed.}
 \label{fig:NGKR_panel}
\end{figure*}

We solve Eq.~\eqref{eq:circle-renormalised} numerically with a dedicated \texttt{python} module, publicly available at \github{mathieukaltschmidt}{NGKR-circular-loop}. Some selected curves are shown in Fig.~\ref{fig:NGKR_panel}, where we see how large tension parameters $\mathcal{T}$ yield solutions close to the NG limit.
We postpone an expansion of the evolution around this limit for large logarithms to Sec.~\ref{sec:compa}, where it will help us to assess our results quantitatively and compare them to our field theory simulations.

\subsection{Axion Spectrum}
\label{subsec:KRspectrum}
To calculate the radiation spectrum in the NGKR model, we follow
Ref.~\cite{Dine:2020pds}. In contrast to an exactly periodic loop, whose
radiation is naturally decomposed into discrete harmonics, see Refs.~\cite{Vilenkin:1986ku,Garfinkle:1987yw,Battye:1993jv,Battye:1994au}, the
finite collapse history produces a continuous spectrum. In this subsection, $t$ and $R(t)$ denote the dimensionless quantities $t/R_0$ and
$R(t)/R_0$, respectively, and
\begin{equation}
 \mathbf X(t,\varphi)=
 \big(R(t)\cos(\varphi),R(t)\sin(\varphi),0\big).
\end{equation}
For radiation propagating along
$\hat{\mathbf{k}}=(\sin(\alpha)\cos(\beta),\sin(\alpha)\sin(\beta),\cos(\alpha))$,
the energy emitted per unit physical angular frequency is
\begin{equation}
 \label{eq:radiation}
 \frac{\dd E_a^{\rm rad}}{\dd \omega}
 =\frac{\omega^2v^2R_0^4}{4\pi}\int \dd\Omega\,
 \left|\int_{t_i}^{t_f}\dd t\int_{-\pi}^{\pi}\dd\varphi\,
 \hat{\mathbf{k}}\cdot(\dot{\mathbf X}\times\mathbf X')
 e^{i\omega R_0(t-\hat{\mathbf{k}}\cdot\mathbf X)}\right|^2 .
\end{equation}
Here a dot and a prime denote derivatives with respect to $t$ and $\varphi$.
The angular measure is $\dd\Omega=\sin(\alpha)\,\dd\alpha\,\dd\beta$.
Performing the $\varphi$ integral along the circular loop gives, 
\begin{equation}
 \label{RadAxionExpr}
 \frac{\dd E_a^{\rm rad}}{\dd \omega}
 =2\pi^2\omega^2v^2R_0^4\int_0^\pi \dd\alpha\,\sin(\alpha)
 \left|\int_{t_i}^{t_f}\dd t\,e^{i\omega R_0t}
 J_0(\omega R_0R\sin(\alpha))\cos(\alpha)\,R\dot R\right|^2 .
\end{equation}

Note that the dimensionless spectrum 
\begin{equation}
 \mathcal P_{\rm KR}(\omega R_0)
 =\frac{1}{v^2R_0^2}\frac{\dd E_a^{\rm rad}}{\dd\omega},
 \label{eq:KR-spectrum-normalisation}
\end{equation}
is a function of the dimensionless frequency $\omega R_0$ and it is most suited for comparison with the field simulations in
Eq.~\eqref{eq:spectrum_normalisation} (for a massless axion $k=\omega$). 

\begin{figure}[t]
\centering
\includegraphics[width=0.65\linewidth]{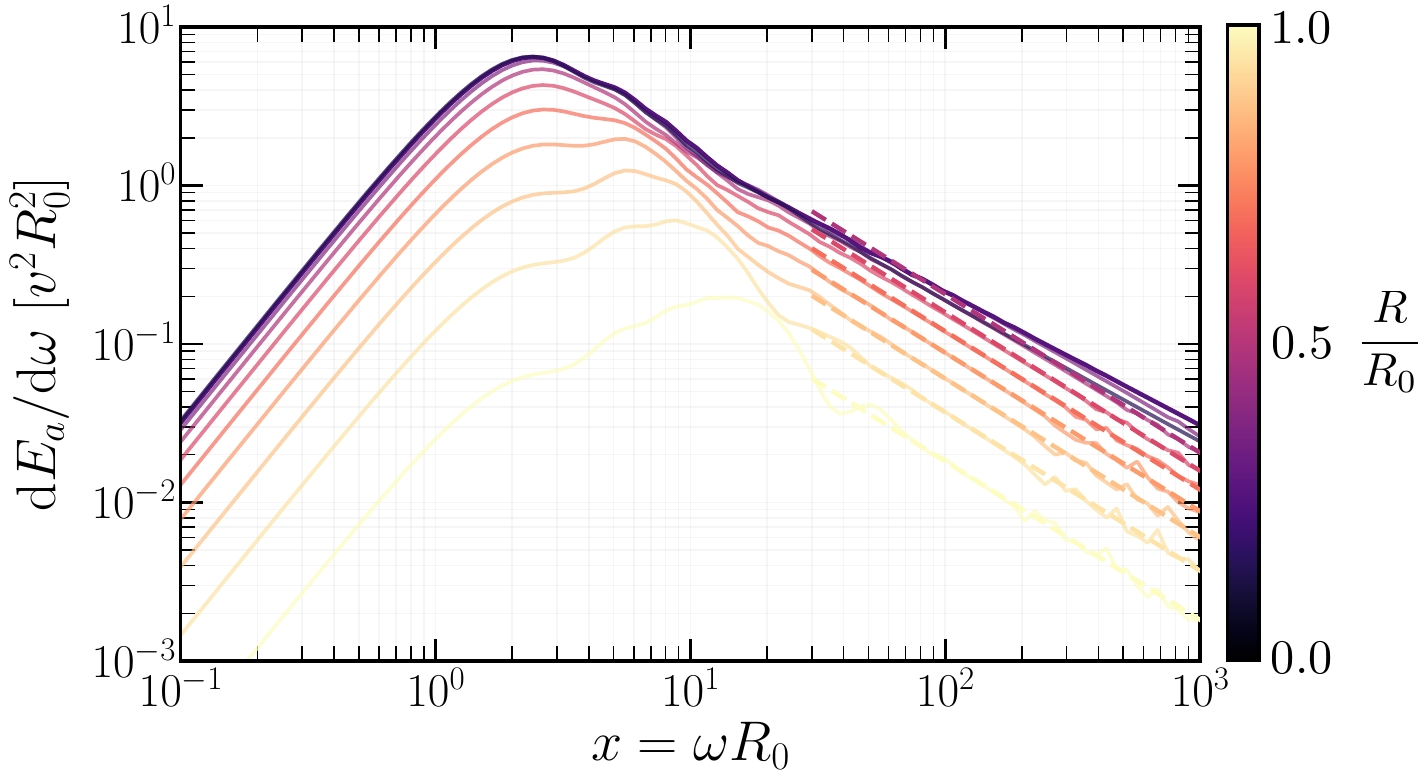}
\caption{Time sequence of the angle-integrated KR spectrum
$\mathcal P_{\rm KR}(x)$ for $\log(m_sR_0)=5.0$. Colour encodes the endpoint
$t_f$ of the emission history, from early (yellow) to late (purple), in terms of the normalised loop radius $R/R_0$ at $t_f$.
Predictions for the ultraviolet (UV) tail from Eq.~\eqref{FinalSpectrum} are
shown as dashed lines before the halving time $R(t_{1/2})=0.5R_0$.}
\label{fig:KR_spectrum}  
\end{figure}

Numerical integration shows that the high-frequency spectrum scales as, 
\begin{equation}
    \frac{\dd E_a^{\rm rad}}{\dd\omega}\propto\omega^{-1}.
\end{equation} 
When the circular loop reaches
collapse, this behaviour can be understood using the stationary phase method~\cite{Dine:2020pds}, or as a limiting case of cusp emission~\cite{Drew:2025txu}, but the same scaling is obtained when the trajectory is
terminated before collapse. In that case it originates from the endpoint of the finite time integral.
To show that, let us assume that $R(t)>0$, $|\dot R(t)|<1$, and that $R(t)$ is sufficiently smooth on $[t_i,t_f]$. We further assume $\dot R(t_i)=0$, so that the initial boundary contribution vanishes. Defining $\lambda\equiv\omega R_0$ and $\Phi(t,\alpha,\beta)\equiv t-R(t)\sin(\alpha)\cos(\beta)$, and using the integral representation of $J_0$ instead of performing the azimuthal integral of Eq.~\eqref{RadAxionExpr}, the remaining amplitude is 
\begin{equation}
\label{IntegrNoSaddl}
\mathcal A=\frac{1}{2\pi}\int_0^{2\pi}\dd\beta
\int_{t_i}^{t_f}\dd t\,e^{i\lambda\Phi(t,\alpha,\beta)}R\dot R
=\frac{1}{2\pi}\int_0^{2\pi}\dd\beta\,\mathcal A_\beta,
\end{equation}
so that the modulus squared in Eq.~\eqref{RadAxionExpr} is
$\cos^2(\alpha)\,|\mathcal A|^2$.
The phase derivative is
\begin{equation}
\dot\Phi(t)=1-\dot R(t)\sin(\alpha)\cos(\beta).
\end{equation}
Since $|\dot R(t)|<1$, we have $|\dot\Phi(t)|\geq1-|\dot R(t)|>0$, and hence the phase has no stationary points with respect to $t$.

Repeated integration by parts therefore gives an expansion in inverse powers of $\lambda$. Defining $B_0\equiv R\dot R/\dot\Phi$ and $B_{n+1}\equiv\dot\Phi^{-1}\dd B_n/\dd t$, we obtain
\begin{equation}
\label{ByPartT}
\mathcal A_\beta=\sum_{k=0}^{n}\frac{(-1)^k}{(i\lambda)^{k+1}}\left[B_k(t)\mathrm{e}^{i\lambda\Phi(t)}\right]_{t_i}^{t_f}+\frac{(-1)^{n+1}}{(i\lambda)^{n+1}}\int_{t_i}^{t_f}\dd t\,\dot\Phi(t)B_{n+1}(t)\mathrm{e}^{i\lambda\Phi(t)}\,.
\end{equation}
Since $\dot R(t_i)=0$, the leading boundary contribution comes from $t=t_f$:
\begin{equation}
\label{LeadingBoundary}
\mathcal A_\beta\simeq\frac{1}{i\lambda}\frac{R(t_f)\dot R(t_f)}{1-\dot R(t_f)\sin(\alpha)\cos(\beta)}\mathrm{e}^{i\lambda[t_f-R(t_f)\sin(\alpha)\cos(\beta)]}.
\end{equation}

The remaining $\beta$ integral has two stationary points, $\beta=0$ and $\beta=\pi$, with $\Phi''(0)=R\sin(\alpha)$ and $\Phi''(\pi)=-R\sin(\alpha)$. Applying the stationary-phase approximation gives
\begin{equation}
\mathcal A\simeq\frac{R\dot R}{2\pi i\lambda}\sqrt{\frac{2\pi}{\lambda R\sin(\alpha)}}\mathrm{e}^{i\lambda t}\left[\frac{\mathrm{e}^{-i\lambda R\sin(\alpha)+i\pi/4}}{1-\dot R\sin(\alpha)}+\frac{\mathrm{e}^{i\lambda R\sin(\alpha)-i\pi/4}}{1+\dot R\sin(\alpha)}\right],
\end{equation}
where $R$ and $\dot R$ are evaluated at $t=t_f$. Consequently,
\begin{equation}
\label{ModulBeta}
|\mathcal A|^2\simeq\frac{R\dot R^2}{2\pi\lambda^3\sin(\alpha)}\left[\frac{1}{(1-\dot R\sin(\alpha))^2}+\frac{1}{(1+\dot R\sin(\alpha))^2}+\frac{2\sin(2\lambda R\sin(\alpha))}{1-\dot R^2\sin^2(\alpha)}\right].
\end{equation}

The first two terms give the leading non-oscillatory contribution to the $\alpha$ integral, while the last term is subleading due to the additional oscillatory suppression and can be neglected. Substituting the leading terms into Eq.~\eqref{RadAxionExpr}, the factor $\sin(\alpha)$ of the angular
measure cancels against the $1/\sin(\alpha)$ of Eq.~\eqref{ModulBeta} and we
obtain
\begin{equation}
\frac{\dd E_a^{\rm rad}}{\dd\omega}\simeq\frac{\pi v^2R_0R\dot R^2}{\omega}\int_0^\pi\dd\alpha\left[\frac{\cos^2(\alpha)}{(1-\dot R\sin(\alpha))^2}+\frac{\cos^2(\alpha)}{(1+\dot R\sin(\alpha))^2}\right].
\end{equation}
The remaining integral can be evaluated exactly for $|\dot R|<1$,
\begin{equation}
\int_0^\pi\dd\alpha\left[\frac{\cos^2(\alpha)}{(1-\dot R\sin(\alpha))^2}+\frac{\cos^2(\alpha)}{(1+\dot R\sin(\alpha))^2}\right]=\frac{2\pi}{\dot R^2}\left(\frac{1}{\sqrt{1-\dot R^2}}-1\right),
\end{equation}
so that, for $|\dot R(t_f)|<1$,
\begin{equation}
\label{FinalSpectrum}
\frac{\dd E_a^{\rm rad}}{\dd\omega}
\simeq\frac{2\pi^2v^2R_0\,R(t_f)}{\omega}
\left(\frac{1}{\sqrt{1-\dot R(t_f)^2}}-1\right),
\end{equation}
or equivalently
$\mathcal P_{\rm KR}(x)\simeq2\pi^2R(t_f)\,[\gamma_R(t_f)-1]/x$ with
$\gamma_R=(1-\dot R^2)^{-1/2}$. Thus, even in the absence of a stationary
point we obtain $\dd E_a^{\rm rad}/\dd\omega\propto\omega^{-1}$, with vanishing amplitude when the endpoint is at rest.

\section{Field Theory Simulations}
\label{sec:simu}

In this section we develop the framework for simulating the global string loop dynamics and discuss the relevant observables, which we will use to compare the outcomes of the simulations to the effective theory described in Sec.~\ref{sec:eft}.

\subsection{Simulation Overview}\label{subsec:sim_overview}

Our strategy is to simulate the same circular loop initial condition with several complementary field-theory codes. This is useful not only for extending the dynamical range, but also for separating physical effects from method-dependent systematics.

\subsubsection{Three-dimensional \texttt{jaxions}}
The baseline calculation uses the ordinary
complex scalar theory in three spatial dimensions, evolved with
\github{veintemillas}{jaxions}~\cite{Jaxions:2026} on a uniform cubic lattice. In Minkowski space
\texttt{jaxions} (and \github{}{axioNyx}~\cite{Schwabe:2020eac}, see~\ref{sec:axionyx} below) solve
\begin{equation}
 \ddot\phi-\nabla^2\phi
 +\lambda\phi\left(|\phi|^2-v^2\right)=0,
 \label{eq:eom_mink_code}
\end{equation}
which is the equation of motion following from Eq.~\eqref{eq:lagrangian}, with $m_s=\sqrt{2\lambda}\,v$ as in Sec.~\ref{subsec:complex-scalar-model}. This is the most general of our setups and can later be applied to non-circular loops and string networks, see Refs.~\cite{Vaquero:2018tib, OHare:2021zrq, Pierobon:2023ozb, Saikawa:2024bta} for a selection of recent works using \texttt{jaxions} in the context of axion string networks. Its principal limitation is the uniform lattice with spacing $\delta=L/N$, for a uniform box of size $L$ with $N$ grid sites per dimension. As the loop accelerates, the Lorentz-contracted core becomes increasingly difficult to resolve. All our 3D domains use periodic boundary conditions.

\subsubsection{Three-dimensional AMR with \texttt{axioNyx}}\label{sec:axionyx}

\begin{figure}[t]
    \centering
    \includegraphics[width=\linewidth]{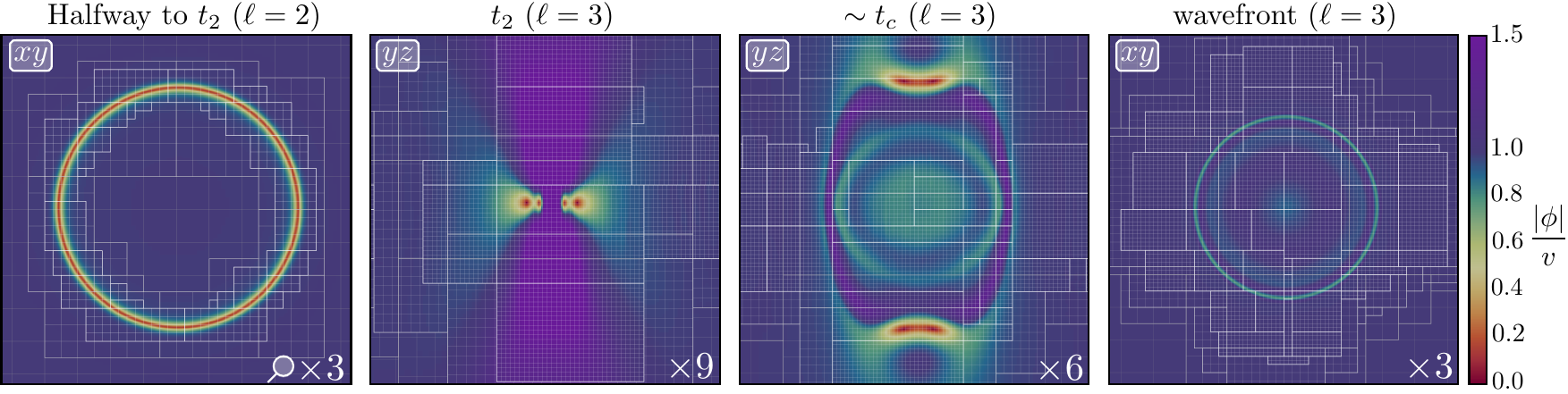}
    \caption{Adaptive mesh refinement in \texttt{axioNyx} for a typical circular loop collapse history, see Sec.~\ref{subsec:collapse_anatomy} for details on the time scales. The (up to) $\ell=3$ refinement levels are overlaid in white and the snapshots are zoomed in by the factor in the lower right of each figure for visualisation purposes.}
    \label{fig:amr_overview}
\end{figure}

The 3D AMR code \texttt{axioNyx} evolves the same equation and
the same initial fields, but dynamically refines the mesh around the string, see Fig.~\ref{fig:amr_overview}. Comparing it with uniform-grid \texttt{jaxions} simulations therefore isolates
discretisation errors, particularly during the relativistic final stage of
collapse, without changing the underlying continuum theory.

\subsubsection{Axisymmetric two-dimensional \texttt{jaxions}}
\label{subsec:2dsims}
For a perfectly circular loop we can exploit axial symmetry
and evolve the complex field on the $(\rho,z)$ half-plane. The corresponding
\texttt{jaxions} equation is
\begin{equation}
 \ddot\phi+\eta\dot\phi
 =\left(\partial_z^2+\partial_\rho^2
 +\frac{1}{\rho}\partial_\rho\right)\phi
 -\lambda\phi\left(|\phi|^2-v^2\right),
 \label{eq:eom_caxion}
\end{equation}
where $\eta$ is a damping rate that is nonzero only in the
absorbing boundary layers.
This reduction makes much larger lattices and scale separations affordable,
and provides our strongest control of the continuum limit. Its price is that
it applies only while axial symmetry is imposed: unlike the 3D codes, it
cannot study generic loop shapes, non-axisymmetric instabilities, or string
networks. Rather than periodic outer boundaries, the 2D implementation uses
damping in radial and axial sponge-layers to absorb outgoing radiation and
prevent it from returning to the loop.
The 2D code also implements a simple form of static regridding: at selected
stages of the collapse we interpolate the fields onto a new uniform grid
covering a smaller domain centered around the shrinking loop. This is not local adaptive refinement, but it concentrates the available resolution around the dynamically relevant region and allows us to follow the late collapse with near-optimal precision. We therefore still refer to it as \enquote{2D AMR} in the following discussion. We use these resampled runs primarily to understand when and how finite
resolution affects the collapse. Agreement in the common regime tests both
the symmetry reduction and the 3D discretisation effects.

\subsubsection{Three-dimensional \texttt{Moore} high-tension code}
\label{subsec:moore}

Finally, we use the 3D two-scalar--vector theory of
Sec.~\ref{subsec:Klaer--Moore-model}, introduced and developed in
Refs.~\cite{Klaer:2017qhr,Klaer:2017ond,Klaer:2019fxc}. In temporal gauge, $A_0=0$, its continuum equations
have the schematic form
\begin{equation}
 \ddot\phi_n=D_iD_i\phi_n-\frac{\partial V}{\partial\phi_n^*}
 \quad(n=1,2),\qquad
 \dot E_i=(\boldsymbol{\nabla}\times\boldsymbol B)_i-J_i,
 \label{eq:eom_moore}
\end{equation}
with $V$ the potential of Eq.~\eqref{eq:KM-lagrangian}, $E_i=-\dot{A}_i$ and
$B_i=(\boldsymbol\nabla\times\boldsymbol A)_i$ the electric and magnetic
fields, and $J_i=\sum_n q_n\,{\rm Im}(\phi_n^*D_i\phi_n)$ the scalar current,
supplemented by Gauss's law. Its massive vector
enhances only the UV core tension while leaving the IR axion field
unchanged. These simulations therefore reach much larger effective tensions
and test whether the collapse depends only on the matched long-distance
parameters rather than on the microscopic core theory.

Together, the four setups form a hierarchy: uniform 3D simulations establish the general problem, 3D AMR controls resolution near collapse, axisymmetric 2D (and 2D AMR) runs push the ordinary scalar theory to much larger scale separations, and the two-scalar--vector runs extend the tension still further.

\subsection{Loop Initial Conditions}
\label{subsec:stringICs}
\begin{figure}[t]
    \centering
    \includegraphics[width=.8\linewidth]{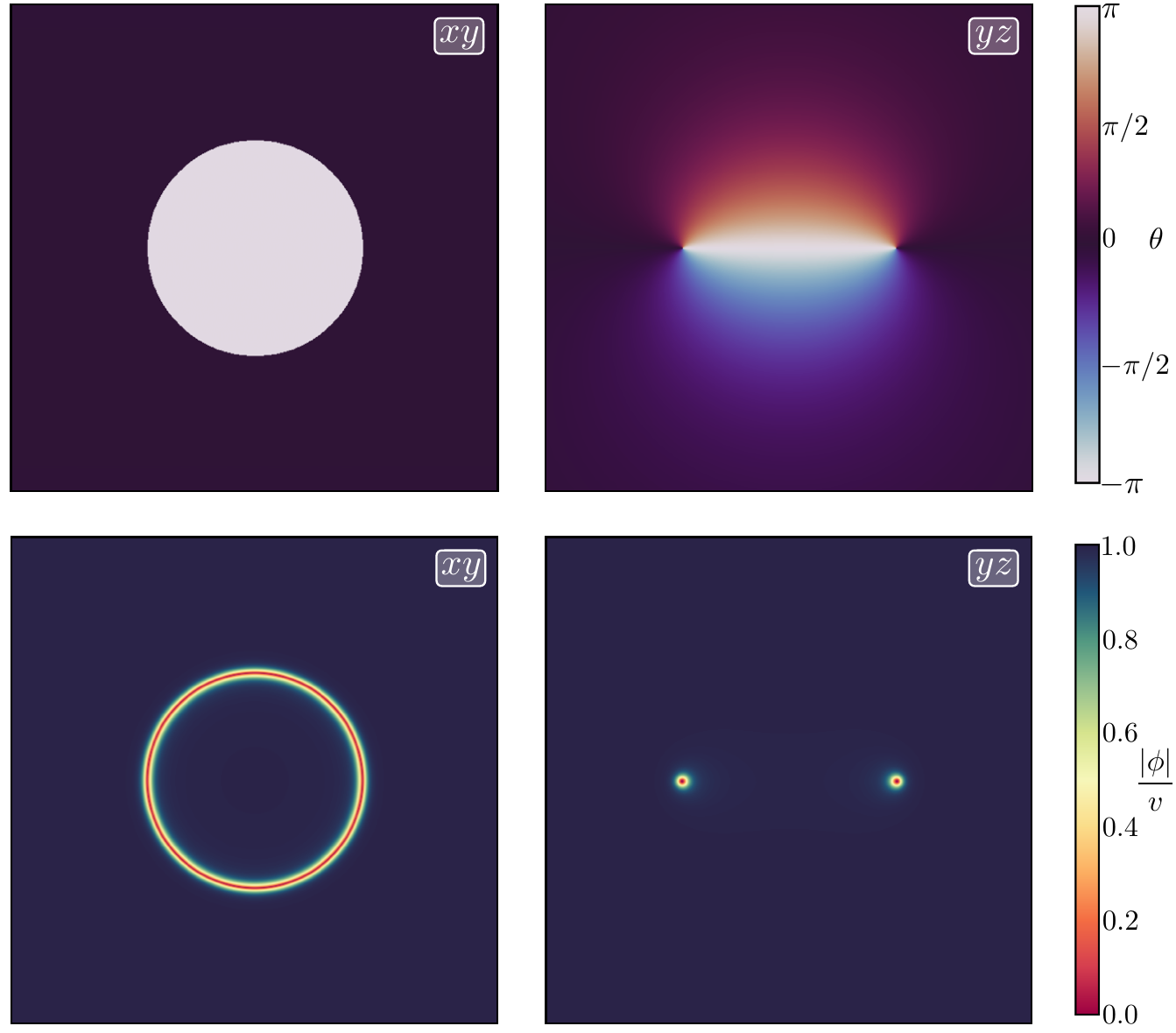}    
    \caption{Snapshot of circular loop initial conditions, showing the axion field $\theta$ (\emph{upper panel}) and the radial field $|\phi|/v$ (\emph{lower panel}) in the $xy$- (\emph{left}) and $yz$-planes (\emph{right}).}
    \label{fig:axionICs}
\end{figure}
All calculations begin with a loop of radius $R_0$ in the
$z=0$ plane, initially at rest. The radial field is initialised with the
straight-string profile evaluated at the shortest distance to the loop. For
the phase we seek the minimum-gradient-energy configuration compatible with
one unit of winding. Away from the core this means solving
$\nabla^2\theta=0$ with the required wrapped boundary condition.

The magnetostatic analogy follows directly from the
Kalb--Ramond-axion duality of Sec.~\ref{subsec:kr-axion-duality}. In the static region outside the core, where $f\simeq1$, the spatial vector
\begin{equation}
 b_\theta^i\equiv\frac{1}{2v}\epsilon^{ijk}H_{0jk}
 =\partial^i\theta,
 \label{eq:KR-magnetic-vector}
\end{equation}
plays the role of a magnetic field. For a static string lying on an oriented
curve $\Gamma$, the Bianchi identity for $H$ together with the KR source
equation~\eqref{eq:Rsource} reduce to
\begin{equation}
 \boldsymbol\nabla\cdot\boldsymbol b_\theta=0,
 \qquad
 \boldsymbol\nabla\times\boldsymbol b_\theta
 =2\pi\boldsymbol j_\Gamma,
 \qquad
 \boldsymbol j_\Gamma(\boldsymbol{x})
 =\oint_\Gamma \dd\boldsymbol\ell\,
 \delta^{(3)}(\boldsymbol{x}-\boldsymbol X).
 \label{eq:KR-magnetostatics}
\end{equation}
These are precisely the magnetostatic equations for the
magnetic field of a filamentary current, with $2\pi$ playing the role of the
current normalisation. Equivalently, the second equation is the
distributional form of the axion winding condition
$\boldsymbol\nabla\times\boldsymbol\nabla\theta
=2\pi\boldsymbol j_\Gamma$. Imposing
$\boldsymbol b_\theta\to0$ at infinity, the unique transverse solution
is the corresponding Biot--Savart field.

Rather than integrating the field numerically, we obtain the phase \emph{geometrically}. If $S$ is any orientable surface with
$\partial S=\Gamma$, and
$\Omega_S(\boldsymbol x)$ is the solid angle subtended by $S$ at $\boldsymbol x$,
then
\begin{equation}
 \theta(\boldsymbol{x})=\frac12\Omega_S(\boldsymbol{x}),
 \label{eq:solid-angle-theta}
\end{equation}
whose gradient reproduces the Biot--Savart field exactly off the surface. Since $\Omega_S$ jumps by $4\pi$ across $S$, the phase jumps by $2\pi$, as required, and reversing the orientation of $\Gamma$ reverses both signs. This
construction is manifestly harmonic away from the string, and therefore
extremises $\frac12v^2\int \dd^3x\,(\boldsymbol\nabla\theta)^2$ at fixed
winding and asymptotic boundary condition; it also requires no path
integration, no gauge fixing, and applies to arbitrary loop shapes. The
identity between Eq.~\eqref{eq:solid-angle-theta} and the Biot--Savart field,
together with the triangulated implementation used in practice, is given in
App.~\ref{app:string_ics}. For snapshots of a typical circular loop initial condition see Fig.~\ref{fig:axionICs}. In the axisymmetric implementation we fix the branch on the loop plane by $\theta(\rho,0)=\pm\pi$ for $\rho<R_0$, approached from $z>0$ ($z<0$), and $\theta(\rho,0)=0$ for $\rho>R_0$, together with $\theta\to0$ far from the loop.

\paragraph{Core Relaxation.}
Fixing the phase does not by itself put the massive radial
field in equilibrium. In the axisymmetric 2D simulations the distance
$d=\sqrt{(\rho-R_0)^2+z^2}$ to the ring is known analytically, so we simply set the radial field to the straight-string profile of Eq.~\eqref{eq:eom_f} evaluated at that distance, $f(m_sd)$. Curvature modifies this local straight-string profile only at relative order $(m_sR_0)^{-1}$. In 3D, extracting a sufficiently smooth distance to the discretised string directly from the phase field is less accurate. We therefore relax the initial configuration by alternately evolving one polar component while holding the other fixed. With artificial damping coefficients $\eta_f,\eta_\theta>0$, the two
projected equations are
\begin{align}
 \ddot f+\eta_f\dot f={}&\nabla^2f-f(\boldsymbol\nabla\theta)^2
 -\frac{m_s^2}{2}f(f^2-1),
 &&\theta\ \text{fixed},
 \label{eq:radial-core-relaxation}\\
 \ddot\theta+\eta_\theta\dot\theta={}&
 \frac{1}{f^2}\boldsymbol\nabla\cdot
 \left(f^2\boldsymbol\nabla\theta\right),
 &&f\ \text{fixed}.
 \label{eq:phase-core-relaxation}
\end{align}
A series of (typically $\sim 30$) alternating damping steps relax the configuration without allowing the loop topology to change.

For the \texttt{Moore} initial conditions we instead evolve the full
two-scalar--vector system through a prepropagation stage with exponential expansion, $a_{\mathrm{FLRW}}(t)=e^{Ht}$. The scalar equations then take the form
\begin{equation}
 \ddot\phi_n+3H\dot\phi_n-\frac{1}{a_{\mathrm{FLRW}}^2}D_iD_i\phi_n
 +\frac{\partial V}{\partial\phi_n^*}=0,
 \qquad n=1,2,
 \label{eq:moore-inflationary-relaxation}
\end{equation}
together with the corresponding expanding-Universe gauge
equations and Gauss constraint. We tune $H$ so that the resulting Hubble friction relaxes the scalar--vector core before the configuration is frozen and then used as input for the (Minkowski) evolution described in Sec.~\ref{subsec:moore}.

None of these procedures constructs a static circular loop,
which does not exist in the unconstrained theory. Rather, we first force
$\theta$ to solve its static equation with the winding and string position
held fixed, and then approximately minimise the core energy for that phase.
Each constrained static problem is well defined even though their freely
evolving combination is not static. Stabilisation of the energy during the
auxiliary evolution, followed by a smooth release with small initial field
accelerations, provides the numerical check that the core has relaxed. These
diagnostics are specified in Appendix~\ref{app:ic-relaxation}.

The remainder of this section is organised as follows. In Sec.~\ref{subsec:collapse_anatomy} we discuss the typical timeline of a circular loop collapse, in Sec.~\ref{subsec:loop_radius} we introduce different estimators for the loop radius, discuss discretisation and finite-volume effects and construct the continuum limit of the simulations, and in Sec.~\ref{subsec:spectrum} we describe the measurement of the axion spectrum from 2D simulations and discuss the relevant features of the pre- and post-collapse spectra. More details on the code-specific setups and the simulations are collected in App.~\ref{app:numerics}.

\subsection{Anatomy of the Collapse}\label{subsec:collapse_anatomy}
The qualitative picture of the collapse can be simulated with the required precision using the
axisymmetric 2D setup introduced in Sec.~\ref{subsec:2dsims}, which allows us to resolve the core throughout the nonlinear
stage.

\begin{figure*}[t]
\centering
\includegraphics[width=\linewidth]{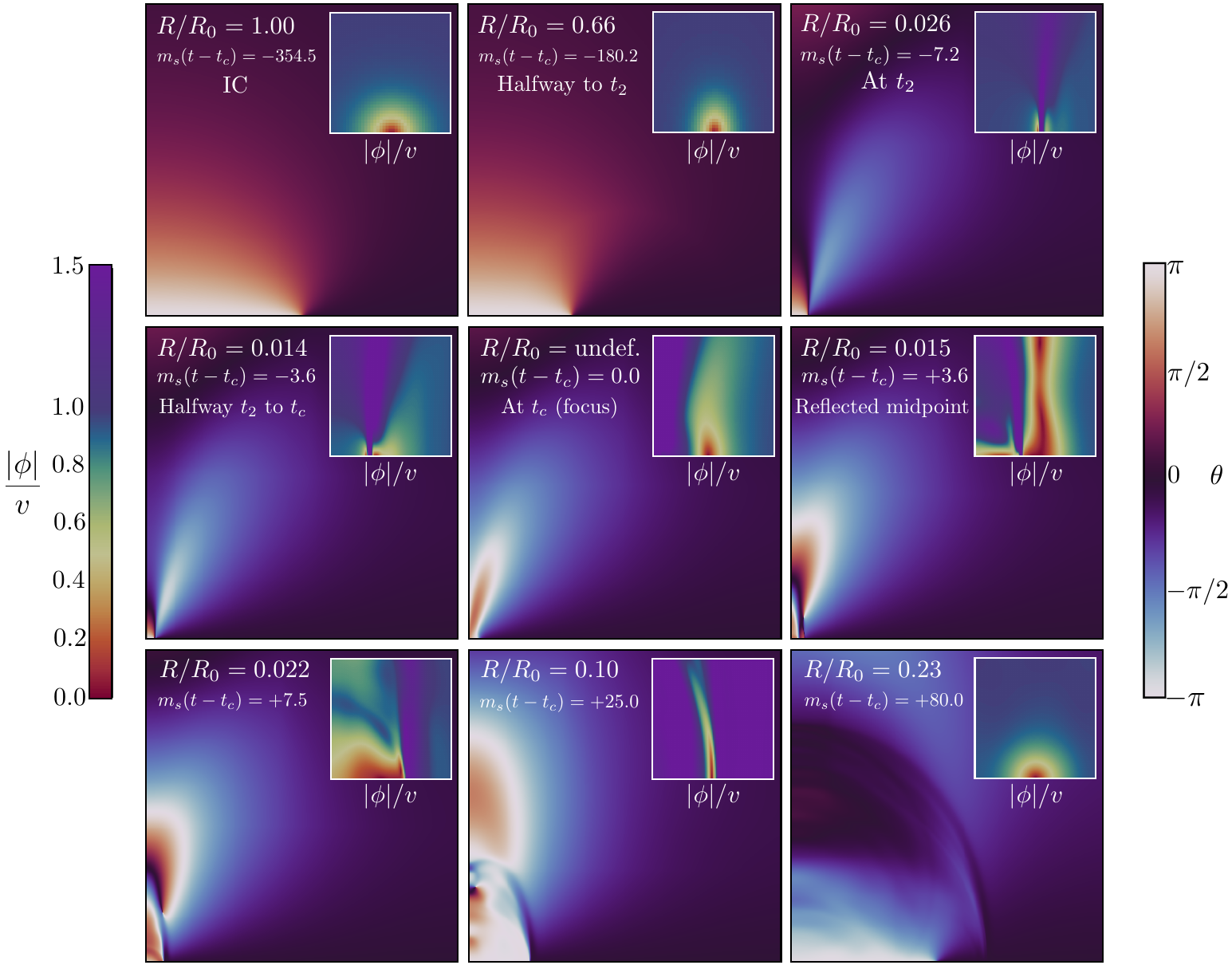}
\caption{Two-dimensional anatomy of a resolved axisymmetric loop collapse.
Only the simulated quarter-plane $\rho,z\geq0$ is shown; the full solution is
obtained using Eq.~\eqref{eq:cylindrical_reflection} and axial symmetry. Main
panels show the periodic phase $\theta(\rho,z)$ from the initial conditions,
through the loss of adiabaticity at $t_2$, Eq.~\eqref{eq:t2_scaling}, and the
focus at $t_c$, to the outgoing configuration. Insets show $|\phi|/v$ within
approximately $10\,m_s^{-1}$ of the core. The spatial window follows the
shrinking loop and returns to the full domain in the final panel.}
\label{fig:collapse_anatomy}
\end{figure*}

We begin with a circular loop of radius $R_0$ at rest in the $z=0$ plane, see the upper-left panel of Fig.~\ref{fig:collapse_anatomy}. Axial symmetry removes any dependence on the azimuthal angle, so the fields
depend only on $(\rho,z)$.  Reflection across the loop plane acts by complex
conjugation,
\begin{equation}
 \phi(\rho,-z)=\phi^*(\rho,z),\qquad
 \theta(\rho,-z)=-\theta(\rho,z)\quad ({\mathrm{mod}}\ 2\pi),
 \label{eq:cylindrical_reflection}
\end{equation}
so it is sufficient to simulate the quarter-plane $\rho\geq0$, $z\geq0$; the
remaining quadrants follow by reflection and complex conjugation.
We identify two characteristic time scales during the collapse. The first, $t_2$, marks the loss of
adiabaticity of the contracting core and is defined in
Eq.~\eqref{eq:t2_scaling} below. The second, $t_c$, is the focus time at which the axion energy density at $\rho=z=0$ is maximal; it can be understood as the \enquote{collapse time} in an energy-transport sense.

For the discussion, we follow the panels of Fig.~\ref{fig:collapse_anatomy} from top left to bottom right.  To reconstruct the full plane, the
individual snapshots must be reflected and complex conjugated.  The string is the point at the lower edge where all phase colours meet; after rotation
around the symmetry axis this point becomes the circular loop.  A contour around it passes from $0$ to $\pi$ in the upper half-plane and from $-\pi$ to $0$ in the lower half-plane, completing the required $2\pi$ winding.

\begin{figure*}[t]
\centering
\includegraphics[width=\linewidth]{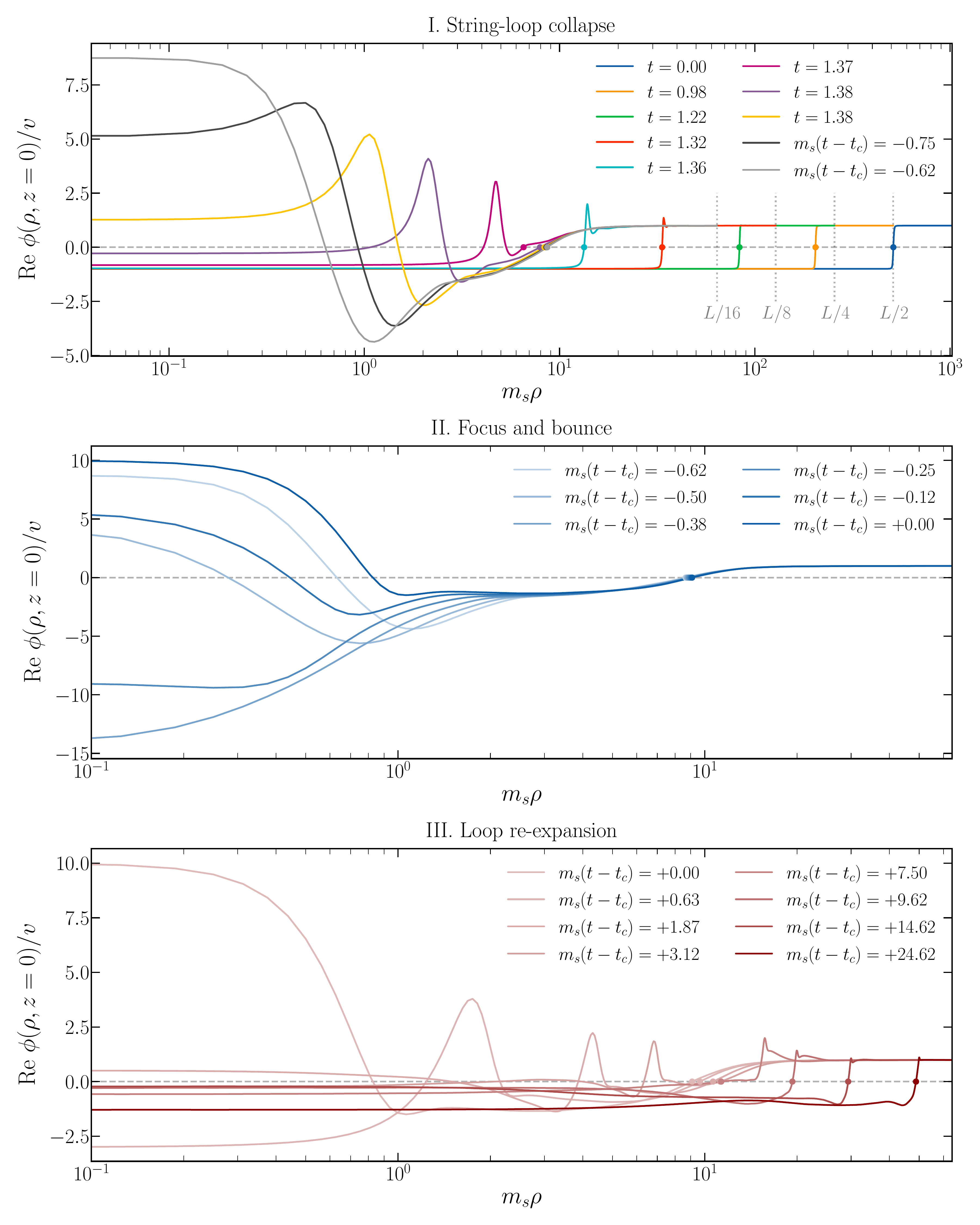}
\caption{Profiles of $\mathrm{Re}\,\phi(\rho,z=0)/v$ against $m_s\rho$ for the
resolved collapse shown in Fig.~\ref{fig:collapse_anatomy}. The three rows
isolate the approximately NG contraction and onset of the trailing crest
(\emph{top}), the nonlinear focus and bounce (\emph{middle}), and the subsequent loop
re-expansion (\emph{bottom}). Dots mark the outermost topological zero used to track
the loop.}
\label{fig:collapse_overview}
\end{figure*}

Equation~\eqref{eq:cylindrical_reflection} constrains the phase on the loop
plane to $\theta=0$ or $\pi\ (\operatorname{mod} 2\pi)$.  Equivalently,
$\mathrm{Im}\,\phi(\rho,0)=0$, and all information about the string positions
in this plane is contained in $\mathrm{Re}\,\phi(\rho,0)$.  A string is a zero
across which $\mathrm{Re}\,\phi$ changes sign, or equivalently the phase jumps
between $0$ and $\pi$.  The resulting profile resembles a one-dimensional
domain wall, although the object is a string after restoring the azimuthal
direction.  Figure~\ref{fig:collapse_overview} uses precisely this
representation: the one-dimensional $z=0$ profiles provide the timeline,
while Fig.~\ref{fig:collapse_anatomy} supplies the complementary
two-dimensional geometry.

At early times the loop contracts adiabatically.  The first three profiles in
Fig.~\ref{fig:collapse_overview} show that the core shape changes mainly by
Lorentz contraction.  The NG trajectory and boost factor,
\begin{equation}
 R(t)=R_0\cos(t/R_0),\qquad \gamma_R(t)=\frac{R_0}{R(t)},
 \label{eq:NG_circular_trajectory}
\end{equation}
provide a useful description, up to the slowly decoupling Goldstone
corrections to the force and inertia.  As the proper acceleration grows like
$a_{\rm prop}\simeq R_0/R^2$, the core eventually cannot readjust on its
natural time scale $m_s^{-1}$.  This yields the estimate
\begin{equation}
 m_sR_{\rm NL}\sim\sqrt{m_sR_0},
 \label{eq:nonlinear_radius_estimate}
\end{equation}
which is parametrically larger than the static core width.

The first sign of this loss of adiabaticity is the small crest left behind the
moving core in the upper row of Fig.~\ref{fig:collapse_overview}.  For a
straight string an arbitrarily boosted profile is an exact solution, but a
curved, accelerating string must continuously readjust its shape.  The field
overshoots the vacuum value and excites a trailing radial oscillation.  We
define $R_2$ as the loop radius when the crest first reaches
$\mathrm{Re}\,\phi=2v$, and call the corresponding time $t_2$.  A fit to simulations in the parameter range $128 \leq m_sR_0 \leq 1024$ gives
\begin{align}
 m_sR_2 &=(0.434\pm0.022)(m_sR_0)^{0.500\pm0.008},
 \label{eq:R2_scaling}\\[.5em]
 \frac{t_2}{R_0} &=\frac{\pi}{2}
 -\frac{1.057\pm0.004}{\log(m_sR_0)}
 -\frac{0.387\pm0.011}{\sqrt{m_sR_0}},
 \label{eq:t2_scaling}
\end{align}
which confirms the NG scaling estimate. Note that the two fits are
consistent with each other: inverting Eq.~\eqref{eq:NG_circular_trajectory} at
$R=R_2$ gives $t_2/R_0=\arccos(m_sR_2/m_sR_0)\simeq\pi/2-0.434/\sqrt{m_sR_0}$,
roughly matching the last term of Eq.~\eqref{eq:t2_scaling} (within $2\sigma$), while the
$1/\log(m_sR_0)$ term is the expected tension correction to the NG
trajectory.  A complete discussion of the measurements and the corresponding fits is presented in App.~\ref{app:collapse_scalings}.

Below $R_2$, nonlinear field dynamics rather than the original string core
control the collapse.  The \enquote{topological zero} moves ahead of the energy-density peak and may have a superluminal phase velocity, without superluminal energy transport.  The energy instead flows inward at nearly the speed of light and is compressed into the focus region.  Here the field amplitude and energy density are so large that crossing $\mathrm{Re}\,\phi=0$ is not especially energetic compared with nearby points.  The zero remains topologically well-defined, but it is no longer a faithful marker of where the energy is located. Indeed, shortly after $t_2$ additional string--antistring pairs appear as zeros of the highly excited field.  They are products of the nonlinear wave dynamics, so identifying any one zero as ``the original loop'' ceases to be meaningful. This is exactly why we adopt the energy-density definition of $t_c$ given above.

The maximum focus density grows rapidly with $m_sR_0$, while its width
decreases. Nevertheless, the focused region is not parametrically narrower
than the Lorentz-contracted core at $R_2$,
\begin{equation}
 \ell_2\sim\frac{1}{m_s\gamma_R(t_2)}
 =\frac{R_2}{m_sR_0}
 \simeq\frac{0.434}{\sqrt{m_sR_0}}\,m_s^{-1},
 \label{eq:NLcorewidth}
\end{equation}
This is the smallest scale that must be resolved through the nonlinear event,
leading to the convergence criterion
\begin{equation}
 m_s\delta\lesssim\frac{1}{\sqrt{m_sR_0}}.
 \label{eq:focus_resolution_requirement}
\end{equation}
The focus-time, density, and width fits are collected in
App.~\ref{app:collapse_scalings}; here we use only their physical trends and
the resulting resolution criterion.

After focusing, energy flows back out of the central region around a new
outgoing zero of $\phi$.  This can be interpreted as a bounce of the original
loop, but it is not an elastic reflection: the field left behind is strongly
excited, and part of the energy remains in nonlinear oscillations and
radiation, so the loop cannot return to its initial radius.  The middle and
lower rows of Fig.~\ref{fig:collapse_overview} show the focus, bounce, and
re-expansion.  Figure~\ref{fig:collapse_anatomy} additionally reveals pairs of
small loops moving away in the $z$ direction; by symmetry an identical set
moves toward negative $z$.  These loops contract and annihilate.  The final
panel is instead dominated by a smooth outgoing phase patch, with smaller
irregularities propagating away from the surviving loop along $z$.  This is
the long-wavelength component of the emitted axion radiation studied in
Sec.~\ref{subsec:spectrum}.

\subsection{Collapse Trajectory}\label{subsec:loop_radius}
The quantity we extract from every simulation is the instantaneous physical
radius $R(t)$ of the circular string loop, with initial value $R_0\equiv R(0)$. The reference against which it is compared is the NG
trajectory of Eq.~\eqref{eq:NG_circular_trajectory},
$R_{\rm NG}(t)=R_0\cos(t/R_0)$, which collapses at $t=\pi R_0/2$. In the following we discuss the different techniques to measure the loop radius and how we obtain the continuum radius from a series of simulations.

\subsubsection{Loop Radius Measurement}
\label{subsec:loop_radius_meas}
The measurement is performed in the two-dimensional slice through the $z=0$ plane where the loop is located. The core appears as a closed ring with $\phi=0$. In practice we work with the real part $\mathrm{Re}\,\phi$. As noted in
Sec.~\ref{subsec:collapse_anatomy}, reflection symmetry fixes the phase in
this plane to $\theta=0$ or $\pi$, so that $\mathrm{Re}\,\phi=\pm v f$ and its
zeros are precisely the core crossings, where $f\to0$.

We first locate the center of the loop, $\mathbf{x}_0$, as the centroid of the
field weighted towards the core by a narrow Gaussian,
\begin{equation}
\label{eq:com}
\mathbf{x}_0 = \frac{\sum_{\mathbf{x}} w(\mathbf{x})\,\mathbf{x}}
                    {\sum_{\mathbf{x}} w(\mathbf{x})}, \qquad
w(\mathbf{x}) = \exp\!\left[-c_w\left(\frac{\mathrm{Re}\,\phi(\mathbf{x})}{v}
\right)^{2}\right],
\end{equation}
with $c_w \simeq 30$, so that $w\simeq 1$ on the core and is exponentially
suppressed in the bulk. 

From $\mathbf{x}_0$ the radius is obtained in two complementary ways, each
evaluated both along the grid axes and along a diagonal.\footnote{The diagonal estimators are available only in the 3D simulations. The axisymmetric runs resolve only a single quarter-plane, so the loop plane is sampled along one line through the core rather than a full plane.} Within angular wedges of half-width
$\Delta\alpha=\pi/8$ centered on a set of directions $\{\alpha_k\}$, we take the $w$-weighted mean distance to $\mathbf{x}_0$,
\begin{equation}
\label{eq:r_weighted}
R_{\{\alpha_k\}} = \frac{\sum_{\mathbf{x}\in\mathcal{W}} w(\mathbf{x})\,
                          |\mathbf{x}-\mathbf{x}_0|}
                        {\sum_{\mathbf{x}\in\mathcal{W}} w(\mathbf{x})},
\qquad
\mathcal{W}=\Big\{\mathbf{x}: \min_k \big|\angle(\mathbf{x}-\mathbf{x}_0)
-\alpha_k\big|<\Delta\alpha\Big\},
\end{equation}
with $\{\alpha_k\}=\{0,\tfrac{\pi}{2},\pi,\tfrac{3\pi}{2}\}$ giving the
axis estimate $R_{\rm axes}$ and $\{\tfrac{\pi}{4},\tfrac{3\pi}{4},\tfrac{5\pi}{4},\tfrac{7\pi}{4}\}$
the diagonal estimate $R_{\rm diag}$. Alternatively, we collect the sign changes of $\mathrm{Re}\,\phi$ along rays
through $\mathbf{x}_0$ in the same two sets of directions, locate each
crossing by linear interpolation, and take half the separation of the
outermost pair,
\begin{equation}
R^{\rm int} = \tfrac{1}{2}\big(s_2-s_1\big),
\label{eq:r_extent}
\end{equation}
where $s_1$ and $s_2$ are the positions of the two outermost core crossings
along the sampled line, so that $R^{\rm int}$ does not
depend on where $\mathbf{x}_0$ lies on it. This yields
$R_{\rm axes}^{\rm int}$ and $R_{\rm diag}^{\rm int}$. Each estimate is
converted to physical units via the lattice spacing $\delta$, and the radius
reported at time $t$ is the set
$\{R_{\rm axes},R_{\rm diag},R_{\rm axes}^{\rm int},R_{\rm diag}^{\rm int}\}$.

The four estimators bracket the radius, and their spread is the measurement
uncertainty that propagates into the continuum analysis of
Sec.~\ref{subsec:continuum}. Two systematics are separated by construction. The axes-vs-diagonal difference quantifies the residual
deformation of the loop due to discretisation effects, since a perfectly circular loop on an isotropic grid would give
$R_{\rm axes}=R_{\rm diag}$; it supplies an uncertainty 
$\sigma_{\rm deform}$ used in the continuum extrapolation. 
The difference between the two independent measurement techniques additionally provides us with a sensitivity estimate to the string core profile and to the soft mask of Eq.~\eqref{eq:com}. 
Nevertheless, both techniques fail once the Lorentz-contracted core is no longer resolved with the required precision.
Its width is $1/(m_s\gamma_R)=R/(m_sR_0)$, which must not be narrower than one cell, giving the floor
\begin{equation}
 m_sR(t)\gtrsim (m_s\delta)\,(m_sR_0),
 \label{eq:radius_resolution_floor}
\end{equation}
or equivalently $R/R_0\gtrsim m_s\delta$. Below this radius the measured
trajectory is no longer under control, and the treatment of the final descent
is discussed in App.~\ref{app:collapse_times}.

\begin{figure}[t]
\centering
\includegraphics[width=\linewidth]{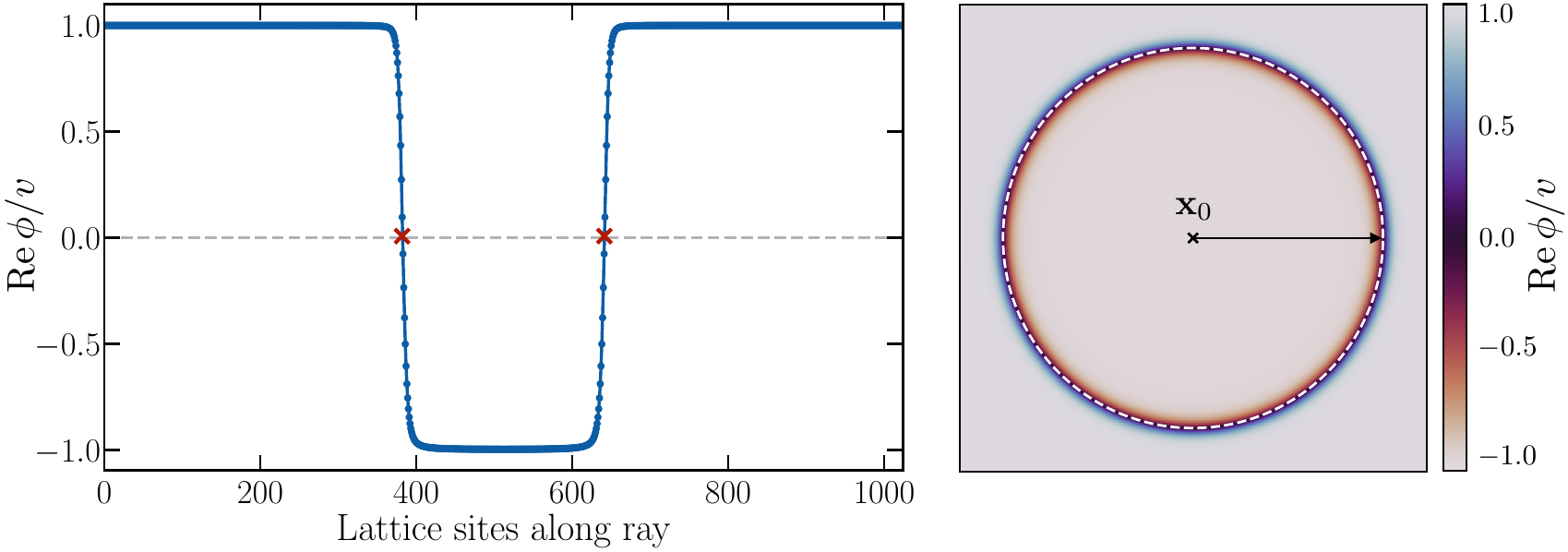}
\caption{Measurement of the loop radius. \emph{Right:} $\operatorname{Re}\phi/v$
on the loop plane, with the center $\mathbf{x}_0$ of Eq.~\eqref{eq:com}. \emph{Left:} $\operatorname{Re}\phi/v$
along a ray through that plane; red crosses mark the two core crossings located by linear interpolation, whose separation gives $R^{\rm int}$ in
Eq.~\eqref{eq:r_extent}. The same rays are used for the weighted estimator of
Eq.~\eqref{eq:r_weighted}, and both are evaluated along the grid axes and
along the diagonals, giving four estimators whose spread sets the measurement
uncertainty.}
\label{fig:meas_radius}  
\end{figure}

\subsubsection{Continuum Extrapolation}\label{subsec:continuum}
The continuum radius is obtained by extrapolating the measured radii to vanishing lattice spacing, $m_s\delta\to 0$. For a fixed scale separation $\log(m_s R_0)$, we run a series of simulations increasing the grid size $N$ and tune both the discretisation effects, controlled via $m_s\delta$, and the effects of simulating in a finite box which is controlled by the size of the loop relative to the box size, $R/L$, simultaneously. Since $L=N\delta$, the two are not independent: they are related by
\begin{equation}
\label{eq:diagonal}
 \frac{R_0}{L}=\frac{m_sR_0}{N\,m_s\delta},
\end{equation}
so a series at fixed $N$-scaling traces a line in the $(m_s\delta,\,R_0/L)$
plane whose origin is the continuum. We choose the one-parameter family
\begin{equation}
\label{eq:diagonal_family}
 m_s\delta=\sqrt{\frac{m_sR_0}{\alpha\,N}},
 \qquad
 \frac{R_0}{L}=\alpha\,m_s\delta=\sqrt{\frac{\alpha\,m_sR_0}{N}},
\end{equation}
with $\alpha$ a fixed slope. Increasing
$N$ then drives $m_s \delta \to 0$ while keeping $R/L$ on the same diagonal, so that the discretisation and the volume effects decrease together.
For loops satisfying $R_0/L\lesssim0.25$ we find the discretisation error to
dominate, which allows the extrapolation to be reduced to a one-parameter
model in $m_s\delta$ along a single diagonal.
\begin{figure}[t]
    \centering
\includegraphics[width=\linewidth]{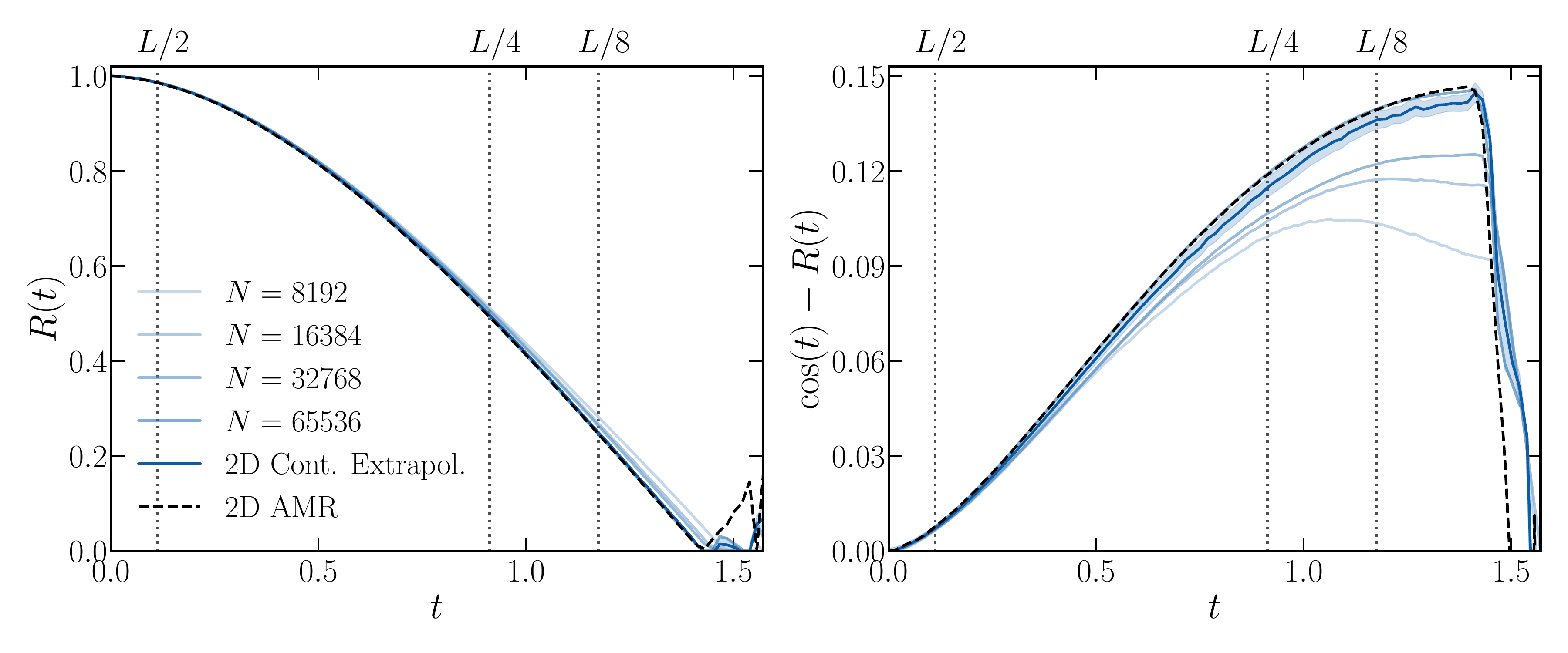}
\includegraphics[width=\linewidth]{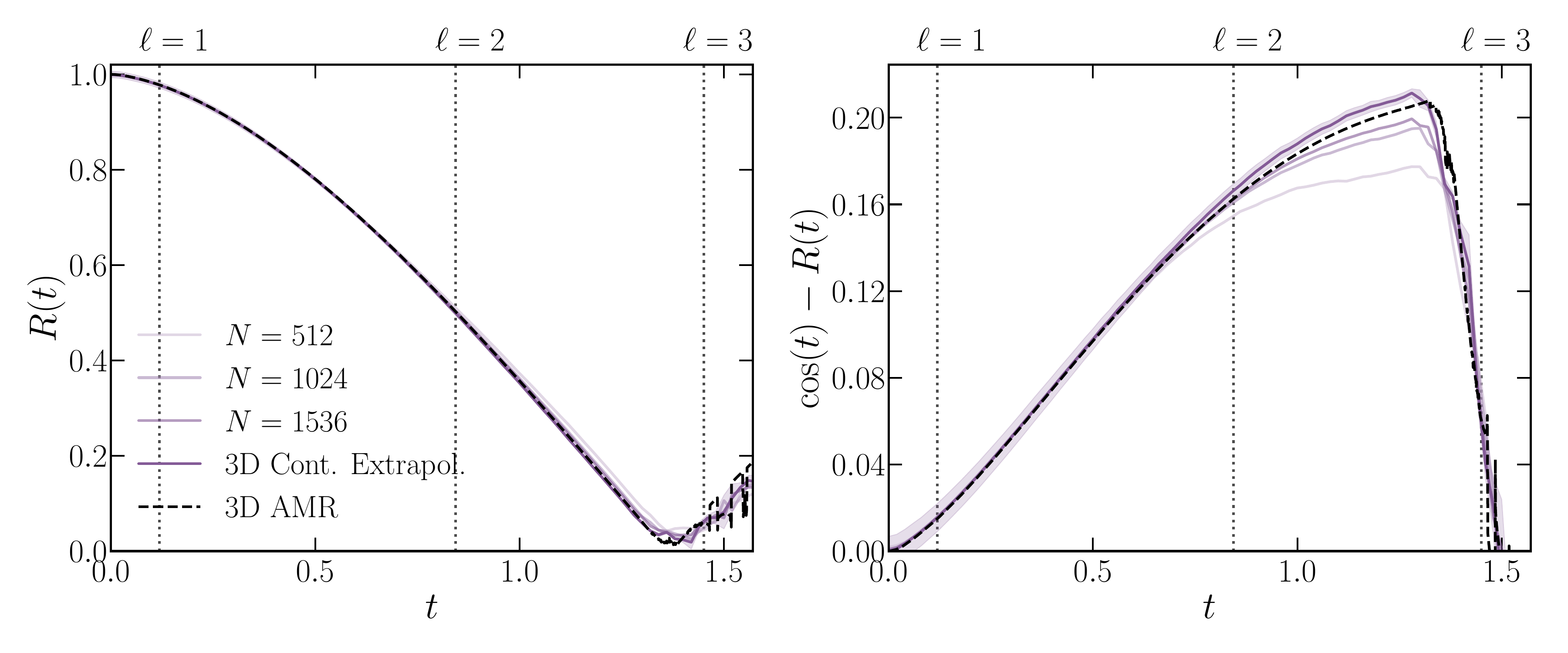}
\caption{Continuum extrapolation of the collapse trajectory in 2D at $\log(m_sR_0)=7.0$
(\emph{top}) and 3D simulations (\emph{bottom}) at $\log(m_sR_0)=5.0$. Thin lines show the individual runs along
the diagonal of Eq.~\eqref{eq:diagonal_family}, the shaded band the
extrapolated continuum curve $c_0(t)$ with the uncertainty of
Eq.~\eqref{eq:band}, and the dashed line a single adaptive-mesh-refinement run
at the same $\log(m_sR_0)$. Vertical dotted lines mark the times at which the resolution is increased: new refinement levels in the AMR runs, and in the 2D case a resampling of the fields onto a box of half the previous size.}
    \label{fig:continuum_vs_amr}
\end{figure}
Having interpolated the runs onto a common time grid, at each
$t$ we model the residual dependence of the measured radius on $m_s\delta$ as 
\begin{equation}
 R( t;\,m_s\delta) =\sum_{n=0}^{\infty} c_{2n}( t)\,(m_s\delta)^{2n}.
\label{eq:cont_model}
\end{equation}
Truncating at second order\footnote{For the \texttt{Moore} simulations we included the fourth order contributions $\sim(m_s\delta)^4$ for better convergence.},
\begin{equation}
 R(t;m_s\delta)=c_0(t)+c_2(t)\,(m_s\delta)^2
 +\mathcal{O}\!\left((m_s\delta)^4\right),
\label{eq:cont_model_truncated}
\end{equation}
and the continuum radius is
$R_{\mathrm{cont}}(t)\equiv c_0(t)=\lim_{m_s\delta\to0}R(t;m_s\delta)$. The coefficients follow from a weighted
least-squares fit performed independently at each time step, the quality of which
we quantify by the reduced $\chi^2$. The weighting is dictated by the physics of the collapse. As the loop shrinks it
becomes relativistic, $|\dd R/\dd t|\to1$ and $\gamma_R\to\infty$, and its core
Lorentz-contracts to a width $1/(m_s\gamma_R)$. The grid then resolves the
core by only $1/(m_s\delta\,\gamma_R)$ cells, which becomes $\mathcal{O}(1)$
already before the collapse, cf.\ Eq.~\eqref{eq:radius_resolution_floor}, so the dynamics in this regime are not properly captured on the lattice. We suppress its influence on the extrapolation by weighting each timestep with a decreasing function of the Lorentz factor,
\begin{equation}
\label{eq:gamma_weight}
 w(t)\propto\gamma_R(t)^{-p},
\end{equation}
with $p=3$, so that the well-resolved early collapse dominates the fit.

The uncertainty band is the sum of three independent contributions,
\begin{equation}
\label{eq:band}
\sigma( t) = \sqrt{\,\sigma_{\rm fit}^2( t) + \sigma_{\rm deform}^2( t) + \sigma_{\rm core}^2( t)\,}.
\end{equation}
The fit term is the regression uncertainty in $c_0$,
scaled by the reduced $\chi^2$ at this timestep,
\begin{equation}
\label{eq:sigma_fit}
 \sigma_{\rm fit}(t)=\sqrt{\,\chi^2_{\rm red}(t)\,
 \big[(A^{\!\top}WA)^{-1}\big]_{00}},
\end{equation}
where $A$ is the design matrix of Eq.~\eqref{eq:cont_model} and $W$ the diagonal weight matrix built from Eq.~\eqref{eq:gamma_weight}, corresponding to $c_0$. The deformation term is the axes-vs-diagonal split of
the radius measurement of Sec.~\ref{subsec:loop_radius_meas},
\begin{equation}
    \sigma_{\rm deform} = \tfrac12| R_{\rm diag} -  R_{\rm axes}|.
\end{equation}
Finally, the core term is the Lorentz-contracted core width
itself,
\begin{equation}
\label{eq:sigma_core}
\sigma_{\rm core}( t) = \frac{1}{m_s\gamma_R( t)},
\end{equation}
which is the fundamental limit at which we can define the concept of the string core size. 

\begin{figure*}
 \centering
 \includegraphics[width=\linewidth]{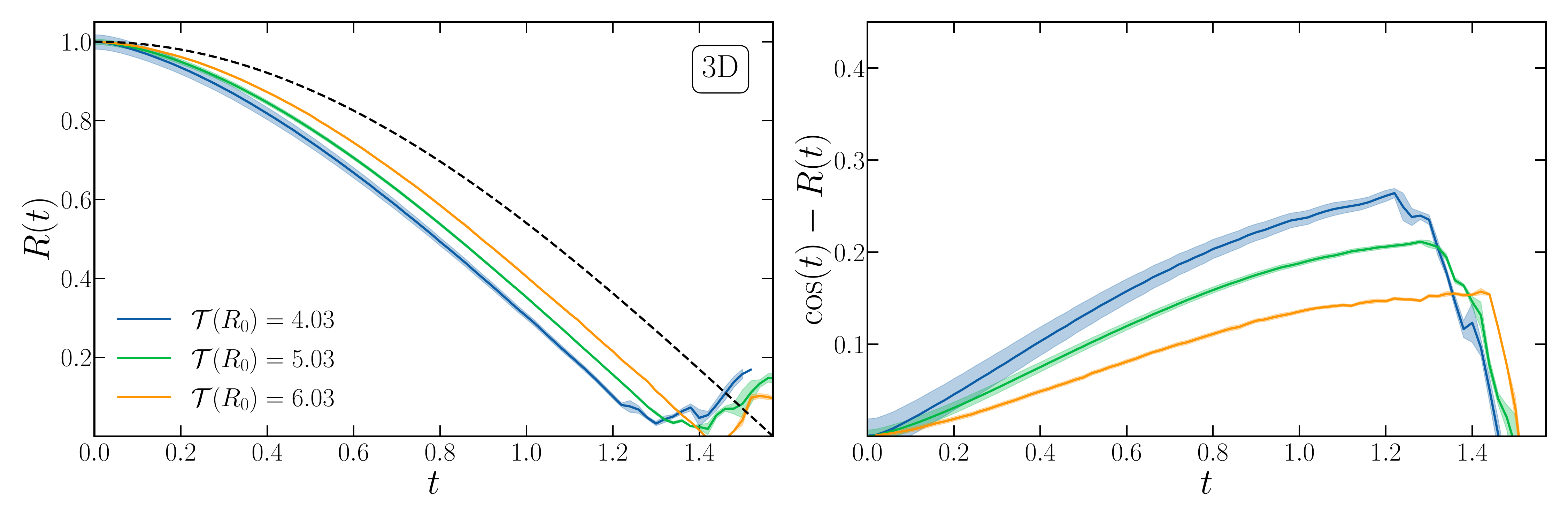}
 \includegraphics[width=\linewidth]{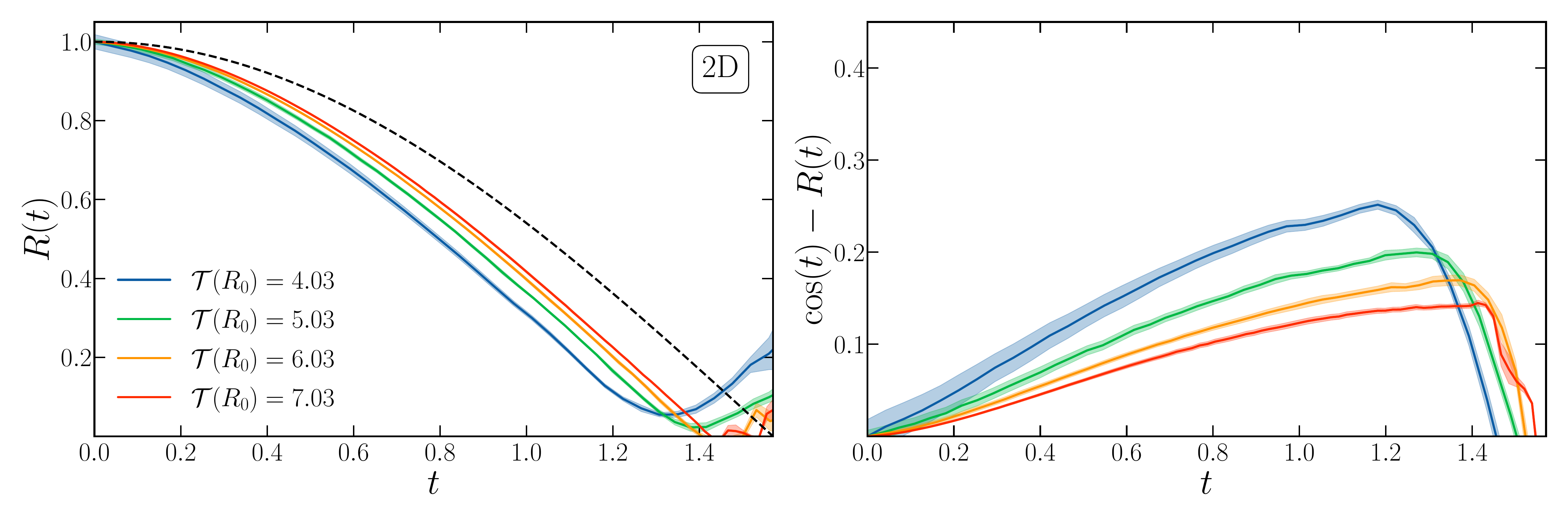}
 \includegraphics[width=\linewidth]{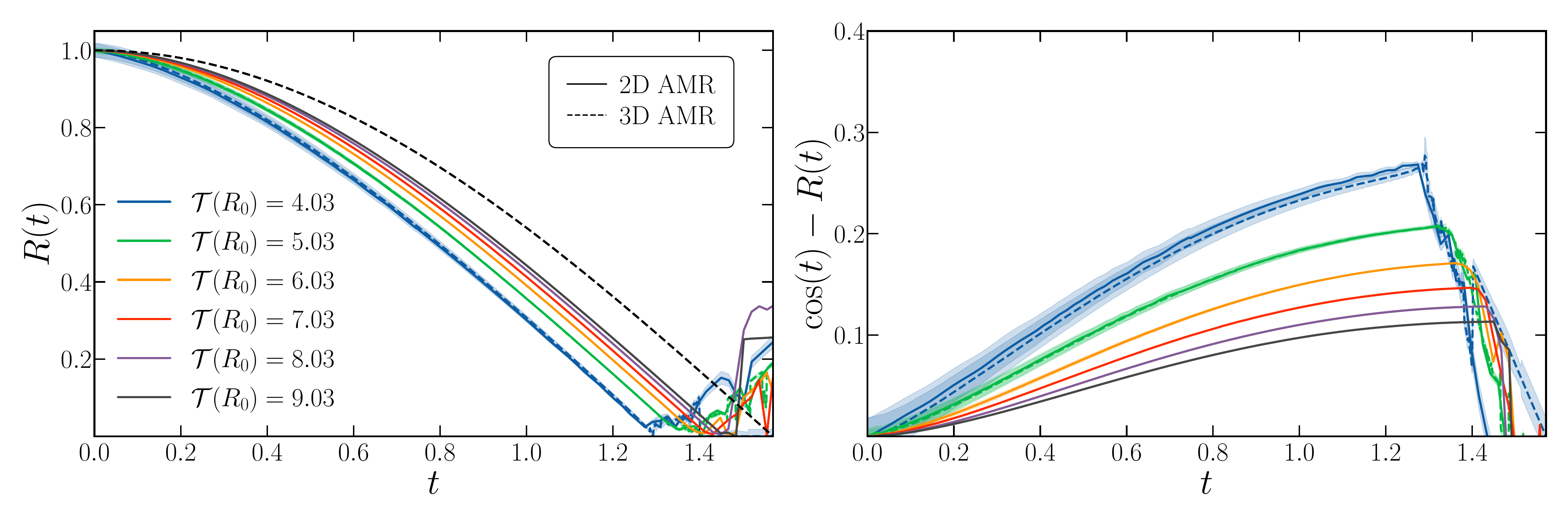}
 \includegraphics[width=\linewidth]{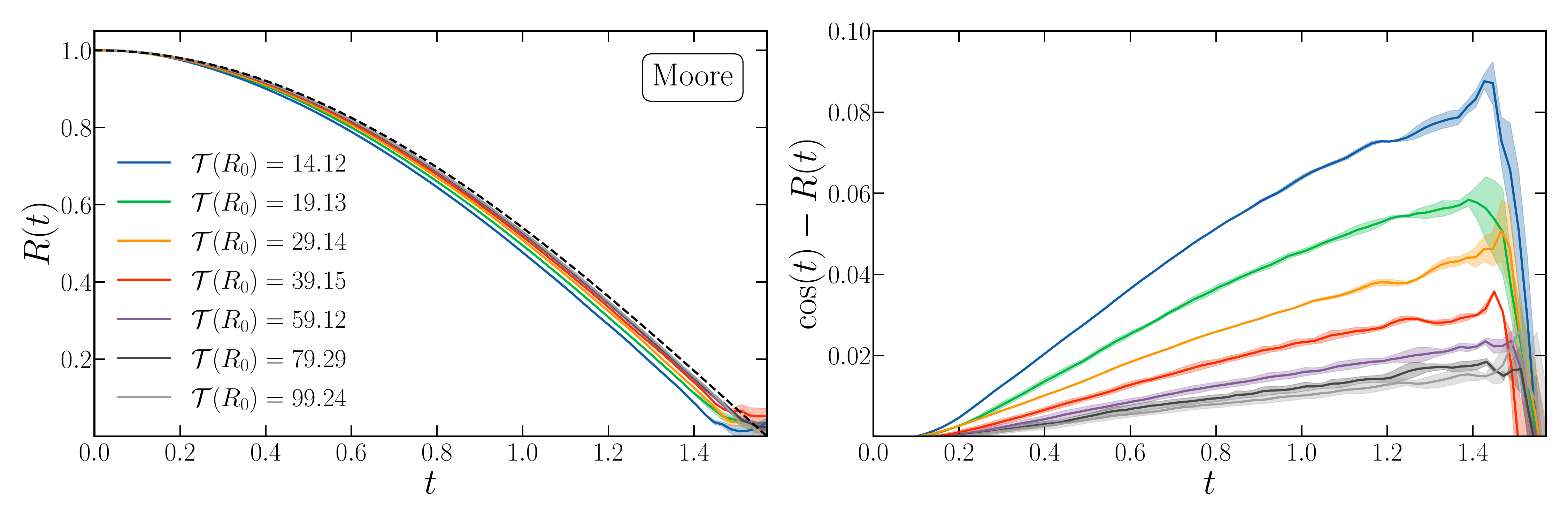}
 \caption{Selected continuum extrapolated loop trajectories for the four simulation setups, labelled in the left panels: \texttt{jaxions} 3D, \texttt{jaxions} 2D, AMR (3D \texttt{axioNyx} and 2D \texttt{jaxions}) and \texttt{Moore}. \emph{Left:} loop radius $R(t)$, with the NG trajectory $R=\cos(t)$ shown as a black dashed line. \emph{Right:} deviation from the NG trajectory, $\cos(t)-R(t)$. Shaded bands are the continuum uncertainties of Eq.~\eqref{eq:band}. Both $t$ and $R$ are given in units of $R_0$. The curves are labelled by their dimensionless tension $\mathcal T(R_0)=c_{\rm FT}+\log(m_sR_0)$, given by Eq.~\eqref{eq:FT-tension-match}.}
 \label{fig:continuum_panel}
\end{figure*}

This continuum extrapolation applies to all simulations in this work, besides the  AMR runs. Since these runs dynamically increase the resolution when required, we only run a single simulation at each value of $\log(m_sR_0)$ and find good agreement with the continuum extrapolation, see Fig.~\ref{fig:continuum_vs_amr}. Some selected continuum curves for the four different approaches are shown in Fig.~\ref{fig:continuum_panel}.

\subsection{Axion Spectrum from Simulations}\label{subsec:spectrum}
Away from the string, the radial mode is close to its vacuum value and the
phase is a free massless field.  We decompose its energy into kinetic and
gradient parts,
\begin{equation}
 E_K=\frac{v^2}{2}\int_{\mathcal{V}}\dd^3x\,\dot\theta^2,
 \qquad
 E_G=\frac{v^2}{2}\int_{\mathcal{V}}\dd^3x\,(\boldsymbol\nabla\theta)^2,
 \qquad E_a=E_K+E_G,
 \label{eq:KG-axion}
\end{equation}
where ${\mathcal{V}}$ is the simulation box volume. With
$\widetilde X_{\boldsymbol k}=\int_{\mathcal{V}}\dd^3x\,
e^{-i\boldsymbol k\cdot\boldsymbol x}X(\boldsymbol x)$, we bin modes in shells
${\mathcal{B}}_k$ of width $\Delta k$ and define the spectral densities
\begin{equation}
 E_K(k)=\frac{v^2}{2{\mathcal{V}}\Delta k}\sum_{\boldsymbol q\in{\mathcal{B}}_k}
 |\widetilde{\dot\theta}_{\boldsymbol q}|^2,
 \qquad
 E_G(k)=\frac{v^2}{2{\mathcal{V}}\Delta k}\sum_{\boldsymbol q\in{\mathcal{B}}_k}
 \sum_i|\widetilde{\partial_i\theta}_{\boldsymbol q}|^2,
 \label{eq:KG-spectra}
\end{equation}
so that $E_K=\sum_kE_K(k)\Delta k$, and likewise for $E_G$.
In the axisymmetric simulations the ordinary Fourier shells are replaced by the
corresponding Fourier--Bessel decomposition with cylindrical measure.  We
have checked both implementations against the direct real-space energy using
Parseval's theorem. Unless stated otherwise, the spectra shown in this
subsection are taken from the 2D axisymmetric runs in the parameter range $16 \leq m_sR_0 \leq 1024$. All the relevant simulation parameters are tabulated in Tab.~\ref{tab:simparams_spectra} in App.~\ref{app:numerics}.

The derivatives are evaluated without explicitly unwrapping the compact phase,
\begin{equation}
 \dot\theta={\rm Im}\!\left(\frac{\dot\phi}{\phi}\right),
 \qquad \partial_i\theta={\rm Im}\!\left(\frac{\partial_i\phi}{\phi}\right),
 \label{eq:theta-derivatives-spectrum}
\end{equation}
and the momentum entering Parseval comparisons is the lattice momentum of the
chosen finite-difference stencil. At points where $|\phi|$ is small these
ratios are singular and evaluating the DFTs produces white noise at a stochastic level depending on the accidental closeness to $|\phi|=0$ of core points.  We avoid this to a large extent by using the smooth mask of Ref.~\cite{Gorghetto:2018myk}
\begin{equation}
 M(\boldsymbol x)=\frac{|\phi(\boldsymbol x)|}{v},
 \label{eq:vil-mask}
\end{equation}
before transforming. In \texttt{jaxions} and in the following discussion we refer to this specific mask as the \enquote{VIL} mask. The masked integrands then involve $f^2(\partial\theta)^2$,
the same combination that appears in Eqs.~\eqref{eq:eom_theta}
and~\eqref{eq:global-string-tension}, so the core singularity is removed
without introducing an external scale. 

For comparisons between loop sizes we use
\begin{equation}
 x\equiv kR_0,\qquad
 {\mathcal{P}}(x)\equiv\frac{1}{v^2}\frac{\dd (E_a/R_0)}{\dd x}
 =\frac{1}{v^2R_0^2}\frac{\dd E_a}{\dd k},
 \label{eq:spectrum_normalisation}
\end{equation}
matching the normalisation used for the NGKR spectrum in
Eq.~\eqref{eq:KR-spectrum-normalisation}.
This removes the leading $v^2R_0$ scaling of the loop energy
and makes spectra at different $m_sR_0$ directly comparable. 

\subsubsection{Initial Self-field}

The initial configuration contains no radiation in the usual asymptotic
sense: its gradient energy is the static axion self-field required by the string winding for a fixed static loop.  In the infinitely thin-ring limit its spectrum can be
calculated analytically up to a one-dimensional integral,
\begin{equation}
 {\mathcal{P}}^{\rm thin}_G(x)=2\pi^2\int_0^\pi \dd\vartheta\,
 \sin\vartheta\,J_1^2(x\sin\vartheta),
 \label{eq:thin_ring_spectrum}
\end{equation}
where $J_n(z)$ are the Bessel functions of the first kind. Its limits are
\begin{equation}
 {\mathcal{P}}^{\rm thin}_G(x\ll1)=\frac{2\pi^2}{3}x^2,
 \qquad
 {\mathcal{P}}^{\rm thin}_G(x\gg1)=\frac{2\pi^2}{x},
 \label{eq:thin_ring_limits}
\end{equation}
the first set by the dipolar far field of a closed loop and the second by the
$1/\rho$ axion gradient near the core.

\begin{figure*}[t]
 \centering
 \includegraphics[width=.65\linewidth]{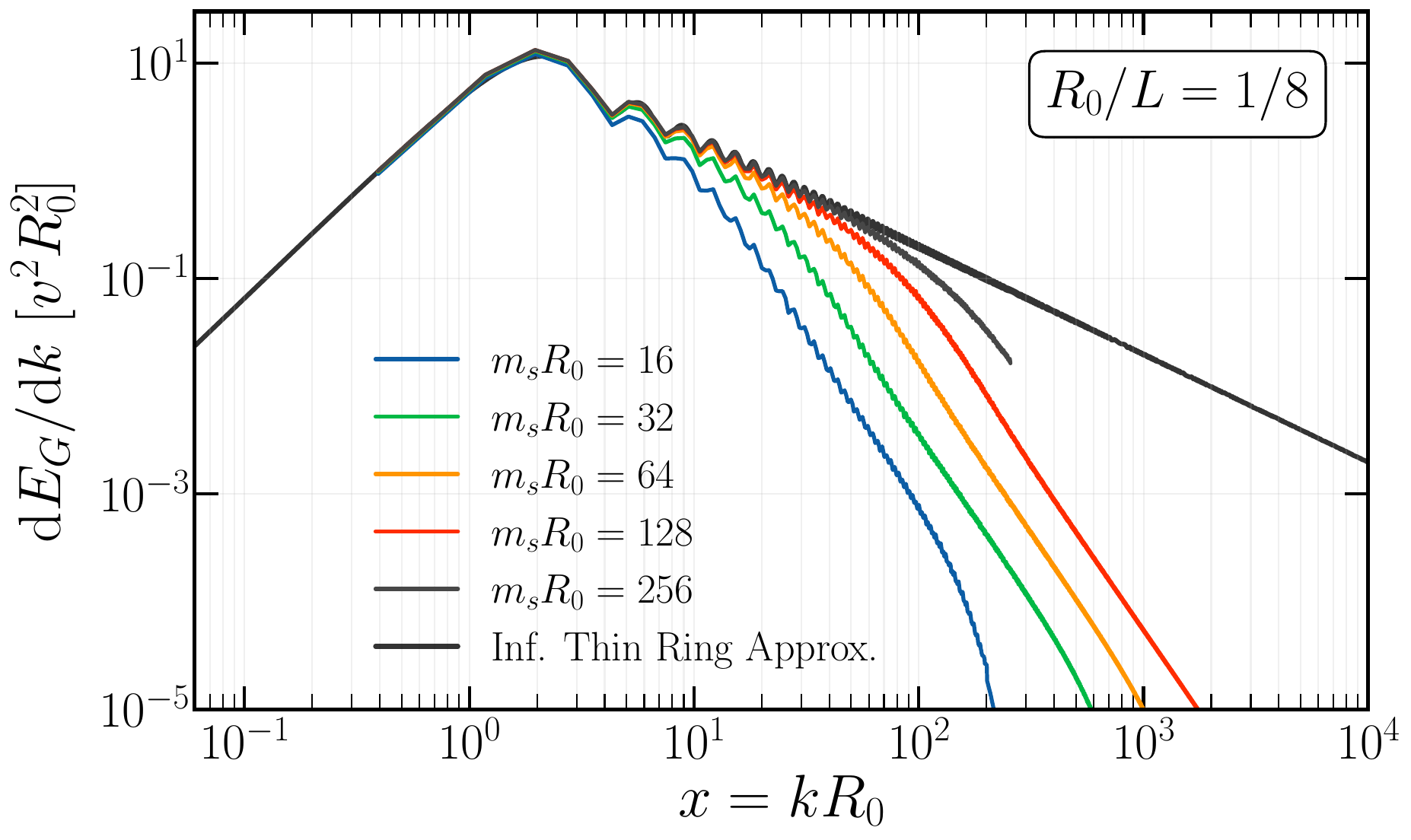}
 \includegraphics[width=.65\linewidth]{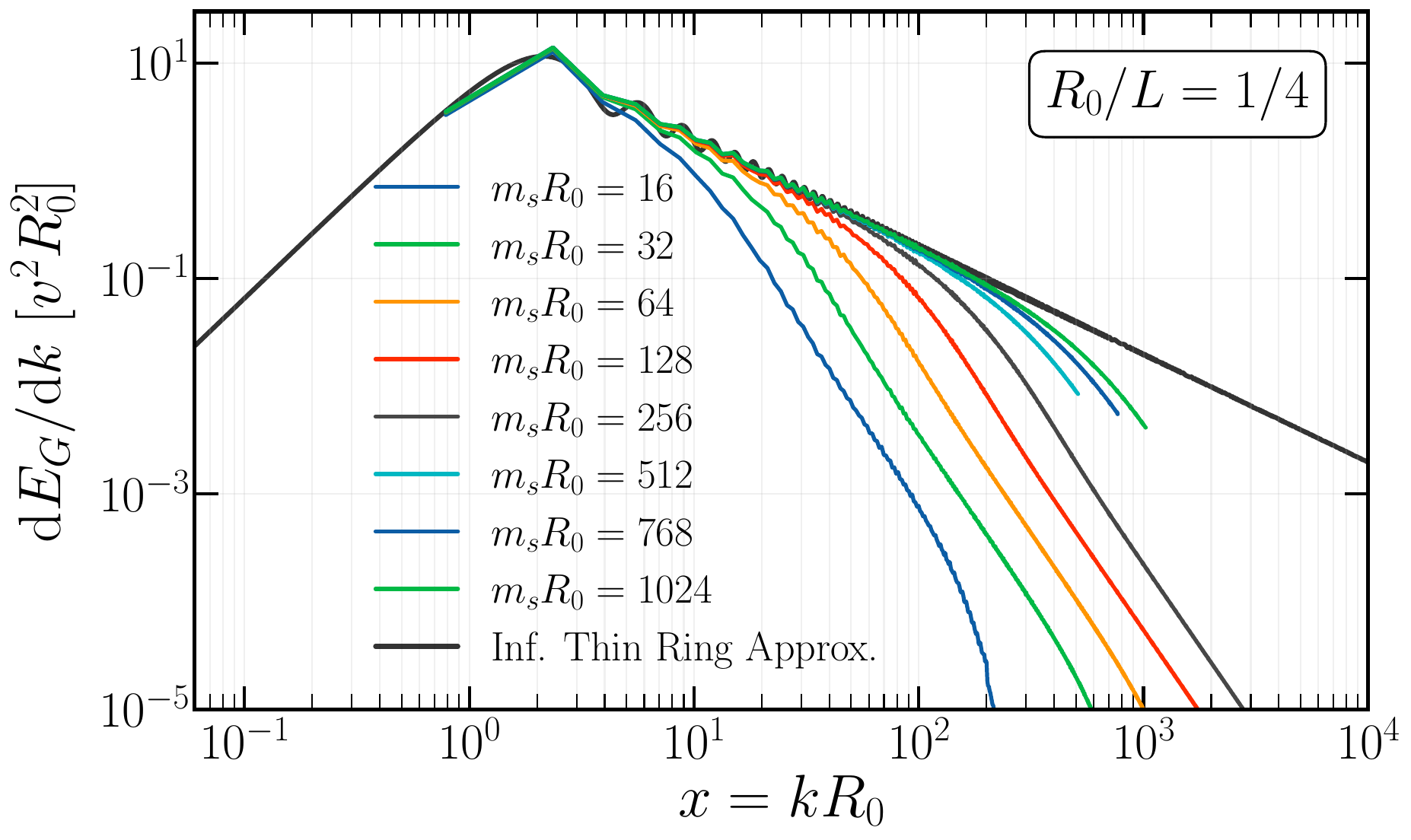}
 \includegraphics[width=.65\linewidth]{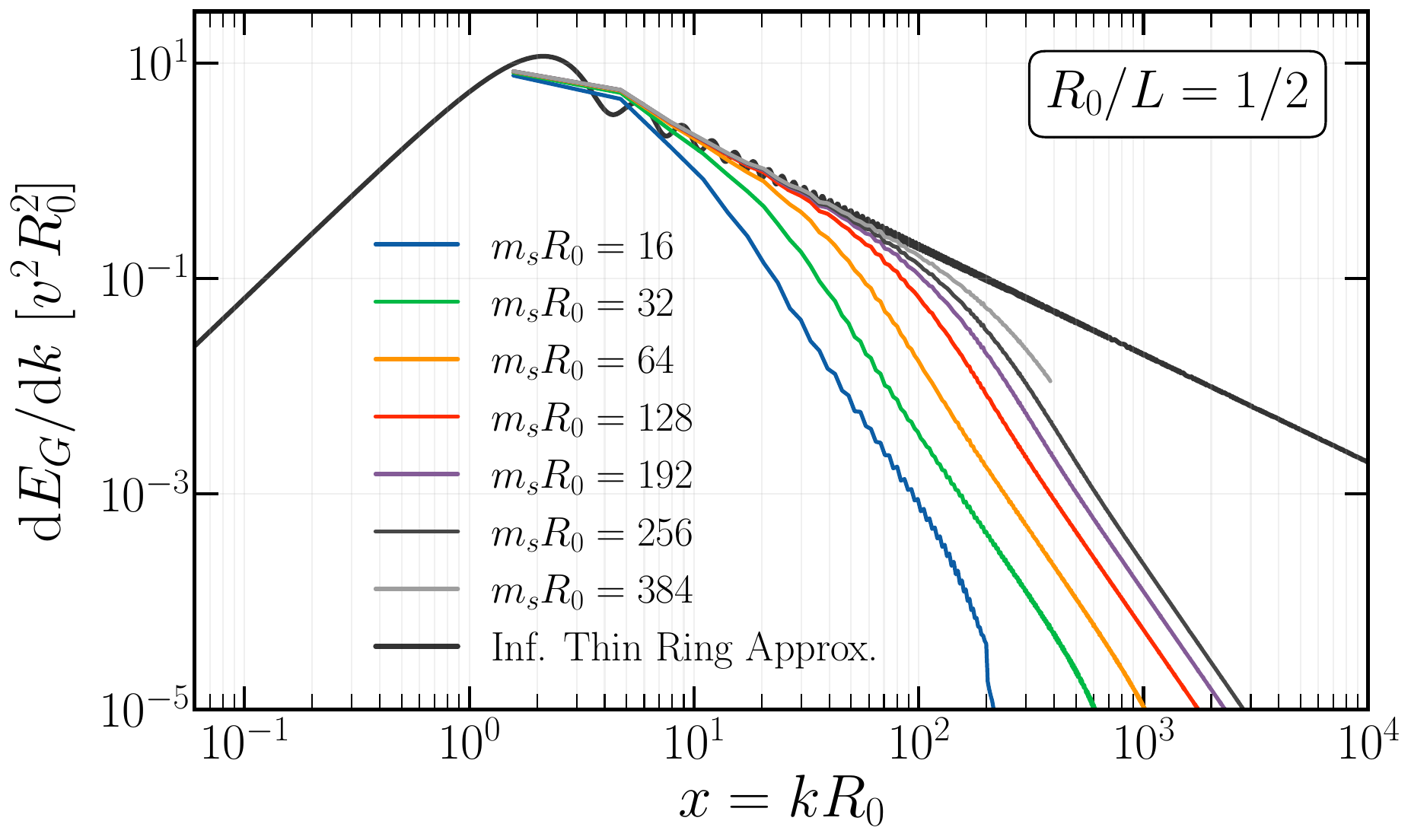}
 \caption{Initial gradient spectra for several $m_sR_0$ and three volume
 ratios.  The black curve is the infinitely thin-ring prediction in
 Eq.~\eqref{eq:thin_ring_spectrum}.  Finite core width regulates the UV,
 whereas increasing $L/R_0$ improves resolution of the dipole IR part of the spectrum.}
 \label{fig:spectrum_initial}
\end{figure*}

Figure~\ref{fig:spectrum_initial} compares this parameter-free result with
the solid-angle initial conditions of Sec.~\ref{subsec:stringICs}.  The
agreement extends to progressively larger $x$ as the hierarchy $m_sR_0$
increases; departures in the UV are the expected regulation by the finite
radial core.  We show the comparison for three values of the radius to the simulation box size ratio $R_0/L=(\frac{1}{8},\frac{1}{4},\frac{1}{2})$, to expose
the sparse IR sampling and finite-volume distortion of the smallest boxes.

The UV limit in Eq.~\eqref{eq:thin_ring_limits} fixes the total
self-field energy without free parameters. Integrating $2\pi^2/x$ up to the
core scale $x\sim m_sR_0$ is equivalent to multiplying the energy per unit
length of Eq.~\eqref{eq:axion-tension} by the loop circumference, and gives
\begin{equation}
 \frac{E_{\theta}^{\rm IC}}{v^2R_0}
 =2\pi^2\left[\log(m_sR_0)+{\rm const}\right],
 \qquad 2\pi^2\approx19.7392\,,
 \label{eq:IC_gradient_energy}
\end{equation}
so that the slope is determined by $\pi v^2$ alone. Fitting our 2D initial
conditions in the $\log(m_sR_0)\sim 3-6$ range yields
\begin{equation}
    \frac{E_{\theta}^{\rm IC}}{v^2R_0}
 =  (19.741 \pm 0.020)\log(m_sR_0) - (9.42\pm 0.10)-\frac{16.0 \pm 0.7}{m_sR_0}.
\end{equation}

The coefficient of the $\log$ is compatible with $2\pi^2$ and the constant part must have a $2\pi^2 \log 0.4725$ from the axion core contribution, but it also needs to contain a $2\pi^2 \log \rho_{\rm eff}/R_0$, where $\rho_{\rm eff}$ is the effective infrared cut-off of the circular loop, a sort of effective radius that gives the axion energy from the straight string formula.  Our result $\sim -9.42$ implies that $\rho_{\rm eff}\simeq 1.313 R_0$.

At large distances the loop is equivalent to a magnetic dipole, whose field
scales with the distance as $d^{-3}$ and whose energy density therefore scales with $d^{-6}$. A finite volume cut-off leads to losses $\sim L^{-3}$. Fitting the data from the simulation with $m_sR_0=64$, we find
\begin{equation}
 \frac{E_{\theta}^{\rm IC}(L)}{v^2R_0}
 =\left.\frac{E_{\theta}^{\rm IC}}{v^2R_0}\right|_{L\to\infty}
 -(7.95\pm0.21)\left(\frac{R_0}{L}\right)^3,
 \label{eq:IC_volume_correction}
\end{equation}
consistent with this expectation. The coefficient is fixed by the
long-range dipole field and should be independent of $m_sR_0$ for $L/R_0>2$.

\subsubsection{From Self-field to Radiation}
Before the nonlinear time $t_2$, the spectrum is well described by an
adiabatically contracting copy of the initial self-field.  If
$ r(t)\equiv R(t)/R_0$, scale invariance predicts
\begin{equation}
 {\mathcal{P}}_G(x,t)=r^2{\mathcal{P}}^{\rm IC}_G(rx),
 \label{eq:adiabatic_spectrum_rescaling}
\end{equation}
which turns out to describe our measurements reasonably well up to a time $t<t_2$, over the range of $x$ that is
unaffected by the finite volume effects in the IR and by the finite core in the
UV.  
This interpretation cannot be extended through
the nonlinear focus: at that stage, the field contains the shrinking
self-field, massive radial excitations, and outgoing axion waves.

We expect and observe that the deepest IR modes cease to follow the contracting loop much earlier than $t_2$. 
To evaluate this quantitatively, we 
follow a simple picture: as the loop accelerates and contracts, modes follow the adiabatic contraction model, until they decouple. Higher-$k$ modes can keep up for longer and decouple when $R/R_0$ is smaller. 
We take the final spectrum of a simulation and fit the mode amplitudes of the first bins to the adiabatic model of Eq.~\eqref{eq:adiabatic_spectrum_rescaling} to calculate at which value of $R/R_0$ they would have decoupled. This yields the effective abandonment
radii
\begin{equation}
 \frac{R_{\rm ab}}{R_0}\simeq
 \begin{cases}
 0.78, & kR_0=\pi/8,\\
 0.73,  & kR_0=\pi/4,\\
 0.55,  & kR_0=\pi/2,
 \end{cases}
 \label{eq:IR_abandonment_radii}
\end{equation}
which at fixed $kR_0$ are nearly independent of $m_sR_0$.  We emphasise that decoupling should not be interpreted as a sharp event.  Yet the results seem quite compatible with an effective
mode-by-mode freeze-out:  shorter wavelengths remain attached to
the self-field longer, whereas the longest modes are already insensitive to
the subsequent collapse while the loop is still large.

Our primary late-time estimator for the axion spectrum is the sum $E_K(k)+E_G(k)$.\footnote{If only
the kinetic piece is available, the same result follows from
\begin{equation*}
 \frac{\dd E_a}{\dd k}=2\overline E_K(k),
 \qquad \overline E_K(k)=\frac{1}{\Delta t}
 \int_{t_1}^{t_1+\Delta t}\dd t\,E_K(k,t),
\end{equation*}
whose fractional finite-window error for coherent standing waves is bounded by
$|2\overline E_K(k)-\dd E_a/\dd k|/(\dd E_a/\dd k)\leq1/(k\Delta t)$. A
fractional accuracy $\epsilon$ thus requires $k\Delta t\gtrsim\epsilon^{-1}$ if
$E_G$ is unavailable.} For a free standing mode
$\theta_{\boldsymbol k}=A_{\boldsymbol k}\cos(kt+\psi_{\boldsymbol k})$,
\begin{equation}
 E_K(\boldsymbol k)\propto k^2|A_{\boldsymbol k}|^2
 \sin^2(kt+\psi_{\boldsymbol k}),
 \qquad
 E_G(\boldsymbol k)\propto k^2|A_{\boldsymbol k}|^2
 \cos^2(kt+\psi_{\boldsymbol k}),
 \label{eq:KG-oscillation}
\end{equation}
so the two pieces oscillate out of phase while their sum is constant,
\begin{equation}
 \frac{\dd E_a}{\dd k}=E_K(k)+E_G(k).
 \label{eq:late-total-spectrum}
\end{equation}
The sum requires no time averaging and provides a direct diagnostic of
whether the post-collapse field is freely propagating.  

\begin{figure*}[t]
 \centering
 \includegraphics[width=0.49\linewidth]{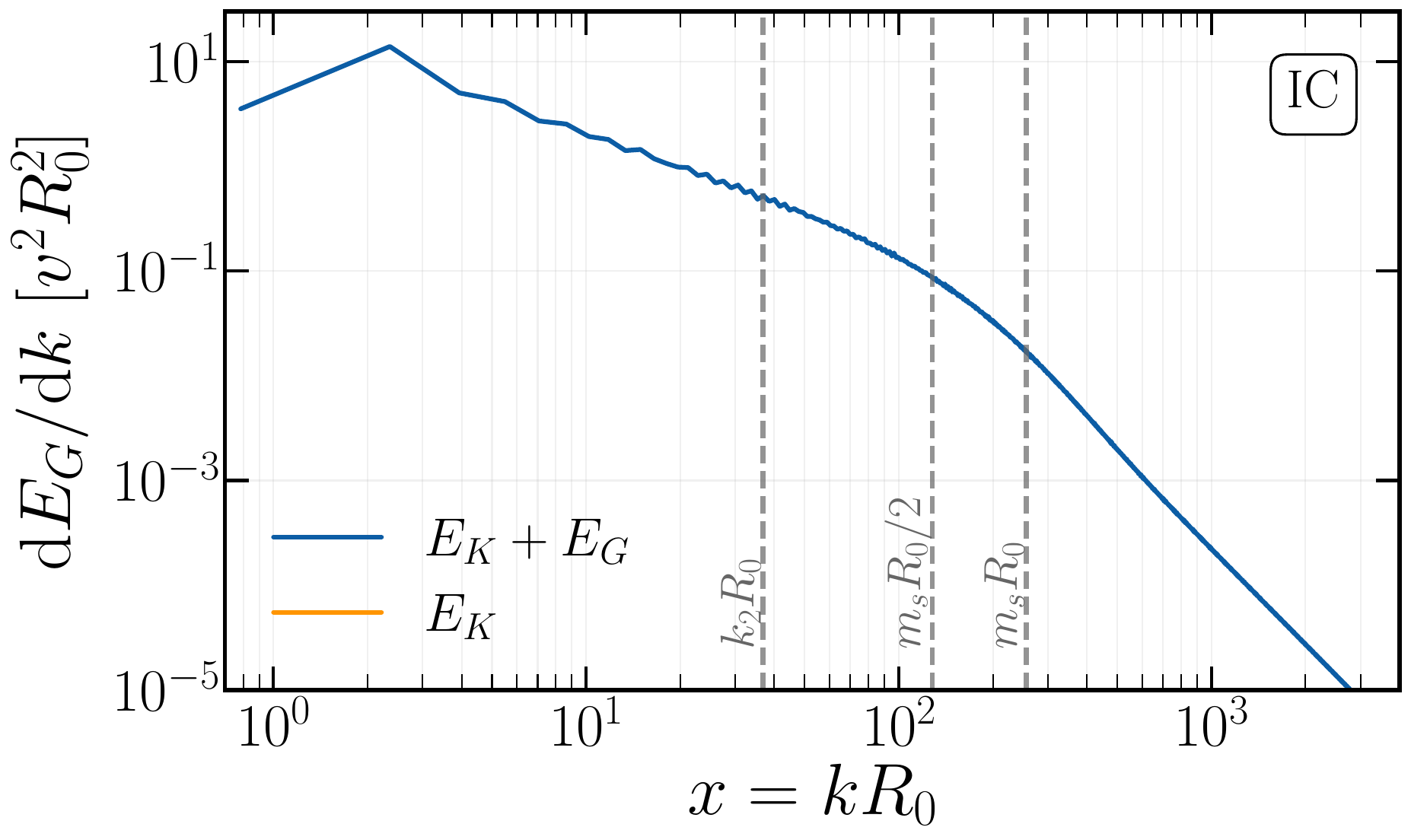}
 \includegraphics[width=0.49\linewidth]{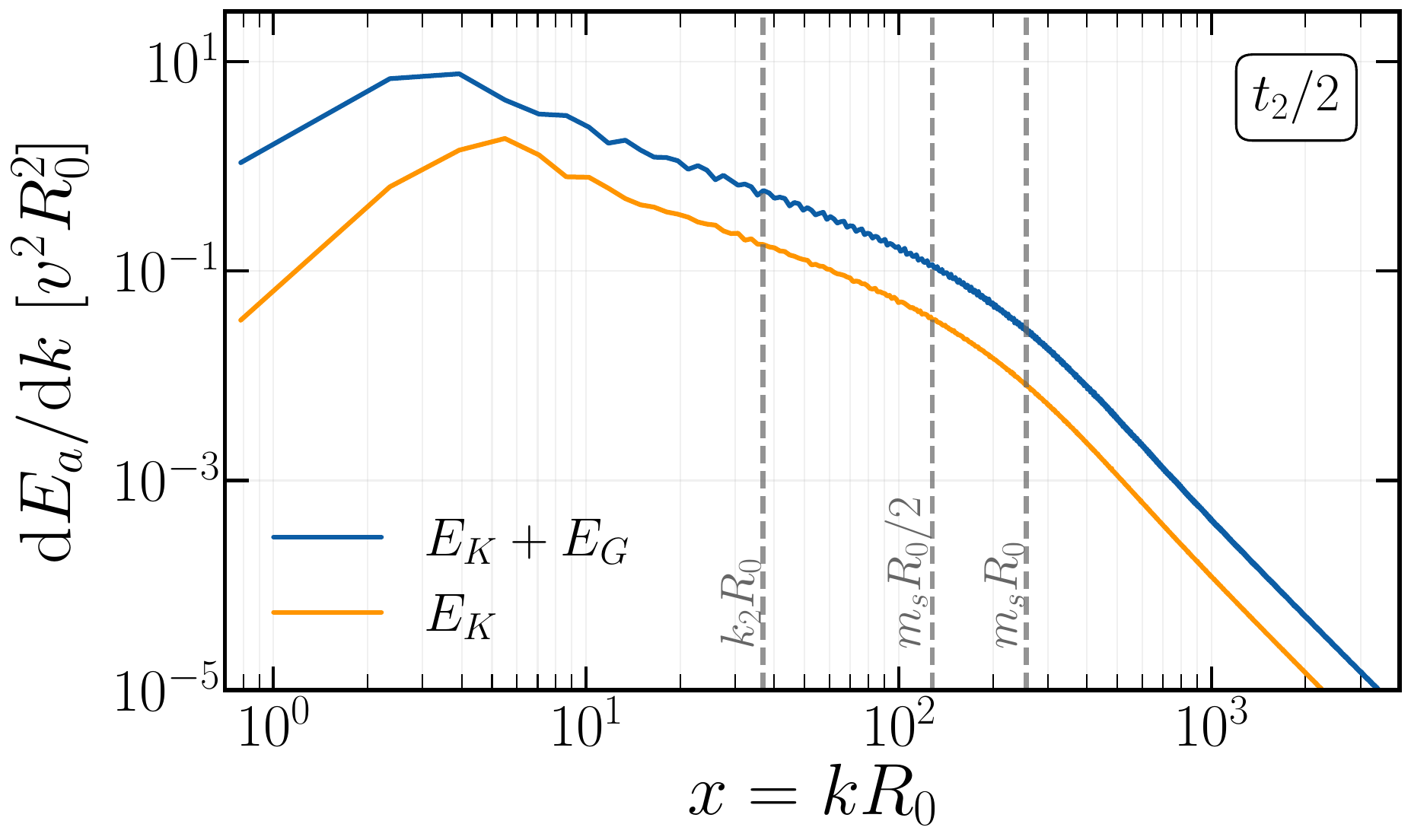}
 \includegraphics[width=0.49\linewidth]{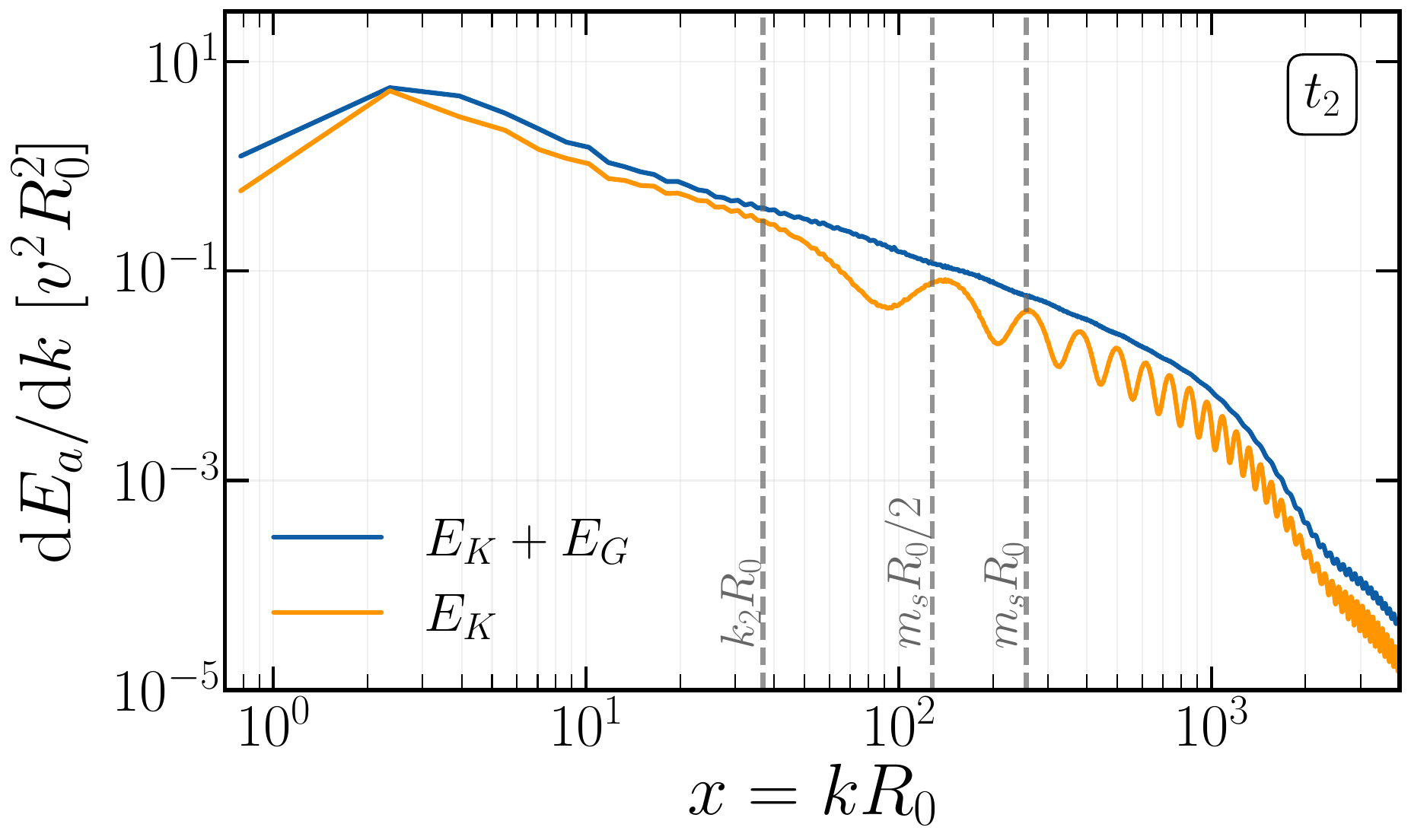}
 \includegraphics[width=0.49\linewidth]{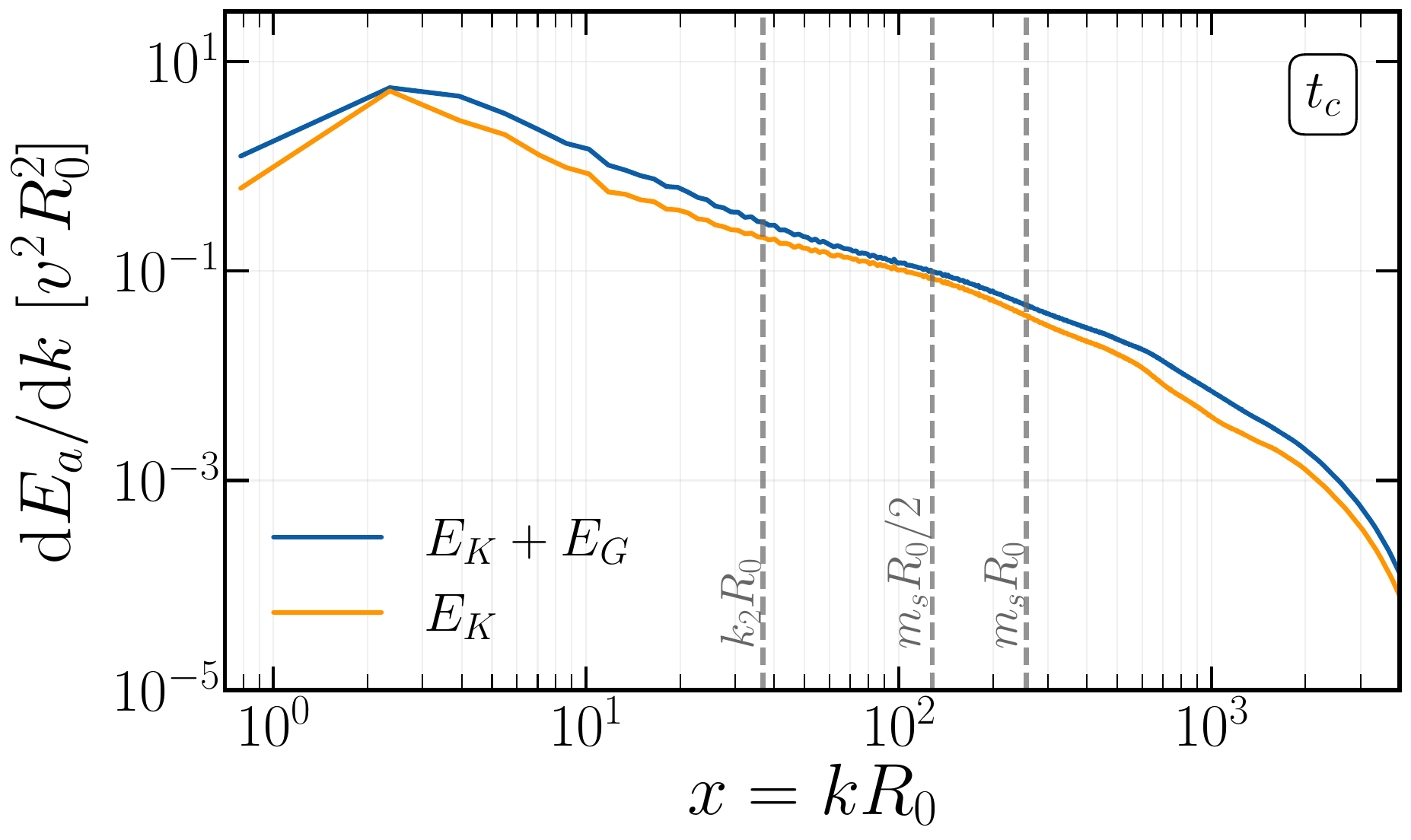}
 \includegraphics[width=0.49\linewidth]{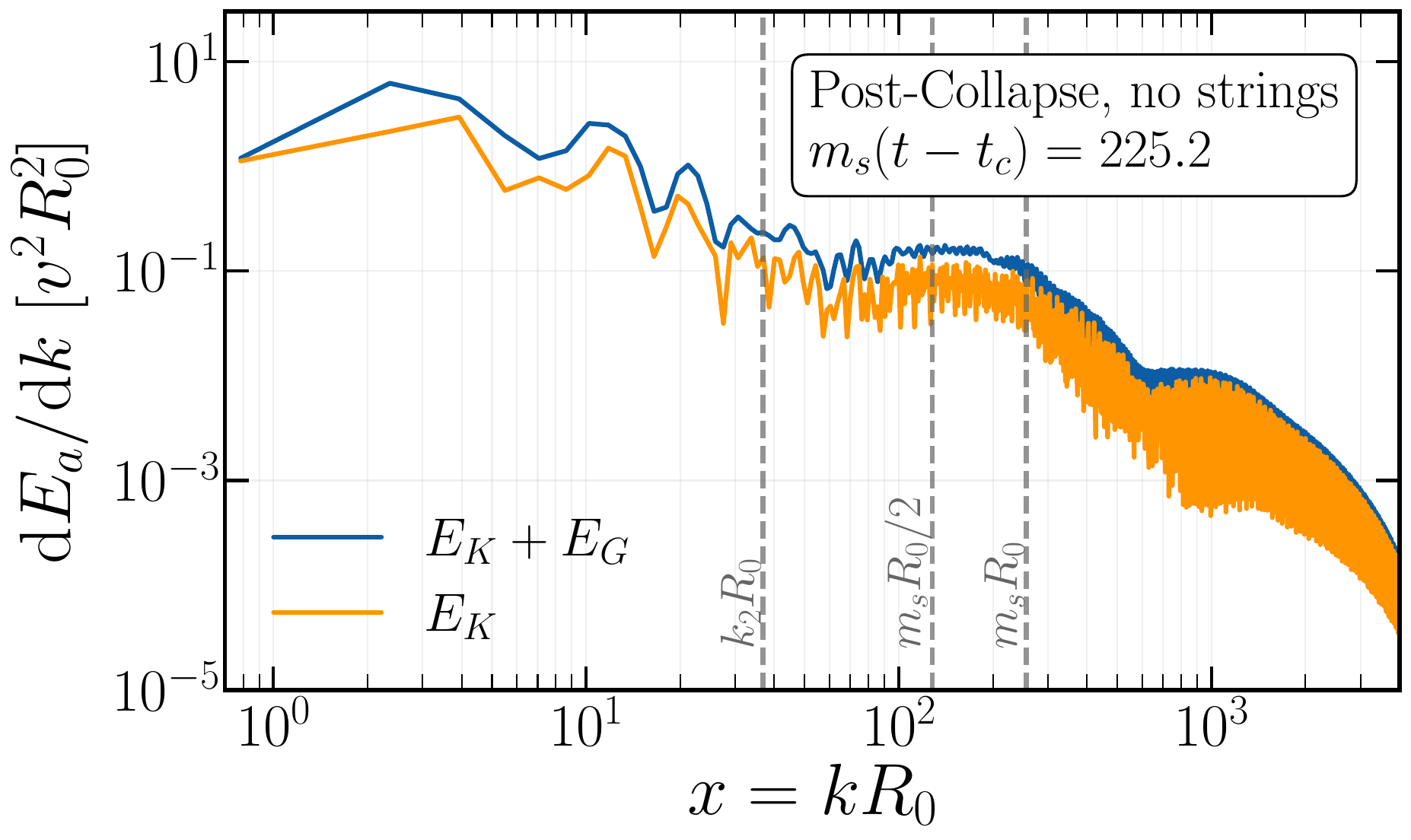}
 \includegraphics[width=0.49\linewidth]{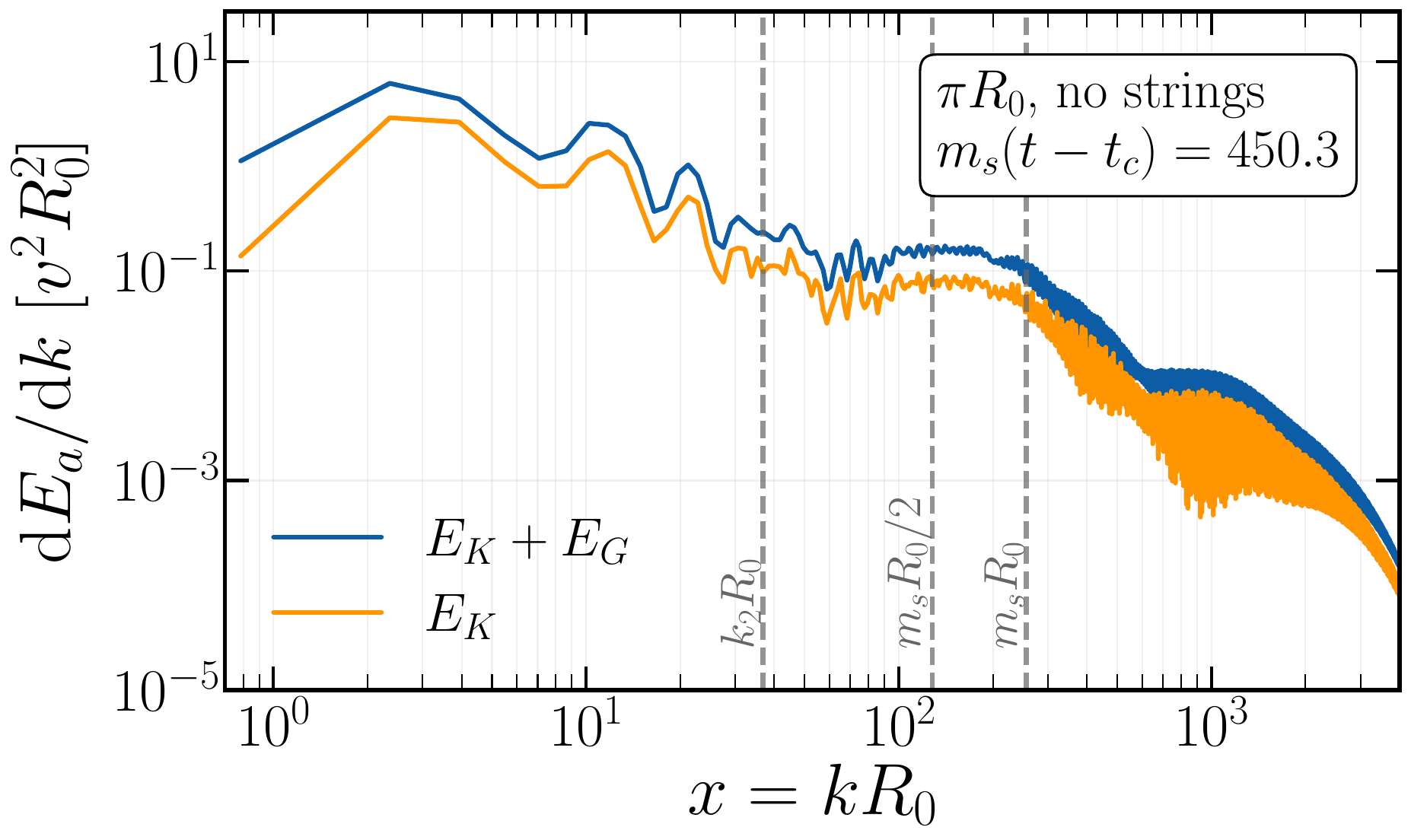}
 \caption{Evolution of the masked $E_K+E_G$ (blue) and $E_K$ (orange)
 spectra for a resolved $m_sR_0=128$, $R_0/L=1/4$ loop collapse.  The last two panels
 are measured after all strings have disappeared and are unmasked.  Vertical dashed
 lines mark $k_2R_0$, $m_sR_0/2$, and $m_sR_0$.}
 \label{fig:spectrum_history}
\end{figure*}

Figure~\ref{fig:spectrum_history} follows one resolved loop through the stages
identified in Sec.~\ref{subsec:collapse_anatomy}. At the initial time the spectrum is bounded by the loop scale $1/R_0$ in the IR and by the core
scale $m_s$ in the UV, and the kinetic part vanishes because the loop
starts at rest. As the loop contracts, the IR peak moves to larger $x$ and smaller amplitude following
Eq.~\eqref{eq:adiabatic_spectrum_rescaling}, while the kinetic component grows as the gradient self-field contracts, as seen in the $t_2/2$ panel. At this time, which is very close to half the collapse time, the gradient energy is still significantly larger than the kinetic component.  
At $t_2$, only slightly before collapse, the peak has moved only slightly further, but the UV cutoff has moved well beyond $m_s$ as a consequence of the Lorentz contraction of the core. 
Oscillatory patterns in $E_K$ indicate a macroscopic feature developing, the wavefront we will describe later.  
Through the nonlinear regime the UV cutoff is pushed even further. After the focus and bounce the spectrum shows a broad wavy-plateau feature with a UV cut-off near
$m_s$ and extending down to
\begin{equation}
 k_2\equiv R_2^{-1},
 \label{eq:k2_definition}
\end{equation}
with $R_2$ from Eq.~\eqref{eq:R2_scaling}. 
The nonlinear processing of the compressed energy seems to have two main spectral effects; it creates the UV plateau and transfers power from the low-$k$ modes $k<k_2$ into the $k>k_2$ nonlinear region. The final spectrum is therefore not the scale invariant spectrum we started with: it has been shifted to the UV by the adiabatic contraction and distorted by the nonlinear period near $k\simeq k_2$, and enriched above $k_2$ by waves generated during the nonlinear stage. Their spatial origin is discussed in more detail in Sec.~\ref{subsec:frontwave_spectrum}.

The spatial and temporal resolution of the focus event has little effect on the instantaneous spectrum below $k_2$ at $t_c$, consistent with the earlier abandonment of those modes.  It is nevertheless essential after the focus: insufficient resolution generates excess waves, produces a strong oscillatory pattern around the $k\sim m_s$ feature, and eventually contaminates the spectrum even for $k<k_2$.

\subsubsection{Late-time Spectrum and Limitations}

\begin{figure*}[h]
 \centering
 \includegraphics[width=.65\linewidth]{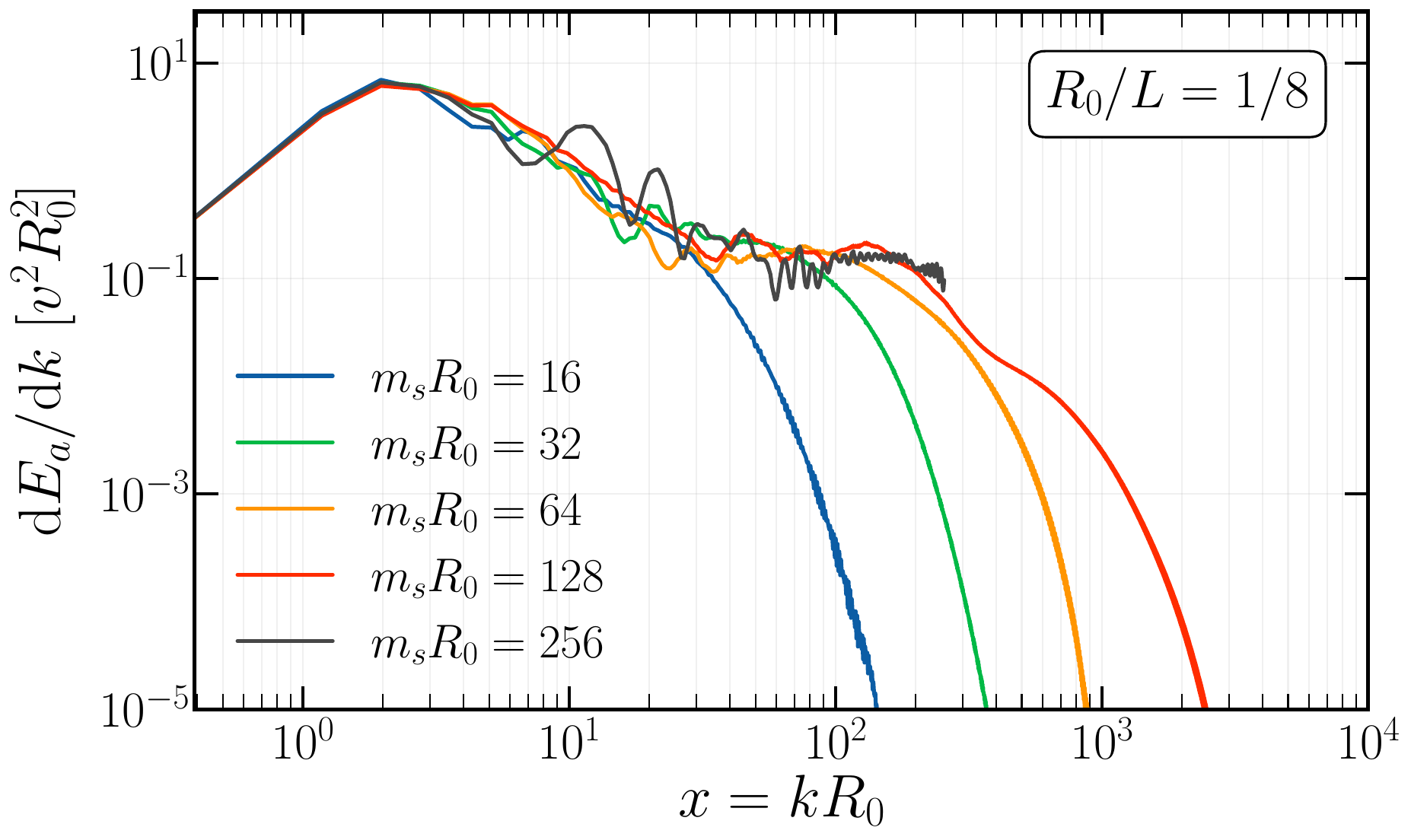}
 \includegraphics[width=.65\linewidth]{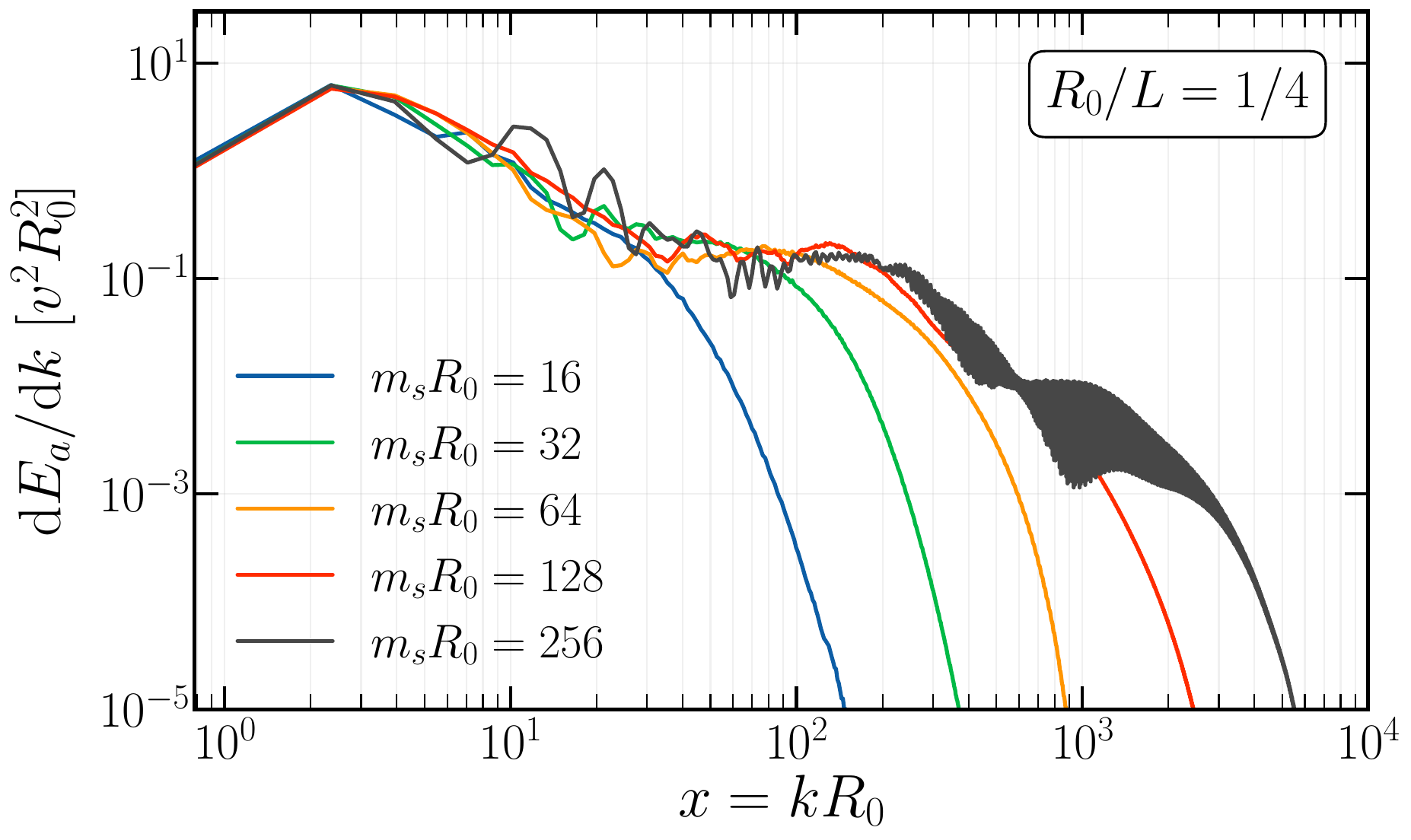}
 \includegraphics[width=.65\linewidth]{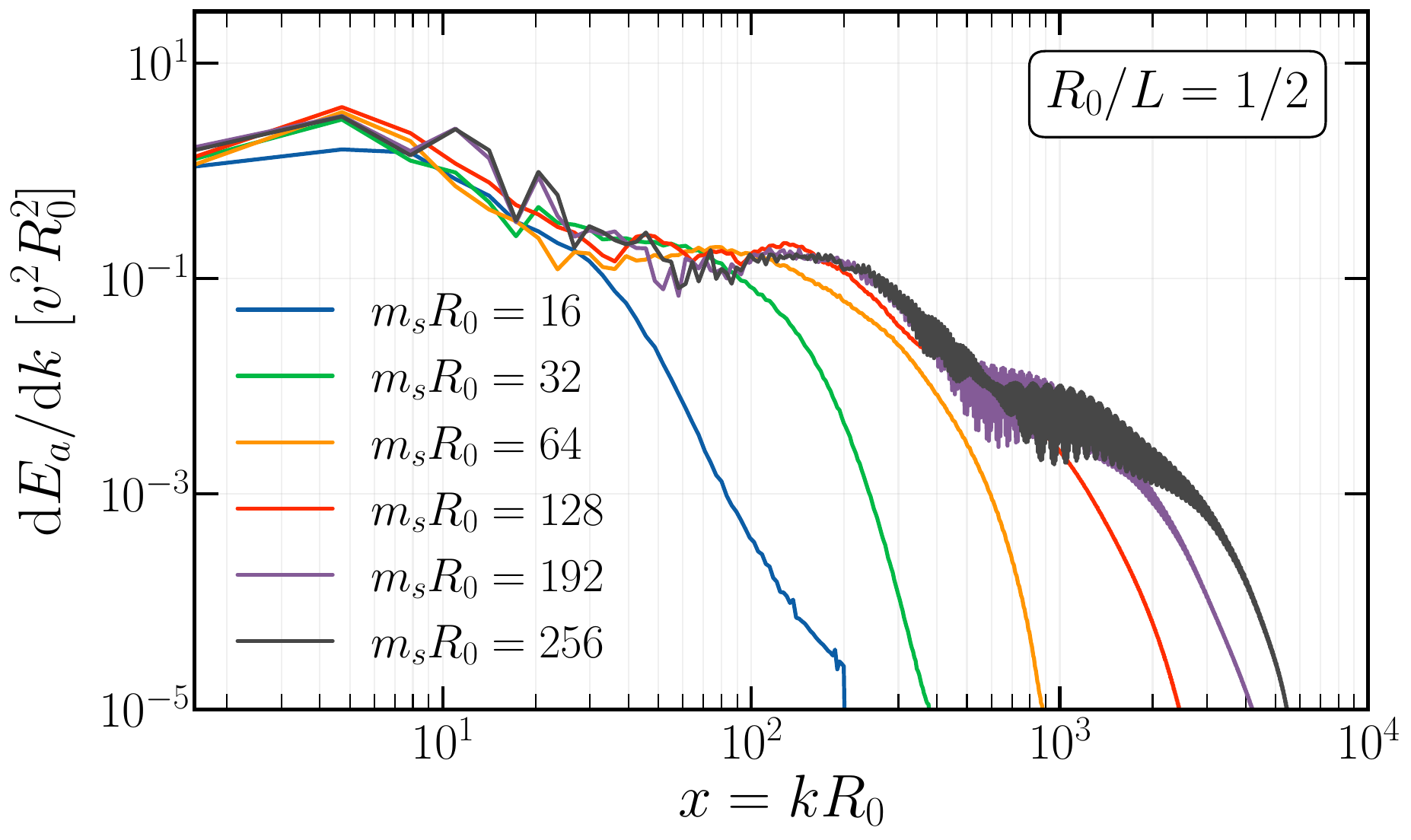}
 \caption{Post-focus, late time $E_K+E_G$ spectra at $t=\pi R_0$ for several $m_sR_0$ evaluated at $t=\pi R_0$. Larger boxes improve the IR reach; the smallest boxes
 display the largest finite-volume and boundary artifacts.}
 \label{fig:spectrum_final}
\end{figure*}

Figure~\ref{fig:spectrum_final} shows the late time spectra at
$t=\pi R_0$, after the string loop has collapsed, bounced and recollapsed, and its non-linear \enquote{daughter} loops disappeared, leaving no string in the box. The spectrum is made of axion and saxion waves with small residual interaction between them. The baseline simulations employ very large volumes $R_0/L=1/8$, which allow us to accurately resolve the $k\sim 1/R_0$ peak and intermediate oscillations but they are very costly in RAM. We find that smaller volumes, $R_0/L=(1/2,1/4)$, reproduce the spectrum very well, albeit with worse resolution with the benefit of being numerically cheaper. We interpret this as a simple early decoupling of the smallest $k$ modes, and the late evolution being very well contained within the volume. If this is firmly established, one could calculate the IR part of the spectrum with large volume simulations, even if the collapse is not fully resolved according to our criterion Eq. \eqref{eq:focus_resolution_requirement}, and the UV part with fully resolved simulations at smaller volumes. 

The data show that the low-IR spectrum, $k<3/R_0$ is largely independent of $\log (m_sR_0)$, the height of the UV plateau does not have a clear $m_sR_0$-dependence (except its $k\sim m_s$ cut-off) and the intermediate regime does not depend monotonically on $m_sR_0$. Indeed, the range between $3/R_0$ and the beginning of the plateau features oscillations and no clear trend allowing for an extrapolation is visible. 

We emphasise that these spectra should not be identified with the asymptotic radiation spectrum, defined as the flux through a surface at infinity, which is the quantity used in the NGKR EFT.  Three limitations must be kept
explicit.  First, the self-field does not vanish globally at $t_c$: a point
at distance $r$ responds only after $t-t_c\gtrsim r$. Second, at $R_0/L=1/4$ a light-speed front emitted near the initial loop reaches the boundary layer after approximately $3R_0$. By $t-t_i=\pi R_0$, its leading edge can therefore begin to be absorbed, although most of the radiation remains in the simulation box. The very good agreement between $R_0/L=1/4$ and $1/2$ data suggests that this is not a significant effect.  The 3D simulations with periodic boundary conditions instead suffer from re-entry and interference effects.
Third, local mesh refinement as implemented for the axisymmetric runs preserves the resolved UV dynamics but removes
access to the IR modes of the original grid.  The mode-dependent
abandonment in Eq.~\eqref{eq:IR_abandonment_radii} offers a controlled way of
reconstructing part of this lost IR contribution, but it should be calibrated with care, which goes beyond the scope of this work.

In the UV, we restrict the interpretation to momenta below the Nyquist frequency $k_{\mathrm{Nyq}}=\pi N/L$, and track the effects of lattice dispersion, the derivative stencil, core masking, and contamination by the massive radial mode.  In the IR the main systematics are sparse shells, the absence of wavelengths larger than $L$, influence of the periodic copies or sponge-layer absorption, and incomplete causal separation of radiation from the initial self-field.  We emphasise once more that the resolution criterion in Eq.~\eqref{eq:focus_resolution_requirement} is therefore necessary for understanding the post-focus spectrum, while a sufficiently large $L/R_0$ is independently required for the IR part.

\subsubsection{Spatial Origin of the Post-focus Ultraviolet Spectrum}
\label{subsec:frontwave_spectrum}

\begin{figure*}[h]
 \centering
 \includegraphics[width=\linewidth]{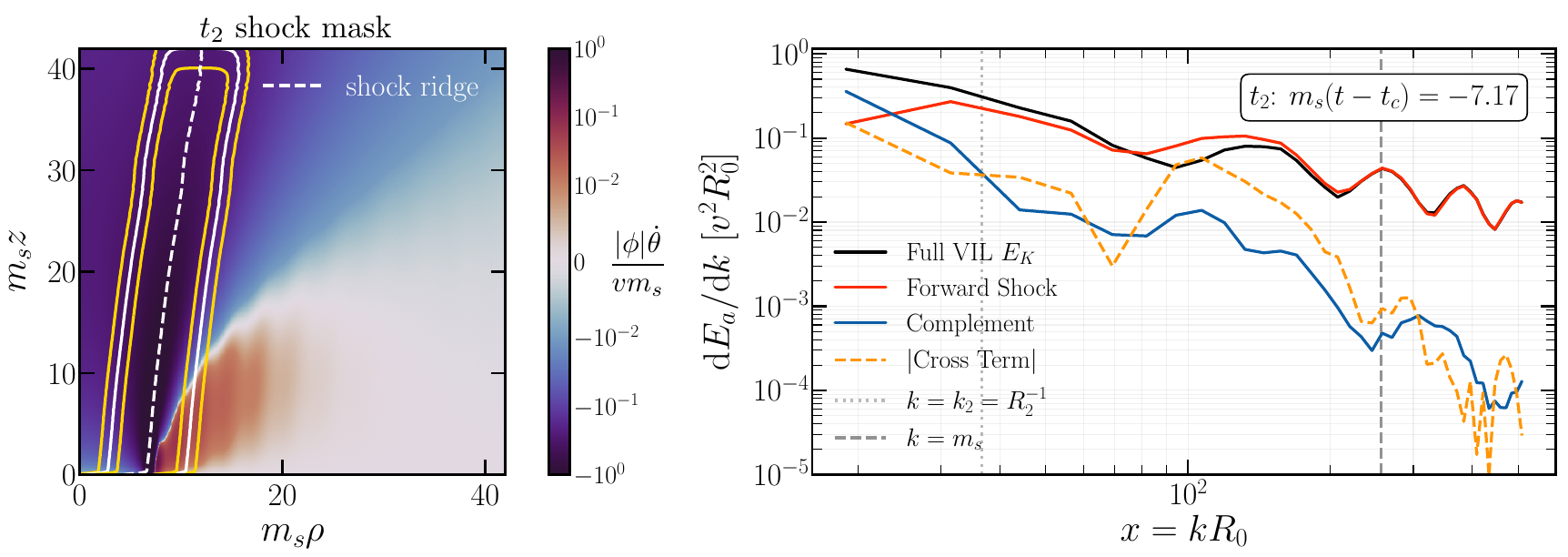}
\caption{Forward nonlinear structure at $t_2$ in a resolved $m_sR_0=256$
 simulation. \emph{Left:} signed phase rate $q_K/m_s=|\phi|\dot\theta/vm_s$, with
 contours of the smooth tube mask and the measured ridge. The disturbance
 extends several $R_2$ ahead of the string. \emph{Right:} full kinetic
 spectrum masked with the VIL method, Eq.~\eqref{eq:vil-mask} (black), the contribution from
 the tube (red), its complement (blue), and the magnitude of their
 interference term (orange dashed). The forward structure already dominates
 above $k_2$.}
 \label{fig:shock_spectrum_t2}
\end{figure*}

Visually, the field maps reveal a spatial separation which is obscured in
the integrated spectrum.  We illustrate it with our resolved
$m_sR_0=256$ simulation.  Already at the nonlinear time $t_2$, the signed
phase velocity $q_K=|\phi|\dot\theta/v$ displays a narrow forward structure
originating from the contracting string.  Figure~\ref{fig:shock_spectrum_t2}
shows that it extends several $R_2$ in the $z$-direction before the final
focus is reached.  A smooth tube of half-width $0.55R_2$ around the measured
ridge isolates this structure without imposing a sharp mask edge.  Its
Fourier transform dominates the kinetic spectrum above
$k_2$, Eq.~\eqref{eq:k2_definition}, and contains nearly all of the power approaching $k\sim m_s$.  The hard component is therefore generated during the approach to the nonlinear scale, rather than appearing instantaneously at $t_c$.

After focusing, the string emerging from the focus initially moves
outward at nearly the speed of light and reaches
$m_sR_{\mathrm{bounce}}^{\mathrm{max}}\simeq61$ before turning around.  This is the
largest bounce observed in our simulations.  Runs which do not satisfy
Eq.~\eqref{eq:focus_resolution_requirement} generally show a much smaller
bounce, or none at all: unresolved nonlinear focusing produces excess
short-wavelength excitations and can erase the outgoing string.

\begin{figure*}[t]
 \centering
 \includegraphics[width=\linewidth]{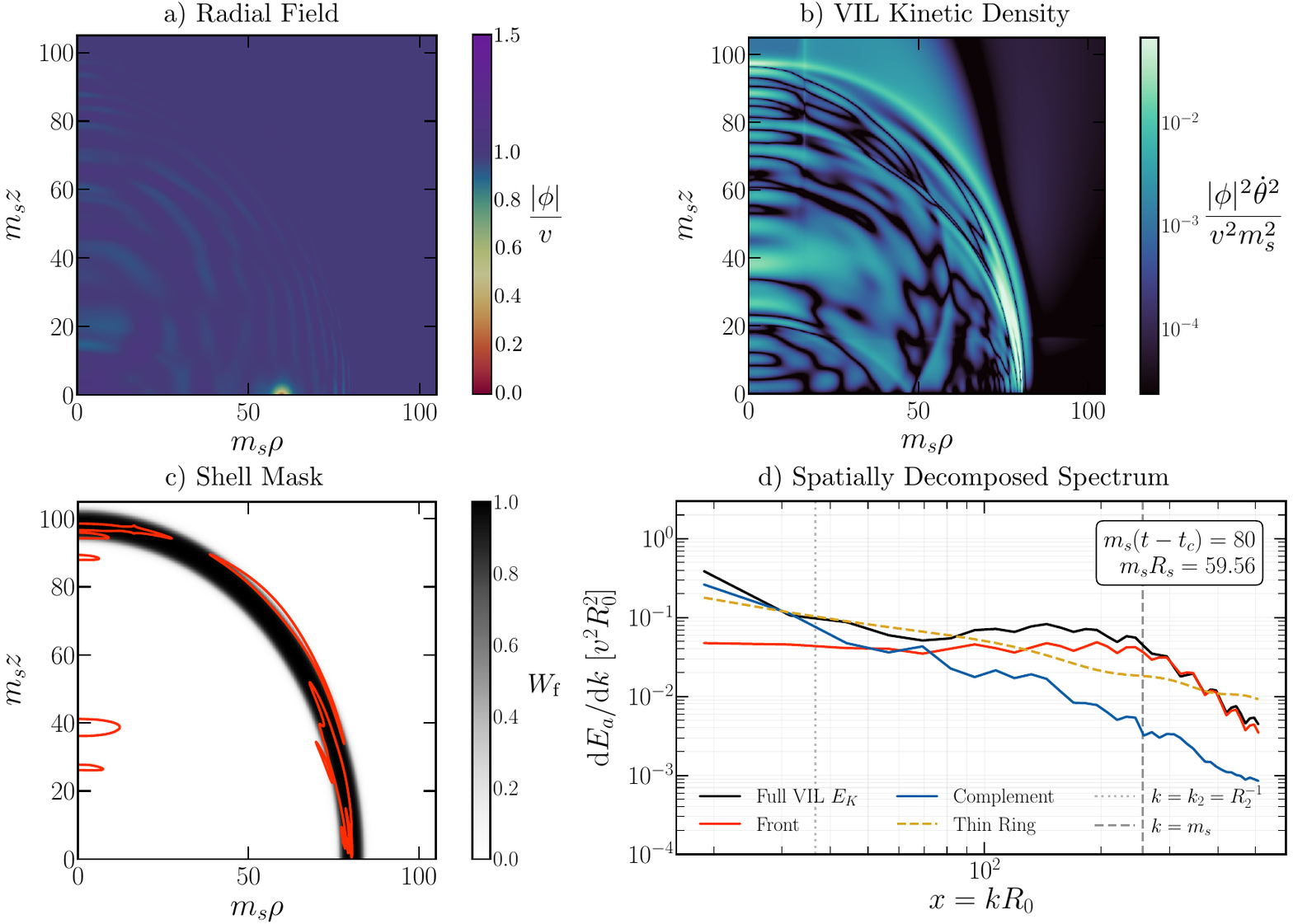}
  \caption{Spatial origin of the post-focus UV spectrum in the
 resolved $m_sR_0=256$ simulation at $m_s(t-t_c)=80$. \emph{Top left:}
 radial field, with the measured string at $m_sR_s\approx 59.6$. \emph{Top right:}
 masked kinetic-energy density, showing the energetic front continuing beyond
 the recollapsing string. \emph{Bottom left:} the smooth diagnostic shell
 $W_{\rm f}$; the red contour traces the kinetic ridge and illustrates that
 the physical front is not spherical. \emph{Bottom right:} full masked kinetic
 spectrum (black), shell contribution (red), complementary contribution
 (blue), and the parameter-free thin-ring spectrum evaluated at the measured
 $R_s$ (orange dashed). Vertical lines mark $k=k_2$ and $k=m_s$. The narrow
 shell contains the hard excess extending from $k_2$ towards and beyond $m_s$,
 whereas the complementary spectrum approximately follows the string
 self-field.}
 \label{fig:postfocus_front_spectrum}
\end{figure*}

At $m_s(t-t_c)\simeq80$ the string has slowed sufficiently that it
can be separated spatially from a second structure: a narrow, energetic
front which continues to propagate away from the focus at approximately the
speed of light. One can see the position of the bounced string\footnote{At this time the zero of
$\operatorname{Re}\phi(\rho,z=0)$ lies at
$m_sR_s=59.565$.} in the $|\phi|/v$-field map in Fig.~\ref{fig:postfocus_front_spectrum} (upper left panel) clearly separated from the kinetic energy maximum which is visible in the upper right panel of the same figure.  
  The front is anisotropic and is not a spherical wave.
Nevertheless, a smooth shell centred on the focus, of total radial thickness $\sim R_2$, provides a useful diagnostic mask $W_{\rm f}(\rho,z)$
over the interval in which the front and string are well separated.  We define
the corresponding real-space axion energies by
\begin{equation}
 E_X^{\rm f}=2\pi\!\int \dd \rho\,\dd z\,\rho\,
 W_{\rm f}(\rho,z)\,\varrho_X(\rho,z),\qquad
 E_X^{\rm c}=2\pi\!\int \dd \rho\,\dd z\,\rho\,
 [1-W_{\rm f}(\rho,z)]\,\varrho_X(\rho,z),
 \label{eq:front_direct_energy}
\end{equation}
where $\varrho_X$ is the energy density of Eq.~\eqref{eq:KG-axion} with
$X=K$, $G$, or $K+G$.  These quantities are evaluated directly on the
lattice and do not rely on Parseval reconstruction.  We find
\begin{equation}
 \frac{E_K^{\rm f}}{E_K^{\rm c}}=1.20,
 \qquad
 \frac{E_G^{\rm f}}{E_G^{\rm c}}=0.576,
 \qquad
 \frac{E_{K+G}^{\rm f}}{E_{K+G}^{\rm c}}=0.773,
 \label{eq:front_energy_ratios}
\end{equation}
so that the shell contains about $44\%$ of the total axion energy and $55\%$ of its kinetic energy. Since it occupies only a small fraction of the
cylindrical volume, its mean energy density is approximately $17$ times that
of the complementary region.  These percentages depend quantitatively on
the diagnostic shell, but the spatial separation and the concentration of
kinetic energy in the front are unambiguous.

Fourier-transforming the masked fields shows that the excess extending from
$k_2\sim R_2^{-1}$ towards $m_s$ is localised in this outgoing front, see the red line in the lower right panel of Fig.~\ref{fig:postfocus_front_spectrum}. Its
spectrum is approximately flat, or mildly increasing, in $\dd E/\dd k$ and is much harder than the complementary spectrum (blue, same figure), which turns out to be softer than the instantaneous thin-ring self-field (dashed-orange, $\propto 1/k$).  For a ring of radius $R_s$, the
parameter-free comparison in the normalisation of
Eq.~\eqref{eq:spectrum_normalisation} is
\begin{equation}
 {\mathcal{P}}^{\rm ring}_G(x;R_s)=
 \left(\frac{R_s}{R_0}\right)^2
 {\mathcal{P}}^{\rm thin}_G\!\left(x\frac{R_s}{R_0}\right),
 \label{eq:instantaneous_ring_spectrum}
\end{equation}
which is Eq.~\eqref{eq:adiabatic_spectrum_rescaling} evaluated at
$r=R_s/R_0$, with its dipolar turnover at $kR_s\sim1$, i.e.\ $x\simeq4.30$ for
the measured radius. 
The agreement of the thin ring and complementary spectrum below $k_2$ supports the interpretation of the complement as the bouncing string self-field plus softer radiation.  A deviation near $x\sim20$ may represent a leading harmonic released from the
abandoned self-field, but this identification is presently only suggestive.

Splitting $q_K=Wq_K+(1-W)q_K$, the corresponding powers obey
\begin{equation}
 P_K=P_K^{W}+P_K^{1-W}+P_K^{\rm cross},
 \qquad
 P_K^{\rm cross}=2\,\mathrm{Re}
 \left[\widetilde{Wq_K}\,\widetilde{(1-W)q_K}^{\,*}\right].
 \label{eq:masked_spectrum_cross}
\end{equation}
Consequently, the two masked spectra need not add to the full spectrum
bin by bin, and either one may locally exceed it when the cross term is
negative.  We use direct real-space integrals such as in
Eq.~\eqref{eq:front_direct_energy} for energy fractions, and the Fourier
decomposition only to diagnose the spatial origin of spectral features.

\begin{figure*}[t]
 \centering
 \includegraphics[width=\linewidth]{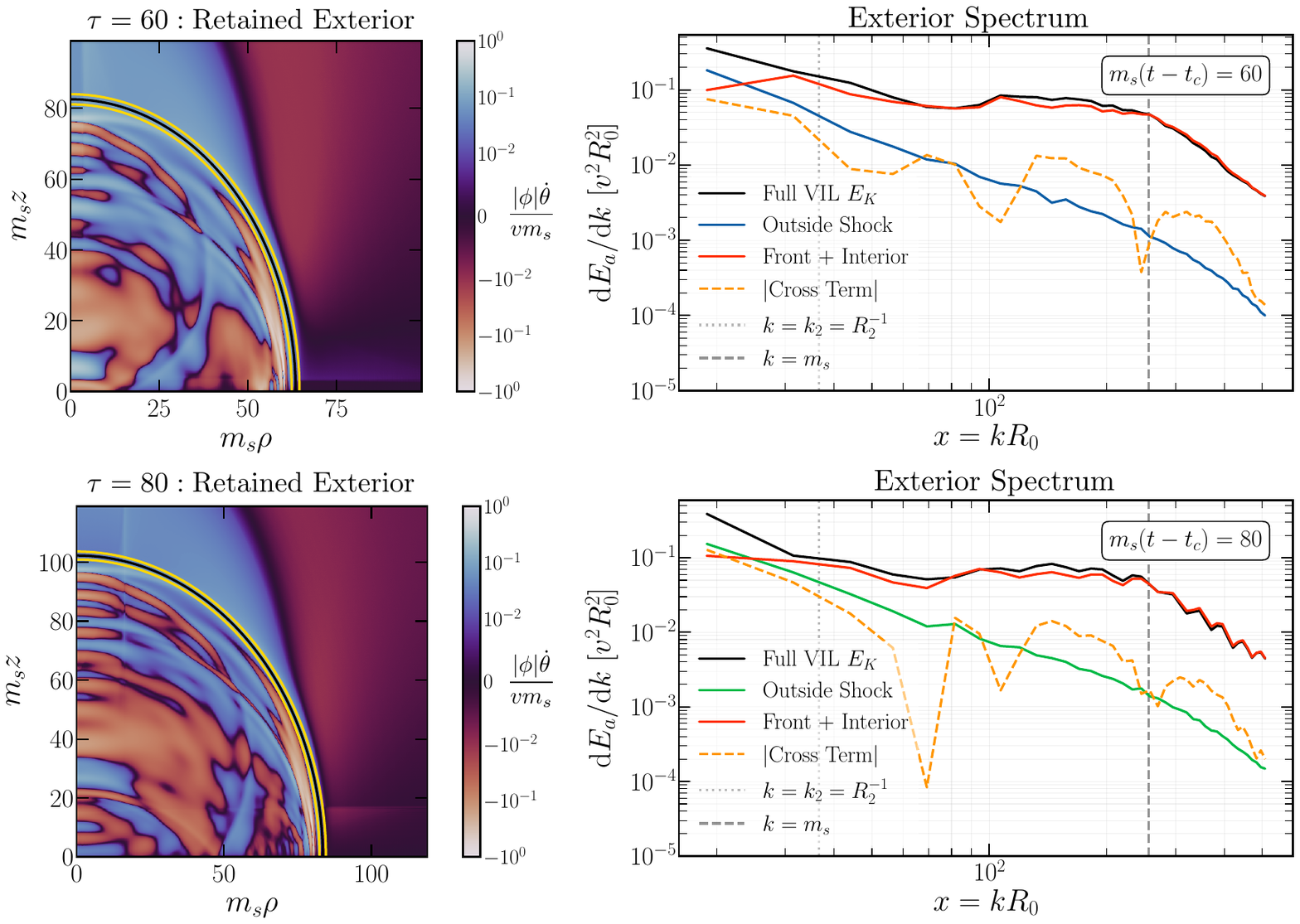}
  \caption{Exterior-field test at $m_s(t-t_c)=60$ (\emph{top}) and $80$
 (\emph{bottom}). \emph{Left:} signed phase velocity with contours marking the
 outer edge of the anisotropic front; only the region outside the contour is
 retained in the exterior transform. \emph{Right:} full masked kinetic
 spectrum (black), exterior contribution (blue/green), front plus interior
 (red), and magnitude of the cross term (orange dashed). The exterior
 contribution is soft and carries negligible power near $m_s$.}
 \label{fig:outside_front_spectrum}
\end{figure*}

As a complementary causal test, we mask both the front and its entire
interior, retaining only the field ahead of its outer edge. Results are shown in Fig. \ref{fig:outside_front_spectrum}.  The anisotropic
front is represented by an elliptical radial coordinate with semiaxes
$m_s(t-t_c)$ and $m_s(t-t_c)+18$ in the $\rho$ and $z$ directions,
respectively. The offset is an empirical description of the longitudinal
lead and is about $2.6R_2$, not an additional physical scale.  At
$m_s(t-t_c)=60$ and $80$, Fig.~\ref{fig:outside_front_spectrum} shows
that the retained exterior contains only a steeply decreasing IR tail.  Its power is already small at $k_2$ and negligible near $m_s$, and decreases
relative to the total between the two snapshots.  The $m_s$-scale excess is
therefore carried by the outgoing front or by the excited field behind it,
not by a precursor propagating ahead of the front. The exterior holds a larger contribution of the low-$k$ modes, which can be attributed both to the early abandoned modes and a simple  interior vs. exterior volume normalisation. 

The late spectra also contain oscillatory structure below $k_2$.  These
fringes are clearest for $m_sR_0=128$, $192$, and $256$ and persist in the
uniform $N^2=8192^2$, $m_sR_0=256$ simulation.  They are thus not generated by the mesh refinement.  Their spacing is substantially coarser than that predicted by treating the instantaneous outgoing front as a single circular or spherical shell.  The most plausible origins are coherent $R_0$-scale structures abandoned at collapse, the bouncing string and its self-field, or interference between radiation emitted at different stages. We have performed 2D numerical simulations with AMR, which allow resolving larger values of $m_sR_0$. The fringes have 
weaker visibility at $m_sR_0=384$ and $512$, but we cannot claim their
physical disappearance: spectral resolution, loss of long modes under mesh refinement, and incomplete focus convergence can all wash them out. Further studies will be dedicated to these features. 

The two characteristic UV positions scale differently,
\begin{equation}
 k_2R_0\sim\sqrt{m_sR_0},
 \qquad
 \frac{m_s}{k_2}\sim\sqrt{m_sR_0},
 \label{eq:front_spectral_scales}
\end{equation}
so the separation between them grows with the scale hierarchy.  A flat $\dd E/\dd k$
plateau does not by itself imply logarithmic energy: its integral is
proportional to its bandwidth.  Determining the asymptotic importance of the
front therefore requires measuring both its height and its energy fraction
as functions of $m_sR_0$.  The analogous hard spectrum in the NGKR
calculation may reflect the generic Fourier response of a sharply localised,
relativistic source.  In particular, the high-frequency NGKR result also
receives temporal endpoint contributions.  Establishing which part is
shared long-distance physics and which part specifically originates in the
resolved radial-mode focus requires the same spatial decomposition at larger
$m_sR_0$.

\subsubsection{Energy Budget}
The analysis of the spectra allows us to decompose the final energy distribution into several distinct components and to study their scaling with the tension parameter $\mathcal{T}$.

The initial energy of the string naturally splits into the core contribution, $E_\mathrm{core} = 2\pi R_0 \times \mu_{\rm core}$ and the regularised axion field contribution
$E_a = 2\pi R_0 \times \pi v^2 \log(m_s R_0)$, where the non-logarithmic contribution is absorbed into the core energy. 
The final spectra are naturally divided into a low-$k$ part, below the non-linear scale $k<k_2=1/R_2=1/R_0\sqrt{m_sR_0}$, and a high-$k$ non-linear region $k>k_2$ reaching $k\gtrsim m_s$.
We note that the axion energy is initially equipartitioned between these low and high momentum components. This happens because the energy spectrum is $\dd E/\dd k \propto 1/k$ and thus, the energies are $\propto \log k_2/k_{\rm min}$ and $\propto \log k_{\rm max}/k_2$, and both of these contributions turn out to be $\sim\log(\sqrt{m_s R_0})=0.5\log (m_s R_0)$ when we take the initial IR and UV cut-offs to be $1/R_0$ and $m_s$, respectively. 

To study the final energy we employ our 2D runs where the focus event is properly resolved; otherwise we cannot guarantee to get the energy budget right. 

The simplest behaviour we find is the low-$k$ axion field energy, which retains $\sim 40\% $ of the total initial energy, independent of $\mathcal{ T}\approx\log(m_sR_0)$. The high-$k$ axion field is most affected by the focus period and its final value increases with $\mathcal{T}$. At small $\mathcal{T}\sim 3$ most of the $\sim 60\%$ remaining energy is taken by the saxion fields, but already at ${\mathcal{T}}=5$ it is only $\sim 10\%$. Our results seem to extrapolate to a vanishingly small saxion fraction (probably suppressed by $1/\mathcal{T}$ as in the initial conditions), a $\sim 40\%$ low-$k$ axion fraction and a $\sim 60\%$ high-$k$ contribution. The initial equipartition  between the high and low-$k$ spectral axion components is almost maintained, only a small $10\%$ fraction migrates from the IR to the UV during the non-linear event. 
This seems to be consistent with the idea that the evolution of the low-$k$ part of the spectrum is largely a trivial abandonment of self-field modes by the string, which are then left to fully propagate. 
\section{Comparison}
\label{sec:compa}

In this section we want to assess how well the  NGKR EFT discussed in Sec.~\ref{sec:eft} reproduces the simulation results presented in the previous Sec.~\ref{sec:simu}. 

The effective string description is expected to be most reliable at early times, during the adiabatic stage of the collapse. In particular, at the half-radius crossing, $R(t)=R_0/2$, the loop is still far away from the nonlinear scale $R_2$, its core is well resolved in the field-theory simulations, and the comparison does not depend on the subsequent focusing event, or how collapse is defined. 

We therefore first focus on the halving time, then extend the comparison to the full collapse time and the axion spectrum after the collapse.  

\subsection{Halving and Collapse Times}
\label{subsec:collapse_time}
Rather than comparing the full continuum collapse trajectories $ R_{\rm cont}( t)$ with the numerical solutions of Eq.~\eqref{eq:circle-renormalised},  we find it more illuminating to focus on two simpler observables: the halving time $t_{1/2}$, at which $R=R_0/2$, and
the collapse time $t_{\rm coll}$, at which $R\to0$. 

We expect the loop collapse to be equivalent to a pure NG trajectory in the limit $\mathcal{T}(R_0) \to \infty$. We thus define the
dimensionless time advances with respect to a NG collapse as,  
\begin{equation}
\label{eq:collapse_advance}
T_{1/2}(\mathcal T)=1-\frac{t_{1/2}(\mathcal T)}{t_{1/2}^{\rm NG}}, \qquad
T_{\mathrm{coll}}(\mathcal T)=1-\frac{t_{\rm coll}(\mathcal T)}{t_{\rm coll}^{\rm NG}}.
\end{equation}
where $ t_{1/2}^{\rm NG}=\pi R_0/3$ and $ t_{\rm coll}^{\rm NG}=\pi R_0/2$ follow from the NG collapse solution $R/R_0=\cos(t/R_0)$.
Both are positive quantities, since the back-reaction of the string accelerates the collapse compared to the NG case. They are expected to vanish only in the large tension limit $\mathcal T\to\infty$. We show our results in Fig.~\ref{fig:collapse_times} as a function of the dimensionless tension parameter $\mathcal T(R_0)$. 
A technical discussion on how we measure $T_{1/2}$ and $T_{\rm coll}$ from simulations is deferred to App.~\ref{app:collapse_times}. 

The halving and collapse times decrease monotonically with $\mathcal{T}$ approaching the NG limit at high $\mathcal{T}$, as expected. At low $\mathcal{T}$, the field-theory simulations and NGKR agree in value near $\mathcal{T}\approx 3-4$, but with different slopes. This is to some extent expected, as high curvature corrections and the effects of the saxion fields can have an impact here and are not represented in the effective theory at leading order as used here. 

The main trend of the difference to NG decreases as $\propto 1/\mathcal{T}$. This is a prediction of the NGKR effective theory, because the renormalised loop evolution equation is the NG equation with a force term decreasing as $1/\mathcal T$, but we also observe it in the field-theory simulations. 
The different complex scalar simulation results are consistent within the uncertainties.

Recall that $\mathcal T(R_0)=c_{\rm FT}+\log(m_sR_0)\approx \log(m_sR_0)$ for the complex scalar simulations and the NGKR effective theory, so the effective tension parameter is essentially the $\log$ setting the hierarchy between core size and string radius. The high-tension simulations are performed at similar values of $\log ( m_s R_0)$ but have much larger core tension, Eq.~\eqref{eq:KM-dimensionless-tension}, so we can extend the $\mathcal{T}$-range up to values of ${\mathcal{T}}\sim 100$. 
Plotting the results as a function of $\mathcal{T}$ shows continuity between the simple complex scalar results and the high-tension model. This proves that indeed, $\mathcal{T}$ is the important parameter of the effective theory, not ${\mathcal{T}}_{\rm core}$ or $\log (m_s R_0)$ individually. This also confirms that exchanging the gradient energy of the axion field near the core for core energy in a vector field, as proposed in Refs.~\cite{Klaer:2017ond,Klaer:2017qhr}, works extremely well. 

Since both NGKR and the field-theory simulations tend towards NG, the agreement between them seems to improve at large $\mathcal{T}$. However, the logarithmic plot reveals a conundrum:
\begin{figure}[t]
    \centering
    \includegraphics[width=0.49\linewidth]{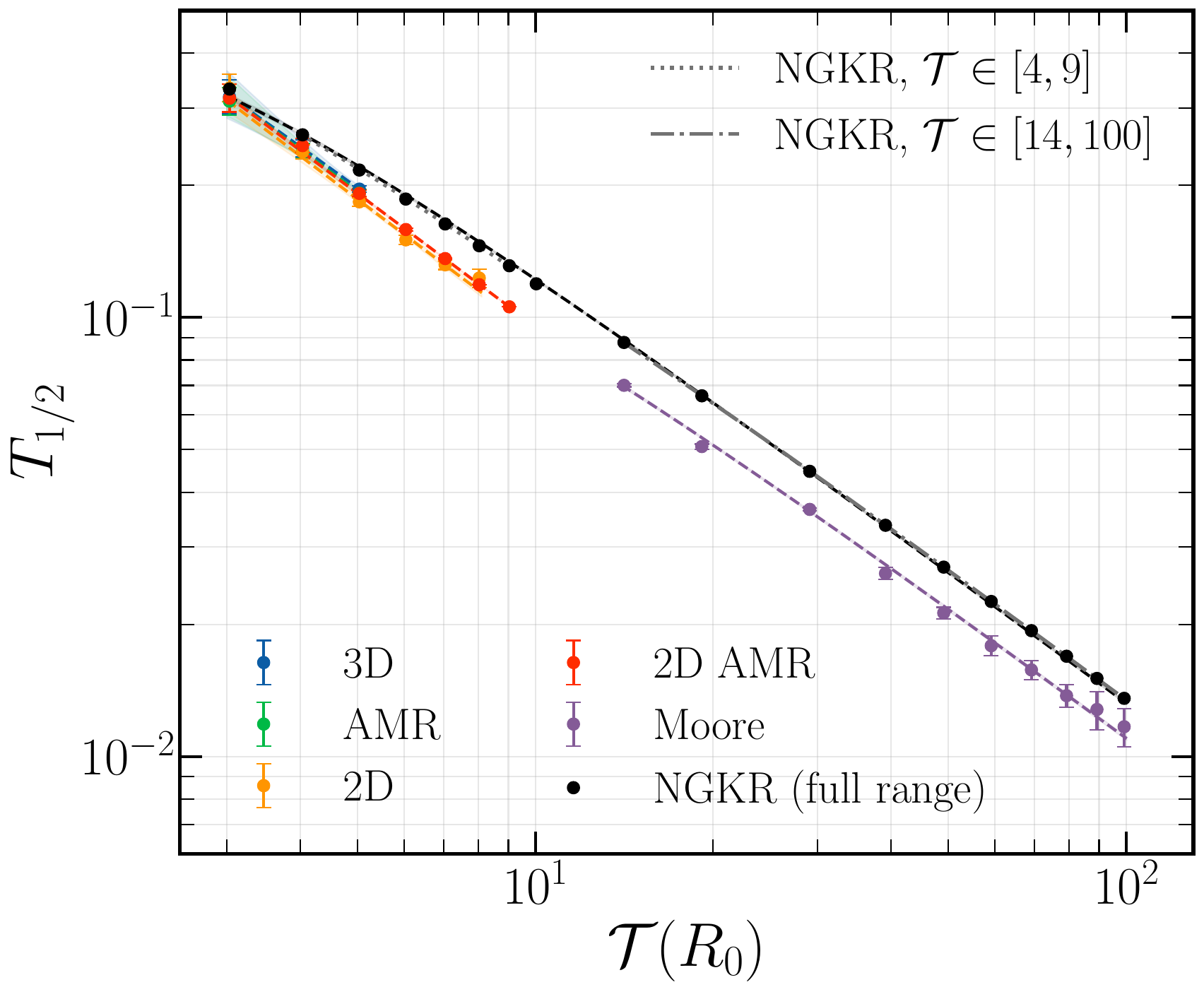}
    \includegraphics[width=0.49\linewidth]{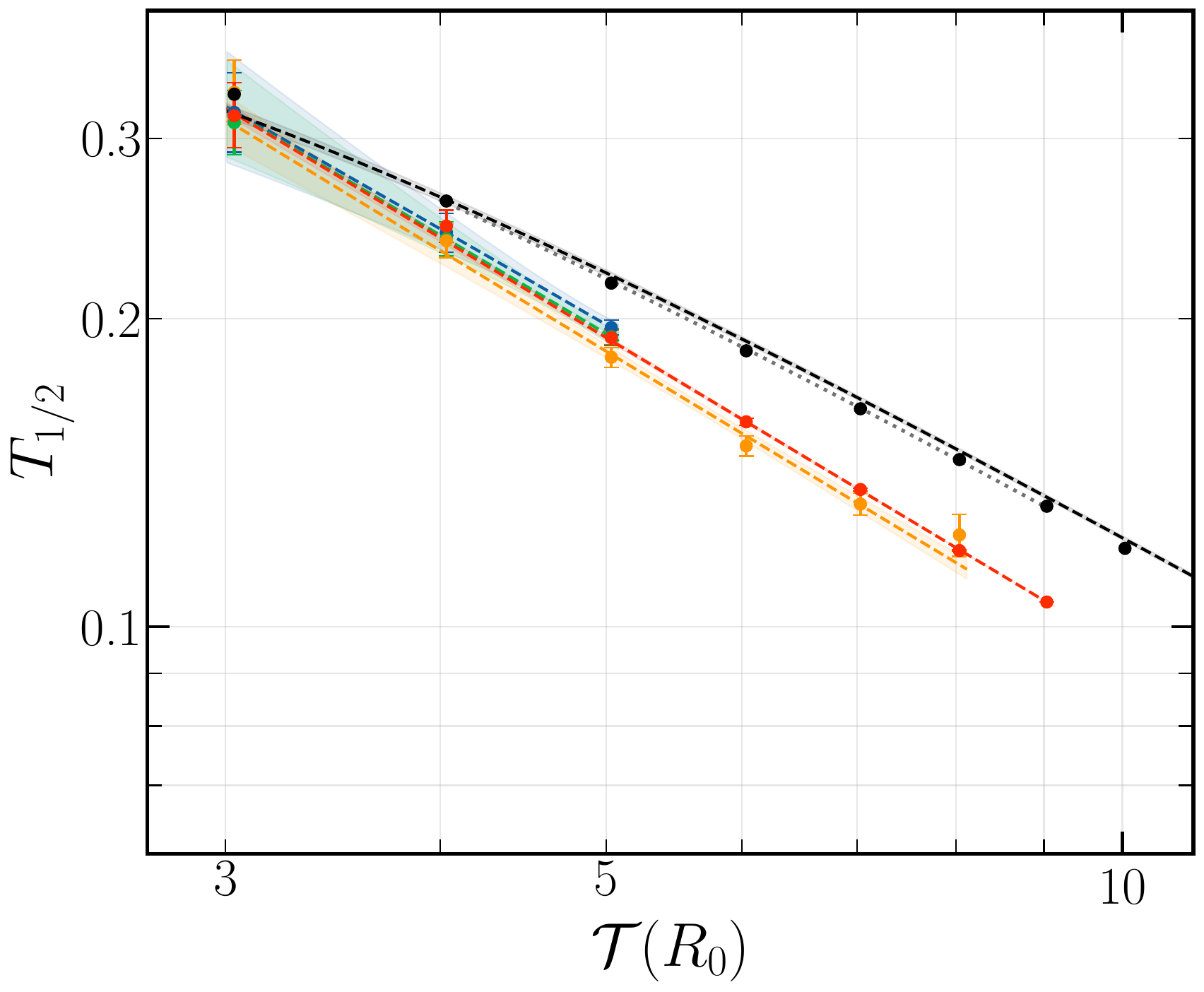}
    \includegraphics[width=0.49\linewidth]{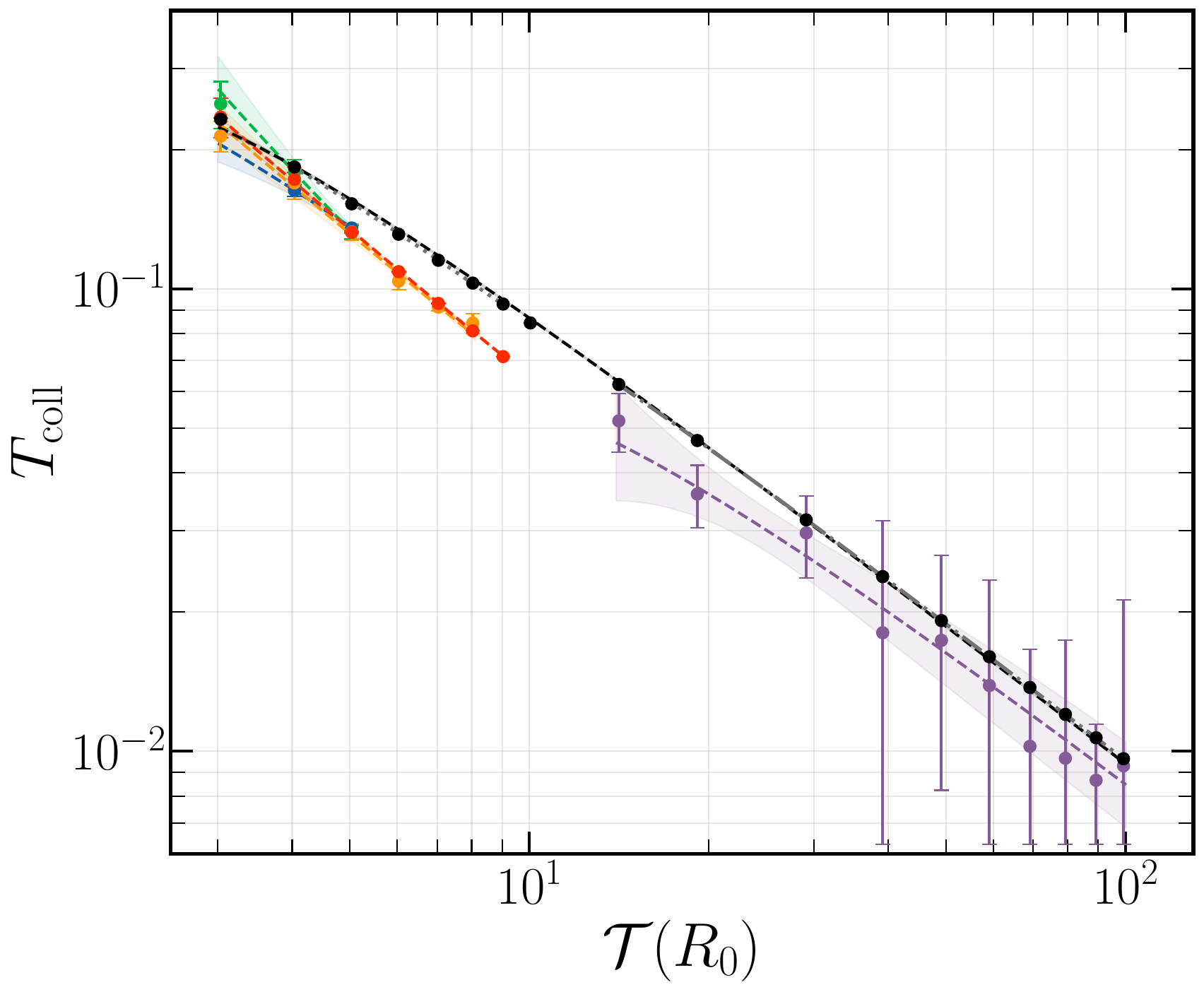}
    \includegraphics[width=0.49\linewidth]{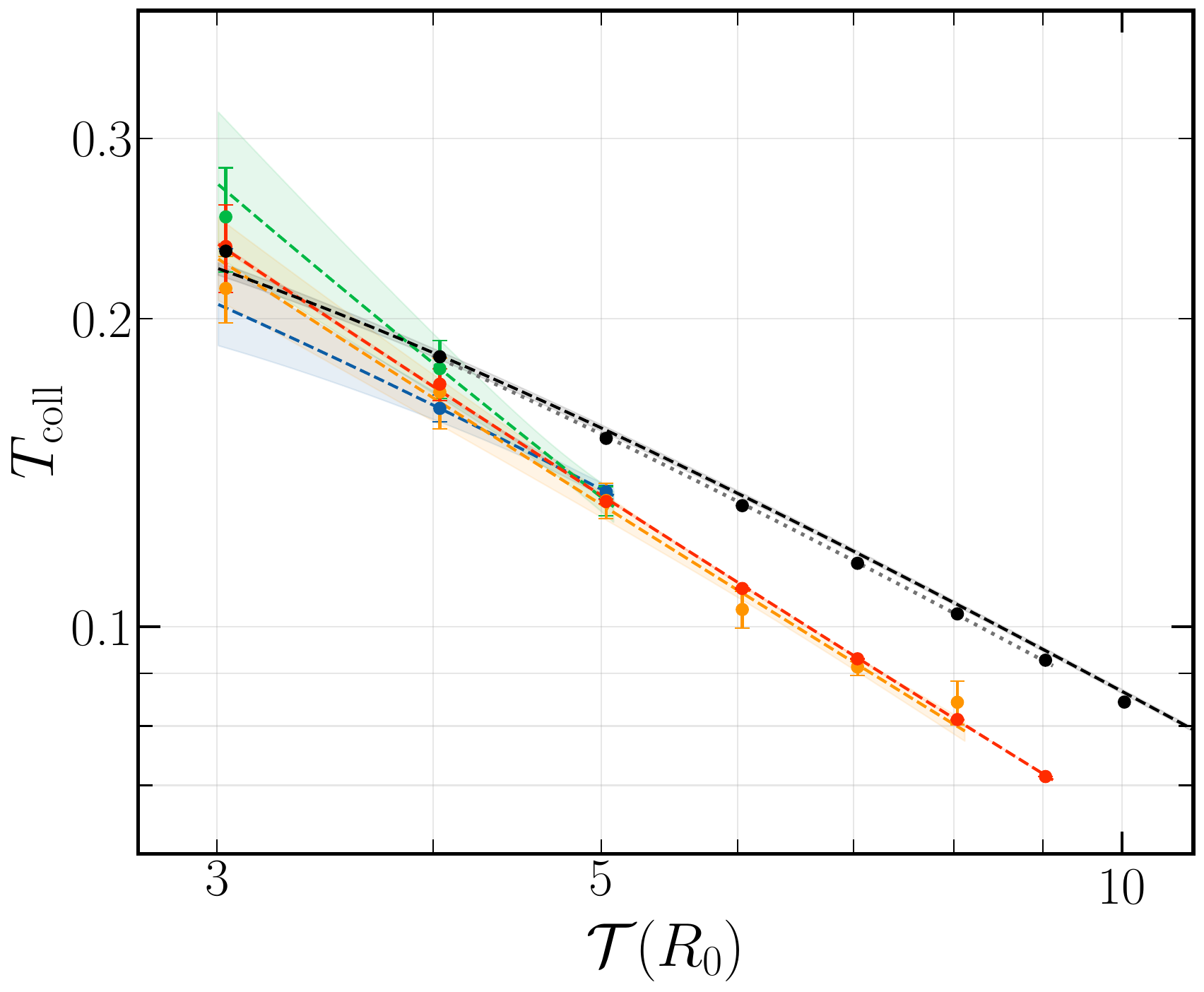}
    
    \caption{Dimensionless time advances with respect to the NG limit, defined in Eq.~\eqref{eq:collapse_advance}, as a function of the dimensionless tension parameter $\mathcal T(R_0)$, for the loop halving time (\emph{top}) and the collapse time (\emph{bottom}). The right panels zoom in on the low-tension region covered by the complex scalar simulations. Dashed lines and shaded bands are fits of $A/\mathcal T+B/\mathcal T^2$ with their $1\sigma$-uncertainty. Additionally, for NGKR we show the same fits over the range relevant for the complex scalar (CS) simulations, $\mathcal{T}\approx 4-9$, and the high-tension Klaer-Moore (KM) simulations, $\mathcal{T}\approx 14-100$ in gray, see Tab.~\ref{tab:half-collapse} for details. Error bars extending below the plotted range are clipped at the axis for visualisation purposes.}
    \label{fig:collapse_times}
\end{figure}
The NGKR results remain visibly different from the simulation results. To exhibit the large-tension behaviour directly, we perform a fit to our different data sets as,  
\begin{equation}
 T_{1/2}=\frac{A}{\mathcal{T}}+\frac{B}{{\mathcal{T}}^2}+\mathcal O({\mathcal{T}}^{-3}), 
 \label{eq:comparison_inverse_log_fit}
\end{equation}
with the results for the coefficients presented in Tab.~\ref{tab:half-collapse}.
\begin{table}[tbp]
\centering
\renewcommand{\arraystretch}{1.2}
\setlength{\arrayrulewidth}{1pt}
\begin{tabular}{l r@{\,$\pm$\,}l r@{\,$\pm$\,}l r@{,\;}l}
\hline\hline
 Code
 & \multicolumn{2}{c}{$A$}
 & \multicolumn{2}{c}{$B$}
 & \multicolumn{2}{c}{$(\mathcal{T}_{\mathrm{min}}$, $\mathcal{T}_{\mathrm{max}})$} \\
\hline

3D      & $1.0$   & $0.2$   & $-0.1$   & $0.9$   & (4  & 6)   \\
AMR     & $0.97$  & $0.17$  & $-0.02$  & $0.80$   & (4  & 5)   \\
2D      & $0.91$ & $0.06$ & $0.07$   & $0.33$  & (4  & 8)   \\ 
2D AMR  & $0.951$ & $0.007$ & $0.04$   & $0.06$  & (4  & 9)   \\
Moore   & $1.120$ & $0.013$ & $-1.95$  & $0.26$  & (14 & 100) \\
\hline
NGKR (full)   & $1.335$ & $0.008$ & $-1.13$  & $0.06$  & (3  & 100) \\
NGKR (CS)   & $1.289$ & $0.007$ & $-0.98$  & $0.04$  & (4  & 9) \\
NGKR (KM)   & $1.3628$ & $0.0008$ & $-1.78$  & $0.02$  & (14  & 100) \\
\hline\hline
\end{tabular}
\caption{Fit coefficients for the ansatz  $T_{1/2}=A/\mathcal{T}+B/\mathcal{T}^{2}$ with the dimensionless tension
$\mathcal{T}(R_0)$. The last column gives the
range of $\mathcal{T}(R_0)$ covered by the different simulations. For the complex scalar simulation data the lowest-tension point, $\mathcal{T}\approx 3$, is excluded, as the thin-string
approximation is not reliable. For NGKR we show the fit result for the extrapolated curves of
App.~\ref{sec:ngkr-evolution} over the full tension range and over the range relevant for the complex scalar (CS) simulations, $\mathcal{T}\approx 4-9$, and the high-tension Klaer-Moore (KM) simulations, $\mathcal{T}\approx 14-100$.}
\label{tab:half-collapse}
\end{table}

The field-theory simulations give $A\simeq0.95$ for $\mathcal T\le9$, in stark contrast to the
NGKR result, $A=1.335\pm0.008$ over the full tension range $3\le\mathcal T\le100$. We find that our results depend on the chosen fit range: for $4\le\mathcal T\le9$, the range of our complex scalar simulations, we find $A=1.289\pm 0.007$, and for the \texttt{Moore} range $14\le\mathcal T\le100$ we find $A=1.3628\pm 0.0008$. The field-theory coefficients may
depend on the exact fit range in a similar way. In summary, we find that, compared over the same range of tensions, the field-theory and NGKR results for the leading coefficient $A$ differ by about $36\%$ for the complex scalar ($1.289$ versus $0.951$) and about $22\%$ for the Moore simulations ($1.36$ versus $1.12$).
  We find no clear reason for this disagreement. We expected the NGKR and field-theory results to agree up to $1/(m_sR_0)$ corrections, but the disagreement at $\mathcal T\simeq\log(m_sR_0)=6$ is $\delta\mathcal T_{1/2}\simeq0.027$, while $1/(m_sR_0)=0.0025$ is a factor $\sim11$ smaller.
Moreover, the 2D AMR simulations yield a very small higher-order coefficient $B$, while a significant $1/m_sR_0$ distortion could have been partially fitted by a larger $B$ and biased $A$. 
The high-tension Moore simulation results lie at intermediate values $A\sim 1.1$. 
On the one hand, this rules out $1/(m_s R_0)$ effects as the sole cause of disagreement, because these simulations are done at $\log(m_sR_0)=4$, which is much larger than the differences between NGKR and the simulations, e.g. $\delta T_{1/2}=0.002$ but $1/m_sR_0= 0.02$ at ${\mathcal{T}}=100$. 
On the other hand, $A\sim1.12$ is much closer to the NGKR result $A\approx 1.363$ so it could be that the field-theory simulations very slowly approach the NGKR results. However, at present, we find no mathematical or physical basis for this extremely slow approach. Finally, note that for the \texttt{Moore} simulations  we find $B\approx -1.95$, which could bias the result towards a larger $A$.

Perhaps not surprisingly, all the aspects discussed here for the halving time are mirrored in our measurements of the collapse time. The trends are correct, but the leading inverse-tension correction to NG behaviour is different for NGKR and field theories. This suggests that whatever causes this mismatch applies already quite early in the evolution. Indeed, we have performed the exercise of calculating the $A$ coefficient not only for the halving and collapsing times but for all values of the $r=R/R_0$ ratio. We observe that $\delta A = A_{\rm NGKR}-A_{\rm FT}$ increases with
$1-r$, stabilising at $\delta A \approx 0.4$ already at  $R/R_0\approx 0.7$. 

Possible reasons for the mismatch of $A$ are either physical or numerical, or both. Recall that field-theory simulations collapse \emph{slower} than predicted by NGKR.
The most obvious physical difference between field-theory simulations and the NGKR effective description is the existence of fields in the string core, i.e. the UV physics. Saxion and vector fields can be radiated from the string, which should accelerate the collapse, since the energy loss in the adiabatic regime must ultimately be supplied by a reduction in the string length, given that we minimised the energy stored in axion gradients. 

Another possible explanation is that the core profiles in the initial conditions could be not completely relaxed, resulting in excess energy in the core. 
At first order, this leads to a positive contribution to the renormalisation of the core tension, which delays the collapse, hence approaching the NG limit. However, a constant shift of $\mathcal{T}$ affects only the $B$ coefficient.  We would need to get the core tension wrong by  a quantity proportional to the tension itself, $\delta {\mathcal{T}}\propto {\mathcal{T}}$. In high-tension Moore simulations this is plausible, because most ${\mathcal{T}}$ is in the core profile. This is a reason why it is really important to get the core relaxation profile right. Given the trend $T_{1/2}\sim 1.12/{\mathcal{T}}$, a mere $1.363/1.12-1\sim 22\%$ excess core energy would reconcile the results to the NGKR effective theory. 
We find that our core relaxation process reduces energy excesses much more than that, see the discussion in App.~\ref{app:ic-relaxation}. 
However, this would not explain the difference in $A$ at intermediate ${\mathcal{T}}\sim 4-9$ seen in the simpler complex-scalar model. There, most of the tension is in the $\log (m_s R_0)$ piece, hence it is distributed logarithmically over the simulation box and getting the core wrong can only produce a constant $
\delta {\mathcal{T}}$ and thus a shift in $B$. Since $A$ for all complex-scalar simulations is compatible this is not a very simple and likely explanation. 

Dine pointed out some difficulties of treating global strings with collective coordinates and the inadequacy of the NGKR description in Ref. \cite{Dine:2021aml}. Our results could be an over-sophisticated example supporting this claim. However, his main point appears to be that the use of collective coordinates or modes leaves the dynamics of certain IR axion modes out of the description. In our NGKR description these modes are not integrated out. Only high-mass fields, non-zero exclusively close to the core, axion fields at the core and the dynamics of the interaction between all these high-frequency modes are deliberately excluded. The NG string moves under its own curvature and the force exerted by the UV axion field around it, which interacts with all the IR modes of the axion field. We therefore conclude that if our description fails, it cannot be due to missing IR modes.

An interesting and possible effect appears also in the interaction between excited core modes and the curvature of the string, see Refs. \cite{Blanco-Pillado:2022axf, Aurrekoetxea:2026xrl}. A displacement of the core profile with respect to the equilibrium could trigger a parametric resonance and produce an excess axion radiation. This effect should be very small in our very relaxed initial conditions and it would accelerate the collapse, not delay it, and is expected to be visible in the axion radiation spectrum. 

Another possible culprit could be an error in solving the evolution equation of the NGKR theory, which is certainly non-trivial. However, the $A$ coefficient can be calculated in the NGKR effective theory by perturbing around the NG trajectory, see 
App.~\ref{app:halving_coefficient} for a detailed derivation. 
We write the trajectory as the NG solution plus a first-order correction
$r_1$ in the large-$\mathcal T$ limit, where in this calculation
$\mathcal T\equiv\mathcal T(R_0)=\bar\mu(R_0)/(\pi v^2)$,
\begin{equation}
 \frac{R(t)}{R_0}=\cos\left(\frac{t}{R_0}\right)
 +\frac{1}{\mathcal T}\,r_1\!\left(\frac{t}{R_0}\right)
 +\mathcal O(\mathcal T^{-2}).
\end{equation}
The circular-loop equation at first order becomes, in units of $R_0$,
\begin{equation}
 \mathcal L r_1={\cal F}_0(t),
 \qquad
 \mathcal L r_1\equiv
 \ddot r_1+2\tan(t)\,\dot r_1-r_1,
 \label{eq:comparison_linearized_eom}
\end{equation}
where ${\cal F}_0 \equiv R_0\bar F(R_0\cos(t);R_0)/(\pi v^2)$ is the dimensionless finite retarded KR force of
Eq.~\eqref{eq:circle-renormalised} evaluated on the NG trajectory. If $G(t,t')$ is the retarded Green function of
$\mathcal L$, expanding the condition $R(t_{1/2})=R_0/2$ gives
\begin{equation}
 A=-\frac{1}{t_*\sin(t_*)}
 \int_0^{t_*}\dd t'\,G(t_*,t'){\cal F}_0(t'),
 \qquad t_*=\frac{\pi}{3}.
 \label{eq:comparison_A_green}
\end{equation}

We note that a finite additive reparametrisation of $\mathcal T$ leaves
Eq.~\eqref{eq:comparison_A_green} invariant and first changes the coefficient
$B$ at order $\mathcal T^{-2}$, as shown in
App.~\ref{app:halving_coefficient}.
Indeed, the local logarithmic counterterms are proportional to the NG curvature operator, $\cal K$,  which vanishes identically
when evaluated on the NG trajectory. 

A direct evaluation of Eq.~\eqref{eq:comparison_A_green}, followed by removal
of the angular endpoint cutoff, yields
\begin{equation}
 A_{\mathrm{NGKR}}=1.368\pm0.004.
 \label{eq:comparison_A_eft_direct}
\end{equation}

This agrees to $0.2\%$ (within $1\sigma$) with the coefficient extracted from the numerical NGKR evolution at large
tension, $A=1.3653\pm0.0003$, obtained from our fit restricted to $29\le\mathcal T\le100$. The level of agreement rules out a problem in the NGKR solver as the origin of the discrepancy.

We note that splitting the retarded
integral at the release time also shows that the result is dominated by the
field laid down while the loop was held static: approximately $1.301$ comes
from the pre-release branch, while only about $0.068$ is generated by the
subsequent dynamical history. This points to a problem in the initial conditions. 

Indeed, the coefficient $A$ must be a time-averaged function of $H_{012}\propto \partial_z \theta$ as this is the quantity sourcing the difference between the NG and NGKR evolution, and the static pre-release field configuration $\partial_z \theta$ is specified differently in the NGKR and field-theory simulations. 
In the latter, the $\theta$-field is initialised geometrically through the solid angle approach of Eq.~\eqref{eq:solid-angle-theta}. We checked that the initial $\boldsymbol\nabla\theta$ agrees with the
field given by the retarded integral of the NGKR calculation in a few points off the string. 
However, as a further direct sensitivity test, we conducted simulations by rescaling the axial coordinate in the initial conditions for $\theta$, $z\to z/ \kappa$, i.e. compressing the field and thus increasing the gradients by a factor of $\kappa$.  Matching the NGKR halving time at $m_sR_0=128$ required $\kappa\simeq 3.2$, rather than a perturbation around $\kappa=1$. Such a large axial distortion makes a modest normalisation of the initial-condition error an implausible explanation.

We therefore do not presently understand why the leading NGKR collapse
advance is about $22$ to $36\%$ larger than in the field-theory simulations. The mismatch is an order-$1/\mathcal{T}$ effect in the observable, and dependences on the finite core matching and regularisation are only relevant at the next order. This discrepancy is theoretically interesting, but its absolute effect on the trajectories is small throughout the range where NGKR is expected to apply, and it does not affect our analysis of the nonlinear focus.

\subsection{Spectrum}
\label{subsec:spectrum_comp}
We want to compare the spectrum of axions radiated during the collapse of a loop in the field theory and the NGKR EFT. This comparison is severely limited by two aspects. In the field theory we would need to have a far-away boundary to separate self-fields, whose energy density decays as $d^{-6}$, and radiation, decaying as $d^{-2}$, where $d$ is a distance. This is computationally very expensive. For NGKR, we can calculate the axions radiated at $d=\infty$, which is impossible to do in the field-theory simulations. Instead of considering an extrapolation of the radiation crossing a spherical shell and $d\to \infty$, we prefer to wait after the loop has collapsed and bounced until there is no string in the box and the axion spectrum stabilises. This takes some time, as the self-fields do not vanish instantaneously. If we control the \enquote{leaking} of axions, which would lead to total energy loss of the box, and keep it under control, the results should be the desired axion spectrum from the loop collapse. 

In the NGKR effective theory, we can take the $d\to \infty$ limit, record the radiation fields as a function of time from the retarded $R=R(t)$ simulated history, and obtain the pure radiation spectrum. Unfortunately, the EFT loses predictability at around the collapse, where the curvature of the loop diverges and the EFT violates the assumptions. Thus, we are forced to stop the computation right before the collapse. 

Interestingly, the $R\to 0$ event can be easily integrated in the NGKR EOM and does not lead to significant complications in the calculation of the spectrum. One could therefore study the increasingly smaller bounce solutions and the interference between them. 
It is well known that the NG evolution, being periodic, produces a (perturbatively) discrete spectrum of radiation, and we expect that at large $\mathcal{T}$ the decreasing bounces will lead to a continuum spectrum with interference fringes. 
This additional study goes beyond the scope of this work, but we will develop these ideas in more detail in a follow-up publication. 

\begin{figure*}[t]
 \centering
 \includegraphics[width=.65\linewidth]{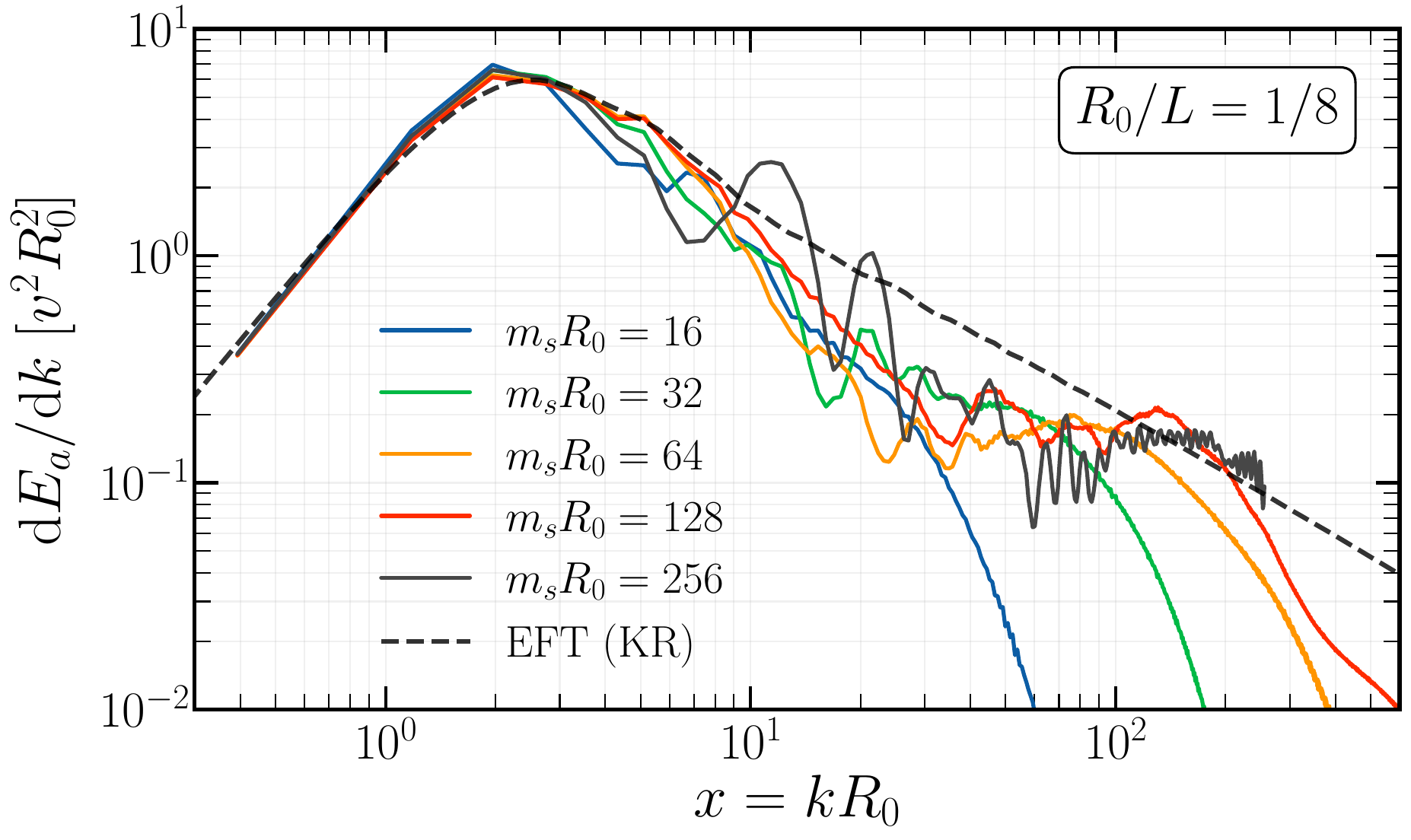}

 \caption{Late-time $E_a=E_K+E_G$ spectra at $t=\pi R_0$ for several $m_sR_0$,
 compared with the NGKR prediction of Eq.~\eqref{RadAxionExpr} evaluated
 close to collapse for $\mathcal{T}\approx 5.0$. Here we show only the result for the $R_0/L=1/8$ simulation, since it resolves the IR part of the spectrum with the required precision for this comparison, cf. Fig.~\ref{fig:spectrum_final}.}
 \label{fig:spectrum_comp}
\end{figure*}

Despite all these complications and the additional constraints we impose, we are nevertheless able to obtain a useful result. Figure \ref{fig:spectrum_comp} shows the final spectrum obtained from 2D simulations fulfilling the focus-resolution requirement of Eq.~\eqref{eq:focus_resolution_requirement} compared to the results of the NGKR radiation spectrum integrated up to the collapse time. 

We observe remarkably good agreement at low $k$, which we attribute to the fact that these axions originate from the non-adiabatic part of the early self-field contraction. As the loop contracts, the self-field initially follows its motion. Eventually, however, the contraction becomes too fast for the low-$k$ modes to adjust on their characteristic time-scales, and they cannot track the dynamics of the loop anymore. These modes are then effectively abandoned by the loop and propagate approximately as free waves. They remain present at the end of the field-theory simulations and are included in the time window used by the NGKR EFT, since they were emitted before the collapse. For $k<2/R_0$, i.e. around the peak and below, the agreement is remarkably good for all simulated values of $\mathcal{T}\approx\log(m_sR_0)$. 

The agreement appears to extend beyond the peak, with the range of agreement depending on $\mathcal{T}$: larger values of $\mathcal{T}$ yield agreement with the EFT up to larger $k$. This trend appears to hold up to $m_sR_0=256$, where the spectrum develops very pronounced oscillations or fringes in the $1/R_0\sim k_2$ range, before the non-linear plateau. These features could be associated with the relatively pronounced bounce we observe in this simulation, and the radiation emitted during the bounce interfering with that produced during the first collapse. Such events are not captured by the NGKR EFT, which only describes the radiation emitted during the initial collapse.

A more systematic study with a wider range of $\mathcal{T}$-values could help to clarify the origin of these fringes and determine whether the system approaches the ever-oscillating NG solution. We defer this study to future work. 

\section{Conclusions}
\label{sec:conclusions}

In this paper, we have studied the dynamics of circular global string loops, with particular emphasis on the comparison between field-theory simulations and the formalism based on the NGKR effective theory.
For the field-theory simulations, we considered two classes of models: the standard single complex scalar model defined by the action~\eqref{eq:lagrangian}, and the Klaer--Moore model~\eqref{eq:KM-lagrangian}. The latter is expected to describe the dynamics of high-tension global strings without introducing a large hierarchy of scales.
For the single complex scalar model, we performed simulations using three different methods: 3D uniform lattice, 3D AMR, and 2D axisymmetric setup.
The results obtained with these three methods agree well with each other up to discretisation effects, demonstrating that the associated numerical systematics are well under control.
The 2D axisymmetric simulations allow us to reach relatively large values of the string tension, $\mathcal{T}(R_0) \approx \log(m_sR_0) \approx 9$.
Combined with the Klaer--Moore simulations, our analysis covers the phenomenologically relevant range of tensions, $3 \lesssim \mathcal{T}(R_0) \lesssim 100$, including $\mathcal{T}\approx 70$ relevant for post-inflationary QCD axion cosmology.

We also developed a numerical scheme for constructing static-loop initial conditions, in which the $\theta$ field is constructed geometrically in terms of the solid angle subtended by the loop 
and the string core is subsequently relaxed.
The ability to construct such initial conditions is crucial for making a precise comparison with the NGKR effective theory.
For the NGKR effective theory, we carefully reviewed the formalism developed by Dabholkar and Quashnock in Ref.~\cite{Dabholkar:1989ju} and computed the loop trajectories by numerically solving the renormalised equation of motion for circular loops.

We find that the loop trajectories predicted by both the field theory and the NGKR effective theory approach those of the NG strings as the string tension is increased (see Figs.~\ref{fig:NGKR_panel} and~\ref{fig:continuum_panel}).
We also confirm that the single complex scalar model and the Klaer--Moore model approach the same NG limit in a consistent manner (see Fig.~\ref{fig:collapse_times}).
This provides further evidence for the validity of the Klaer--Moore construction~\cite{Klaer:2017qhr}, in which parametrically large string tensions are realised using two scalar fields and a vector field.

Our analysis also reveals an unexplained discrepancy between the loop trajectories obtained from the field-theory simulations and those predicted by the NGKR effective theory.
This discrepancy manifests itself as a difference of up to $36\%$ in the coefficient of the leading $1/\mathcal{T}$ correction to the NG result when both are fitted over the same
range of tensions (see Eq.~\eqref{eq:comparison_inverse_log_fit} and Tab.~\ref{tab:half-collapse}).
The NGKR effective theory consistently predicts slightly earlier loop collapse than is observed in the field-theory simulations.
Although we carefully examined possible sources of this discrepancy (see Sec.~\ref{subsec:collapse_time}), we were unable to identify its origin conclusively.
Further scrutiny is needed to determine whether this discrepancy arises from unaccounted systematics in the simulations, the effective theory, or both, or instead calls for a more fundamental reassessment of the theoretical framework (see Appendix~\ref{app:limitations} for a discussion of the limitations of the effective theory).

In addition to the loop trajectories, we also compared the axion spectra obtained from the field-theory simulations and the NGKR effective theory (see Fig.~\ref{fig:spectrum_comp}).
We find good quantitative agreement in the IR part of the spectrum, including the peak and the $1/k$ power law.
This component arises from an adiabatically contracting copy of the initial self-field released as radiation.
However, around or after the stage of the nonlinear focus, additional features appear in the UV part of the spectrum. 
Our analysis shows that a significant fraction of the energy is carried by the outgoing front, whose spectrum is flat from $k \approx k_2$ (given by Eqs.~\eqref{eq:R2_scaling} and~\eqref{eq:k2_definition}) to a cutoff at $k \approx m_s$. 
We emphasise that adequate numerical resolution is essential for an accurate determination of this spectrum at the nonlinear focus.
The convergence criterion we obtained is given in Eq.~\eqref{eq:focus_resolution_requirement}.

In Ref.~\cite{Dabholkar:1989ju}, it was argued that the axion spectrum should become IR dominated in the large-tension limit.
The underlying reasoning was that the loop trajectory approaches the NG solution, radiative effects become negligible, and the loop is therefore expected to undergo many oscillations before losing its energy.
Our results for the IR part of the spectrum lend partial support to this picture, but nonlinear collapse dynamics prevent us from drawing a definitive conclusion about IR dominance in the large-tension limit. 
These nonlinear dynamics transfer energy from the loop into a burst of UV axions. 
Such wave fronts are also seen in network simulations and might explain why some of them yield spectral indices $q<1$.\footnote{
All large-scale simulations performed to date have yielded values of $q$ close to, or somewhat below, unity. Although several studies have found a tendency for $q$ to increase with the string tension~\cite{Gorghetto:2018myk,Gorghetto:2020qws,Kim:2024wku,Saikawa:2024bta,Correia:2025nns}, no simulation has yet observed a value of $q$ that clearly exceeds unity. Claims of $q \gg 1$ (e.g., Refs.~\cite{Gorghetto:2020qws, Kim:2024wku}) rely on extrapolations based on the assumption that $q$ continues to grow logarithmically with the scale separation.}
The presence of features in the UV part of the spectrum associated with the wave fronts, which begin to appear at a scale $k_2$ parametrically smaller than the cutoff scale $k \sim m_s$, 
also calls into question the validity of the single power-law ansatz, $\propto k^{-q}$, commonly adopted in the literature.

Our analysis can be improved and extended in several directions.
To assess the importance of the UV features generated by the nonlinear focus for high-tension strings, their dependence on $m_sR_0$ should be examined systematically.
The post-focus dynamics of the string also warrant closer investigation.
Our field-theory simulations show that the string core rebounds after the loop collapses (see Fig.~\ref{fig:collapse_anatomy} and accompanying discussions).
The existence of such a bounce indicates that the loop does not lose all of its energy in a single collapse.
Whether this behaviour becomes more pronounced in the large-tension limit is an interesting open question.
If loops undergo many oscillations as the NG limit is approached, the predictions for massless radiation could differ substantially from those inferred from current field-theory simulations (see Refs.~\cite{Battye:1994au,Wantz:2009it} for the analysis of the axion relic abundance based on this assumption).
It would also be valuable to extend the comparison with the NGKR effective theory to more general configurations, such as perturbed straight strings~\cite{Drew:2019mzc,Battye:2026whd,Battye:2026iuy} and non-circular loops~\cite{Hagmann:2000ja,Saurabh:2020pqe,Baeza-Ballesteros:2023say}.
Clarifying the validity and limitations of the effective theory through such studies would enable more robust predictions for axion radiation from global strings and could help resolve the long-standing controversy.

\begin{acknowledgments}
This article is based upon work from COST Action COSMIC WISPers (CA21106), supported by COST (European Cooperation in Science and Technology).
We thank Atish Dabholkar, Amelia Drew and Jean Quashnock for helpful discussions and comments on this work. We are grateful to Guy Moore and to the authors of \texttt{axioNyx}, particularly Benedikt Eggemeier and Bodo Schwabe, for sharing their codes with us. M.K. also thanks Amelia Drew and the Centre for Theoretical Cosmology at the Department of Applied Mathematics and Theoretical Physics (DAMTP), University of Cambridge, UK, for their hospitality during a research stay in the initial phase of this project. This stay was funded by the STSM COST Grant CA21106-e89ce97c: \enquote{AMR for Axion String Simulations}.
M.\,K. is funded by the Italian Ministero dell'Università e della Ricerca through the FIS 2 project FIS-2023-01577 (DD n. 23314 10-12-2024, CUP C53C24001460001) and
by the Istituto Nazionale di Fisica Nucleare (INFN) through
the Theoretical Astroparticle Physics (TAsP) project.
The work of M.\,K., J.\,R., I.\,R. and I.\,S.-G. was supported by the grants PGC2022-\-126078NB-\-C21 and PID2024-\-160228NB-I00 funded by MCIN/\-AEI/\-10.13039\-501100011033 and ERDF - A way of making Europe, and DGA-\-FSE 2023-\-E21-\-23R by the Government of Arag\'{o}n, Spain, and the European Union NextGenerationEU Recovery and Resilience Program on Astrof\'{i}sica y F\'{i}sica de Altas Energ\'{i}as CEFCA-\-CAPA-\-ITAINNOVA. 
Additionally, M.\,K. was supported by the Government of Arag\'{o}n, Spain, with a PhD fellowship as specified in DGA-ORDEN-CUS/702/2022. 
I.\,R. is supported by IBS under the project code IBS-R018-D3. 
The work of K.\,S. is supported by JSPS KAKENHI Grant Numbers JP24K07015 and JP26K00695.
Parts of the simulations were performed on the HPC system Raven at the Max Planck Computing and Data Facility.

The authors take full responsibility for the scientific content, analysis, and conclusions presented
in this work. Large language models (LLMs), including Claude Opus 5.5 and ChatGPT 5.5, were used to
improve the readability and performance of the NGKR code, as well as to check the consistency of the draft. The final code was carefully checked by the authors against earlier independent versions of the numerical calculations, and all text and results were explicitly verified by the authors.
\end{acknowledgments}

\newpage
\appendix
\renewcommand{\theequation}{A.\arabic{equation}}
\setcounter{equation}{0}

\section{Numerical Details}
\label{app:numerics}
The four simulation setups are introduced in Sec.~\ref{subsec:sim_overview}. Here we collect more technical implementation details and the parameters of the runs used in this work.

\subsection{\texttt{jaxions}: 3D and Axisymmetric 2D}\label{app:jaxions}

Both the 3D and the axisymmetric branch of \github{veintemillas}{jaxions}~\cite{Jaxions:2026} decompose Eq.~\eqref{eq:eom_mink_code} into a system of first-order equations for $\phi$ and $\Pi=\dot\phi$ and evolve them on a uniform lattice with spacing $\delta=L/N$. The Laplacian is built from an $N_g$-neighbour stencil,
\begin{align}
(\nabla^2 \phi)_{{\bm i}} = \frac{1}{\delta^2}\sum_{u}\sum_{n=1}^{N_g}
  C_n\big(\phi_{{\bm i}+n{\bm n}_u} + \phi_{{\bm i}-n{\bm n}_u} - 2\phi_{{\bm i}}\big),
\end{align}
where ${\bm n}_u$ are unit vectors representing displacements of one lattice spacing along the coordinate direction $u$, running over $(x,y,z)$ in 3D and over $(\rho,z)$ in the axisymmetric case. In our simulations we use at least $N_g=2$.

The time step is set by the fastest lattice mode, $\omega_{\rm max}=\sqrt{m_s^2+12/\delta^2}$, as
\begin{equation}
 \dd t = \frac{1.5\,\delta}{\sqrt{(m_s\delta)^2+12}} < 0.433\,\delta,
\end{equation}
which satisfies the Courant condition. The 3D runs are propagated with a RK4 integrator, the 2D runs with a staggered kick--drift leapfrog,
\begin{align}
\Pi^{n+1} = \Pi^n + \Delta t\,\mathcal{F}_\phi, \qquad
\phi^{n+1} = \phi^n + \Delta t\,\Pi^{n+1},
\end{align}
with $\mathcal{F}_\phi$ the right-hand side of Eq.~\eqref{eq:eom_caxion}.

The 3D \texttt{jaxions} lattice is cubic with $N^3$ grid sites and periodic boundary conditions. The parameters of all runs are listed in Table~\ref{tab:simparams_jaxions3D}.

\begin{table}[tbp]
\centering
\renewcommand{\arraystretch}{1.2}
\setlength{\arrayrulewidth}{1pt}
\begin{tabular}{c c c c c c}
\hline \hline
$N$ & $\log(m_sR_0)$ & $\mathcal{T}(R_0)$ & $L$ & $R_0/L$ & $m_s\delta$ \\
\hline
\multirow{2}{*}{128}  & 3.0 & 3.03 & 4.61  & 0.22 & 0.72 \\
                       & 4.0 & 4.03 & 3.06  & 0.33 & 1.31 \\
\hline
\multirow{2}{*}{256}  & 3.0 & 3.03 & 6.52  & 0.15 & 0.51 \\
                       & 4.0 & 4.03 & 4.33  & 0.23 & 0.92 \\
\hline
\multirow{4}{*}{512}  & 3.0 & 3.03 & 9.22  & 0.11 & 0.36 \\
                       & 4.0 & 4.03 & 6.13  & 0.16 & 0.65 \\
                       & 5.0 & 5.03 & 4.80  & 0.21 & 1.39 \\
                       & 6.0 & 6.03 & 2.91  & 0.34 & 2.29 \\
\hline
\multirow{4}{*}{1024} & 3.0 & 3.03 & 13.04 & 0.08 & 0.26 \\
                       & 4.0 & 4.03 & 8.66  & 0.12 & 0.46 \\
                       & 5.0 & 5.03 & 6.78  & 0.15 & 0.98 \\
                       & 6.0 & 6.03 & 4.11  & 0.24 & 1.62 \\
\hline
\multirow{2}{*}{1536} & 5.0 & 5.03 & 8.31  & 0.12 & 0.80 \\
                       & 6.0 & 6.03 & 5.04  & 0.20 & 1.32 \\
\hline
\multirow{2}{*}{2048} & 5.0 & 5.03 & 9.59  & 0.10 & 0.70 \\
                       & 6.0 & 6.03 & 5.82  & 0.17 & 1.15 \\
\hline \hline
\end{tabular}
\caption{Numerical parameters for the \texttt{jaxions} 3D simulations. $N$ is the number of lattice sites per dimension, $L$ the box size, $R_0/L$ the initial loop radius in units of the box size and $m_s\delta$ the lattice spacing in units of the inverse saxion mass. The dimensionless tension at the initial radius is $\mathcal T(R_0)=c_{\rm FT}+\log(m_sR_0)$ with $c_{\rm FT}=0.0347$, see Eq.~\eqref{eq:FT-tension-match}.}
\label{tab:simparams_jaxions3D}
\end{table}

In the axisymmetric branch the fields are evolved on the $(\rho,z)$ half-plane, where a circular loop of radius $R$ appears as a single vortex at $(\rho,z)=(R,0)$, promoted to a ring by the symmetry, and the collapse corresponds to this vortex drifting towards the axis, $R\to0$. The $1/\rho$ term of Eq.~\eqref{eq:eom_caxion} is given a dedicated treatment on the axis $\rho=0$, where the radial part of the discrete operator is doubled; away from the axis the same stencil as in 3D is used. The symmetry is enforced through a conjugate reflection $\phi(-z)=\phi^*(z)$ across the loop plane and an even reflection $\phi(-\rho)=\phi(\rho)$ across the axis, with the axis row handled separately. The outer radial and axial edges are surrounded by absorbing sponge layers, in which the momentum update is replaced by the damped form
\begin{align}
 \Pi^{n+1}=d_2\,\Pi^n + d_1\,\Delta t\,\mathcal{F}_\phi,
 \qquad
 d_1=\Big(1+\tfrac12\eta\,\Delta t\Big)^{-1},
 \qquad
 d_2=\Big(1-\tfrac12\eta\,\Delta t\Big)\,d_1,
\end{align}
with $\eta$ the local damping strength of Eq.~\eqref{eq:eom_caxion}, non-zero only in the sponge layers.

At selected radii during the collapse the field configuration is interpolated onto a new uniform grid centered on the string core. Keeping the number of sites while reducing the physical domain increases the resolution of the shrinking loop; in contrast to genuine AMR there is only one resolution at each stage, and the far-field part of the old domain is not retained, with consequences for the spectrum measurements as discussed in Sec.~\ref{subsec:spectrum}.

We ran three dedicated series of simulations using the axisymmetric 2D code to study the final state of the string collapse and the radiation spectrum.  The parameters of the collapse runs are listed in Tab.~\ref{tab:simparams_jaxions2D} (\enquote{classic} 2D runs) and Tab.~\ref{tab:simparams_AMR} (2D AMR, as discussed in this section). For the dedicated spectrum runs of Secs.~\ref{subsec:spectrum} and~\ref{subsec:spectrum_comp} we report the simulation parameters in Tab.~\ref{tab:simparams_spectra}. For these runs, we used $N_g=4$ and the other parameters were chosen to satisfy the focus-resolution requirement of Eq.~\eqref{eq:focus_resolution_requirement} . For  values of $m_sR_0>256$, this was realised by employing the 2D AMR technique.

\begin{table}[tbp]
\centering
\renewcommand{\arraystretch}{1.2}
\setlength{\arrayrulewidth}{1pt}
\begin{tabular}{c c c c c c}
\hline \hline
$N$ & $\log(m_sR_0)$ & $\mathcal{T}(R_0)$ & $L$ & $R_0/L$ & $m_s\delta$ \\
\hline
\multirow{3}{*}{256}
  & 3.0 & 3.03 & 6.52 & 0.15 & 0.51 \\
  & 4.0 & 4.03 & 4.33 & 0.23 & 0.92 \\
  & 5.0 & 5.03 & 3.39 & 0.29 & 1.97 \\
\hline
\multirow{3}{*}{512}
  & 3.0 & 3.03 & 9.22 & 0.11 & 0.36 \\
  & 4.0 & 4.03 & 6.13 & 0.16 & 0.65 \\
  & 5.0 & 5.03 & 4.80 & 0.21 & 1.39 \\
\hline
\multirow{4}{*}{1024}
  & 3.0 & 3.03 & 13.04 & 0.08 & 0.26 \\
  & 4.0 & 4.03 & 8.66  & 0.12 & 0.46 \\
  & 5.0 & 5.03 & 6.78  & 0.15 & 0.98 \\
  & 6.0 & 6.03 & 4.11  & 0.24 & 1.62 \\
\hline
\multirow{4}{*}{1536}
  & 3.0 & 3.03 & 15.97 & 0.06 & 0.21 \\
  & 4.0 & 4.03 & 10.61 & 0.09 & 0.38 \\
  & 5.0 & 5.03 & 8.31  & 0.12 & 0.80 \\
  & 6.0 & 6.03 & 5.04  & 0.20 & 1.32 \\
\hline
\multirow{4}{*}{2048}
  & 3.0 & 3.03 & 18.44 & 0.05 & 0.18 \\
  & 4.0 & 4.03 & 12.25 & 0.08 & 0.33 \\
  & 5.0 & 5.03 & 9.59  & 0.10 & 0.70 \\
  & 6.0 & 6.03 & 5.82  & 0.17 & 1.15 \\
\hline
\multirow{5}{*}{4096}
  & 3.0 & 3.03 & 26.07 & 0.04 & 0.13 \\
  & 4.0 & 4.03 & 17.32 & 0.06 & 0.23 \\
  & 5.0 & 5.03 & 13.56 & 0.07 & 0.49 \\
  & 6.0 & 6.03 & 8.23  & 0.12 & 0.81 \\
  & 7.0 & 7.03 & 4.69  & 0.21 & 1.25 \\
\hline
\multirow{3}{*}{8192}
  & 5.0 & 5.03 & 19.18 & 0.05 & 0.35 \\
  & 6.0 & 6.03 & 11.64 & 0.09 & 0.57 \\
  & 7.0 & 7.03 & 6.63  & 0.15 & 0.89 \\
\hline
\multirow{3}{*}{16384}
  & 6.0 & 6.03 & 16.45 & 0.06 & 0.41 \\
  & 7.0 & 7.03 & 9.38  & 0.11 & 0.63 \\
  & 8.0 & 8.03 & 5.69  & 0.18 & 1.03 \\
\hline
\multirow{2}{*}{32768}
  & 7.0 & 7.03 & 13.26 & 0.08 & 0.44 \\
  & 8.0 & 8.03 & 8.04  & 0.12 & 0.73 \\
\hline
\multirow{2}{*}{65536}
  & 7.0 & 7.03 & 18.75  & 0.05 & 0.31 \\
  & 8.0 & 8.03 & 11.37  & 0.09 & 0.52 \\
\hline
\multirow{1}{*}{131072}
  & 8.0 & 8.03 & 16.08  & 0.06 & 0.37 \\
\hline \hline
\end{tabular}
\caption{Numerical parameters for the \texttt{jaxions} 2D simulations, with the same notation as in Table~\ref{tab:simparams_jaxions3D}.}
\label{tab:simparams_jaxions2D}
\end{table}

\begin{table}[t]
\centering
\renewcommand{\arraystretch}{1.2}
\setlength{\arrayrulewidth}{1pt}
\begin{tabular}{c c c c c}
\hline \hline
$m_sR_0$ & $\mathcal{T}(R_0)$ & $N$ & $R_0/L$ & $m_s\delta$ \\
\hline
\multirow{3}{*}{16}   & \multirow{3}{*}{2.81} & 128   & 1/2 & 0.250 \\
                      &                       & 256   & 1/4 & 0.250 \\
                      &                       & 512   & 1/8 & 0.250 \\
\hline
\multirow{3}{*}{32}   & \multirow{3}{*}{3.50} & 512   & 1/2 & 0.125 \\
                      &                       & 1024  & 1/4 & 0.125 \\
                      &                       & 2048  & 1/8 & 0.125 \\
\hline
\multirow{3}{*}{64}   & \multirow{3}{*}{4.19} & 1024  & 1/2 & 0.125 \\
                      &                       & 2048  & 1/4 & 0.125 \\
                      &                       & 4096  & 1/8 & 0.125 \\
\hline
\multirow{3}{*}{128}  & \multirow{3}{*}{4.89} & 4096  & 1/2 & 0.062 \\
                      &                       & 8192  & 1/4 & 0.062 \\
                      &                       & 16384 & 1/8 & 0.062 \\
\hline
192                   & 5.29                  & 8192  & 1/2 & 0.047 \\
\hline
\multirow{3}{*}{256}  & \multirow{3}{*}{5.58} & 8192  & 1/2 & 0.062 \\
                      &                       & 16384 & 1/4 & 0.062 \\
                      &                       & 32768 & 1/8 & 0.062 \\
\hline
*384                  & 5.99                  & 8192  & 1/2 & 0.094 \\
*512                  & 6.27                  & 8192  & 1/4 & 0.250 \\
*768                  & 6.68                  & 8192  & 1/4 & 0.375 \\
*1024                 & 6.97                  & 8192  & 1/4 & 0.500 \\
\hline \hline
\end{tabular}
\caption{Numerical parameters of the spectrum runs of Sec.~\ref{subsec:spectrum}, with $\mathcal T(R_0)=c_{\rm FT}+\log(m_sR_0)$ as in Table~\ref{tab:simparams_jaxions3D}. The box size is fixed by $m_sL=N\,m_s\delta$. Rows with an asterisk use the regridding procedure of App.~\ref{app:jaxions}, so that their resolution increases during the collapse and the quoted $m_s\delta$ refers to the initial grid.}
\label{tab:simparams_spectra}
\end{table}

\subsection{\texttt{axioNyx}: Adaptive Mesh Refinement}\label{app:axionyx}
\texttt{axioNyx}~\cite{Schwabe:2020eac} is an extension of the \texttt{Nyx} cosmological simulation code~\cite{Almgren:2013sz}, built on the \texttt{AMReX} framework~\cite{Zhang:2021abc}.  It uses block-structured Adaptive Mesh Refinement (AMR) to locally refine the grid around the string cores. In the context of axion strings AMR was successfully used for network and individual string simulations in  Refs.~\cite{Buschmann:2021sdq, Benabou:2024msj} and Refs.~\cite{Drew:2019mzc, Drew:2022iqz,Drew:2023ptp}, respectively. It solves the same Eq.~\eqref{eq:eom_mink_code} as \texttt{jaxions}, again with a Laplacian built with $N_g=2$, and a kick--drift leapfrog for the time integration on the coarse grid and all $\ell$ refinement levels.

The mesh hierarchy, with refinement ratio $r_\ell=2$ between levels, is managed by \texttt{AMReX}. Ghost zones are filled by coarse-to-fine interpolation and the Laplacian is evaluated level by level. Cells are tagged for refinement following the plaquette-detection method of Ref.~\cite{Fleury:2015aca}, together with regions of large field gradients $\boldsymbol\nabla\phi$. An additional refinement level $\ell$ is unlocked whenever
\begin{equation}
 (m_s\delta)_{\rm finest}\,\gamma_R(t) \gtrsim 0.5,
 \label{eq:amr-criterion}
\end{equation}
so that the Lorentz-contracted core stays resolved as the loop accelerates, cf.\ Eq.~\eqref{eq:radius_resolution_floor}.
\begin{table}[tbp]
\centering
\renewcommand{\arraystretch}{1.3}
\setlength{\arrayrulewidth}{1pt}
\begin{tabular}{c c c c c c c c c}
\hline \hline
$\log(m_sR_0)$ & $\mathcal{T}(R_0)$ & $\ell_{\rm max}$ & $N_{\mathrm{coarse}}$ & $m_s\delta_{\mathrm{coarse}}$ & $N_{\mathrm{finest}}^{\mathrm{eff}}$ & $m_s\delta_{\mathrm{finest}}$ & $L$ & $R_0/L$ \\
\hline
*3.0 & 3.03 & 2 & 256 & 0.51 & 1024 & 0.13 & 6.52 & 0.15 \\
*4.0 & 4.03 & 3 & 256 & 0.92 & 2048 & 0.12 & 4.33 & 0.23 \\
*5.0 & 5.03 & 3 & 512 & 1.15 & 4096 & 0.15 & 3.96 & 0.25 \\
\phantom{*}6.0 & 6.03 & 3 & 2048 & 1.05 & 16,384 & 0.13 & 5.31  & 0.19 \\
\phantom{*}7.0 & 7.03 & 3 & 4096 & 1.06 & 32,768 & 0.13 & 3.95  & 0.25 \\
\phantom{*}8.0 & 8.03 & 3 & 16,384 & 1.01 & 131,072 & 0.13 & 5.53  & 0.18 \\
\phantom{*}9.0 & 9.03 & 3 & 32,768 & 1.02 & 262,144 & 0.13 & 4.11  & 0.24 \\
\hline \hline
\end{tabular}
\caption{Simulation parameters for the AMR simulations, with the same notation as in Table~\ref{tab:simparams_jaxions3D}. $\ell_{\rm max}$ is the maximum number of refinement levels, and the lattice spacing is given on the coarse and the effective finest grid. Rows with an asterisk indicate simulations that were performed with both 3D \texttt{axioNyx} and 2D \texttt{jaxions}, whereas the other rows were only performed with 2D AMR.}
\label{tab:simparams_AMR}
\end{table}
To ensure that the initial conditions are identical across codes, we generate them once in \texttt{jaxions}, following Sec.~\ref{subsec:stringICs}, and convert them into the required \texttt{boxlib} format using \texttt{MUSIC}~\cite{Hahn:2011}.

The parameters of all AMR runs, including the effective resolution on the finest level, are listed in Table~\ref{tab:simparams_AMR}.

\subsection{\texttt{Moore} Simulations}\label{app:Mooresims}
The model, its straight-string solution and its tension are described in Sec.~\ref{subsec:Klaer--Moore-model}; here we focus on the implementation and the parameter choices for the simulations.

The fields are evolved on a cubic lattice with a kick--drift leapfrog integrator. Scalar gradients use the compact covariant difference $D_i\phi(x)\simeq U_i(x)\,\phi(x+\hat{\imath})-\phi(x)$ with link variables $U_i=e^{iA_i}$, improved through the next-to-nearest neighbours, so that $N_g=2$ as in \texttt{axioNyx}. The gauge field is updated from square and rectangular plaquettes, with $E_i$ the gauge electric field, $B_i$ the plaquette magnetic field and $J_i$ the covariant scalar current of Eq.~\eqref{eq:eom_moore}, supplemented by Gauss's law.

The code sets $v_2=1$ and adjacent charges $q_2=q_1-1$ and additionally we impose the BPS condition $m_1=m_2=m_V=m_s$, which fixes the gauge coupling in terms of the charges and vacuum expectation values as
\begin{equation}
 e=\frac{m_s\delta}{\left(q_1^2v_1^2+q_2^2v_2^2\right)^{1/2}}
 \label{eq:moore-lattice-coupling}
\end{equation}
in lattice units. The remaining parameter $v_1$ is then chosen, for each charge assignment, to realise a given core tension $\mathcal T_{\rm core}$; the resulting values are listed in Table~\ref{tab:simparams_moore}, together with the dimensionless tension $\mathcal T(R_0)$ of Eq.~\eqref{eq:KM-dimensionless-tension}. All runs use the same \texttt{jaxions} initial conditions at $\log(m_sR_0)=4$, prepared as described in Sec.~\ref{subsec:stringICs}, so that the runs differ only in their core tension.

\begin{table}[tbp]
\centering
\renewcommand{\arraystretch}{1.3}
\setlength{\arrayrulewidth}{1pt}
\begin{minipage}{0.46\textwidth}
\centering

\begin{tabular}{c c c c}
\hline \hline
$\mathcal{T}_{\rm core}$ & $\mathcal{T}(R_0)$ & $v_1^2$ & $(q_1,q_2)$ \\
\hline
11.27 & 14.12 & 1.31 & $(2,1)$ \\
16.28 & 19.13 & 2.66 & $(2,1)$ \\
26.28 & 29.14 & 1.00 & $(3,2)$ \\
36.29 & 39.15 & 2.37 & $(3,2)$ \\
46.26 & 49.11 & 3.54 & $(3,2)$ \\
56.26 & 59.12 & 1.58 & $(4,3)$ \\
66.28 & 69.14 & 2.32 & $(4,3)$ \\
76.43 & 79.29 & 3.01 & $(4,3)$ \\
86.41 & 89.26 & 1.32 & $(5,4)$ \\
96.39 & 99.24 & 1.86 & $(5,4)$ \\
\hline \hline
\end{tabular}
\end{minipage}
\hfill
\begin{minipage}{0.48\textwidth}
\centering
\begin{tabular}{c c c c c}
\hline \hline
$\log(m_sR_0)$ & $N$ & $L$ & $m_s\delta$ & $R_0/L$ \\
\hline
\multirow{4}{*}{4.0} & 256  & 4.33 & 0.92 & 0.23 \\
                      & 512  & 6.13 & 0.65 & 0.16 \\
                      & 768  & 7.50 & 0.53 & 0.13 \\
                      & 1024 & 8.66 & 0.46 & 0.12 \\
\hline \hline
\end{tabular}
\end{minipage}%

\caption{\emph{Left:} \texttt{Moore}-specific parameters for each high-tension run. The charges $(q_1,q_2)$ and $v_1^2$, at fixed $v_2^2=1$, were chosen using the approximate core tension of Eq.~\eqref{eq:KM-core-tension}, keeping $v_1^2$ at moderate values. We list the core tension $\mathcal T_{\rm core}$ obtained from the full numerical straight-string solution at the BPS point, and the resulting dimensionless tension $\mathcal T(R_0)=\mathcal T_{\rm core}+\log(m_sR_0/\pi)$ of Eq.~\eqref{eq:KM-dimensionless-tension} for the \texttt{jaxions} initial conditions with $\log(m_sR_0)=4.0$. \emph{Right:} Numerical parameters of the \texttt{jaxions} (3D) initial conditions at $\log(m_sR_0)=4.0$ used for all \texttt{Moore} runs, with the same notation as in Table~\ref{tab:simparams_jaxions3D}. All \texttt{Moore} runs are performed at the BPS point, $m_1=m_2=m_V=m_s$, so the gauge coupling depends on the charges and the vacuum expectation values, $e = m_s\delta\,/\,(q_1^2v_1^2+q_2^2v_2^2)^{1/2}$ in lattice units. It ranges from $e\simeq0.37$ for $(q_1,q_2)=(2,1)$, $v_1^2=1.31$ at $N=256$ down to $e\simeq0.06$ for $(q_1,q_2)=(5,4)$, $v_1^2=1.86$ at $N=1024$.}
\label{tab:simparams_moore}
\end{table}

For completeness, we present the results of the fit of the Klaer--Moore string and gauge field profiles as discussed in Sec.~\ref{subsec:Klaer--Moore-model} for the parameter range of interest for this work in Tab.~\ref{tab:KM_fit_coefficients}.

\subsection{Renormalised Nambu--Goto--Kalb--Ramond Loop}
\label{sec:ngkr-evolution}
This section provides the details of the  \github{mathieukaltschmidt}{NGKR-circular-loop} module.
We solve Eq.~\eqref{eq:circle-renormalised} using the complete stored causal
history of the loop. At every observation time, the past-light-cone
condition Eq.~\eqref{eq:lightcone-function} determines the interval
$[t_\pi,t_\delta]$ and the retarded angle. The derivative in
Eq.~\eqref{eq:forceKR} is evaluated before imposing the light-cone condition,
so all explicit $\dot R(t_r)$ and $\ddot R(t_r)$ terms are retained.

\paragraph{Regulator and matching.}
The code keeps the regulator
$\delta_\sigma$ finite and implements it as the angular cutoff $|\varphi|>\delta_\varphi\equiv\delta_\sigma/R_0$.
The matched local coefficient and the running tension are
\begin{equation}
 \frac{\mu_0}{\pi v^2}=c_{\rm FT}+\log(4m_s\delta_\sigma),
 \qquad
 \bar\mu(\Delta)=\mu_0+\pi v^2
 \log \left( \frac{\Delta}{4 \delta_{\sigma}} \right),
 \label{eq:ngkr-code-matching}
\end{equation}
where in the code the logarithm is replaced by the exact form of the bracket in Eq.~\eqref{eq:muKR-exact}.
The code can evolve either the bare equation
\begin{equation}
 \mu_0(\delta_\sigma)\mathcal K=
 -\frac{2\pi v}{\gamma_R^3}H_{012}(R;\delta_\sigma)
 \label{eq:ngkr-code-bare}
\end{equation}
or the algebraically identical finite-$\Delta$ form
\begin{equation}
 \bar\mu(\Delta)\mathcal K=
 -\frac{2\pi v}{\gamma_R^3}H_{012}
 +\pi v^2 \log \left( \frac{\Delta}{4 \delta_{\sigma}} \right) \mathcal K.
 \label{eq:ngkr-code-finite-Delta}
\end{equation}
In practice both formulations agree to $10^{-5}$, and varying $\Delta$ leaves the trajectories unchanged. The imposed cutoff needs to be
resolved, $\delta_\sigma/R_0\gtrsim\Delta t$, and therefore the cost of a run scales as
$(R_0/\delta_\sigma)^2\propto e^{2\log(m_sR_0)}$.

\begin{table}[t]
\centering
\renewcommand{\arraystretch}{1.2}
\setlength{\arrayrulewidth}{1pt}
\setlength{\tabcolsep}{10pt}
\small
\begin{tabular}{c | c c c | c c c | c c}
\hline \hline
 & \multicolumn{3}{c}{$f_1$} & \multicolumn{3}{c}{$f_2$} & \multicolumn{2}{c}{$A_\varphi/A_\varphi(\infty)$} \\
\cline{2-9}
$\mathcal{T}_{\rm core}$ & $p_1$ & $p_2$ & $p_3$ & $p_1$ & $p_2$ & $p_3$ & $b$ & $c$ \\
\hline
11.27 & $0.6188$ & $0.0555$ & $0.0040$ & $0.4994$ & $0.0128$ & $-0.0747$ & $1.2929$ & $0.5788$ \\
16.28 & $0.6166$ & $0.0545$ & $0.0250$ & $0.4938$ & $0.0127$ & $-0.0729$ & $1.3037$ & $0.5955$ \\
26.28 & $0.6201$ & $0.0561$ & $0.0070$ & $0.5473$ & $0.0320$ & $-0.0277$ & $1.3107$ & $0.6092$ \\
36.29 & $0.6166$ & $0.0541$ & $0.0258$ & $0.5270$ & $0.0181$ & $-0.0672$ & $1.3143$ & $0.6143$ \\
46.26 & $0.6144$ & $0.0522$ & $0.0278$ & $0.5232$ & $0.0164$ & $-0.0714$ & $1.3164$ & $0.6177$ \\
56.26 & $0.6176$ & $0.0548$ & $0.0246$ & $0.5652$ & $0.0419$ & $\phantom{-}0.0021$  & $1.3177$ & $0.6203$ \\
66.28 & $0.6155$ & $0.0530$ & $0.0271$ & $0.5597$ & $0.0393$ & $-0.0063$ & $1.3187$ & $0.6216$ \\
76.43 & $0.6142$ & $0.0519$ & $0.0279$ & $0.5563$ & $0.0374$ & $-0.0120$ & $1.3194$ & $0.6227$ \\
86.41 & $0.6172$ & $0.0544$ & $0.0253$ & $0.5792$ & $0.0464$ & $\phantom{-}0.0179$  & $1.3199$ & $0.6239$ \\
96.39 & $0.6155$ & $0.0530$ & $0.0271$ & $0.5756$ & $0.0455$ & $\phantom{-}0.0144$  & $1.3203$ & $0.6245$ \\
\hline \hline
\end{tabular}
\caption{Fit coefficients of the Klaer--Moore string profiles for the parameter range relevant for this work.}
\label{tab:KM_fit_coefficients}
\end{table}

\paragraph{Time stepping.}
The retarded force depends only on accepted history. We first predict the
new $(R,\dot R)$, evaluate the history integral with the predicted endpoint,
and then correct the step. In particular, the code does not infer a new
instantaneous inertia by varying the observation acceleration. Such a
procedure would tie the UV matching to the discontinuity in $\ddot R$ when
the external force holding the initial loop is released. Instead, the Wilson
coefficient is fixed once through Eq.~\eqref{eq:ngkr-code-matching}; retarded
accelerations remain inside the causal integral. The retarded integral is evaluated with the trajectory interpolated to the light-cone endpoints, and the integrable square-root caustic at the antipodal endpoint is integrated analytically after fitting the integrand to $A/\sqrt{s}+B\sqrt{s}+Cs^{3/2}$, with $s=t_r-t_{\rm min}$. The loop is released abruptly from rest at $R=R_0$ from a static history. Collapse times are obtained by extending the trajectory linearly from the
crossing of $R=10^{-3}R_0$ to $R=0$, cf.\ App.~\ref{app:collapse_times}.

\paragraph{Small and intermediate tensions ($\mathcal{T}< 10$).}
At finite $\delta_\sigma$ the trajectories receive corrections of order $\sim(\delta_\sigma/R)$, see Eq.~\eqref{eq:muKR-exact}. For each $\log(m_sR_0)< 10$ we run the code at four values\footnote{At $\log(m_sR_0)=9$, $\{1.5,2,3,4\}$, to keep
$\delta_\sigma/R_0\ge2.5\,\Delta t$.}
$m_s\delta_\sigma\in\{0.75,1,1.5,2\}$ with $\Delta t=5\times10^{-5}R_0$ and extrapolate
$\mathcal T_{1/2}$, $\mathcal T_{\rm coll}$ and $R(t)$ up to quadratic order in $m_s\delta_\sigma$ to
$\delta_\sigma\to0$ at fixed $m_sR_0$. Relative to $m_s\delta_\sigma=1$, we find corrections to
$\mathcal T_{1/2}$ of $0.032$ for $\log(m_sR_0)=3$ to $\le9\times10^{-4}$ for $\log(m_sR_0)=6$ (and larger). As a cross-check we ran the same simulations with the smallest regulator at $\Delta t/2$ and find that $\mathcal T_{1/2}$ changes by $\lesssim 10^{-4}$, which we include as an uncertainty.

\paragraph{Large tensions ($\mathcal{T}\geq 10$).}
Since the dynamics depend on the UV only through the sum $\mathcal T=c_{\rm FT}+\log(m_sR_0)$, up to power
corrections in $\delta_\sigma/R$, we can equivalently reach $\mathcal T\ge10$ at fixed $\log(m_sR_0)\equiv X$ by setting $c_{\rm FT}=\mathcal T-X$.

We perform calculations for $X=7$
(with $\delta_\sigma/R_0\simeq10^{-3}$) at two different temporal resolutions, $\Delta t/R_0=5\times10^{-5}$ and $2.5\times10^{-5}$, and perform a linear extrapolation to
$\Delta t\to0$. As a cross-check, setting $X=8$ instead of
$X=7$ we find agreement to $0.06\%$ for the halving time $\mathcal T_{1/2}$. The comparison with the field-theory simulations is presented in Sec.~\ref{sec:compa}.

\subsection{Prepropagation}
\label{app:ic-relaxation}

The auxiliary evolutions described in Sec.~\ref{subsec:stringICs} are used
only to prepare the initial state. Their purpose is to reduce the energy at
fixed loop topology and position, not to approximate the physical collapse. Here we specify the diagnostics used to decide when a configuration is relaxed.

For the complex scalar model we monitor the static energy
\begin{equation}
 E_{\mathrm{static}}=\frac{v^2}{2}\int \dd^3x\left[
 (\boldsymbol\nabla f)^2+f^2(\boldsymbol\nabla\theta)^2
 +\frac{m_s^2}{4}(f^2-1)^2\right],
 \label{eq:relaxation-static-energy}
\end{equation}
which decreases under the damped evolution until it reaches a plateau. We quantify saturation over one relaxation interval $\Delta t_{\rm rel}$ through
\begin{equation}
 \Delta_E(t)=
 \frac{|E_{\mathrm{static}}(t+\Delta t_{\mathrm{rel}})-E_{\mathrm{static}}(t)|}
 {E_{\mathrm{static}}(t)}.
 \label{eq:relaxation-energy-saturation}
\end{equation}
The analogous diagnostic in the Klaer--Moore model includes the scalar-gradient,
potential, electric, and magnetic energies. During the de Sitter preparation
stage these energies are evaluated in the corresponding physical volume and we
again stop only after the localised core contribution has reached a plateau.

\begin{figure}[t]
    \centering
\includegraphics[width=0.49\linewidth]{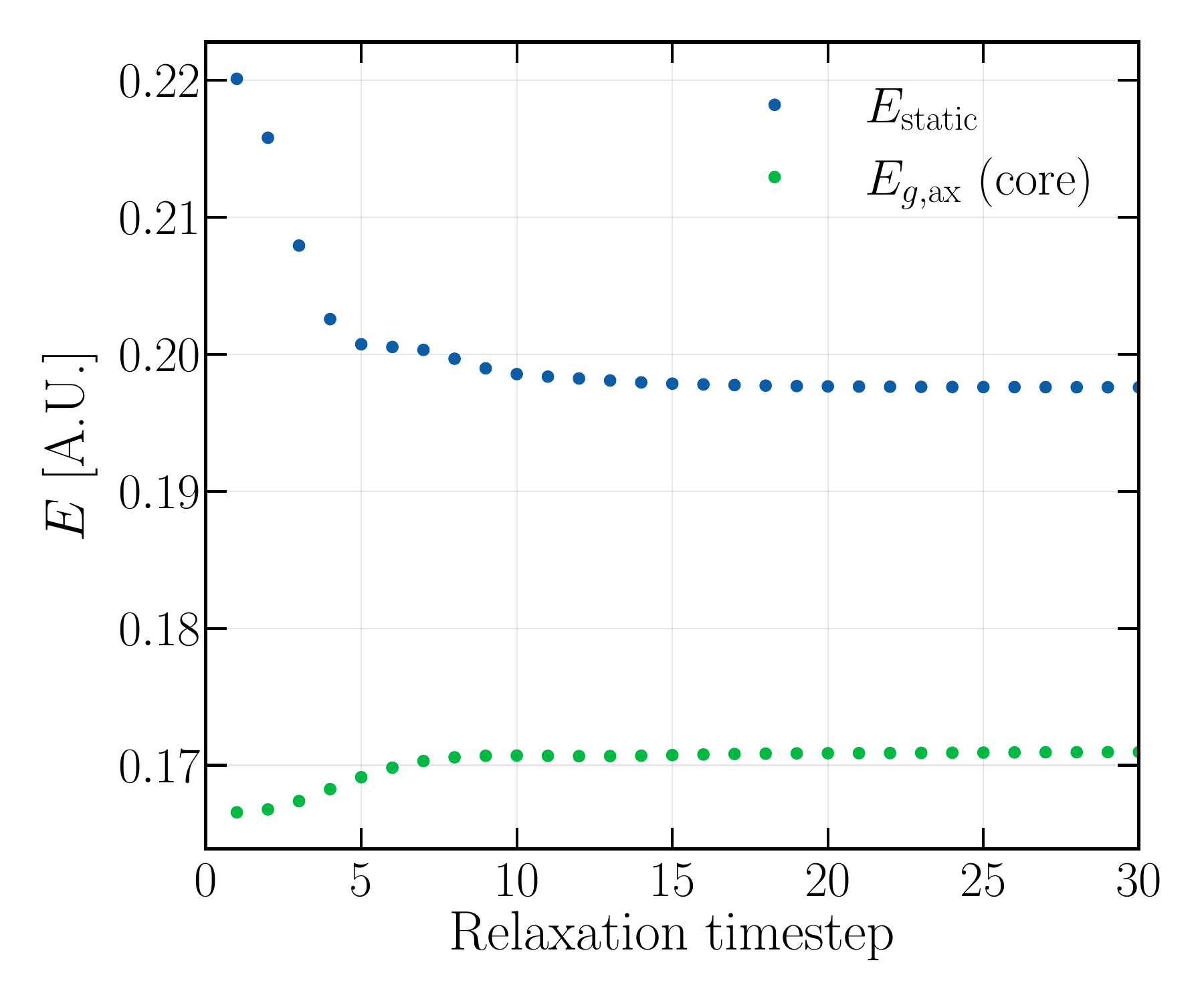}
\includegraphics[width=0.49\linewidth]{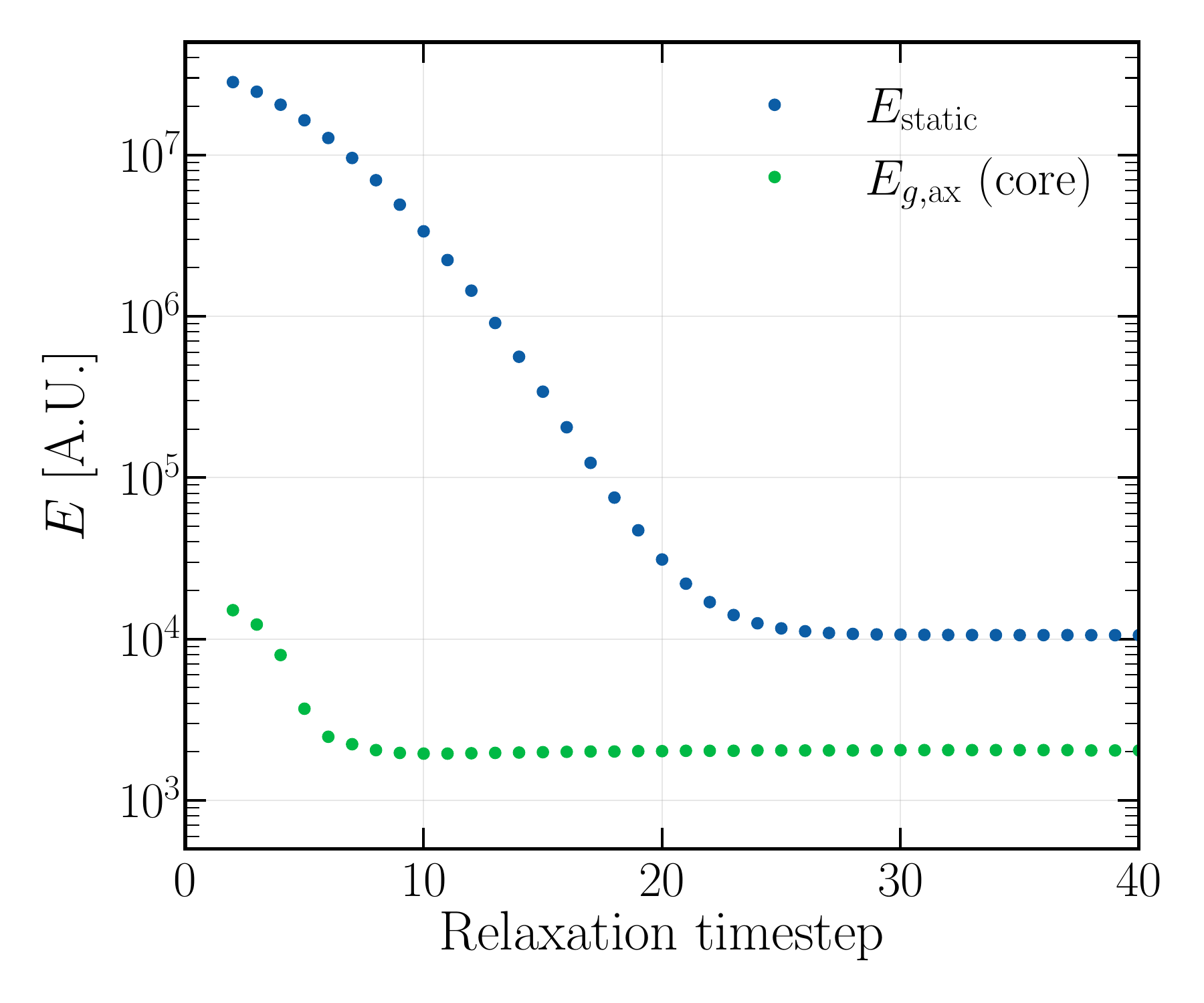}

    \caption{Relaxation of the initial conditions. \emph{Left:} damped relaxation of typical \texttt{jaxions} initial conditions. \emph{Right:} \texttt{Moore} prepropagation in the de Sitter stage. Shown are the static energy of Eq.~\eqref{eq:relaxation-static-energy} and the axion gradient energy in the core region. Both quantities saturate far before the last relaxation timestep.}
    \label{fig:relaxation}
\end{figure}

A complementary local test is supplied by the residuals of the two
constrained static equations,
\begin{align}
 \mathcal R_f&=\nabla^2f-f(\boldsymbol\nabla\theta)^2
 -\frac{m_s^2}{2}f(f^2-1),\\
 \mathcal R_\theta&=\boldsymbol\nabla\cdot
 \left(f^2\boldsymbol\nabla\theta\right).
 \label{eq:relaxation-residuals}
\end{align}
We require their core-weighted norms to stabilise together with the energy.
After switching off the artificial damping or expansion, a well-relaxed
configuration begins the physical evolution smoothly: the field velocities
contain no relaxation transient and the field accelerations are small apart
from the genuine curvature-driven acceleration of the loop. Energy
saturation and the absence of a short initial burst therefore provide two
independent checks of the initial conditions. Throughout the relaxation the loop radius is conserved to better than one percent and the loop velocity is damped to a value consistent with zero, so that the released configuration is the intended loop at rest, see Fig.~\ref{fig:relaxation}. Note that in \texttt{jaxions} the string core is initialised according to the profile of Eq.~\eqref{eq:solution_field}, whereas in \texttt{Moore} one has $f_{1,2} \approx 1$ everywhere at the initial time, and the gauge fields must build up during the initialisation procedure. This can explain why we need shorter and less relaxation for the \texttt{jaxions} initial conditions.

\renewcommand{\theequation}{B.\arabic{equation}}
\setcounter{equation}{0}

\section{Collapse Scales and Continuum Resolution}
\label{app:collapse_scalings}

This appendix collects the quantitative measurements underlying the physical
picture in Sec.~\ref{subsec:collapse_anatomy}.  We simulate loops with several
values of $m_sR_0$ and repeat each case at multiple lattice spacings and refinements.  Observables are extrapolated to the continuum following
Sec.~\ref{subsec:continuum}, by fitting the quadratic correction in the
lattice spacing of Eq.~\eqref{eq:cont_model_truncated} using at least three
resolutions whenever available.  The plotted error bars include the uncertainty of this
extrapolation; widths are extracted from the finest resolved profiles.

\begin{figure}[t]
\centering
\includegraphics[width=0.49\linewidth]{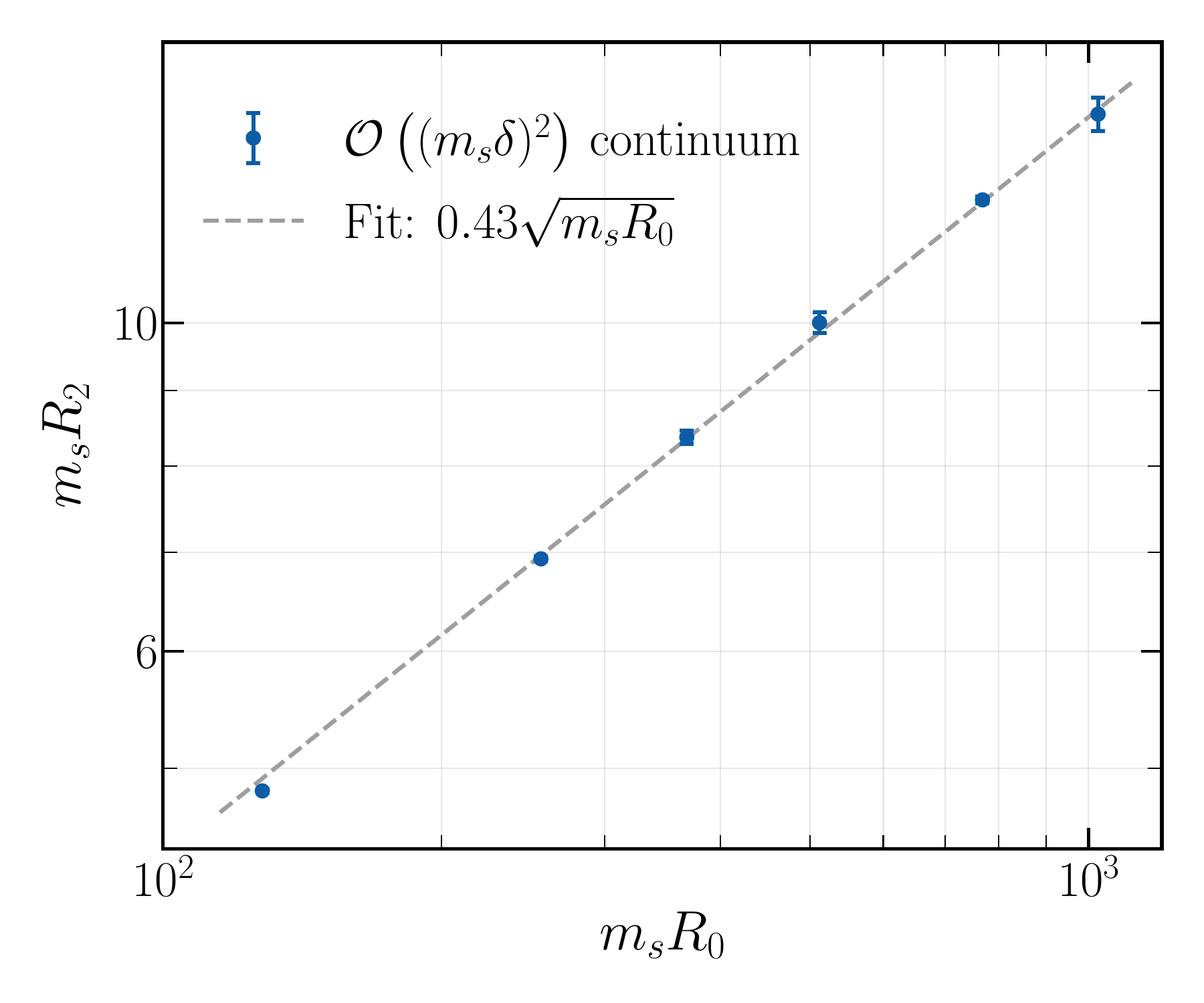}
\includegraphics[width=0.49\linewidth]{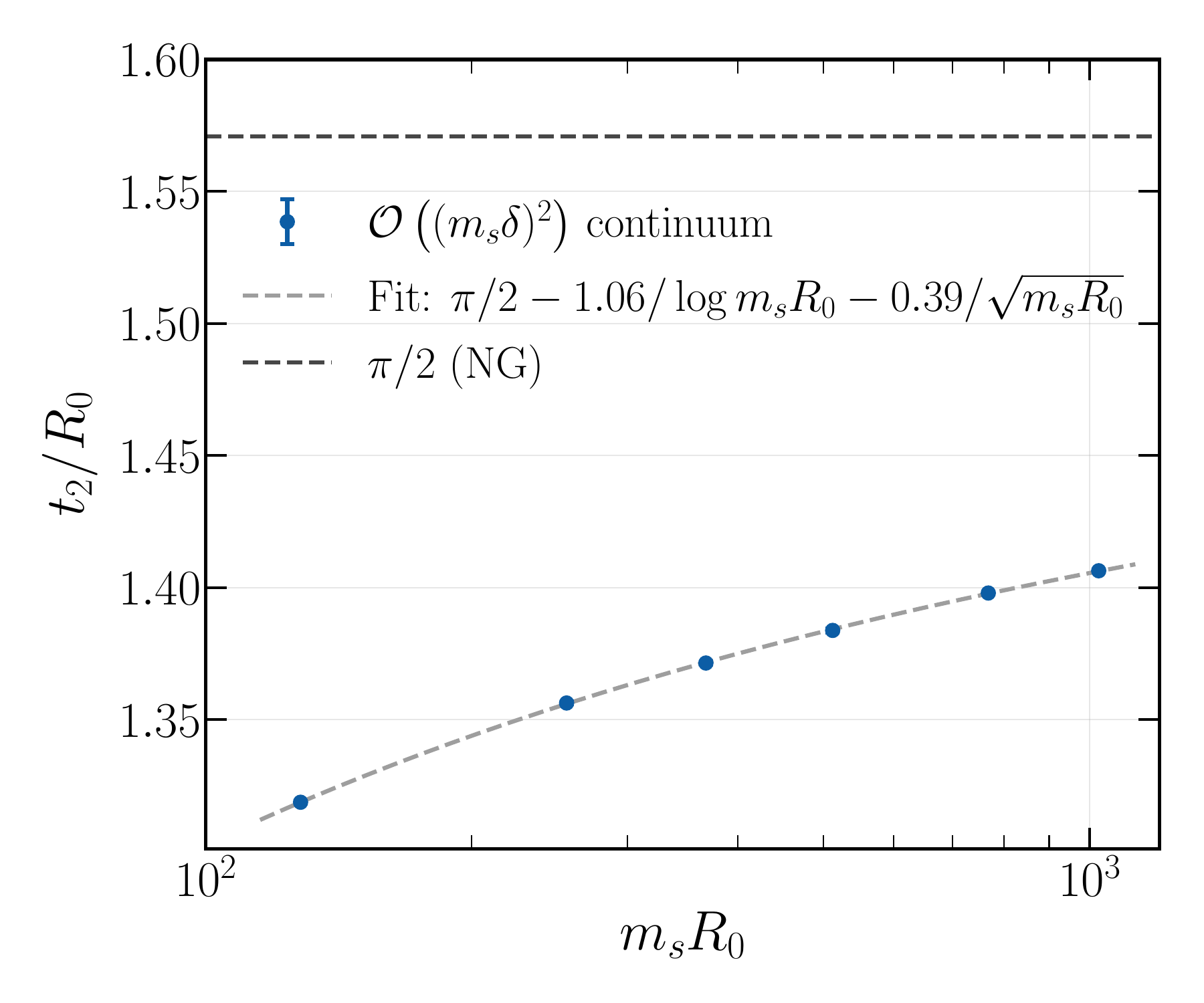}
\caption{Radius $R_2$ at which the trailing radial-field crest first reaches
$\mathrm{Re}\,\phi/v=2$ (\emph{left}), and the corresponding time $t_2$ (\emph{right}),
as functions of $m_sR_0$.}
\label{fig:r2t2fits}
\end{figure}

We define $R_2$ as the radius of the outermost string zero when the trailing
radial-field crest first reaches $\mathrm{Re}\,\phi/v=2$, and $t_2$ as the
corresponding time.  The measurements in Fig.~\ref{fig:r2t2fits} give
Eqs.~\eqref{eq:R2_scaling} and \eqref{eq:t2_scaling}.  In particular,
$R_2\propto\sqrt{R_0/m_s}$ confirms the acceleration estimate
of Eq.~\eqref{eq:nonlinear_radius_estimate}.  The $1/\log(m_sR_0)$ correction
to $t_2$ describes the slowly decoupling global self-field, while the
$1/\sqrt{m_sR_0}$ term accounts for the finite radius at which nonlinear
evolution begins.

\begin{figure}[t]
\centering
\includegraphics[width=0.49\linewidth]{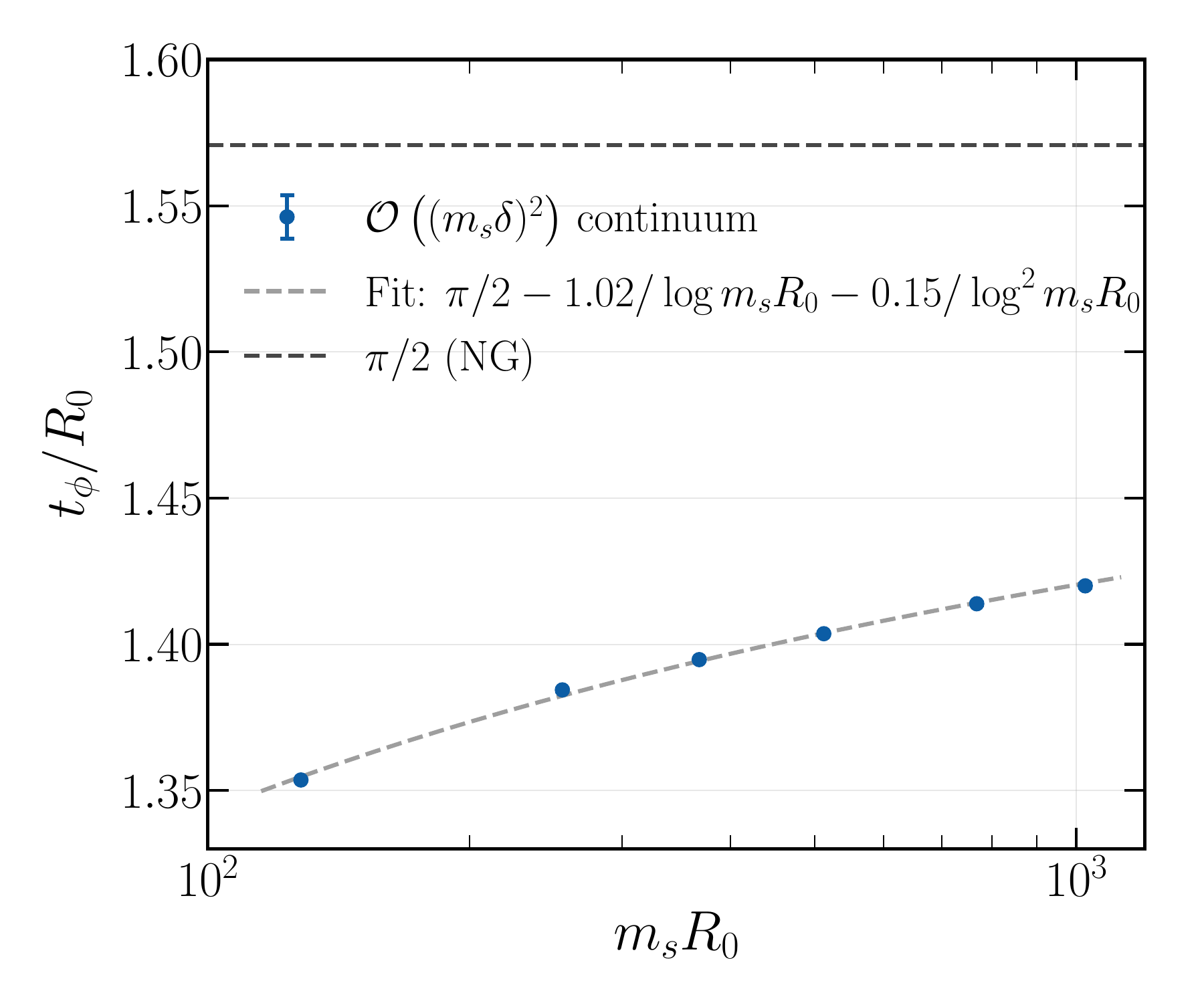}
\includegraphics[width=0.49\linewidth]{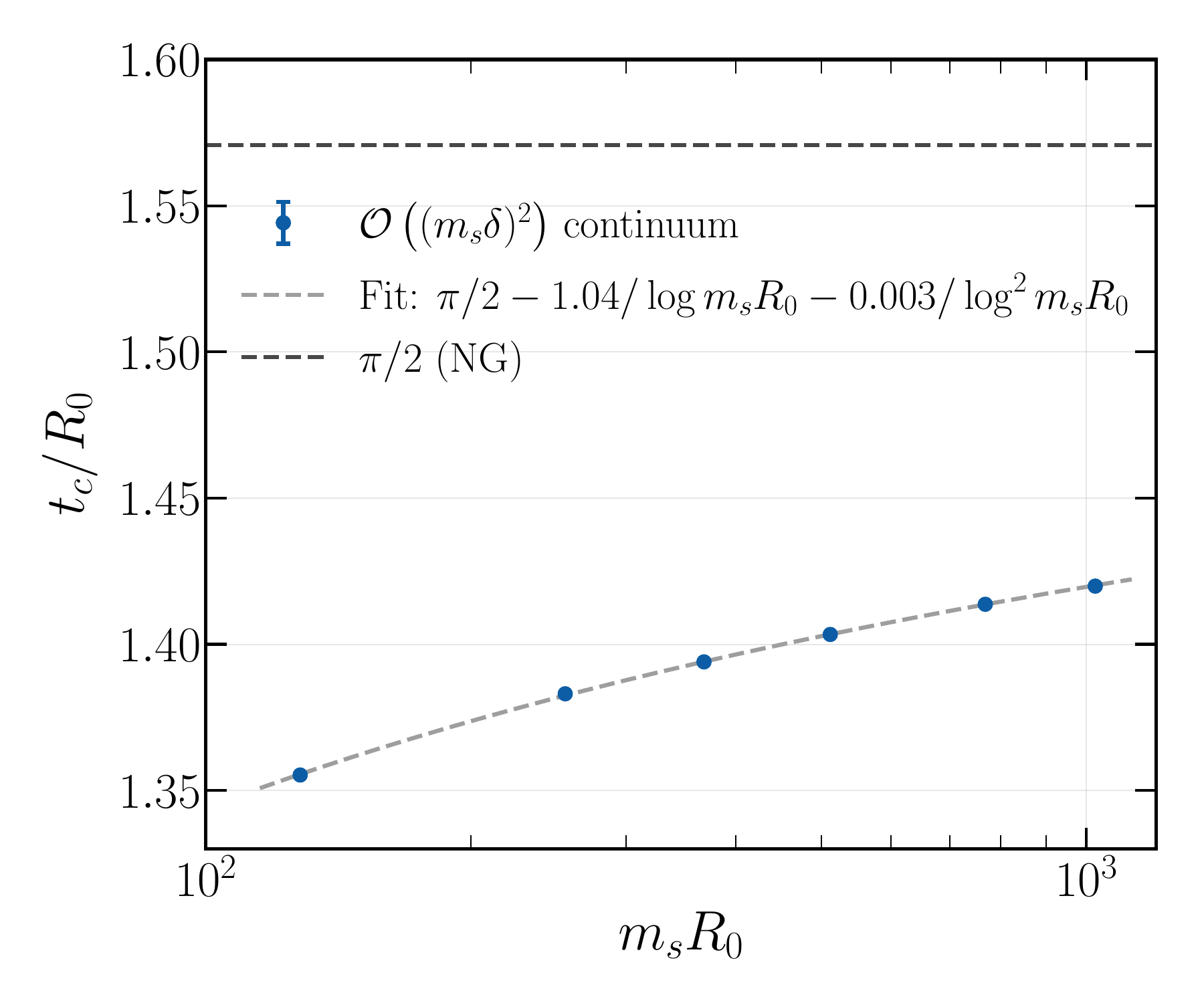}
\caption{Focus times defined by the maximum central field amplitude
(\emph{left}) and by the maximum central energy density (\emph{right}), as
functions of $m_sR_0$, together with the fits of
Eqs.~\eqref{eq:t_field_focus} and~\eqref{eq:t_energy_focus}. Both approach
the NG value $\pi R_0/2$ (dashed) from below and agree with each
other to better than $0.2\%$. We adopt the latter as the focus time $t_c$
used in the main text.}
\label{fig:tcollapses2D}
\end{figure}

Once the evolution becomes nonlinear, several distinct events could be called
the \enquote{collapse}. Figure~\ref{fig:tcollapses2D} compares the time $t_{\phi}$ at which
$|\phi(0,0)|$ is maximal with the focus time $t_c$ of Sec.~\ref{subsec:collapse_anatomy}, at which the axion energy density at $\rho=z=0$ is maximal. Both approach the NG value $\pi R_0/2$ from below,
\begin{align}
 \frac{t_{\phi}}{R_0}&=\frac{\pi}{2}-\frac{1.02}{\log(m_sR_0)}
 -\frac{0.15}{\log^2(m_sR_0)},
 \label{eq:t_field_focus}\\[.5em]
 \frac{t_{c}}{R_0}&=\frac{\pi}{2}-\frac{1.04}{\log(m_sR_0)}
 -\frac{0.003}{\log^2(m_sR_0)},
 \label{eq:t_energy_focus}
\end{align}
with the expected slowly varying global-string correction. The two definitions
agree to better than $0.2\%$ across the whole range, and the energy-density definition is described almost entirely by a single $1/\log(m_sR_0)$ term. We adopt $t_c$ in the main text because it follows energy transport and is
insensitive to the superluminal motion or proliferation of field zeros.

\begin{figure}[t]
\centering
\includegraphics[width=0.49\linewidth]{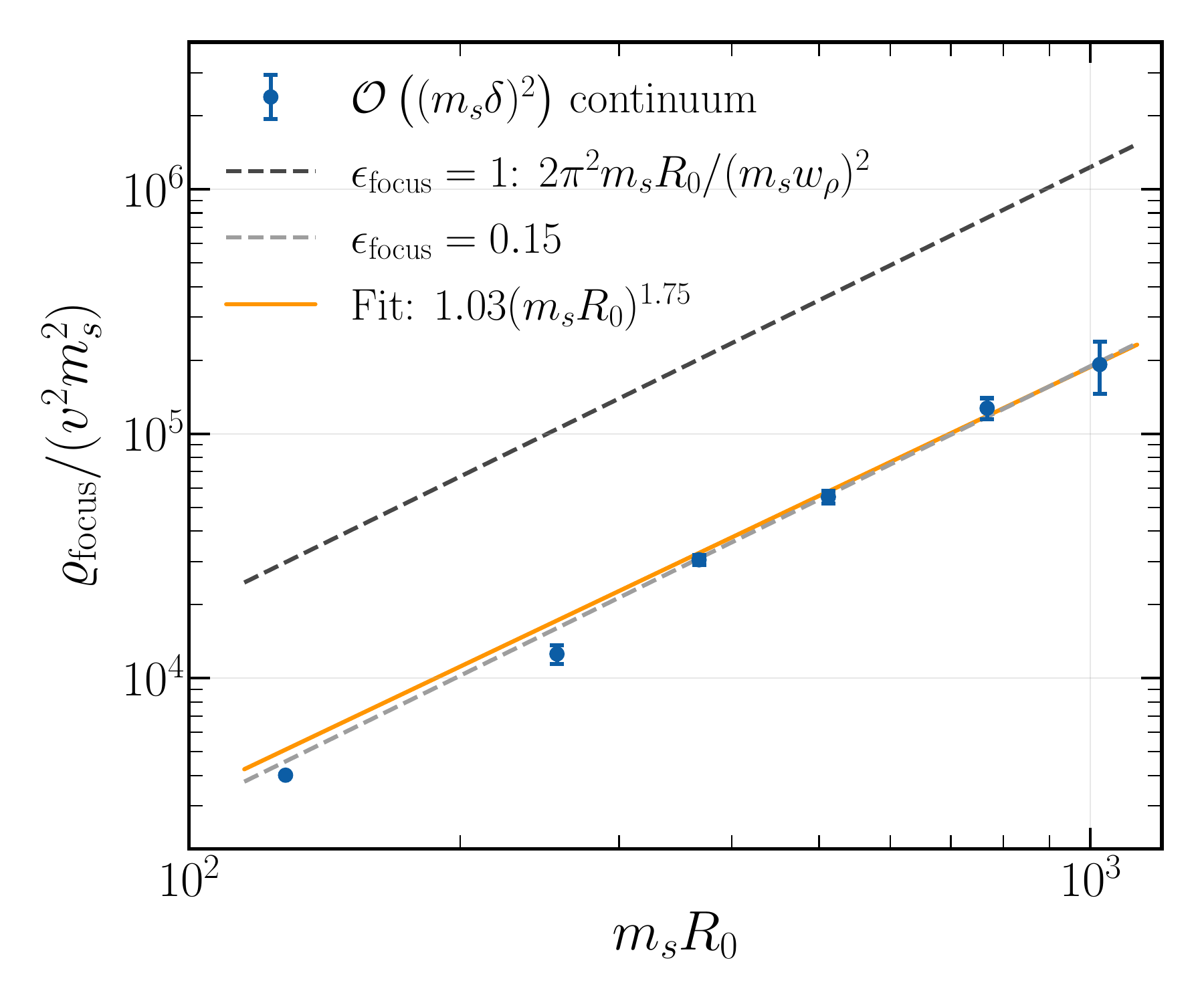}
\includegraphics[width=0.49\linewidth]{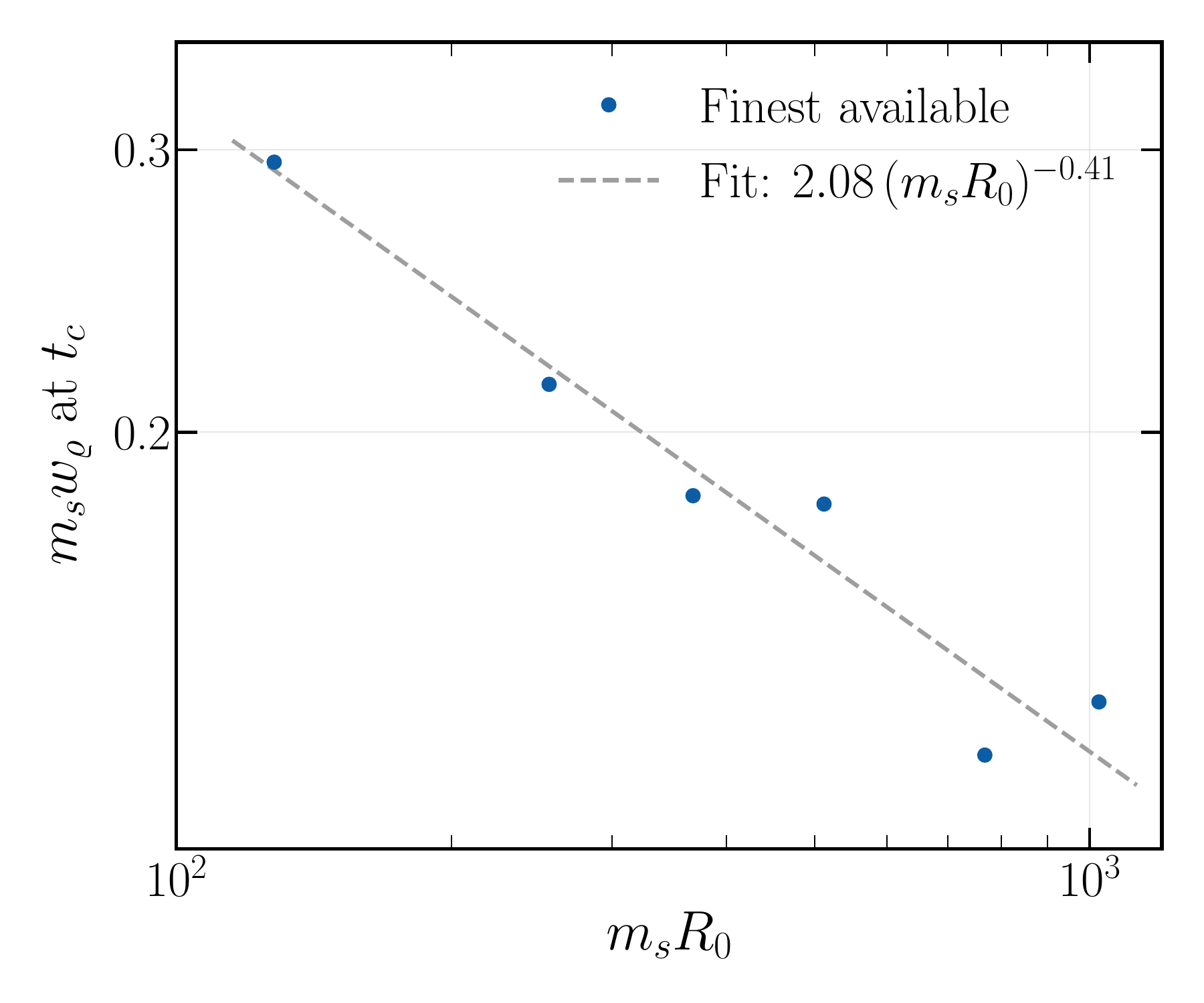}
\caption{Maximum central energy density at $t_c$ (\emph{left}) and the radial
FWHM $w_\varrho$ of the central energy-density peak at the same time
(\emph{right}), as functions of $m_sR_0$. The dashed lines in the left panel
show Eq.~\eqref{eq:focus_density_scaling} for $\epsilon_{\rm focus}=1$ and
$0.15$, evaluated with the fitted width of
Eq.~\eqref{eq:focus_width_scaling}; the solid line is the direct fit
Eq.~\eqref{eq:focus_density_fit}. Widths are taken from the finest available
resolution rather than extrapolated, and therefore the data is more
scattered.}
\label{fig:energyfocus}
\end{figure}

At $t_c$ we measure both the maximum central energy density
$\varrho_{\rm focus}$ and the radial FWHM $w_\varrho$ of its central peak; the results are shown in
Fig.~\ref{fig:energyfocus}.  Compressing the loop
self-field energy of Eq.~\eqref{eq:IC_gradient_energy} into a region of this
width gives the estimate
\begin{equation}
 \frac{\varrho_{\rm focus}}{v^2m_s^2}
 \simeq \epsilon_{\rm focus}\,\frac{2\pi^2\,m_sR_0}{(m_sw_\varrho)^2},
 \qquad \epsilon_{\rm focus}=0.15,
 \label{eq:focus_density_scaling}
\end{equation}
where $\epsilon_{\rm focus}$ is the fraction of that energy participating in
the central focus. The measured width follows
\begin{equation}
 m_sw_\varrho\simeq2.08\,(m_sR_0)^{-0.41},
 \label{eq:focus_width_scaling}
\end{equation}
so that Eq.~\eqref{eq:focus_density_scaling} predicts a density growing as
$(m_sR_0)^{1+2\times0.41}=(m_sR_0)^{1.82}$, consistent with the direct fit
\begin{equation}
 \frac{\varrho_{\rm focus}}{v^2m_s^2}\simeq1.03\,(m_sR_0)^{1.75}.
 \label{eq:focus_density_fit}
\end{equation}
The focus is thus substantially denser than a static straight-string core, but
its width follows the same parametric scale as the Lorentz-contracted core at
$R_2$, Eq.~\eqref{eq:NLcorewidth}; it is not parametrically narrower.

These convergence studies lead to Eq.~\eqref{eq:focus_resolution_requirement}.
When this condition is violated, the focus amplitude overshoots, excess
short-wavelength radiation is produced, and the post-focus string bounce can
be displaced or erased.  Resolving the scale already present at $R_2$ is
therefore sufficient in our simulations to obtain continuum-stable focusing
and subsequent expansion.

\renewcommand{\theequation}{C.\arabic{equation}}
\setcounter{equation}{0}

\section{Halving and Collapse Times from Simulations}
\label{app:collapse_times}

Here we describe the details of the analysis used to extract halving and collapse times for our comparison in Sec. \ref{sec:compa}.

For all simulations we restrict the analysis to the \emph{first} descent of the continuum curve, from $R(0)=R_0$ to its first minimum, which we identify as the point after which $R$ rises again by more than $0.02\,R_0$. Any subsequent rebound or re-collapse is discarded.

The halving time $t_{1/2}$ is obtained as the first crossing of $R=R_0/2$ on this descent, by linear interpolation between neighbouring measurements. At this stage the loop is still mildly relativistic, $\gamma_R\sim\mathcal O(1)$, and the string core is well resolved in all our simulations. 

The collapse time $t_{\rm coll}$ requires more care. First, the measured radius of the topological string (i.e. where $|\phi|=0$) does not reach $R=0$ in all of the simulations due to finite measurement intervals and the finite lattice resolution. Moreover, for $R<R_{\rm NL}\sim R_0/\sqrt{m_sR_0}$ the core becomes highly distorted and the topological string gets displaced from the peak of the energy density, distorting the whole concept of the string collapse and the non-linear core dynamics are not described by the thin-string effective theory. However, at this point, the loop energy and the topological string already move very close to the speed of light and we can define $t_{\rm coll}$ as an extrapolation from the last part of the descent. We employ $R\geq0.1\,R_0$ and extrapolate to $R=0$ with the following caveats:
\begin{itemize}
 \item If the minimum radius satisfies $R_{\min}\leq0.05\,R_0$ and the loop is still descending steeply, $\dd R/\dd t\leq-0.3$, at $0.1\leq R/R_0\leq0.3$, we fit a straight line to the descent in this interval and take its root. The fitted slopes lie between $-0.98$ and $-1.01$, as expected for an ultra-relativistic final collapse.
 \item Otherwise we fit a linear and a quadratic polynomial to the descent at \\ $\operatorname{max}\!\big(0.1,\,2/(m_sR_0)\big)\leq R/R_0\leq0.4$ and take the mean of their roots, with their difference as an additional model uncertainty $\sigma_0$.
\end{itemize}
In the first case, we attribute negligible error to the collapse time, $\sigma_0=0$. The lower limit $0.1\,R_0$ keeps the fits within the thin-string regime for all our runs. Including radii down to $0.05\,R_0$ instead lets the final approach to the core bias individual collapse times by up to $0.05\,R_0$, beyond the NG value in one case.

The NGKR trajectories, in contrast, always reach $R\lesssim10^{-3}R_0$. Their collapse times are
obtained by extrapolating the trajectory linearly from the crossing of $R=10^{-3}R_0$ to $R=0$, yielding a
correction of $\simeq10^{-3}R_0$, since $|\dot R|\to1$.

The uncertainties of the simulation results are propagated from the width
of the continuum band, cf. Eq.~\eqref{eq:band}, at the respective crossing through the
local slope of $R(t)$,
\begin{equation}
\label{eq:sigma_tc}
 \sigma_{t}=\sqrt{\left(\frac{\sigma_R}{|\dd R/\dd t|}\right)^{\!2}+\sigma_{0}^2},
 \qquad
 \sigma_R=\tfrac12\big(R_{\rm hi}-R_{\rm lo}\big),
\end{equation}
with both $\sigma_R$ and $\dd R/\dd t$ evaluated at $t=t_{1/2}$ or $t=t_{\rm coll}$, respectively. For $t_{\rm coll}$, $\dd R/\dd t$ is the slope of the fit. The halving-time uncertainties are much smaller, since the crossing is measured directly and $\sigma_0=0$.

\renewcommand{\theequation}{D.\arabic{equation}}
\setcounter{equation}{0}

\section{Validity and Limitations of the Effective Description}
\label{app:limitations}
To describe a cosmic string analytically we must assume that the string thickness is infinitely  small. We introduce coordinates adapted to the worldsheet, 
\begin{equation}
 x^{\mu}\equiv X^{\mu}(\sigma^{a})+n^{\mu}_{A}(\sigma^a)\,x_{\perp}^{A},
 \label{eq:app_adapted_coords}
\end{equation}
where $a=\{0,1\}$ label the tangent directions, $A=\{2,3\}$ the two orthogonal
directions, $n_A$ are orthonormal vectors perpendicular to the worldsheet, and
$\sigma^{a}$ parametrise the worldsheet as in Eq.~\eqref{eq:EFT_NG+KR}. 

The metric in these coordinates follows from
$\bar{g}_{\mu\nu}\equiv A_{\mu}^{\eta}A_{\nu}^{\rho}g_{\eta\rho}$ with
$A^{\mu}_{a}=\partial X^{\mu}/\partial\sigma^a
+x_{\perp}^A\,\partial n^{\mu}_{A}/\partial\sigma^a$ and
$A^{\mu}_{A}=n^{\mu}_A$, and reads (see Sec.~1.5 of
Ref.~\cite{Anderson:2003gg})
\begin{equation}
\begin{aligned}
\label{MetrNGCoord}
\bar{g}_{ab} &= \gamma_{ab}+\left[2K_{A\,ab}x_{\perp}^A\right]
+\left[K^{c}_{A\,a}K_{B\,bc}x_{\perp}^Ax_{\perp}^B
-\omega_a\omega_b\left(x_{\perp}^A\right)^2\right],\\[.5em]
\bar{g}_{aA} &= \omega_a\varepsilon_{AB}x_{\perp}^B,\\[.5em]
\bar{g}_{AB} &= -\delta_{AB},
\end{aligned}
\end{equation}
where $\gamma_{ab}\equiv X^{\mu}_{,a}X^{\nu}_{,b}g_{\mu\nu}$ is the induced
metric of Eq.~\eqref{eq:EFT_NG+KR}, $X_{,a}\equiv\partial X/\partial\sigma^a$,
$K_{A\,ab}\equiv n^{\mu}_{A\,,a}X^{\nu}_{,b}g_{\mu\nu}$ is the extrinsic
curvature, $\omega_a=n^{\mu}_{2}n^{\nu}_{3\,,a}g_{\mu\nu}$ the normal
fundamental form, and $\varepsilon_{AB}$ the Levi-Civita tensor with
$\varepsilon_{23}=1$. 
Quantities tangent to the worldsheet vary over the local curvature radius
$\mathcal R$, whereas the perpendicular coordinates $x_{\perp}^A$ vary over
the core scale $m_s^{-1}$. The terms in Eq.~\eqref{MetrNGCoord} are therefore
of the orders \cite{Anderson:1997ip}
\begin{equation}
\begin{aligned}
 \gamma_{ab}&\sim\mathcal{O}(1), \\[.5em]
 K_{A\,ab}x^A_{\perp}&\sim\mathcal{O}\!\left((m_s\mathcal R)^{-1}\right),\\[.5em]
 \omega_a\varepsilon_{AB}x_{\perp}^B&\sim
 \mathcal{O}\!\left((m_s\mathcal R)^{-1}\right),\\[.5em]
 K^{c}_{A\,a}K_{B\,bc}x_{\perp}^Ax_{\perp}^B&\sim
 \mathcal{O}\!\left((m_s\mathcal R)^{-2}\right), \\[.5em]
 \omega_a\omega_b\left(x_{\perp}^A\right)^2&\sim
 \mathcal{O}\!\left((m_s\mathcal R)^{-2}\right),
\end{aligned}
\label{eq:app_orders}
\end{equation}
with the same expansion parameter as in Sec.~\ref{sec:eft}. For
$m_s\mathcal R\gg1$ the invariant volume element is
\begin{equation}
\label{NGcoordinates}
 \dd^4x\,\sqrt{-g}=\dd^2\sigma\,\dd^2x_{\perp}\,\sqrt{-\gamma}
 +\mathcal{O}\!\left((m_s\mathcal R)^{-1}\right).
\end{equation}
Applying Eq.~\eqref{NGcoordinates} to the action of
Eq.~\eqref{eq:lagrangianJ}, the leading contribution is
\begin{equation}
\label{NG_KR}
 S\approx-\mu_0\int\dd^2\sigma\,\sqrt{-\gamma}
 + \frac{1}{12}\int\dd^4x\,\frac{H^2}{f^2}
 +\pi v\int\dd^4x\,B_{\mu\nu}J^{\mu\nu},
\end{equation}
where $\mu_0$ and the sign of the $H^2$ term follow the conventions of
Eq.~\eqref{eq:EFT_NG+KR}.

The current $J^{\mu\nu}$ is fixed by the winding of the phase. Integrating
around the string core gives, for unit winding \cite{Vilenkin:1986ku},
\begin{equation}
\label{Winding}
 \int\dd^2x_{\perp}\,\varepsilon^{\mu\nu}\nabla_{\mu}\nabla_{\nu}\theta=2\pi,
\end{equation}
so that, with the definition in Eq.~\eqref{eq:Rsource},
\begin{equation}
 \int\dd^4x\,J^{\mu\nu}
 =\frac{1}{2\pi}\int\dd^2x_{\perp}\dd^2\sigma\,
 \varepsilon^{\lambda\mu\nu\rho}\nabla_{\lambda}\nabla_{\rho}\theta
 \approx \int\dd\sigma^{\mu\nu}
 +\mathcal{O}\!\left((m_s\mathcal R)^{-1}\right),
 \label{eq:app_current_integrated}
\end{equation}
where $\dd\sigma^{\mu\nu}\equiv X^{\mu}_{,a}X^{\nu}_{,b}\varepsilon^{ab}
\dd^2\sigma$ is the directed area element of Eq.~\eqref{eq:EFT_NG+KR}. Within
the thin-string approximation the current is therefore given by Eq.~\eqref{eq:worldsheet-current}.

The term $H^2/f^2$ in Eq.~\eqref{NG_KR} is finite when integrated over the string core, since $H$ vanishes there together with $f$. Setting $f\to1$ everywhere, as required for an infinitely thin string, reintroduces a UV divergence which must be regulated with a cutoff at a transverse radius $\delta_\rho$ chosen such that
\begin{equation}
\label{UVcut}
 \int^{r_{\rm cut}}_{0}\dd^4x\,\frac{H^2}{f^2}
 =\int^{r_{\rm cut}}_{\delta_\rho}\dd^4x\,H^2\big|_{f=1}.
\end{equation}
This $\delta_\rho$ is the transverse cutoff, which is not in general equal to the worldsheet
regulator $\delta_\sigma$ of Sec.~\ref{sec:eft}, nor should it be
identified with a physical core radius.
With these definitions, the effective action is given by (for the massive Kalb-Ramond case, see \cite{Rybak:2023bky})
\begin{equation}
\label{EFT_NG+KR_app}
 S_{\mathrm{EFT}}=-\mu_0\int\dd^2\sigma\sqrt{-\gamma}
 +\frac{1}{12}\int\dd^4x\sqrt{-g}\,H^2
 +\pi v\int\dd\sigma^{\mu\nu}\,B_{\mu\nu},
\end{equation}
evaluated with $f=1$ and with the last integral excluding a region of size
$\delta_\rho$ around the string core, which is
Eq.~\eqref{eq:EFT_NG+KR}.

\paragraph{Initial times.}
The physical scenario we consider is an initially stationary
circular loop whose long-range axion field is the constrained static solution
for a prescribed ring.  We then ``release'' the loop, allowing it to collapse
under its own tension and the self-interaction mediated by the KR
field.  Initially stationary, approximately relaxed circular loops are widely
used in field-theory simulations, see Refs.~\cite{Hagmann:1990mj,Hagmann:2000ja,Gorghetto:2018myk}, although their detailed
constructions differ as reviewed in Sec.~\ref{sec:simu} and App.~\ref{app:context}; a related release
prescription is used in the effective-model calculations of Ref.~\cite{Dabholkar:1989ju}.  This language should not be taken to imply that an unconstrained static circular-loop solution exists.  Rather, the string
position is prescribed while the exterior field is constructed or relaxed.
Consequently, a substantial part of the KR field that subsequently
influences the evolution belongs to a constrained static configuration which,
once released, does not satisfy the freely evolving loop equation of motion.
This limitation is specific to such release initial data and
does not apply in the same way to loops produced dynamically from a string
network or through string collisions, as for example in
Ref.~\cite{Baeza-Ballesteros:2023say}.

\paragraph{Curvature corrections for the circular loop.}
For the circular loop of Eq.~\eqref{eq:loop}, two unit vectors orthogonal to
the worldsheet may be chosen as \cite{Anderson:1997ip}
\begin{equation}
\begin{gathered}
 n_2^{\mu}=\left\{\dot R,\,\cos(\varphi),\,\sin(\varphi),\,0\right\}
 /\sqrt{1-\dot R^2}, \qquad n_3^{\mu}=\left\{0,\,0,\,0,\,1\right\},
\end{gathered}
\label{eq:app_normals}
\end{equation}
leaving a single non-trivial extrinsic curvature with diagonal entries
\begin{equation}
 K_{2\,aa}=\left\{\ddot R,\,-R\right\}/\sqrt{1-\dot R^2}
 \quad\text{(no summation, $a=0,1$)},
 \label{eq:app_K2}
\end{equation}
so that the leading correction to Nambu--Goto, proportional to
$K_{2\,ab}K_2^{ab}$, is
\begin{equation}
 K_{2\,ab}K_2^{ab}=\frac{\left(1-\dot R^2\right)^2+R^2\ddot R^2}
 {R^2(1-\dot R^2)^3},
 \label{eq:app_curvature_correction}
\end{equation}
from which one can estimate the curvature radius,
\begin{equation}
\mathcal{R} = \sqrt{\frac{2}{K_{2\,ab}K_2^{ab}}}.
\label{eq:app_curvature_radius}
\end{equation}
The quantity $K_{2\,ab}K_2^{ab}$ given by Eq.~\eqref{eq:app_curvature_correction} grows as $R$ decreases or $|\ddot R|$ increases. Evaluated on the NG trajectory of Eq.~\eqref{eq:NG_circular_trajectory} it reduces to
$2\gamma_R^4/R_0^2$, so the local curvature radius scales as $\mathcal R\sim R_0/\gamma_R^2$ and the expansion parameter $(m_s\mathcal R)^{-1}$ of Eq.~\eqref{eq:app_orders} grows accordingly during the collapse. Furthermore, starting from a static loop and switching to dynamical evolution produces a large $|\ddot R|$ at the moment of release. If it becomes sufficiently large, higher-order curvature terms might be relevant.

\begin{figure*}[t]
 \centering
 \includegraphics[width=.65\linewidth]{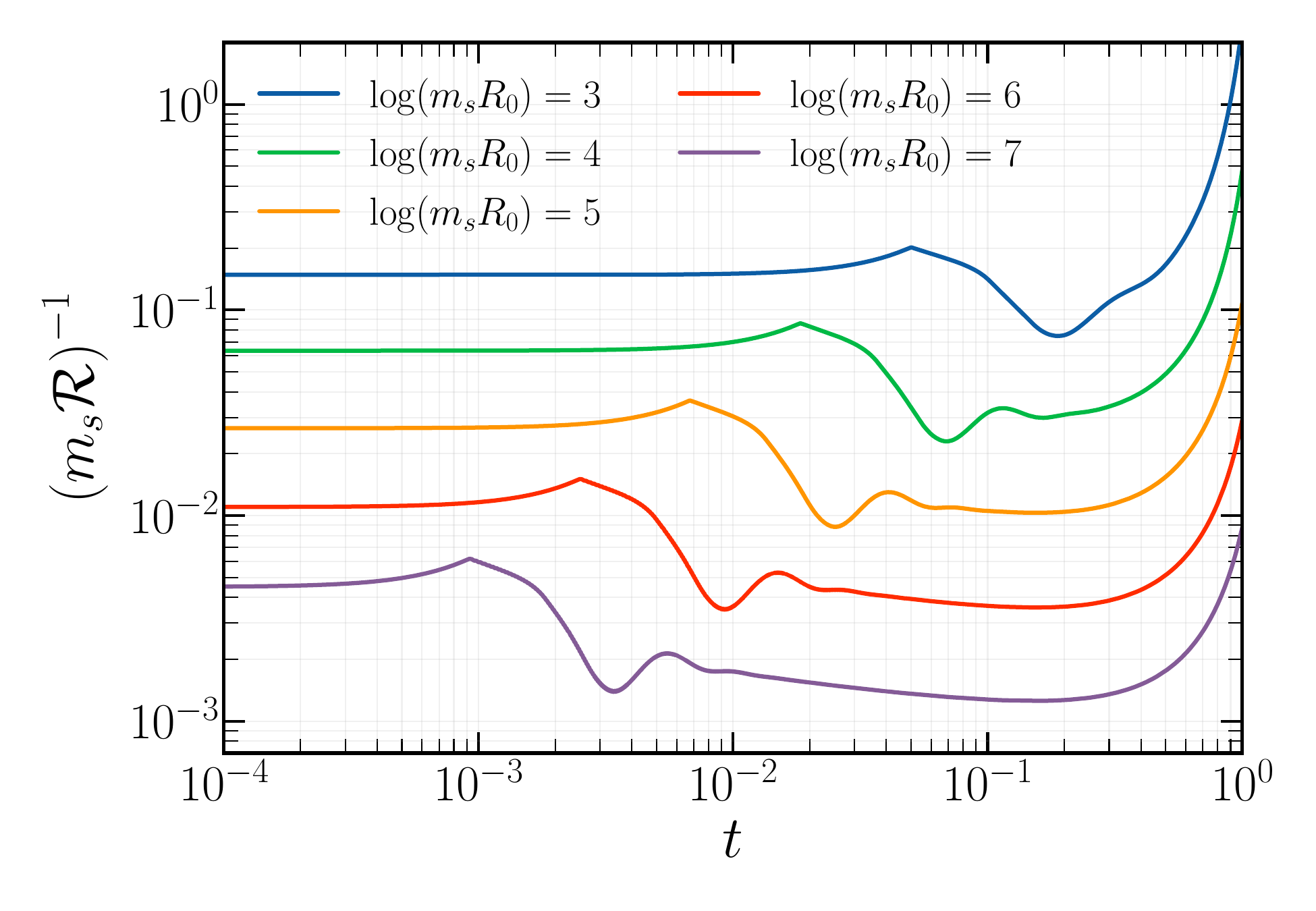}
 \caption{Time evolution of the expansion parameter for corrections to the Nambu--Goto action evaluated from Eqs.~\eqref{eq:app_curvature_correction} and~\eqref{eq:app_curvature_radius} for NGKR loop trajectories with different values of $\mathcal{T}(R_0)\approx\log(m_sR_0)$.}
 \label{fig:curvature_correction}
\end{figure*}

In Fig.~\ref{fig:curvature_correction}, we show the expansion parameter $(m_s\mathcal{R})^{-1}$ along the NGKR loop trajectories obtained by numerically solving Eq.~\eqref{eq:circle-renormalised}. We see that $(m_s\mathcal{R})^{-1}$ is relatively large shortly after the release ($t \lesssim 0.1$) and grows as the loop approaches collapse at late times. Nevertheless, its overall magnitude is suppressed by a factor of $(m_sR_0)^{-1}$. The figure shows that $(m_s\mathcal{R})^{-1}$ remains below 0.1 for most of the evolution when the tension satisfies $\mathcal{T}(R_0) \approx \log(m_sR_0) \gtrsim 4$. We therefore expect higher-order terms to be safely neglected in our analysis.

\renewcommand{\theequation}{E.\arabic{equation}}
\setcounter{equation}{0}

\section{Leading Inverse-tension Correction to the Halving Time}
\label{app:halving_coefficient}

In this appendix we derive the coefficient $A$ in Eqs.~\eqref{eq:comparison_A_green} and~\eqref{eq:comparison_A_eft_direct} controlling the leading
large-tension correction to the halving time. 
\subsection{Derivation}
We normalise lengths and times by $R_0$. Throughout this appendix, the
dimensionless tension without an explicit argument denotes its value at the initial 
loop scale,
\begin{equation}
 \mathcal T\equiv\mathcal T(R_0)
 =\frac{\bar\mu(R_0)}{\pi v^2}
 =c_{\rm FT}+\log(m_sR_0).
 \label{eq:app_tension_parameter}
\end{equation}
At subtraction scale $\Delta=R_0$, the renormalised circular-loop equation is
therefore
\begin{equation}
 \mathcal T\,\mathcal K[R]=\mathcal F[R],\qquad
 \mathcal K[R]\equiv\ddot R+\frac{1-\dot R^2}{R},
 \label{eq:app_large_tension_eom}
\end{equation}
where $\mathcal K$ is the curvature of Eq.~\eqref{eq:circle-curvature}.
Here $\mathcal F [R]= R_0 \bar F(R;R_0)/(\pi v^2)$ is the finite retarded KR force of Eq.~\eqref{eq:circle-renormalised} evaluated at $\Delta=R_0$ and including a normalisation factor.
The NG solution released
from rest and its prediction for the halving time are given by
\begin{equation}
 R_{\rm NG}(t)=\cos(t),\qquad
 t_*\equiv\frac{t_{1/2}^{\rm NG}}{R_0}=\frac{\pi}{3}.
 \label{eq:app_ng_solution}
\end{equation}
Assuming $\mathcal T \gg 1$, let us consider the expansion from the NG solution:
\begin{align}
 R(t)&=\cos(t)+\frac{r_1(t)}{\mathcal T}+\frac{r_2(t)}{\mathcal T^2}
 +\mathcal O(\mathcal T^{-3}),\\[0.5em]
 t_{1/2}&=t_*+\frac{\tau_1}{\mathcal T}+\frac{\tau_2}{\mathcal T^2}
 +\mathcal O(\mathcal T^{-3}).
 \label{eq:app_large_tension_expansion}
\end{align}
Linearising the NG operator gives
\begin{equation}
 \mathcal K[\cos(t)+\epsilon r]
 =\epsilon\,\mathsf L r+\mathcal O(\epsilon^2),\qquad
 \mathsf L r=\ddot r+2\tan(t)\,\dot r-r.
 \label{eq:app_linearized_ng}
\end{equation}
Because $\mathcal K[R_{\rm NG}]=0$, the leading equation is obtained by
evaluating the finite force on the NG history. With
\begin{equation}
 f_0(t)\equiv\mathcal F[R_{\rm NG}](t),
\end{equation}
the leading response obeys
\begin{equation}
 \mathsf Lr_1=f_0(t),\qquad r_1(0)=\dot r_1(0)=0.
 \label{eq:app_first_order_response}
\end{equation}
Two homogeneous solutions and their Wronskian are
\begin{equation}
 y_1(t)=\sin(t),\qquad y_2(t)=\cos(t)+t\sin(t),
 \qquad W(t)=-\cos^2(t).
 \label{eq:app_homogeneous}
\end{equation}
The causal Green function is consequently
\begin{equation}
 G(t,s)=
 \frac{\sin s\,[\cos(t)+t\sin(t)]
 -[\cos(s)+s\sin(s)]\sin(t)}{-\cos^2(s)},
 \qquad 0\leq s\leq t,
 \label{eq:app_green_function}
\end{equation}
which satisfies $\mathsf L_tG=0$, $G(s,s)=0$ and
$\partial_tG|_{t=s}=1$, and
\begin{equation}
 r_1(t)=\int_0^t \dd s\,G(t,s)f_0(s).
 \label{eq:app_r1_green}
\end{equation}
Expanding the crossing condition $R(t_{1/2})=1/2$ gives
$\tau_1=r_1(t_*)/\sin(t_*)$. Therefore, writing
\begin{equation}
 T_{1/2}\equiv1-\frac{t_{1/2}}{t_{1/2}^{\rm NG}}
 =\frac{A}{\mathcal T}+\frac{B}{\mathcal T^2}+\cdots,
 \label{eq:app_T_expansion}
\end{equation}
we obtain
\begin{equation}
 A=-\frac{1}{t_*\sin(t_*)}
 \int_0^{t_*}\dd s\,G(t_*,s)f_0(s).
 \label{eq:app_A_functional}
\end{equation}
Thus, a multiplicative force error
$f_0^{\rm calc}=qf_0^{\rm physical}$ produces
$A_{\rm calc}=qA_{\rm physical}$. 
The KR force perturbs an order-one NG trajectory rather than generating the complete motion.

\subsection{Universality and Next-to-leading Order}

The finite matching contribution $c_{\rm FT}$ is already included in
$\mathcal T$. Equation~\eqref{eq:app_A_functional} therefore contains no
separate dependence on it. At next order,
\begin{equation}
 \mathsf L r_2=\delta\mathcal F[r_1]+\dots,
 \label{eq:app_second_order_response}
\end{equation}
where $\delta\mathcal F$ is the linear response of the nonlocal force and
the ellipsis represents the terms quadratic in $r_1$ and $\dot{r}_1$.
The crossing condition yields
\begin{equation}
 \tau_2=\frac{r_2(t_*)+\tau_1\dot r_1(t_*)
 -\tfrac12\cos(t_*)\,\tau_1^2}{\sin(t_*)},
 \qquad B=-\frac{\tau_2}{t_*}.
 \label{eq:app_tau2}
\end{equation}
The coefficient $A$ is also invariant under a finite additive redefinition of
the tension. Indeed, if $\mathcal T'=\mathcal T+\delta c$, then
\begin{equation}
 \frac{A}{\mathcal T}+\frac{B}{\mathcal T^2}
 =\frac{A}{\mathcal T'}
 +\frac{B+A\delta c}{\mathcal T'^2}
 +\mathcal O(\mathcal T'^{-3}).
 \label{eq:app_finite_tension_redefinition}
\end{equation}
Hence $A$ is invariant under such a finite reparametrisation, whereas $B$
contains the nonlinear response and transforms as in
Eq.~\eqref{eq:app_finite_tension_redefinition}. This bookkeeping dependence is
not a physical regulator dependence: at fixed matched $\mathcal T(R_0)$,
regulator conversion leaves the trajectory unchanged.

\subsection{Force on the Nambu--Goto History}

The force in Eq.~\eqref{eq:app_A_functional} reduces to finite
one-dimensional quadratures, as it is given by the integral over the retarded time, see Eq.~\eqref{eq:forceKR}. 
Here we evaluate it on the NG history, $R(t) =R_{\rm NG}(t)$.

Let $t$ be the observation time, $s$ the
retarded source time, and define
\begin{equation}
\begin{aligned}
 u&=t-s,\quad a=\cos(t),\quad b=\cos(s),\quad w=-\sin(s),\\
 \quad c&=\frac{a^2+b^2-u^2}{2ab},\quad S=\sqrt{1-c^2}.
 \label{eq:app_retarded_variables}
 \end{aligned}
\end{equation}
We decompose the integral kernel into dynamical ($0 < s < t$) and static ($s < 0$) contributions.
After combining the apparently singular terms, the dynamical kernel is found to be
\begin{equation}
 P_{\rm dyn}(t,s)=-\frac{2a^2M}{bSD^2},\qquad
 \lim_{s\to t^-}P_{\rm dyn}(t,s)=-\frac23\sin(t),
 \label{eq:app_dynamic_kernel}
\end{equation}
where
\begin{equation}
\begin{aligned}
M&=-wQD-2ub^3D+bQU, \quad
D=2ub+w(b^2+u^2-a^2),\\
Q&=a^2-b^2-u^2-2ubw, \quad
U=2b(w^2-1)-b^3-b(u^2-a^2).
\label{eq:app_dynamical_kernel_functions}
\end{aligned}
\end{equation}
On the other hand, before the release the prescribed source history is static. 
On this branch, $b=1$, $w=\ddot R=0$, and we obtain
\begin{equation}
 P_{\rm stat}(t,s)=
 -\frac{a^2(u^2+1-a^2)}{u^2
 \sqrt{1-[(a^2+1-u^2)/(2a)]^2}}.
 \label{eq:app_static_kernel}
\end{equation}
The earliest retarded point is the antipode, $s_{\min}=t-a-1$. 
The force on the NG history is therefore given by
\begin{equation}
 f_0(t)=\int_{t-a-1}^{0}\dd s\,P_{\rm stat}(t,s)
 +\int_0^t \dd s\,P_{\rm dyn}(t,s).
 \label{eq:app_force_split}
\end{equation}
The static contribution may equivalently be written
\begin{equation}
 -2a^3\int_t^{1+a}\dd u\,
 \frac{u^2+1-a^2}{u^2
 \sqrt{[(1+a)^2-u^2][u^2-(1-a)^2]}},
 \label{eq:app_static_elliptic}
\end{equation}
an incomplete elliptic integral, whose endpoints $u=1\pm a$ are the zeros of
$S$ in Eq.~\eqref{eq:app_retarded_variables}. The dynamical contribution remains a smooth
quadrature because the retarded geometry contains both $t-s$ and
trigonometric functions of $s$.

\begin{table}[t]
\centering
\renewcommand{\arraystretch}{1.2}
\setlength{\arrayrulewidth}{1pt}
\begin{tabular}{S[table-format=1.4]c}
\hline\hline
{$\delta_\varphi$} & $A$ \\
\hline
0.02   & $1.32865$ \\
0.01   & $1.34752$ \\
0.005  & $1.35730$ \\
0.0025 & $1.36317$ \\
\hline\hline
\end{tabular}
\caption{Leading coefficient $A$ of Eq.~\eqref{eq:app_A_functional} for different values of $\delta_\varphi=\delta_\sigma/R_0$.}
\label{tab:A_cutoff}
\end{table}

Inserting Eq.~\eqref{eq:app_force_split} into
Eq.~\eqref{eq:app_A_functional} yields the results presented in Tab.~\ref{tab:A_cutoff}. 
A linear endpoint extrapolation gives
\begin{equation}
A =1.368\pm0.004
 \qquad(\delta_\varphi\rightarrow0),
 \label{eq:app_A_numerical}
\end{equation}
where the uncertainty quantifies the spread of the result for comparing a linear and quadratic extrapolation in $\delta_\varphi\rightarrow0$.
The force is negative throughout the interval; representative values are
$f_0(0.10)=-4.14$, $f_0(0.50)=-2.06$, and
$f_0(\pi/3)=-0.716$. 

Evaluating the two terms in Eq.~\eqref{eq:app_force_split} separately shows
that the constrained static history supplies approximately $1.301$ of $A$,
whereas the post-release branch supplies only about $0.068$. The coefficient
is therefore controlled mainly by the field present at release. This is why
an additive finite change in the definition of $\mathcal T$
cannot reconcile the leading field-theory result
with $A_{\rm FT}\sim 0.951\pm 0.007$.

\renewcommand{\theequation}{F.\arabic{equation}}
\setcounter{equation}{0}

\section{Loop Initial Conditions via Solid Angle}
\label{app:string_ics}
Axion strings are topological line defects, described by certain configurations of complex scalar fields,  whose phase $\theta$ winds by $2\pi$ upon encircling
the string core. As a consequence the phase field must satisfy
\begin{equation}
\oint_C \dd\boldsymbol{\ell}\cdot\nabla\theta  = 2\pi m,\qquad m\in\mathbb{Z},
\end{equation}
for any closed loop $C$ linking a string of winding number $m$.
Away from the core, $\nabla\theta$ is curl-free except on the string itself.
Constructing $\theta$ by integrating a discretised Biot--Savart field leads
to gauge ambiguities and accumulation of numerical error, since $\theta$ is
globally multi-valued.
\subsection{Solid Angle Construction}
The construction we employ in this work proceeds \emph{geometrically}. Let $\Gamma$ denote the spatial
curve describing a closed string loop, and choose any orientable surface $S$
such that $\partial S = \Gamma$. The phase field is obtained directly from the
\emph{solid angle} $\Omega_S(\boldsymbol{x})$ under which $S$ is seen from the point
$\boldsymbol{x}$ as
\begin{equation}
\theta(\boldsymbol{x}) = \frac{1}{2}\,\Omega_S(\boldsymbol{x}),
\label{eq:solid_angle_theta}
\end{equation}
which is Eq.~\eqref{eq:solid-angle-theta} of the main text. Crossing the surface $S$ changes $\Omega_S$ by $4\pi$, and therefore $\theta$ by
$2\pi$, implementing the required branch structure of the axion field.
Furthermore,
\begin{equation}
\nabla\theta(\boldsymbol{x}) = \frac{1}{2}\,\nabla\Omega_S(\boldsymbol{x}),
\end{equation}
which yields the characteristic $1/r$ azimuthal circulation of the vortex near its core without any explicit integration. An explicit proof of the relation Eq.~\eqref{eq:solid_angle_theta} is given in
Sec.~\ref{app:solid_angle_proof}.

\begin{figure*}[t]
    \centering
    \includegraphics[width=.85\textwidth]{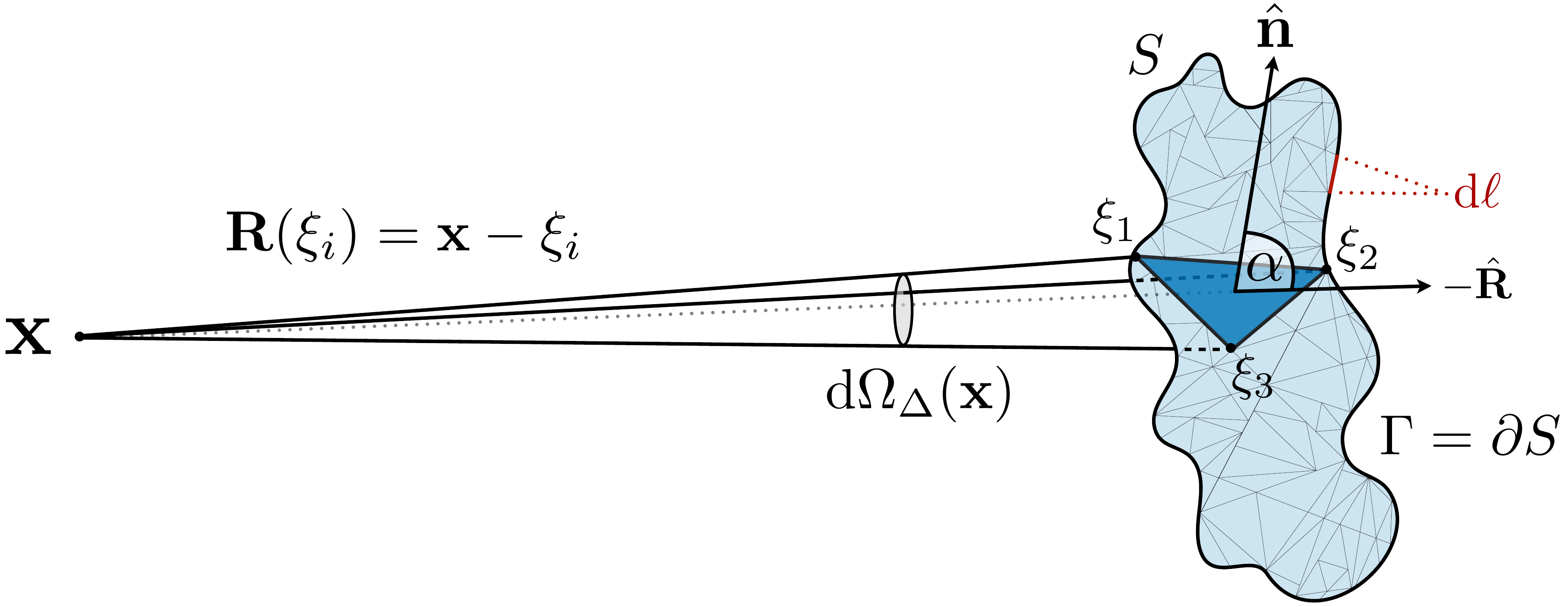}
    \caption{Construction of the axion field around a closed string loop described by the curve $\Gamma = \partial S$ via the solid angle approach, Eq.~\eqref{eq:solid_angle_theta}. In practice, the computation is realised by triangulating the surface $S$ according to Eq.~\eqref{eq:triangles}.}
    \label{fig:solid_angle}
\end{figure*}

For numerical implementation the surface $S$ is discretised into oriented
triangles with vertices $(\boldsymbol{v}_1, \boldsymbol{v}_2, \boldsymbol{v}_3)$, see Fig.~\ref{fig:solid_angle} for a visualisation. With $\boldsymbol{r}_i = \boldsymbol{v}_i - \boldsymbol{x}$ and $\hat{\boldsymbol{r}}_i = \boldsymbol{r}_i / \|\boldsymbol{r}_i\|$, a robust expression for the solid angle of a single triangle is
\begin{equation}
 \Omega_{\triangle}(\boldsymbol{x})=
 2\operatorname{atan2}\!\left(
 \hat{\boldsymbol{r}}_1\cdot(\hat{\boldsymbol{r}}_2\times\hat{\boldsymbol{r}}_3),
 \,
 1+\hat{\boldsymbol{r}}_1\!\cdot\!\hat{\boldsymbol{r}}_2
 +\hat{\boldsymbol{r}}_2\!\cdot\!\hat{\boldsymbol{r}}_3
 +\hat{\boldsymbol{r}}_3\!\cdot\!\hat{\boldsymbol{r}}_1
 \right),
 \label{eq:triangles}
\end{equation}
and summing over all triangles gives the complete phase field,
\begin{equation}
\theta(\boldsymbol{x}) = \frac{1}{2}\sum_{\triangle\in S}
\Omega_{\triangle}(\boldsymbol{x}).
\end{equation}
This guarantees all the required properties: exact $2\pi$-winding around each string core, a single-valued $\theta(\boldsymbol{x})$ away from the chosen branch surface, no gauge-fixing and no integration-induced drift, a trivial parallelisation and applicability to arbitrary curved string loops and networks. The axion phase field is thus constructed directly from topology, through the geometry of the solid angle, ensuring numerical stability and physical correctness. We implemented this computation in \texttt{jaxions} and it serves as the standard tool for computing all initial conditions for the simulations throughout this work. 

\subsection{$\nabla\Omega_S$ equals the Biot-Savart Integral}
\label{app:solid_angle_proof}
Let $\Gamma$ be a closed string loop and $S$ any orientable surface with
$\partial S=\Gamma$, as visualised in Fig.~\ref{fig:solid_angle}. For an observation point $\mathbf x$ \emph{not} on $S$ or $\Gamma$, define $\mathbf R(\boldsymbol\xi) = \mathbf x-\boldsymbol\xi$, where $\boldsymbol\xi$ denotes the source point. A standard representation of the solid angle seen from $\mathbf x$ is the \enquote{monopole flux} through $S$, 
\begin{equation}
\Omega_S(\mathbf x)\;=\;\int_S \dd S(\boldsymbol\xi)\,\frac{\mathbf R(\boldsymbol\xi)}{\|\mathbf R(\boldsymbol\xi)\|^3}\cdot
\mathbf n(\boldsymbol\xi)\,,
\label{eq:app_monopole_flux}
\end{equation}
where $\mathbf n\,\dd S$ is the oriented area element on $S$.

\paragraph{Claim.} For $\mathbf x\notin S$,
\begin{equation}
\nabla_{\!x}\,\Omega_S(\mathbf x)\;=\;\oint_\Gamma
\dd\boldsymbol\ell(\boldsymbol\xi)\times
\frac{\mathbf x-\boldsymbol\xi}{\|\mathbf x-\boldsymbol\xi\|^3}\,.
 \label{eq:app_claim}
\end{equation}

\paragraph{Proof.} 
Let $\mathbf a$ be an arbitrary, constant vector. We consider the directional
derivative
\begin{equation}
\mathbf a\cdot \nabla_{\!x}\Omega_S(\mathbf x)
= \int_S \dd S\, \mathbf n \cdot \left[\,\mathbf a\cdot\nabla_{\!x}\,
\frac{\mathbf R}{\|\mathbf R\|^3}\,\right]\,.
\end{equation}
where the derivative may be taken under the integral sign because
$\mathbf x\notin S$ keeps the integrand smooth on the domain of integration.
For $\mathbf R=\mathbf x-\boldsymbol\xi$, derivatives with respect to $\mathbf x$ are  minus the derivatives with respect to $\boldsymbol\xi$, 
\begin{equation}
\mathbf a\cdot\nabla_{\!x}\,\frac{\mathbf R}{\|\mathbf R\|^3}
= -\,\mathbf a\cdot\nabla_{\!\xi}\,\frac{\mathbf R}{\|\mathbf R\|^3}.
\end{equation}
Since $\mathbf a$ is constant, we can apply the vector identity, 
\begin{equation}
\nabla_{\!\xi}\times\!\big(\mathbf a\times \mathbf F\big)
= \mathbf a\,(\nabla_{\!\xi}\!\cdot \mathbf F)\;-\;(\mathbf a\cdot\nabla_{\!\xi})\,\mathbf F,
\qquad \mathbf F(\boldsymbol\xi)=\frac{\mathbf R}{\|\mathbf R\|^3}.
 \label{eq:app_vector_identity}
\end{equation}
Away from $\boldsymbol\xi=\mathbf x$ we have $\nabla_{\!\xi}\!\cdot \mathbf F=0$,
hence
\begin{equation}
(\mathbf a\cdot\nabla_{\!\xi})\,\mathbf F
= -\,\nabla_{\!\xi}\times\!\big(\mathbf a\times \mathbf F\big),
 \label{eq:app_identity_reduced}
\end{equation}
and therefore
\begin{equation}
\mathbf a\cdot \nabla_{\!x}\Omega_S(\mathbf x)
= \int_S\dd S\, \mathbf n \cdot \nabla_{\!\xi}\times\!\big(\mathbf a\times \mathbf F\big)\,.
 \label{eq:app_before_stokes}
\end{equation}
By Stokes' theorem, applied with respect to the source variable $\boldsymbol\xi$),
\begin{equation}
\int_S\dd S\, \mathbf n \cdot \nabla_{\!\xi}\times \mathbf G
= \oint_\Gamma\dd\boldsymbol\ell\cdot\mathbf G,
\qquad \mathbf G=\mathbf a\times \mathbf F,
\end{equation}
which yields
\begin{equation}
 \mathbf a\cdot\nabla_{\!x}\Omega_S(\mathbf x)
 =\oint_\Gamma\dd\boldsymbol\ell\cdot(\mathbf a\times\mathbf F)
 =\oint_\Gamma\big(\dd\boldsymbol\ell\times\mathbf F\big)\cdot\mathbf a,
 \label{eq:app_triple_product}
\end{equation}
where we used the scalar triple product identity
$(\mathbf a\times\mathbf F)\cdot\dd\boldsymbol\ell
=\mathbf a\cdot(\dd\boldsymbol\ell\times\mathbf F)$.
Since $\mathbf a$ is arbitrary, Eq.~\eqref{eq:app_claim} follows.
\hfill$\square$

\paragraph{Consequence.}
With the axion-string convention of Eq.~\eqref{eq:KR-magnetostatics},
\begin{equation}
 \boldsymbol b_\theta(\mathbf x)=\frac{1}{2}\oint_\Gamma
 \dd\boldsymbol\ell(\boldsymbol \xi)\times
 \frac{\mathbf x-\boldsymbol \xi}{\|\mathbf x-\boldsymbol \xi\|^3}
 =\frac{1}{2}\,\nabla_{\!x}\Omega_S(\mathbf x)
 =\nabla\theta(\mathbf x),
 \qquad
 \theta(\mathbf x)=\frac{1}{2}\Omega_S(\mathbf x).
 \label{eq:app_consequence}
\end{equation}
The Biot-Savart construction and the solid-angle potential are
therefore \emph{identical} off the surface $S$ and the $2\pi$-branch jump of
$\theta$ arises from the $4\pi$ discontinuity of $\Omega_S$ across $S$.

\renewcommand{\theequation}{G.\arabic{equation}}
\setcounter{equation}{0}

\section{Retarded Kalb--Ramond Self-field of a Circular Loop}
\label{app:KR-force-derivation}

In this appendix we derive the retarded field used in
Eq.~\eqref{eq:forceKR} and make explicit the origin and coefficient of its
short-distance singularity.  In Lorenz gauge,
$\partial_\mu B^{\mu\nu}=0$, the retarded solution of
Eq.~\eqref{eq:Rsource} is \cite{Poisson:2011nh}
\begin{equation}
 \begin{split}
 B_{\rm ret}^{\mu\nu}(x)
 &=-2\pi v\int\dd^4x'\,G_{\rm ret}(x-x')J^{\mu\nu}(x')\\
 &=-2\pi v\int_\Sigma \dd\sigma^{\mu\nu} G_{\rm ret}(x-X),
 \qquad \Box G_{\rm ret}(x)=\delta^{(4)}(x).
 \end{split}
 \label{eq:retarded-B}
\end{equation}
We evaluate the field at the point $(t,R(t),0,0)$ of the circular loop.
Writing $R_r\equiv R(t_r)$, the support of the retarded Green function is
specified by the light-cone function in Eq.~\eqref{eq:lightcone-function}.
Direct differentiation of Eq.~\eqref{eq:retarded-B} in
\begin{equation}
 H_{012}=\partial_0B_{12}+\partial_1B_{20}+\partial_2B_{01}
 \label{eq:H012-definition}
\end{equation}
gives
\begin{equation}
 H_{012}
 =v\int\dd t_r\dd\varphi\,
 Q(t,t_r,\varphi)\,\delta'\!\left(\mathfrak f\right),
 \qquad
 Q\equiv2R_r\left[\dot R_r(t-t_r)+R_r-R(t)\cos\varphi\right].
 \label{eq:H012-delta-prime}
\end{equation}
The derivative of the delta function can be removed by integrating by
parts in retarded time,
\begin{equation}
 \int\dd t_r\,Q\,\delta'(\mathfrak f) =\left[\frac{Q}{\partial_{t_r}\mathfrak f}
 \delta(\mathfrak f)\right]_{-\infty}^{\infty}
 -\int\dd t_r\,\delta(\mathfrak f)
 \frac{\dd}{\dd t_r}
 \left(\frac{Q}{\partial_{t_r}\mathfrak f}\right).
 \label{eq:by_parts}
\end{equation}
The boundary term vanishes for the retarded solution
\cite{Copeland:1990qu,Battye:1995hw,Buonanno:1998kx}.  For every
$t_r\in(t_\pi,t)$, the equation $\mathfrak f=0$ has two roots with
opposite values of $\varphi$.  Since the remaining integrand is even,
\begin{equation}
 \int_{-\pi}^{\pi}\dd\varphi\,
 \delta(\mathfrak f)\,g(\cos\varphi)
 =\frac{2g(\cos\varphi_r)}
 {|\partial_\varphi\mathfrak f|_{\varphi_r}}.
 \label{eq:angular-delta-jacobian}
\end{equation}
Equations~\eqref{eq:H012-delta-prime}--\eqref{eq:angular-delta-jacobian}
then give Eq.~\eqref{eq:forceKR}, including its overall factor of two.
The worldsheet regulator removes the interval around the coincidence
point and fixes its upper limit as described below
Eq.~\eqref{eq:lightcone-function}.

We next determine the local behaviour of the integrand.  Let
$u\equiv t-t_r>0$, and let quantities without an $r$ subscript be
evaluated at the observation time $t$.  We denote by
$\varphi_r=\varphi(t_r;t)>0$ the positive solution of the light-cone
condition; the two roots are $\pm\varphi_r$.  For a trajectory that is
smooth at the observation time,
\begin{equation}
 R_r=R-\dot Ru+\frac12\ddot Ru^2+\mathcal O(u^3),
 \qquad
 \dot R_r=\dot R-\ddot Ru+\mathcal O(u^2).
 \label{eq:coincidence-expansion}
\end{equation}
On the light cone,
\begin{equation}
 \cos\varphi_r=\frac{R^2+R_r^2-u^2}{2RR_r},
 \qquad
 |\partial_\varphi\mathfrak f|
 =\sqrt{\big[(R+R_r)^2-u^2\big]
 \big[u^2-(R-R_r)^2\big]}
 =\frac{2Ru}{\gamma_R}+\mathcal O(u^2).
 \label{eq:angular-jacobian-local}
\end{equation}
Thus, the angular delta-function Jacobian is the only factor that is
singular at coincidence:
\begin{equation}
 \frac{1}{|\partial_\varphi\mathfrak f|}
 =\frac{\gamma_R}{2Ru}\left[1+\mathcal O(u)\right].
 \label{eq:angular-jacobian-pole}
\end{equation}

For completeness, the remaining factor can be expanded directly.  Define
$D\equiv\partial_{t_r}\mathfrak f$ and introduce the radial projection of
the observer-source separation,
\begin{equation}
 \Delta_\parallel\equiv R\cos\varphi-R_r,
 \qquad
 \left.\Delta_\parallel\right|_{\mathfrak f=0}
 =\frac{R^2-R_r^2-u^2}{2R_r}.
 \label{eq:radial-separation}
\end{equation}
After taking all $t_r$ derivatives at fixed $\varphi$ and only then
imposing $\mathfrak f=0$, one finds
\begin{align}
 D&=-2u+2\dot R_r\Delta_\parallel
   =-2\gamma_R^{-2}u+\mathcal O(u^2),
 \nonumber\\
 Q&=2R_r(\dot R_r u-\Delta_\parallel)
   =(\gamma_R^{-2}-R\ddot R)u^2+\mathcal O(u^3),
 \nonumber\\
 \partial_{t_r}Q&=2R\ddot R\,u+\mathcal O(u^2),
 \nonumber\\
 \partial_{t_r}D&=2\gamma_R^{-2}+\mathcal O(u).
 \label{eq:regular-factor-expansion}
\end{align}
Consequently, the apparently indeterminate differentiated quotient is
regular:
\begin{align}
 \left[\frac{\dd}{\dd t_r}\left(\frac{Q}{D}\right)\right]_{\mathfrak f=0}
 &=\frac{D\,\partial_{t_r}Q-Q\,\partial_{t_r}D}{D^2}
 \nonumber\\
 &=-\frac{\gamma_R^{-2}+R\ddot R}{2\gamma_R^{-2}}+\mathcal O(u)
 =-\frac{R\gamma_R^2}{2}\,\mathcal K(t)+\mathcal O(u),
 \label{eq:regular-factor-limit}
\end{align}
where Eq.~\eqref{eq:circle-curvature} was used in the last equality.
Combining Eqs.~\eqref{eq:angular-jacobian-pole} and
\eqref{eq:regular-factor-limit} gives the integrand quoted in the main
text,
\begin{equation}
 \frac{1}{|\partial_\varphi\mathfrak f|}
 \left[\frac{\dd}{\dd t_r}
 \left(\frac{Q}{\partial_{t_r}\mathfrak f}\right)\right]_{\mathfrak f=0}
 =-\frac{\gamma_R^3\mathcal K(t)}{4(t-t_r)}+\mathcal O(1).
 \label{eq:KR-integrand-local}
\end{equation}
The light-cone relation gives
$\varphi_r\simeq u/(R\gamma_R)$.  Hence the proper-distance cutoff
$R|\varphi_r|>\delta_\sigma$ is equivalent to
$u>\gamma_R\delta_\sigma$.  The pole in
Eq.~\eqref{eq:KR-integrand-local} therefore produces the logarithmic
cutoff dependence
\begin{equation}
 -\frac{2\pi v}{\gamma_R^3}H_{012}
 =\pi v^2\mathcal K(t)\log\delta_\sigma
 +\text{terms finite as }\delta_\sigma\to0,
 \label{eq:radial-force-local-divergence}
\end{equation}
which is cancelled by the counterterm in
Eq.~\eqref{eq:circle-renormalised}.

\renewcommand{\theequation}{H.\arabic{equation}}
\setcounter{equation}{0}

\section{Literature Comparison}
\label{app:context}

The collapse of circular cosmic string loops has previously been investigated in several studies using numerical simulations and the effective NGKR approximation. In this appendix, we compare our results with those obtained in the existing literature. Since the numerical data from these simulations are not publicly available, we digitised the published curves to enable a quantitative comparison. 

\begin{figure}[t]
    \centering
    \includegraphics[width=0.49\linewidth]{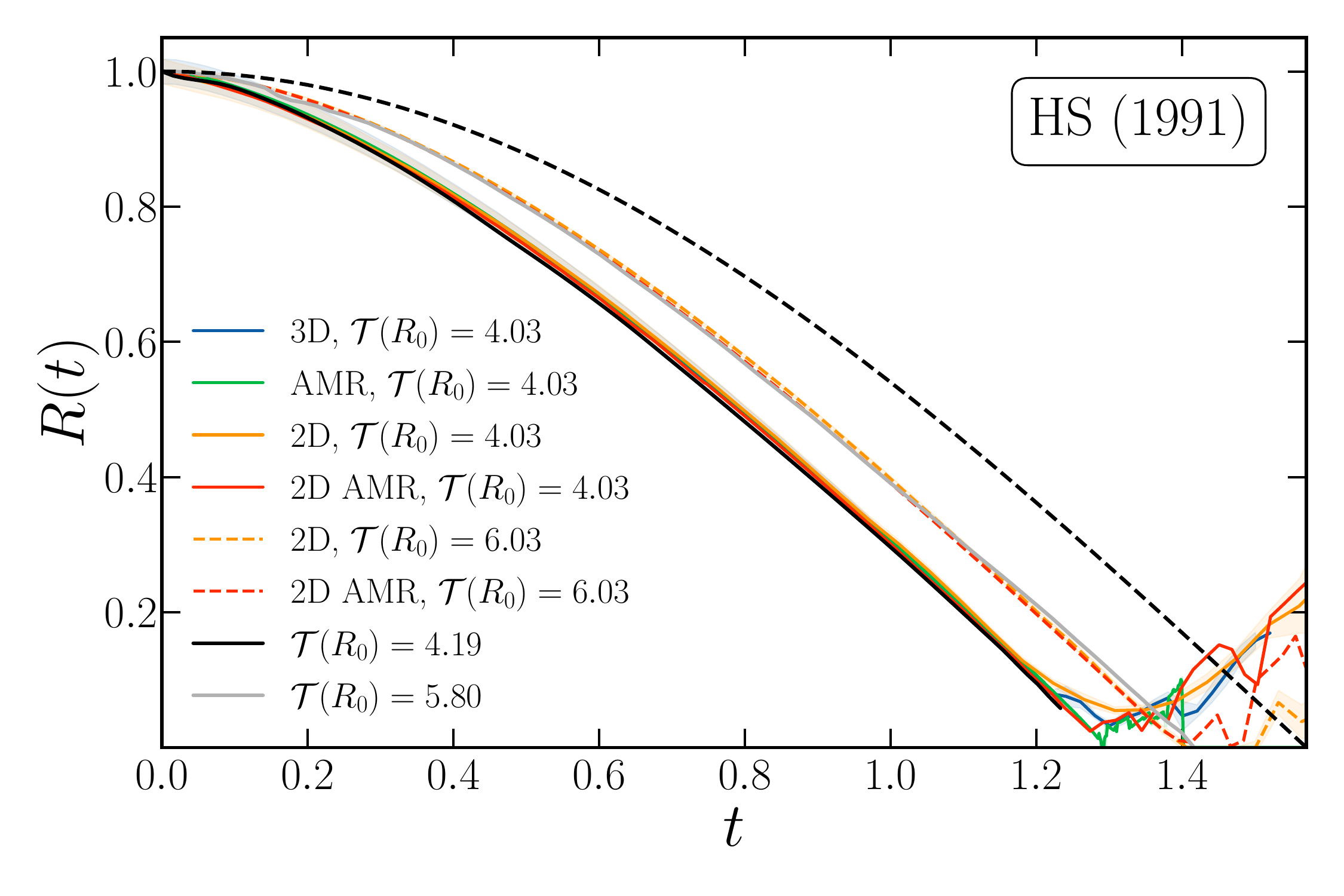}
    \includegraphics[width=0.49\linewidth]{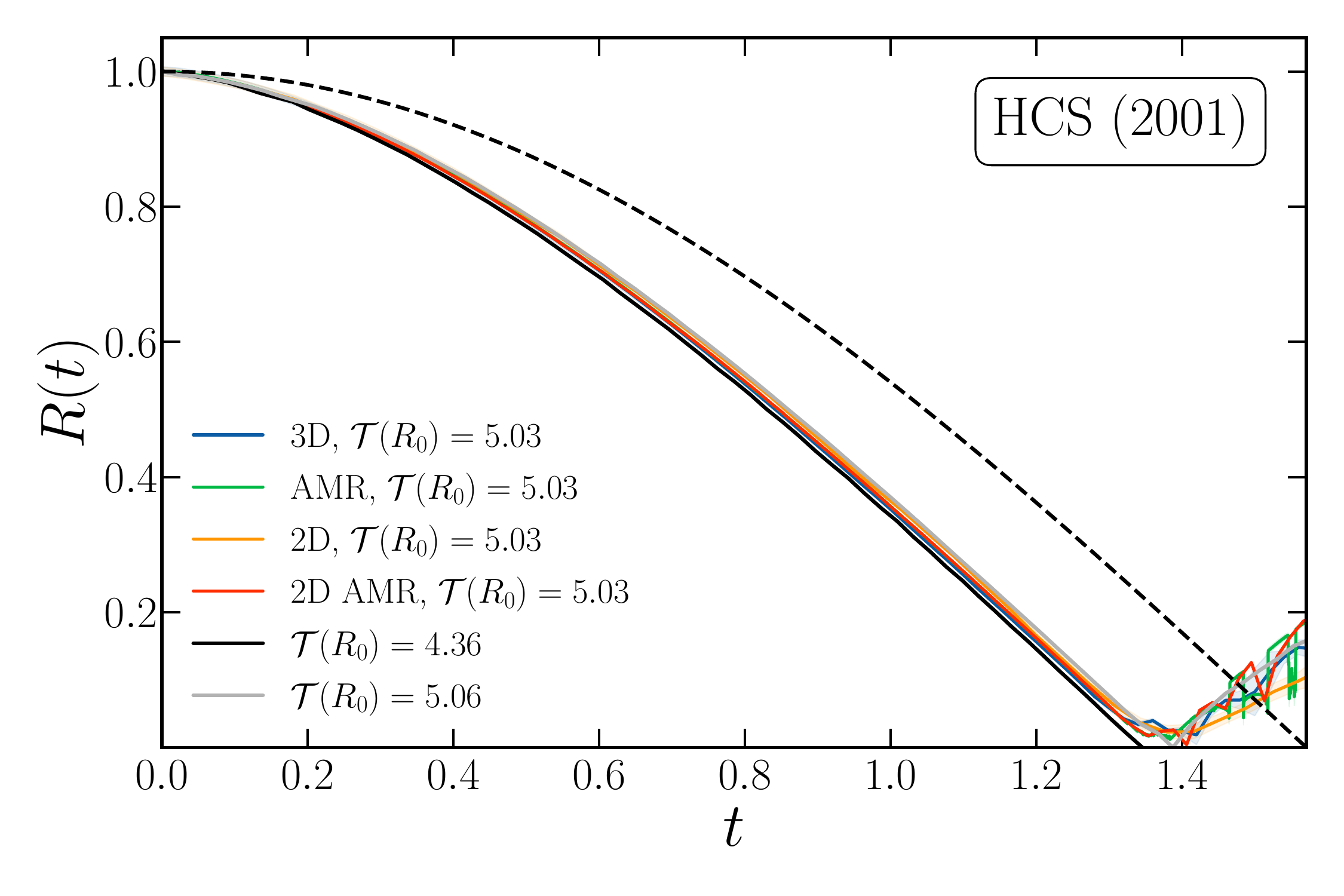}
    \includegraphics[width=0.49\linewidth]{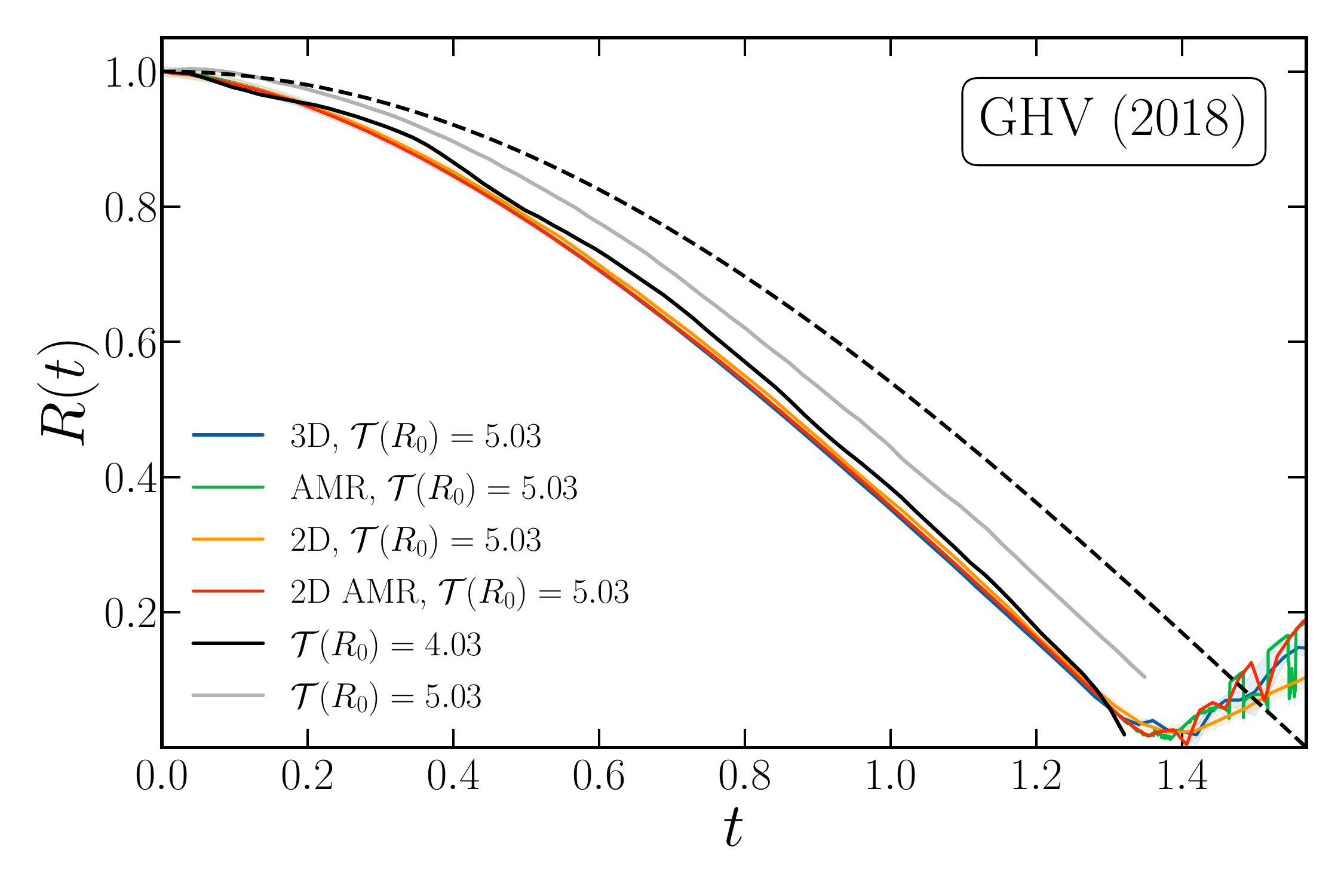}
    \includegraphics[width=0.49\linewidth]{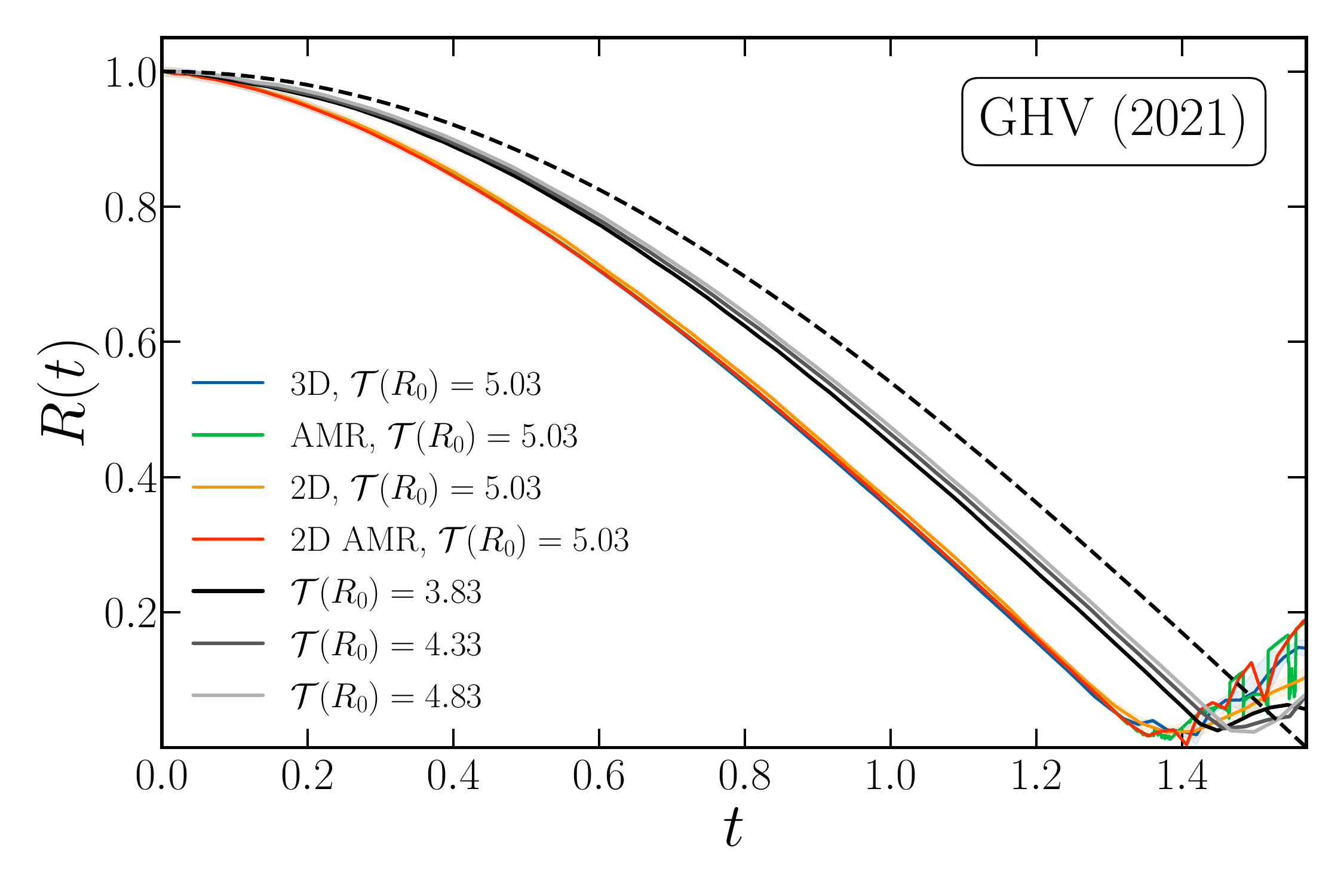}
    \caption{Continuum extrapolated loop trajectories compared with digitised results of Refs.~\cite{Hagmann:1990mj,Hagmann:2000ja,Gorghetto:2018myk,Gorghetto:2020qws}, one panel per reference. Grey lines are the digitised curves, labelled by the dimensionless tension $\mathcal T(R_0)=c_{\rm FT}+\log(m_sR_0)$ corresponding to $\log(m_sR_0)$ as quoted in the respective reference, from black at the lowest tension to light grey at the highest. Coloured lines are our continuum extrapolated trajectories at the closest available tension, with the same colours as in Fig.~\ref{fig:collapse_times}. Where two tensions are shown, the larger one is dashed. The black dashed line is the NG trajectory $R=\cos(t)$, and both $t$ and $R$ are given in units of $R_0$. The results of Refs.~\cite{Hagmann:1990mj,Hagmann:2000ja} agree with ours, whereas those of Refs.~\cite{Gorghetto:2018myk,Gorghetto:2020qws} collapse much later at the same quoted tension. We were unable to identify the source of this discrepancy.}
    \label{fig:radius_comparison}
\end{figure}

We compare our numerical calculation for the effective NGKR field with the result reported in Ref.~\cite{Dabholkar:1989ju}. Ref.~\cite{Dabholkar:1989ju} integrates by parts in the worldsheet coordinate $\sigma$, which generates boundary terms at $\sigma=1/2$ and at $\sigma=0$, but keeps only the former. In particular, we demonstrate that omitting the $\sigma=0$ boundary term, Eq.~\eqref{eq:boundaryMissed}, as done in Ref.~\cite{Dabholkar:1989ju}, allows us to reproduce their result, see Fig.~\ref{fig:DQ_comparison}. For the numerical field-theory simulations, the resulting comparisons are shown in Figs.~\ref{fig:radius_comparison} and~\ref{fig:collapse_times_lit}. Note that the digitised data are restricted to $\mathcal T(R_0)\lesssim 10$, so we do not include our \texttt{Moore} results in this comparison.

\subsection{Comparison with NGKR}
\label{app:DQ}
First, we compare our results with the ones obtained in Ref.~\cite{Dabholkar:1989ju} for which we adopt, only in this section, their notation to clarify the differences.

The first difference is that the integrations in Eq.~\eqref{eq:force_renorm}  and Eq.~\eqref{eq:forceKR} include a UV cutoff. In particular, it is never the case that $t_{\rm f} = t$ when the string has non-zero thickness, $m_s < \infty$. We define the UV-cutoff scale by matching the renormalised string tension to the tension obtained from our numerical simulations.
\begin{figure}[t]
\centering
\includegraphics[width = 0.495\linewidth]{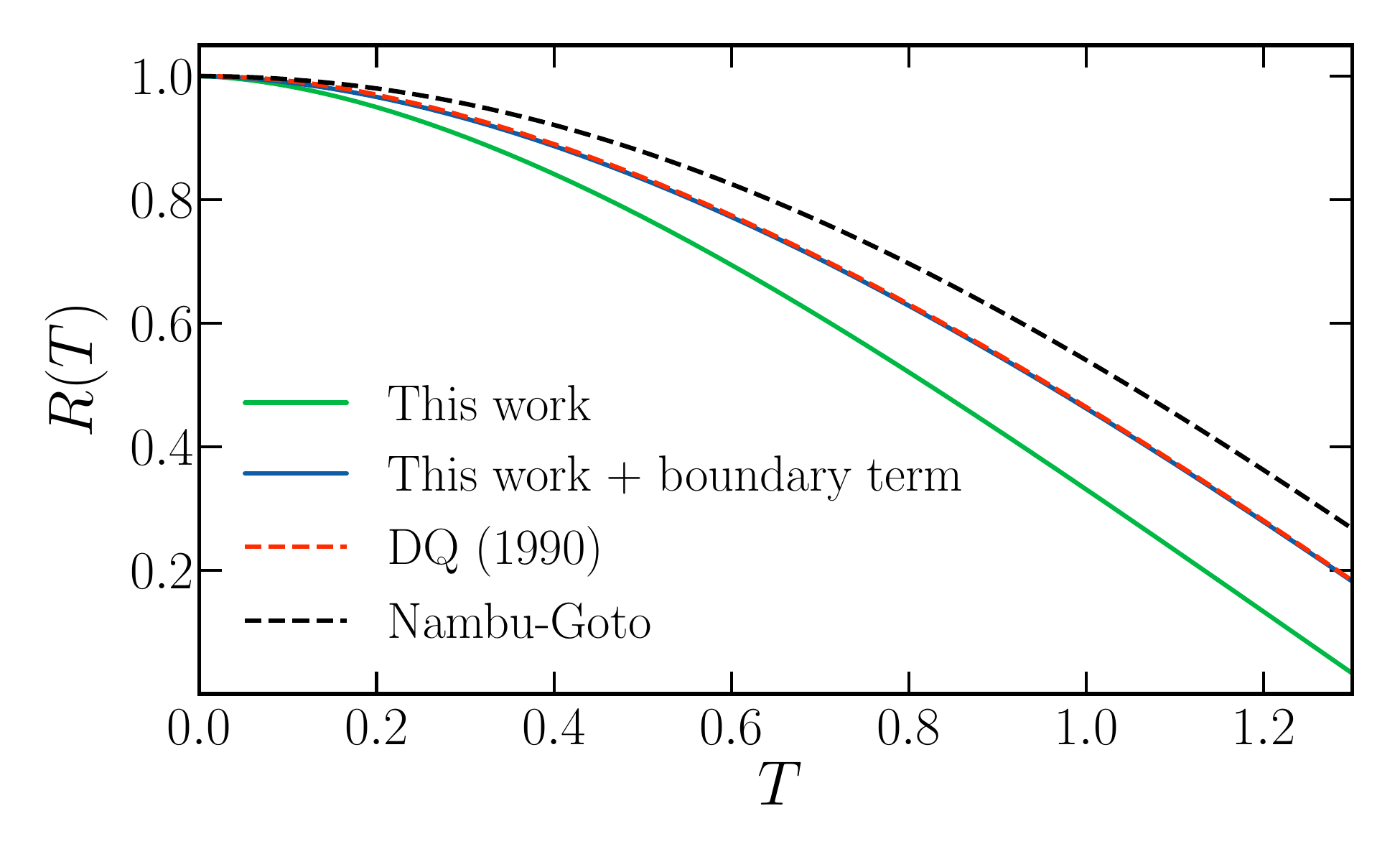}
\includegraphics[width = 0.495\linewidth]{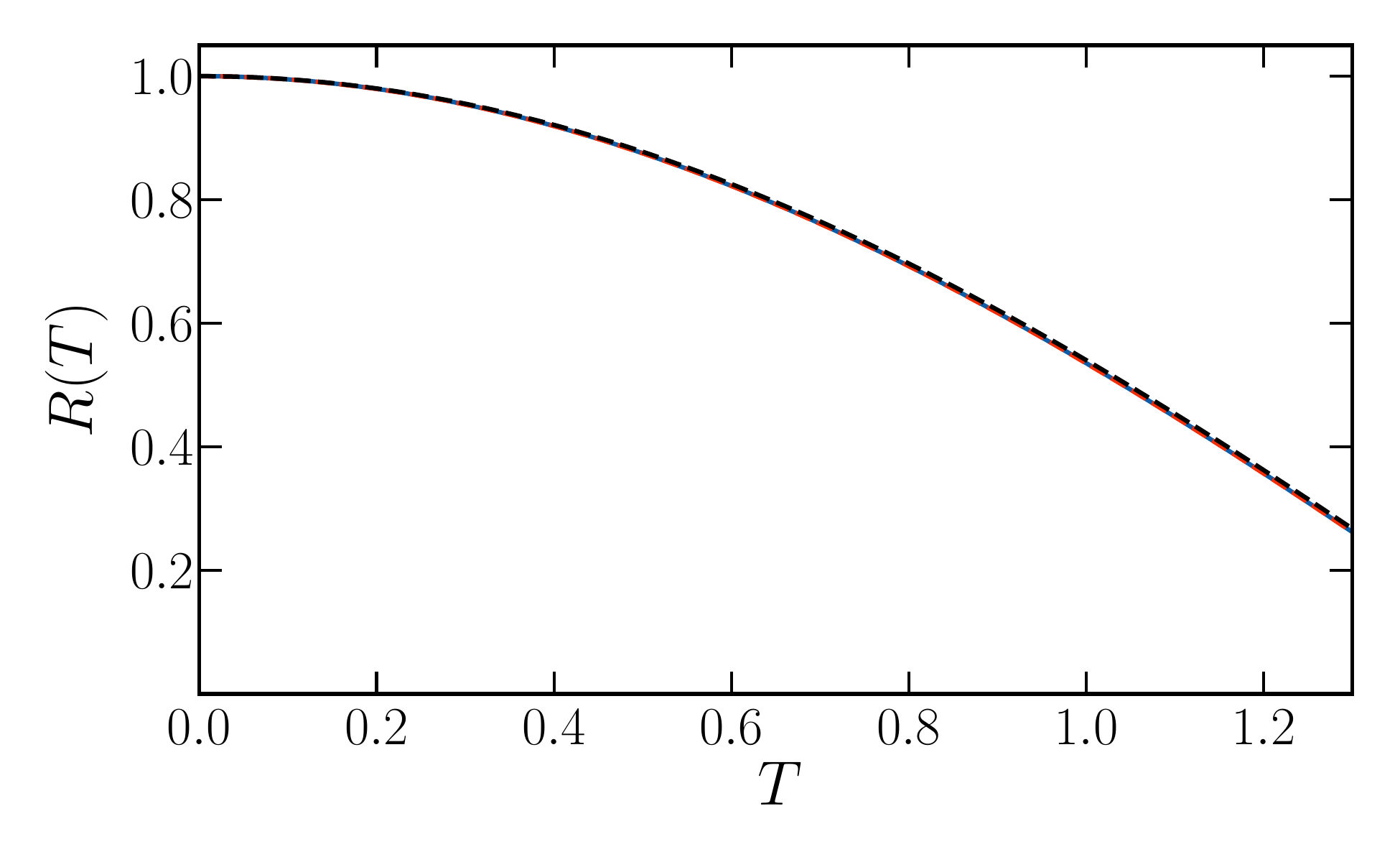}\\
\includegraphics[width = 0.495\linewidth]{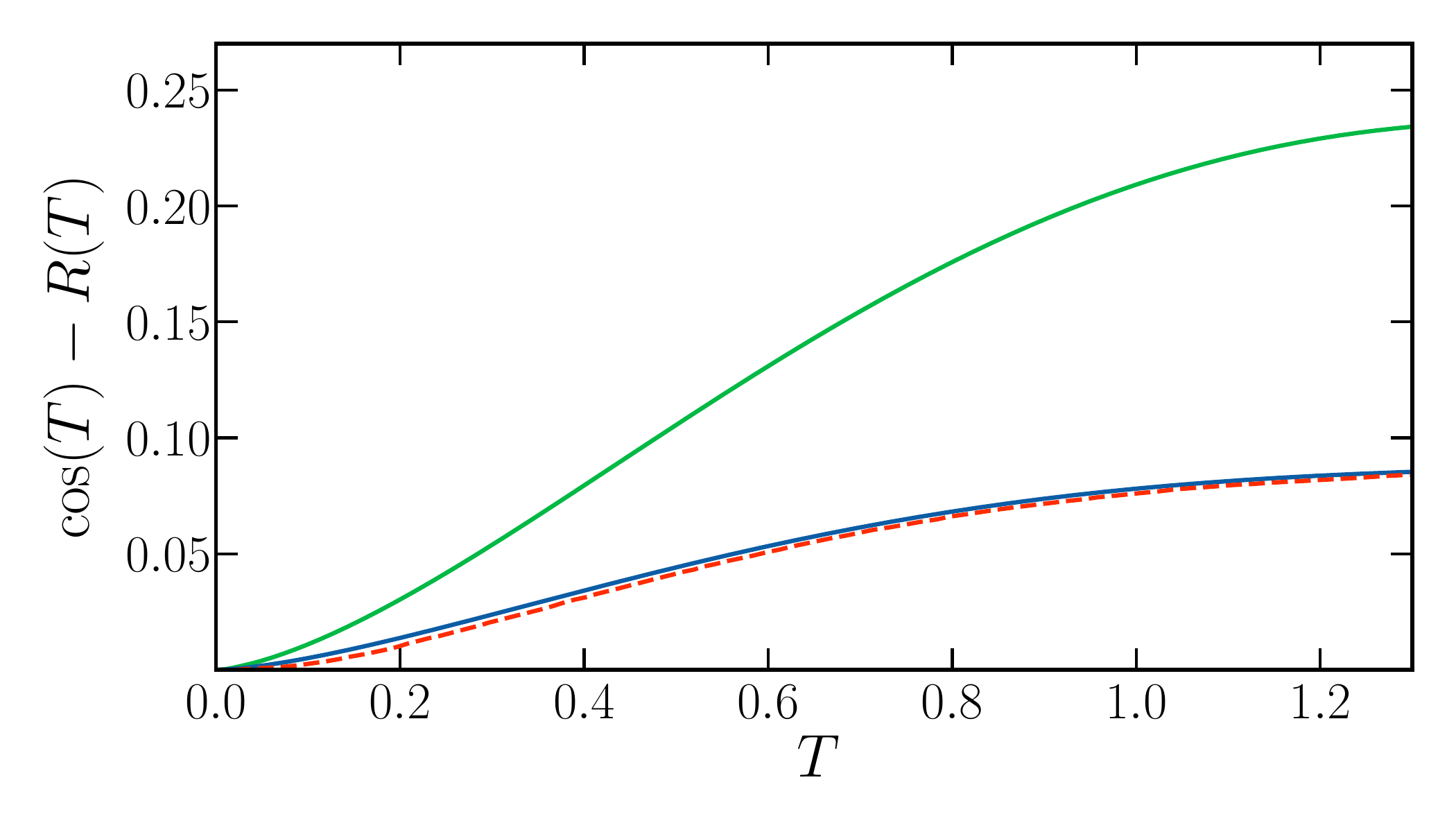}
\includegraphics[width = 0.495\linewidth]{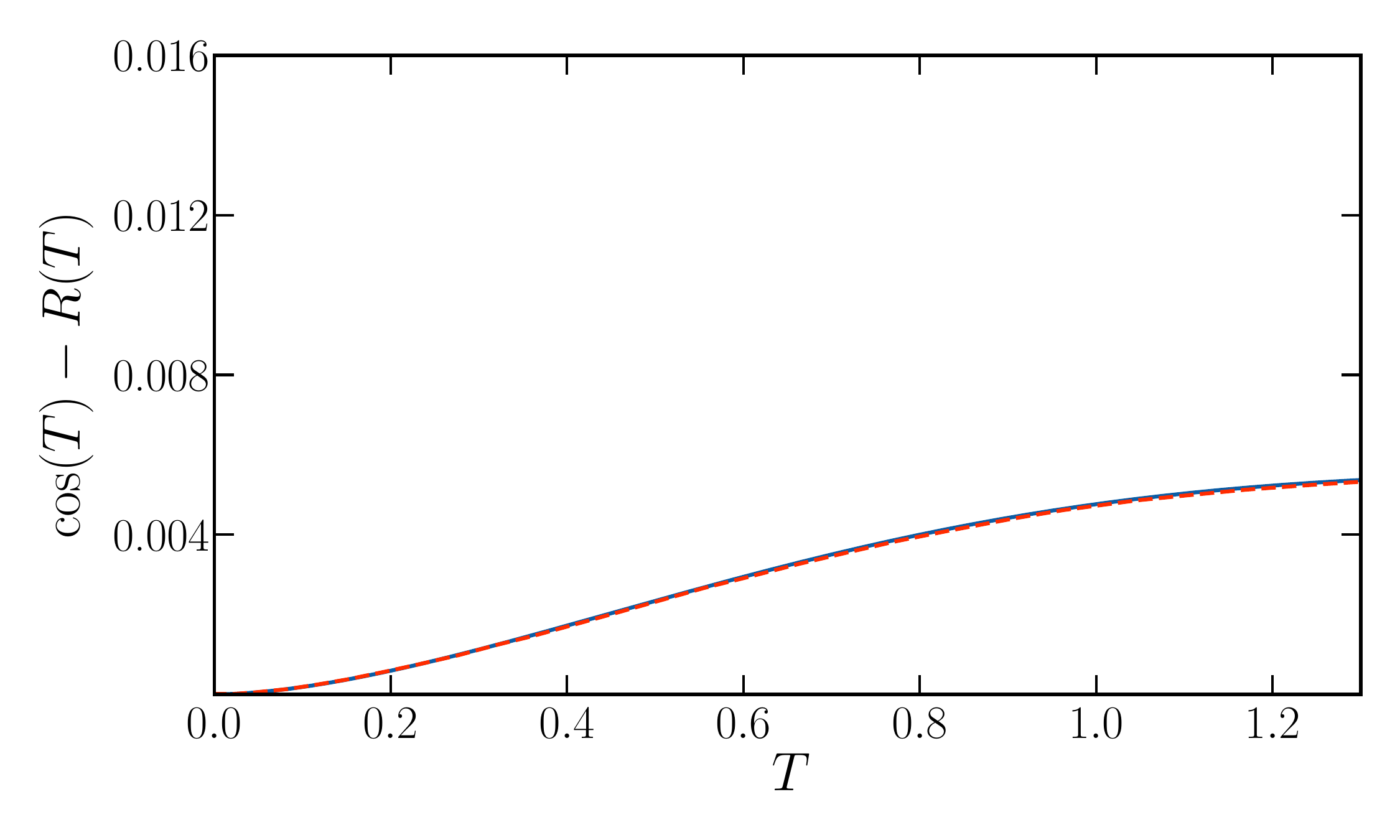}
\caption{Collapse of circular loops with back-reaction. Green: our result, obtained by integrating by parts in $t$, where no boundary terms arise. Blue: the same calculation with the contribution of the $\sigma=0$ boundary term, Eq.~\eqref{eq:boundaryMissed}, removed, i.e.\ reproducing its omission in Ref.~\cite{Dabholkar:1989ju}. Both are compared with the NG trajectory (black dashed) and with digitised results of Ref.~\cite{Dabholkar:1989ju} (red dashed). \emph{Top:} loop radius $R(T) \equiv r(t)/r_0$ as a function of $T=t/r_0$. \emph{Bottom:} deviation from the NG trajectory, $\cos(T)-R(T)$. The left column corresponds to $\log(r(0)/\delta)=5$ and the right column to $\log(r(0)/\delta)=100$.}
\label{fig:DQ_comparison}
\end{figure}
The second, and more significant difference with respect to Ref.~\cite{Dabholkar:1989ju} is that we perform the integration by parts in the variable $t$ rather than $\sigma$. Because of parametrisation invariance, the integration by parts can be carried out in different ways, allowing one to work with either $t(\sigma)$ or $\sigma(t)$. More explicitly, we may write 
\begin{equation}
\begin{gathered}
    \label{eq:A}
    A = \int_{-\infty}^{\infty}  \int_{0}^{1/2} \dd\sigma   \dd t\,   Q(\sigma,t)\delta'(f) = \cdots\\ 
   \text{(3.12) in \cite{Dabholkar:1989ju} } = \int_{-\infty}^{\infty} \dd t \left( \frac{Q(\sigma,t)}{\partial_{\sigma} f} \delta( f) \right) \Bigg|^{\sigma=1/2}_{\sigma=0}  - \int_{-\infty}^{\infty} \dd t\, \frac{\partial_{\sigma} \left( \frac{Q(\sigma,t)}{\partial_{\sigma} f} \right)_{\sigma=\sigma(t)}}{|\partial_\sigma f|} \\[.5em]
   \text{\eqref{eq:by_parts} or \cite{Battye:1995hw, Copeland:1990qu, Buonanno:1998kx}:} = \int_{0}^{1/2}  \dd\sigma\, \left( \frac{Q(\sigma,t)}{\partial_{t} f} \delta( f) \right) \Bigg|^{t=\infty}_{t=-\infty} - \int_{-\infty}^{\infty} \dd t \frac{ \partial_{t}\left( \frac{Q(\sigma,t)}{\partial_{t} f} \right)_{\sigma=\sigma(t)}}{|\partial_\sigma f|}, 
\end{gathered}
\end{equation}
where we performed the integration over $\sigma$ using the $\delta$-function, leaving an integral over $t$ in both cases. Alternatively, one may first integrate over $t$, leaving the $\sigma$ integration. We have verified that both procedures yield the same result.

The boundary contributions from the first term in the third line of \eqref{eq:A} should vanish, since they correspond to contributions from the infinite past and future, see Refs.~\cite{Copeland:1990qu, Battye:1995hw, Buonanno:1998kx}. Our result therefore contains no boundary terms. In contrast, when integrating by parts in $\sigma$, as in Ref.~\cite{Dabholkar:1989ju}, the boundary terms in the second line of \eqref{eq:A} do not vanish and are in fact required to cancel a coordinate divergence arising in the integral, see the discussion between Eqs.~(3.12) and (3.13) in Ref.~\cite{Dabholkar:1989ju}. However, in Ref.~\cite{Dabholkar:1989ju} the boundary conditions were kept only at $\sigma = 1/2$, and not at $\sigma = 0$. Note that here we follow their convention of allowing integration down to $\sigma = 0$. Moreover, if one performs the integration over $\sigma$ from $-1/2$ to $1/2$ without invoking the symmetry under $\sigma \to -\sigma$, there is still a boundary contribution at $\sigma = 0$. This is because a single time value $t$ corresponds to two values of $\sigma$ with opposite signs, both of which satisfy $ f=0$, and thus each contributes a boundary term at $\sigma=0$ as well. More explicitly, the expression (3.13) of Ref.~\cite{Dabholkar:1989ju} should be written as 
\begin{equation}
\begin{aligned}
    \label{eq:hDQ}
    h^{012}_{\rm ret} &= \int^{t_{\rm max}}_{t_{\rm min}+\epsilon_t} \frac{\dd t}{r_0} \left( \frac{(\dot{r}(t_0-t)/r + 1) \cos(2 \pi \sigma) - r_0/r }{|\sin^3(2 \pi \sigma)|} \right)_{\mathfrak f=0} \\[0.5em]
    &+ \frac{1}{\sin (2 \pi \epsilon)} \frac{\dot{r}(t_0-t)/r+1+(r_0/r) \cos(2 \pi \epsilon)}{(t_0-t)/r+\dot{r}(1+(r_0/r) \cos(2 \pi \epsilon)) } \\[0.5em]
    &- \frac{2 \pi \epsilon}{(1+(r/r_0))^2} \left(\frac{r/r_0 + \dot{r}}{1+\dot{r}}\right)_{\sigma=-1/2} + \mathcal{O}(\epsilon^2)   \\[0.5em]
    &+ \left(\frac{1}{\sin(2 \pi \sigma)} \frac{r_0 \cos(2 \pi \sigma) - r - (t_0 - t) \dot{r}}{t_0 - t + \dot{r}\left(  r - \cos(2 \pi \sigma)  r_0 \right)} \right)_{\sigma = \delta}, 
\end{aligned}
\end{equation}
where $\epsilon \ll 1$ is a parameter that helps to treat the singularity, and the last line is the $\sigma=0$ boundary term missing in Ref.~\cite{Dabholkar:1989ju}. In the limit $\delta \rightarrow 0$, the additional boundary term takes the simple form
\begin{equation}
     \lim_{\sigma \rightarrow 0} \left(\frac{1}{\sin(2 \pi \sigma)} \frac{r_0 \cos(2 \pi \sigma) - r - (t_0 - t) \dot{r}}{t_0 - t + \dot{r}\left(  r - \cos(2 \pi \sigma)  r_0 \right)} \right) 
 = \frac{\gamma_0^3}{2} \left(r_0 \ddot{r}_0 - \left( 1-\dot{r}_0^2 \right)  \right),
 \label{eq:boundaryMissed}
\end{equation}
with $\gamma_0^{-1} = \sqrt{1-\dot{r}_0^2}$.

Meanwhile, if we integrate by parts with respect to $t$, we obtain the result given in Eq.~\eqref{eq:forceKR}, with no boundary terms. The two procedures yield the same result only if the $\sigma=0$ boundary term is kept in the $\sigma$-integration. Equivalently, the result of Ref.~\cite{Dabholkar:1989ju} equals ours with the contribution of Eq.~\eqref{eq:boundaryMissed} removed.

For a direct comparison with the digitised results of Ref.~\cite{Dabholkar:1989ju}, we repeat our calculation of circular loop dynamics with back-reaction with the contribution of the $\sigma=0$ boundary term removed, thereby reproducing its omission in Ref.~\cite{Dabholkar:1989ju}, see Fig.~\ref{fig:DQ_comparison}. One can see small differences for $\log(r(0)/\delta)=5.0$ between this calculation and Ref.~\cite{Dabholkar:1989ju}. This deviation can be caused by numerical errors or by inaccuracy of the digitisation; however, this difference is much smaller than present uncertainties when compared to numerical simulations. 

To our knowledge, the results regarding the collapse of circular loops obtained using the NGKR effective description, including those presented in Refs.~\cite{Hagmann:1990mj,Gorghetto:2018myk,Dine:2020pds}, rely on Ref.~\cite{Dabholkar:1989ju}, where the boundary term at $\sigma=0$ is omitted. Our analysis indicates that this boundary contribution can affect the dynamics of circular loops noticeably, see Fig.~\ref{fig:DQ_comparison}. Therefore, some conclusions of these works concerning the NGKR effective description should be interpreted with caution.

\subsection{Comparison with Field Theory Simulations}
\begin{figure}[t]
    \centering
    \includegraphics[width=0.49\linewidth]{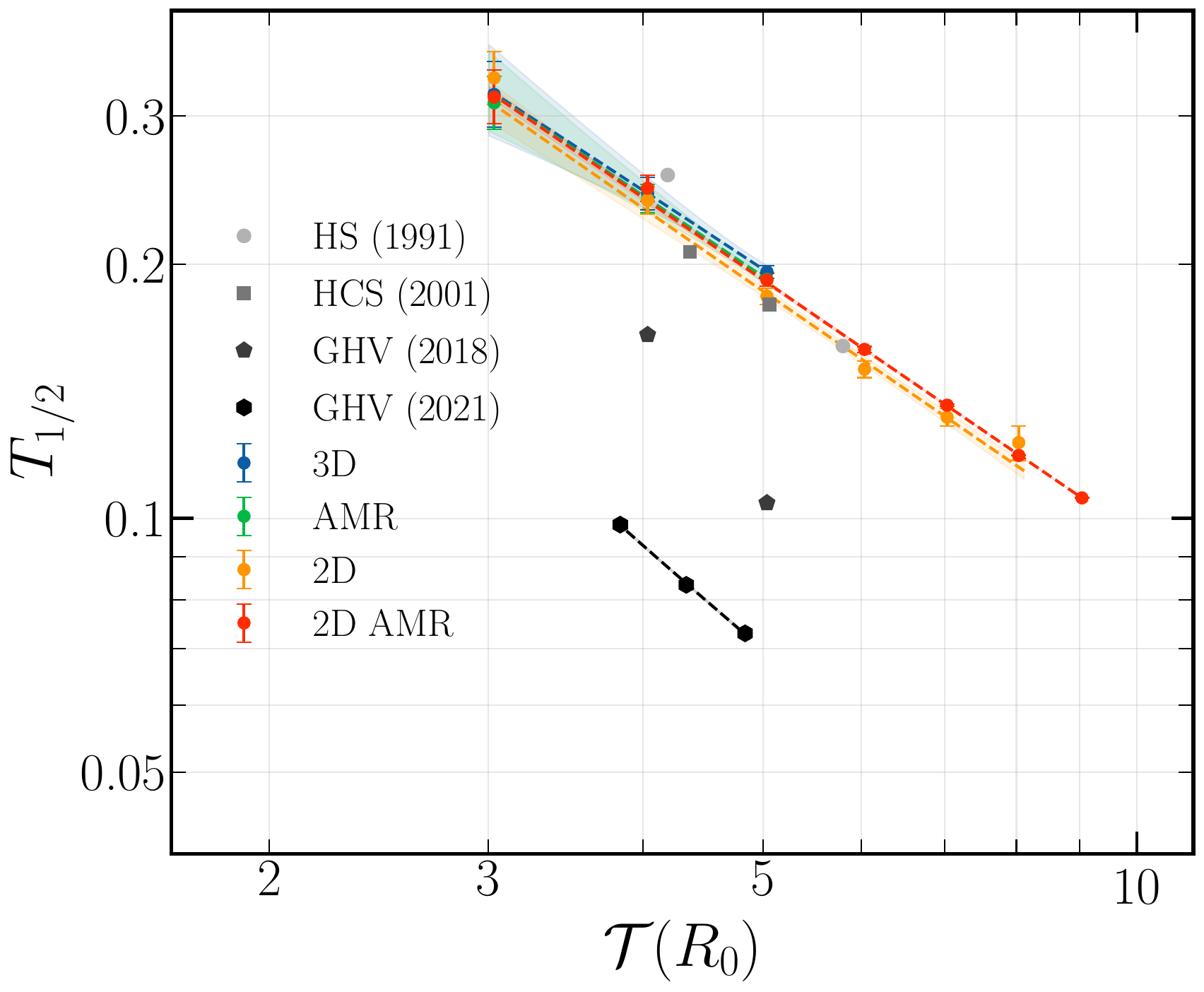}
    \includegraphics[width=0.49\linewidth]{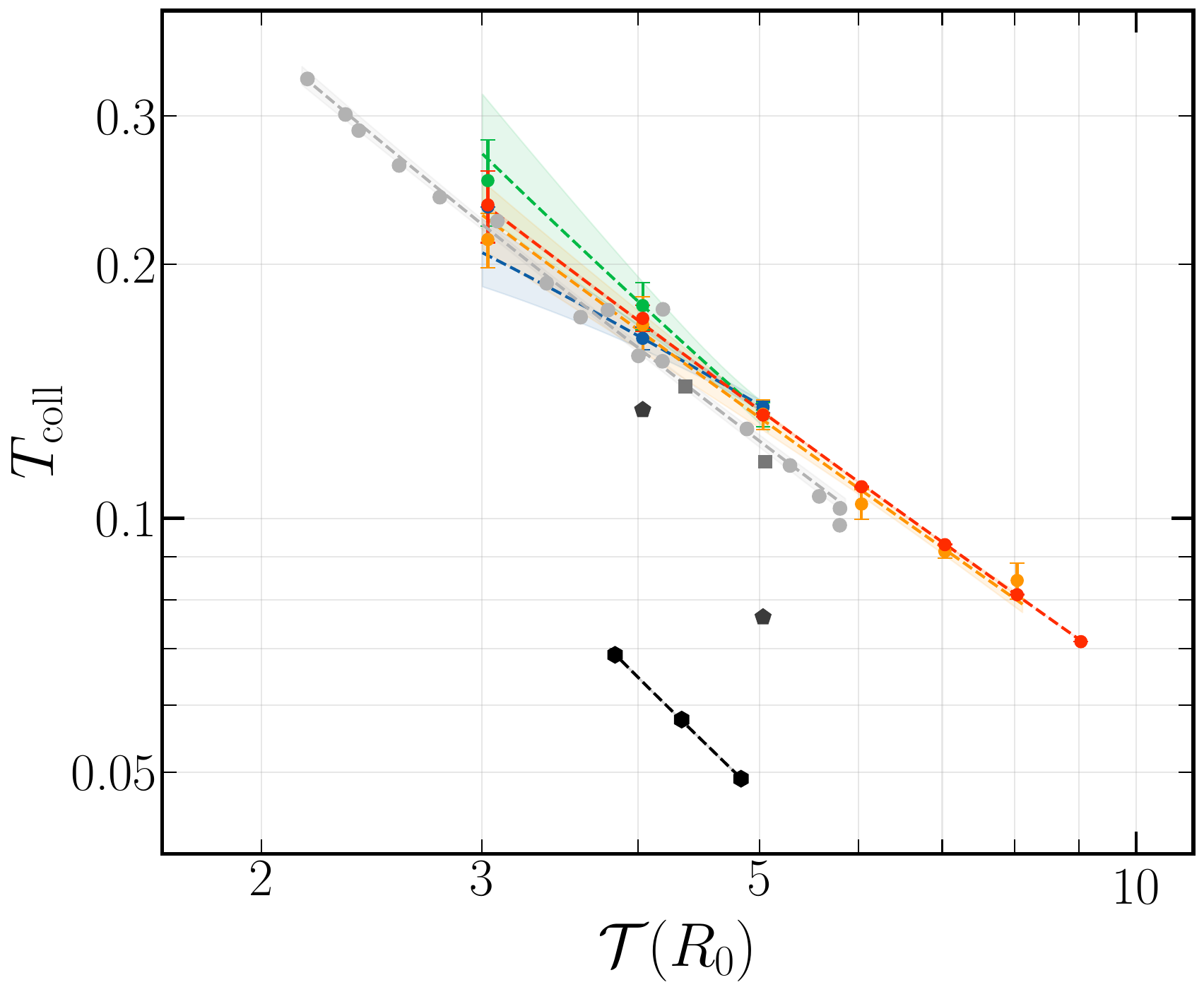}
    \caption{The simulation data of Fig.~\ref{fig:collapse_times} compared to
digitised results of Refs.~\cite{Hagmann:1990mj, Hagmann:2000ja,
Gorghetto:2018myk, Gorghetto:2020qws}, for the halving time (\emph{left}) and
the collapse time (\emph{right}) against $\mathcal T(R_0)$ on logarithmic axes.
Fits are only shown for datasets with more than two points. The data for both
the halving and the collapse times were extracted from the digitised collapse
curves of Fig.~\ref{fig:radius_comparison} using the methods described in
App.~\ref{app:collapse_times}, with the exception of the collapse times for
Ref.~\cite{Hagmann:1990mj}, which reported the relevant data (only for
$T_{\mathrm{coll}}$) in a separate figure.}
    
    \label{fig:collapse_times_lit}
\end{figure}
The initial study of circular-loop collapse using field-theory simulations over a wide range of the ratio between the string core size and the initial loop radius, for which we use the notation $\mathcal{T}(R_0)$, see Eq.~\eqref{eq:FT-tension-match}, was performed in Ref.~\cite{Hagmann:1990mj} using both 2D and 3D simulations. Later studies investigating circular-loop collapse for several values of $\mathcal{T}(R_0)$ include Refs.~\cite{Hagmann:2000ja,Gorghetto:2018myk,Gorghetto:2020qws}. From the plots presented in these studies, we extracted the evolution of the loop radius $R(t)$ for different values of $\mathcal{T}(R_0)$ and compare their results with our simulations in Fig.~\ref{fig:radius_comparison}. The corresponding halving and collapse times are shown in Fig.~\ref{fig:collapse_times_lit}. We find good agreement with Refs.~\cite{Hagmann:1990mj,Hagmann:2000ja}, whereas the loops of Refs.~\cite{Gorghetto:2018myk,Gorghetto:2020qws} collapse considerably later at the same quoted tension. We were unable to identify the origin of this discrepancy.

Some aspects of the collapse of global string loops were also studied in Refs.~\cite{Battye:1993jv,Saurabh:2020pqe,Baeza-Ballesteros:2023say,Benabou:2023klr}, but the evolution of the circular-loop radius was not explicitly reported in these studies.

The comparison is complicated by the fact that the initial conditions differ between these works. Ref.~\cite{Gorghetto:2018myk} uses a product of oppositely wound straight-string fields in a meridional plane, whereas more recent studies select loops from networks or form them by colliding four boosted straight strings, see Ref.~\cite{Baeza-Ballesteros:2023say}, or generate a circular loop dynamically through a PQ transition~\cite{Benabou:2023klr}. None of these procedures needs to begin with the minimum-energy axion field: product, collision and transition initial conditions can carry radiation inherited from their preparation. Even the relaxed solid-angle construction of Sec.~\ref{subsec:stringICs} has an unavoidable release transient, because no free static circular solution exists, but it removes avoidable excess radiation. We therefore use this construction, derived from the magnetostatic analogy for the KR field, and check energy saturation, static-EOM residuals and spectral stability under changes of the relaxation and of the analysis start time, see App.~\ref{app:ic-relaxation}. 

We finally note that Ref.~\cite{Gorghetto:2018myk} reports agreement of its simulations with the effective description of Ref.~\cite{Dabholkar:1989ju}, which we reproduce only when the $\sigma=0$ boundary term of Eq.~\eqref{eq:boundaryMissed} is omitted, cf. App.~\ref{app:DQ}. Both results therefore deviate from ours in the same direction, but we cannot conclude that this is more than a coincidence.

\newpage
\bibliography{bib/strings}
\bibliographystyle{bib/utphys}

\end{document}